\documentclass[twoside,12pt]{article}
\usepackage{epsfig}
\usepackage{rotating}
\usepackage{textcomp}
\usepackage{amstext}
\usepackage{graphicx}
\usepackage{marvosym}
\def\Journal#1#2#3#4{{#1} {#2} (#4) #3 }
\def\NATURE{\em Nature}
\def\NIM{{\em Nucl. Instrum. Methods Phys. Res.}}
\def\NIMA{{\em Nucl. Instrum. Methods Phys. Res.} A}
\def\CPC{{\em Chinese Physics} C}
\def\JINST{\em JINST}

\def\NPB{{\em Nucl. Phys.} B}

\def\PLB{{\em Phys. Lett.} B}

\def\PRL{\em Phys. Rev. Lett.}

\def\PRA{{\em Phys. Rev.} A}
\def\PRD{{\em Phys. Rev.} D}
\def\PRC{{\em Phys. Rev.} C}

\def\P{\vec P}

\newcommand{\be}{\begin{equation}}
\newcommand{\ee}{\end{equation}}
\newcommand{\bea}{\begin{eqnarray}}
\newcommand{\eea}{\end{eqnarray}}

\begin{document}

\title{ \vspace{1cm} Experimental Aspects of Geoneutrino Detection: Status and Perspectives}
\author{O.\ Smirnov,$^{1}$ \\
$^1$JINR, Joint Institute for Nuclear Research, Dubna, Russian Federation}
\maketitle
\begin{abstract}
Neutrino geophysics, the study of the Earth's interior
by measuring the fluxes of geologically produced neutrino at its surface,
is a new interdisciplinary field of science, rapidly developing as
a synergy between geology, geophysics and particle physics. Geoneutrinos,
antineutrinos from long-lived natural isotopes responsible for
the radiogenic heat flux, provide valuable information for the
chemical composition models of the Earth. The calculations of the expected
geoneutrino signal are discussed, together with experimental aspects
of geoneutrino detection, including the description of possible backgrounds
and methods for their suppression. At present, only two detectors,
Borexino and KamLAND, have reached sensitivity to the geoneutrino.
The experiments accumulated a set of $\sim$190 geoneutrino events and continue the data acquisition. The detailed description of the experiments, their
results on geoneutrino detection, and impact on geophysics are presented.
The start of operation of other detectors sensitive to geoneutrinos
is planned for the near future: the SNO+ detector is being filled
with liquid scintillator, and the biggest ever 20 kt JUNO detector
is under construction. A review of the physics potential of these experiments
with respect to the geoneutrino studies, along with other proposals,
is presented. New ideas and methods for geoneutrino detection
are reviewed.
\end{abstract}
\tableofcontents
\section{Introduction}

The Earth's surface heat flow is, in part, fed
by the energy released in the radioactive decay of long-lived isotopes
naturally present in its crust and mantle. Geoneutrinos, antineutrinos
emitted in these decays, would provide useful geological information
if detected. But the neutrino detection is not a simple task as they almost do not interact with matter. Interest
in the geoneutrinos significantly rose due to the development of
large volume detectors, able to detect their exiguous interaction
rates. These measurements are expected to shed light on abundances
and distributions of radioactive elements inside the Earth beyond
the reach of direct measurements by sampling. The natural radioactivity
of the Earth is a powerful source of heat influencing the thermal
history of the planet. The knowledge of the radioactive content in
the Earth's depths is essential for many problems in geoscience.

The importance of radioactivity in the Earth's heat balance was understood
quite early after its discovery. In 1905, Irish physicist J. Joly
was the first who commanded attention to rock radioactivity as a source
of the Earth's thermal energy. His early calculations showed that
the distribution of radioactive elements is not uniform throughout
the Earth's volume, as the concentration of radioactive elements in
the crust extrapolated to the entire volume would result in significantly
higher heat production than the known total heat flux from the Earth.
In other words, these calculations indicated that the concentration
of radioactive elements in the mantle and the core are lower than in the
crust \cite{Joly1909}. In the 1970s, uranium and thorium concentrations were
measured in mantle rock samples extracted from the ocean floor, and
they indeed turned out to be low. Hence, natural radioactive isotopes
are concentrated mainly in the continental crust. The energy from
the decay of radioactive isotopes is an important source of heat in
the Earth, which is estimated to account for roughly half of the Earth's total 
heat release. The Earth emitted more radiogenic heat at the
initial stage of its existence than it emits now, primarily because
rocks at that time had more radioactive elements. Decays of the $^{26}$Al isotope with a half-life of 0.717 Myr were the primary heat source in the early Earth.

There are different types of radioactive decays among the radioactive
isotopes heating the Earth. The energy of $\alpha$-decays is totally
converted into heat, while the energy released in $\beta$-decays
is partially taken away by antineutrinos. Historically, the study of $\beta$-decay
provided the first physical evidence for the neutrino existence. The continuous
spectrum of the $\beta$-decay was a mystery, as there was no apparent
third particle to provide the energy and angular momentum balance.
In a famous letter written in 1930, Pauli suggested the existence
of a light neutral particle, which he called the neutron. This "neutron"
emitted during $\beta$-decay would account for the missing energy and angular momentum. In 1931 Enrico Fermi renamed the particle the neutrino, which literally means "little neutral one". The neutrino's
interaction with matter is extremely weak, and its detection was an
experimental challenge for a quarter of a century, accomplished only
in 1956 through an experiment by Reines and Cowan~\cite{Cowan1956}.
Because each $\beta$-decay is accompanied by antineutrino emission,
detecting these antineutrinos would mean registering the corresponding
$\beta$-decays in the Earth's depths.

The first mention of geologically produced neutrinos in modern physics
can be found in the Gamow letter to Reines (see e.g.~\cite{Fiorentini2007}).
Gamow suggested that unidentified background observed in the nuclear
reactor experiment by Reines and Cowan (devoted to the reactor
antineutrino search) could be due to antineutrinos from $\beta$-decaying
members of the U and Th families. In his teletype answer Reines reported
the first estimate of the 1~MeV antineutrino flux of $10^{8}$ cm$^{-2}$s$^{-1}$
at the Earth's surface based on the heat loss at
the surface of 50~erg$\cdot$cm$^{-2}$ s$^{-1}$, which was not enough to
explain the excess despite the fact that the antineutrino flux was significantly
overestimated in this calculation.

In 1960 Marx and Menyh\'{a}rd \cite{Marx1960} were the first to obtain a realistic
estimate of the geoneutrino flux at the Earth surface of 6.7$\times10^{6}$
s$^{-1}$cm$^{-2}$. Geoneutrinos were discussed by Markov in
his monograph on neutrinos~\cite{Markov1964}, later
Eder~\cite{Eder1966} was the first who pointed out the importance of geoneutrino\footnote{Eder used the term "terrestrial neutrinos".} studies
in the context of the problem of heat release in the Earth, and Marx's
discussion of the geophysics by neutrinos in \cite{Marx1969} closed
the decade.

The modern theoretical study of this problem is attributed to the
work of Krauss, Glashow, and Schramm \cite{Krauss1984}. The first
paper in a geophysical journal appeared only in the 1990s~\cite{Kobayashi1991}.
The potential of the Borexino and KamLAND detectors for geoneutrino
detection was first pointed out by Raghavan at al.~\cite{Raghavan1998}\footnote{A revolutionary 50~kt Borex detector was also proposed by Raghavan~\cite{Raghavan1988}; the idea was materialized as a smaller version, a 0.3~kt detector called Borexino.} and Rothschild et al.~\cite{Rothschild1998}, respectively.
The interest in geoneutrinos in the geoscience community exploded
after the first experimental indication of their presence by the KamLAND~\cite{KamLAND2005} and Borexino collaborations~\cite{Borexino2010}.
At present one can state that we are witnessing the formation of a
new field of geophysics: neutrino geoscience, merging the fine art
of neutrino detection with geophysics.

Geoneutrino flux measurements will give answers to the still debatable
questions about the natural radioactivity of our planet: what is the
radiogenic contribution to the total heat generated by the Earth?
how much uranium, thorium and potassium are distributed in the Earth's
crust and mantle? what are the planetary Th/U and K/U mass ratios? how
uranium and thorium are distributed in the Earth's crust and mantle?
is the mantle chemically uniform, layered, or more complicated? are
there mantle reservoirs enriched in U and Th? is there any georeactor
or hidden excessive $^{40}$K in the Earth's core, as suggested
by some theoreticians? are the geochemical Bulk Silicate Earth (BSE)
models consistent with geoneutrino data? and which of the proposed
Earth models will fit the observed geoneutrino flux?

An important note: in this review we are concentrated on the geologically
produced neutrinos. The use of other natural or artificial neutrino
sources for the purpose of geological research have been actively
studied in parallel with the development of particle acceleration
techniques and neutrino detectors. There were proposals to use neutrino
beams to search for deposits of oil and gas or high-Z ores at large
distances from the accelerator. Other projects were aiming the exploration of
the vertical density profile of the Earth and especially its core.
All of these interesting projects are beyond the scope of this review.

\section{Geoneutrinos and the Earth's heat}

The natural radioactivity of the present Earth is associated with
long-lived radioactive isotopes (radionuclides) with half-lives of
the order of billions of years, comparable with the Earth's and Solar
system's age. The complete list of these isotopes (including both
$\alpha$ and $\beta$ radioactive nuclides) contains 28 nuclides, and
another six nuclides have half-lives longer than 80 million years,
long enough for some of them to survive in noticeable quantities
up to the present day. These 34 naturally occurring radioactive nuclides
comprise the primordial nuclides with different abundances and life-times,
but the major contribution to the Earth's radioactivity comes from very
few of them, namely from decays of radioactive elements in chains
of decays started with $^{238}$U and $^{232}$Th, and from decays
of $^{40}$K, with some contribution from the $^{235}$U chain and from
decays of $^{87}$Rb and $^{147}$Sm. 

Electron antineutrinos ($\overline{\nu_{e}}$) are produced in radioactive
$\beta$-decays
\begin{equation}
_{Z}^{A}X\rightarrow{}_{Z+1}^{A}Y+e^{-}+\overline{\nu_{e}},
\end{equation}
where A and Z are the mass and atomic number of the parent nucleus
X. The daughter nucleus Y has the same A and Z increased by 1. The energy released 
in the reaction corresponds to mass
difference $Q_{0}$ between the parent and the daughter nuclei. The
energy release $Q_{0}$ is distributed
between the electron and the antineutrino. Because of the
tiny cross section of interaction with matter the antineutrino carries away
a fraction of the total energy release. The energy of electrons emitted
in $\beta$-decays, as well as the energy of the accompanying $\gamma$-radiation
and energy produced in $\alpha$-decays, is converted into the heat,
called radiogenic. Another source of heat is primordial heat accumulated during the formation of the Earth. In accordance with modern understanding, radioactive decays are responsible for roughly a half of the total heat flow of the Earth, with the other half coming from the secular cooling of
the Earth. The energy budget of the Earth is schematically shown in Fig.~\ref{fig:EarthBudget}.

\subsection{Long-lived radiogenic elements}

We start with reviewing the radiogenic elements. The most important
decays, from the geophysical point of view, are those in the U and
Th decay chains, and the $^{40}$K decay.

The uranium chain (historically called the ``radium series'') begins
with naturally occurring $^{238}$U. It includes isotopes of astatine,
bismuth, lead, polonium, protactinium, radium, radon, thallium, and
thorium, providing that all are present in any uranium-containing
sample. The total energy released per decay of $^{238}$U into the stable $^{206}$Pb
is 51.773$\pm0.017$ MeV on the assumption of secular equilibrium~\footnote{The
life-time of the $^{238}$U and $^{232}$Th is much longer than life-time of any daughter 
isotope in the corresponding chain, therefore on the long time-scale the decay rates of daughter radioactive isotopes coincide with the decay rate of the parent isotope. In nuclear physics this situation is called secular equilibrium.}. Natural
uranium, along with $^{238}$U and $^{235}$U, contains the decay
product $^{234}$U, which is in secular equilibrium with its parent
$^{238}$U, and its long lifetime provides 5.5$\cdot10^{-5}$ isotopic
abundance in natural uranium. Uranium-234 and its daughter decays are
included in the calculations as a part of the parent $^{238}$U chain. In 
the course of transformation of $^{238}$U into $^{206}$Pb  8 $\alpha$- and 6 $\beta$-decays occur, the latter are accompanied by 6 antineutrinos
\begin{equation}
_{92}^{238}\text{U}\rightarrow_{82}^{206}\text{Pb}+8\alpha+6e^{-}+6\overline{\nu_{e}}+51.77\;\text{MeV.}
\end{equation}
Another chain starts with the less abundant uranium isotope, $^{235}$U
("actinium series"). This decay series includes isotopes of actinium,
astatine, bismuth, francium, lead, polonium, protactinium, radium,
radon, thallium, and thorium. The total energy released per decay
of the parent $^{235}$U into the stable $^{207}$Pb is 46.398$\pm0.010$
MeV on the assumption of secular equilibrium. The transformation of $^{235}$U into $^{207}$Pb  occurs through 7 $\alpha$- and 4 $\beta$-decays accompanied by 4 antineutrinos
\begin{equation}
_{92}^{235}\text{U}\rightarrow_{82}^{207}\text{Pb}+7\alpha+4e+4\overline{\nu_{e}}+46.40\;\text{MeV.}
\end{equation}
The last chain is the "thorium series" beginning with
naturally occurring $^{232}$Th. This series includes isotopes of
actinium, bismuth, lead, polonium, radium, and radon. The total energy
released per decay of the parent $^{232}$Th into the stable $^{208}$Pb is $42.6575\pm0.0053$ MeV. The transformation of $^{232}$Th into $^{208}$Pb  occurs through 6 $\alpha$- and 4 $\beta$-decays accompanied by 4 antineutrinos
\begin{equation}
_{90}^{232}\text{Th}\rightarrow_{82}^{208}\text{Pb}+6\alpha+4e^{-}+4\overline{\nu_{e}}+42.66\;\text{MeV.}
\end{equation}

\begin{figure}
\includegraphics[angle=90,scale=0.8]{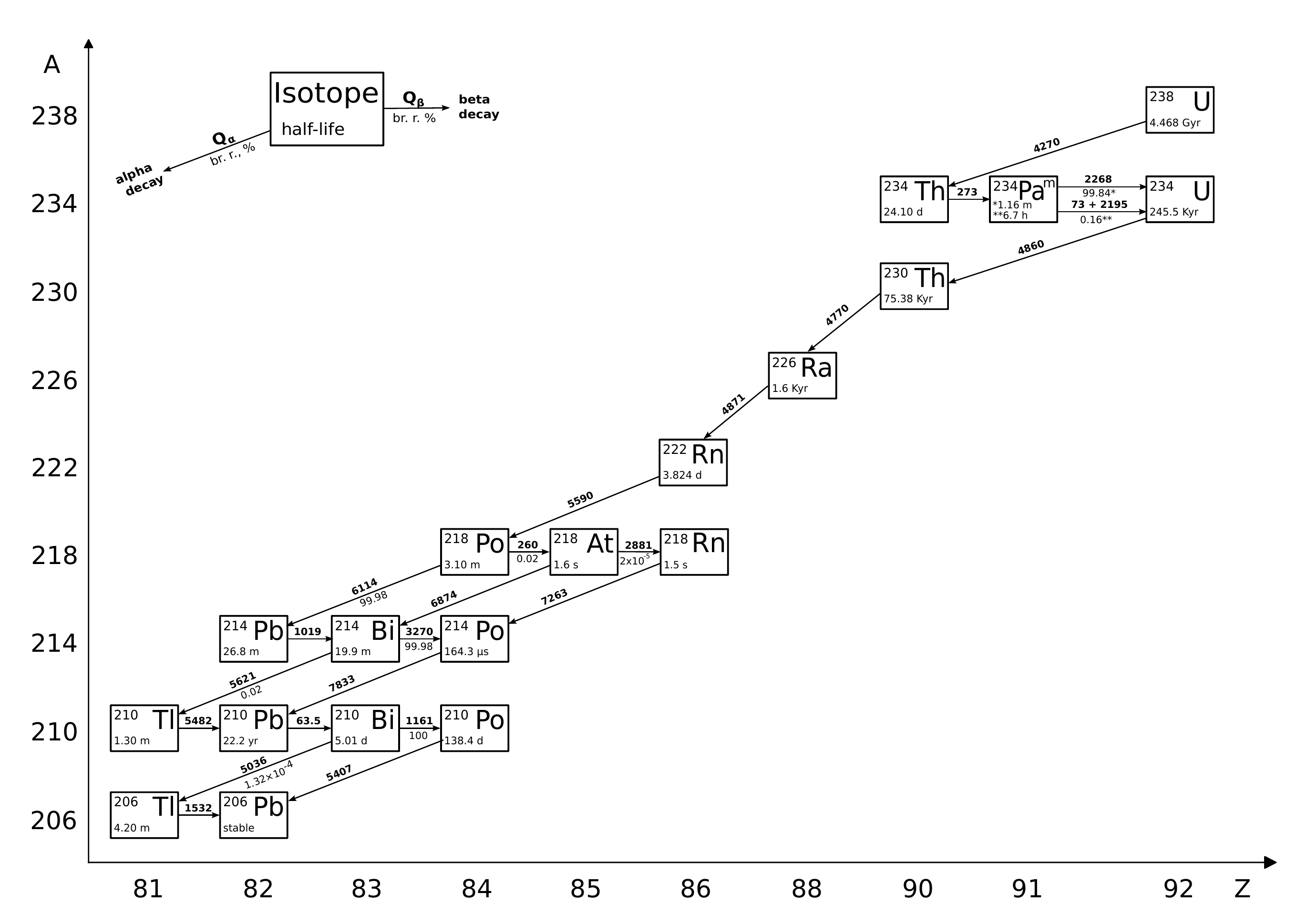}
\protect\caption{\label{fig:U-decay-chain}$^{238}$U decay chain. The scheme reproduces the
original one from Fiorentini et al.~\cite{Fiorentini2007} using the updated nuclear data.}
\end{figure}

\begin{figure}
\includegraphics[angle=90,scale=0.8]{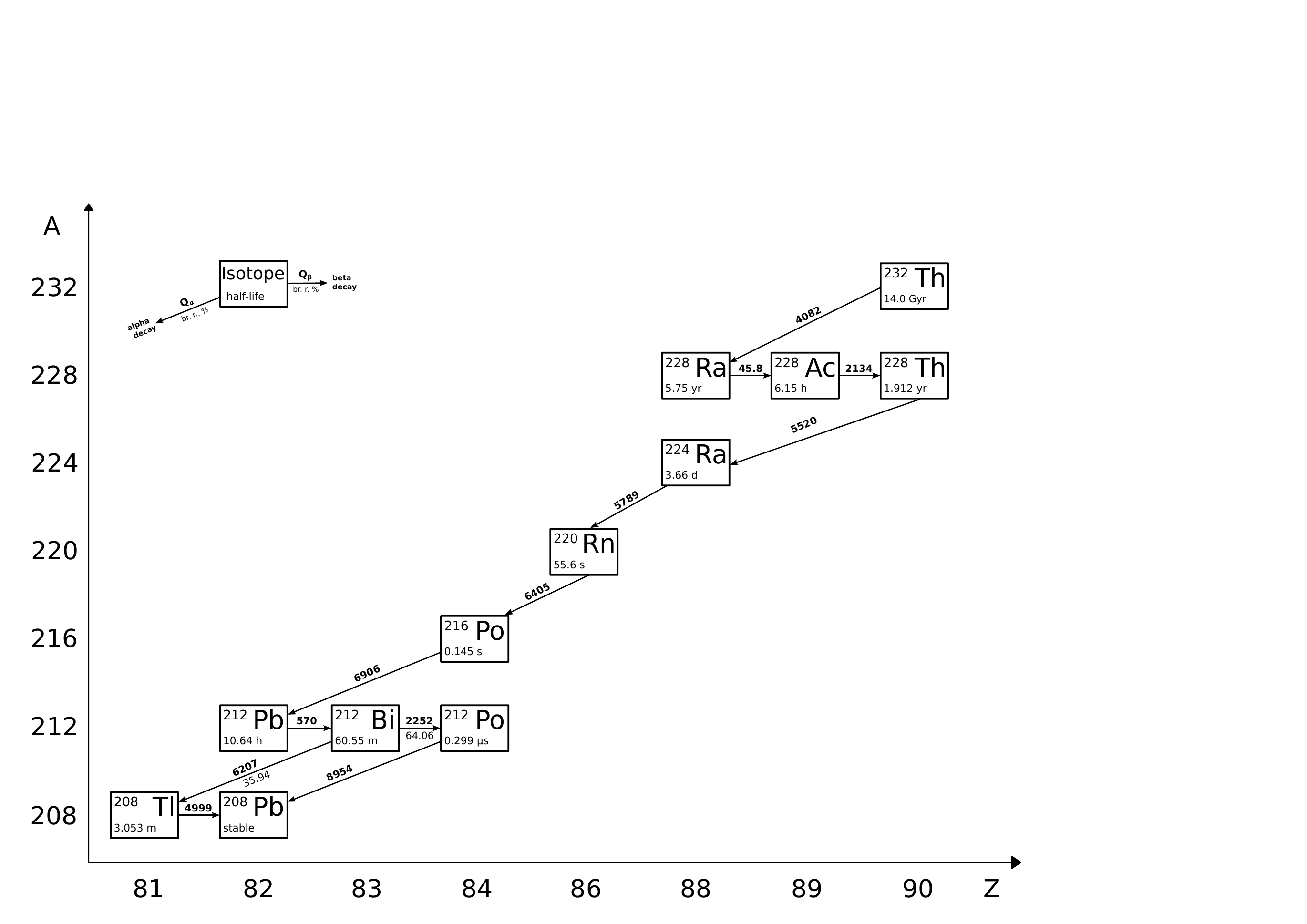}
\protect\caption{\label{fig:Th-decay-chain}$^{232}$Th decay chain. The scheme reproduces the
original one from Fiorentini et al.~\cite{Fiorentini2007} using the updated nuclear data.}
\end{figure}

The main route of the $^{238}$U and $^{232}$Th decay chains are presented schematically in Fig.~\ref{fig:U-decay-chain} and Fig.~\ref{fig:Th-decay-chain}. The less important branches are excluded in these charts for the sake of simplicity.

Potassium-40 ($^{40}$K) is a radioactive isotope of natural potassium with a  half-life long enough to survive in noticeable amount until the present day, the modern isotopic abundance of $^{40}$K is $\sim$117 ppm. $^{40}$K decays through two channels, $\beta$-decay 
\begin{equation}
_{19}^{40}\text{K}\rightarrow{}_{20}^{40}\text{Ca}+e^{-}+\overline{\nu_{e}}+1.311\;\text{MeV,}\quad(89.3\%),
\end{equation}
and electron capture
\begin{equation}
_{19}^{40}\text{K}+e^{-}\rightarrow{}_{18}^{40}\text{Ar}+\nu_{e}+1.505\;\text{MeV,}\quad(10.72\%).
\end{equation}
A fraction of the EC decays occurs into the excited state of $^{40}$Ar (1460 MeV).
In accordance with available data \cite{ENSDF} the intensity of the 1460.82 MeV $\gamma$-line
is (10.66$\pm$0.17)\%, and the total EC branch has (10.72$\pm$0.13)\%
branching ratio, leaving for EC into the ground state (0.06$\pm$0.04)\%.
This decay is anticorrelated with 10.66\% decay into the excited
state. The EC into the ground state slightly reduces the heating
energy. The $^{40}$K $\beta^{-}$decay is unique third forbidden, our estimate of the effective energy release (average energy of the $\beta$-
spectrum) $E_{eff}=$583.6 keV for $^{40}$K is in a good agreement
with both the experimentally obtained value $\left\langle E\right\rangle $=583.98
keV~\cite{Leutz1965} and the recent calculations $\left\langle E\right\rangle $=583.27
keV \cite{Mougeot2018}.

The last two decays of minor importance are those of $^{87}$Rb and
$^{147}$Sm. Rubidium is the 23d most abundant element in the Earth. So, despite
its relatively low specific heat yield, it could be important for
precision calculations
\begin{equation}
_{37}^{87}\text{Rb}\rightarrow{}_{38}^{87}\text{Sr}+e^{-}+\overline{\nu_{e}}+0.2833\;\text{MeV.}
\end{equation}
According to McDonough and Sun~\cite{McDonough1995}, natural Rb is
$\sim$400 times less abundant than K with specific heat yield 2.4 times
greater than that of natural K, thus adding to the total heat
about 0.6\% of the potassium contribution. 

Samarium is just the 40th among the most abundant elements. The isotopic
abundance of natural Sm is lower than that of $^{87}$Rb, $\sim$600 times less abundant compared to K~\cite{McDonough1995}, but its $\alpha$-decay is more energetic, $E_{0}=2.3$~MeV, providing specific heat yield 14 times greater than that of natural K.
As a result, $^{147}$Sm contributes to the total radiogenic heat even more than $^{87}$Rb, adding $\sim$2.3\% of the potassium contribution.
The total heat from Rb and Sm is about 0.1~TW, well below the uncertainty
of the total heat flow measurement. 

Heat production of other elements
from the list of long-lived isotopes is calculated in~\cite{Enomoto2006}, and
all the contributions are at least an order of magnitude lower than the
aforementioned ones.

\begin{table}
\caption{\label{tab:Isotopes}Characteristics of the main contributors to the
radiogenic heat production. Following the ENSDF database notations,
the numeric uncertainty, shown in parentheses, denotes an uncertainty
in the last significant figure(s)}

\centering{}%
\begin{tabular}{ccccccc}
\hline 
 & {\scriptsize{}$^{238}$U } & {\scriptsize{}$^{235}$U } & {\scriptsize{}$^{232}$Th } & \multicolumn{1}{c}{{\scriptsize{}$^{40}$K}} & {\scriptsize{}$^{87}$Rb} & {\scriptsize{}$^{147}$Sm}\tabularnewline
\hline 
{\scriptsize{}Decay(s) type} & {\scriptsize{}$\beta+\alpha$ } & {\scriptsize{}$\beta+\alpha$ } & {\scriptsize{}$\beta+\alpha$ } & {\scriptsize{}$\beta$ \& EC ($\nu$)} & {\scriptsize{}$\beta$ } & {\scriptsize{}$\alpha$ }\tabularnewline
{\scriptsize{}Natural isotopic abundance } & {\scriptsize{}$0.992742$} & {\scriptsize{}$0.007204$} & {\scriptsize{}$1.0000$} & \multicolumn{1}{c}{{\scriptsize{}$1.1668(8)\cdot10^{-4}$}} & {\scriptsize{}$0.2783(2)$} & {\scriptsize{}$0.1499(18)$}\tabularnewline
{\scriptsize{}Life-time, T$_{1/2}$, Gy } & {\scriptsize{}$4.468(6)$} & {\scriptsize{}$0.7038(5)$ } & {\scriptsize{}$14.0(1)$ } & \multicolumn{1}{c}{{\scriptsize{}$1.248(3)$}} & {\scriptsize{}$49.7(3)$} & {\scriptsize{}$106.0(11)$}\tabularnewline
{\scriptsize{}Isotope atomic mass, g/mol } & {\scriptsize{}$238.051$} & {\scriptsize{}$235.044$} & {\scriptsize{}$232.038$} & \multicolumn{1}{c}{{\scriptsize{}$39.9640$}} & {\scriptsize{}$86.909$} & {\scriptsize{}$146.915$}\tabularnewline
{\scriptsize{}Total decay energy, $Q_{tot}$, MeV } & {\scriptsize{}$51.773(17)$} & {\scriptsize{}$46.398(10)$} & {\scriptsize{}$42.6575(53)$} & \multicolumn{1}{c}{{\scriptsize{}$1.3116(11)$}} & {\scriptsize{}$0.283$} & {\scriptsize{}$2.311(1)$}\tabularnewline
{\scriptsize{}Heating energy, $Q_{H}$, MeV } & {\scriptsize{}$47.678(10)$} & {\scriptsize{}$44.383(5)$} & {\scriptsize{}$40.438(5)$} & \multicolumn{1}{c}{{\scriptsize{}$0.6767(17)$}} & {\scriptsize{}$0.0567(2)$} & {\scriptsize{}$2.311(1)$}\tabularnewline
{\scriptsize{}Specific heat yield, $\mu$W$\cdot$kg$^{-1}$ } & {\scriptsize{}$94.32(13)$ } & {\scriptsize{}$4.097(3)$} & {\scriptsize{}$26.4(2)$} & \multicolumn{1}{c}{{\scriptsize{}$3.356(8)\cdot10^{-3}$ }} & {\scriptsize{}$7.74(14)\cdot10^{-3}$} & {\scriptsize{}$4.72(6)\cdot10^{-2}$}\tabularnewline
{\scriptsize{}Specific $\bar{\nu_{e}}$ luminosity, kg$^{-1}$s$^{-1}$ } & {\scriptsize{}$7.41(1)\cdot10^{7}$ } & {\scriptsize{}$2.305(2)\cdot10^{6}$ } & {\scriptsize{}$1.63(1)\cdot10^{7}$ } & {\scriptsize{}$2.763(7)\cdot10^{4}$ } & {\scriptsize{}$8.5(2)\cdot10^{5}$} & {\scriptsize{}-}\tabularnewline
{\scriptsize{}Specific $\nu_{e}$ luminosity, kg$^{-1}$s$^{-1}$ } & {\scriptsize{}-} & {\scriptsize{}-} & {\scriptsize{}-} & {\scriptsize{}$0.332(1)\cdot10^{4}$ } & {\scriptsize{}-} & {\scriptsize{}-}\tabularnewline
\hline 
\end{tabular}
\end{table}

The properties of the main contributors to the radiogenic heat production
are presented in Table~\ref{tab:Isotopes}. $Q_{tot}$ is the
total energy released in the corresponding decay or chain of decays including
the energy taken away by antineutrinos. Heating energy $Q_{H}$
is the "visible" (dissipated for heating) energy release, namely the
total released energy, excluding the energy fraction taken away by
antineutrinos. Specific heat yield $\epsilon_{H}$ and specific antineutrino
(or neutrino in the case of electron capture on $^{40}$K) luminosities
$\epsilon_{\nu}$ per 1 kg of naturally occurring element are provided
in the last two rows with account taken of the natural abundance of the
corresponding radioactive isotope fraction. In the case of uranium,
the corresponding values for two isotopes should be summed to obtain the total
heat and antineutrino flux per 1 kg of natural U.

The data for the evaluation of the energy release and its uncertainty were taken from the
last available version of the ENSDF database \cite{ENSDF} \footnote{Evaluated Nuclear Structure Data File (ENSDF). The ENSDF database
contains evaluated nuclear structure and decay information for over
3000 nuclides, continuously updated with new evaluations published
in the Nuclear Data Sheets. We used the database version of December
2018 in the calculations. The ENSDF is compiled by the International Network of Nuclear
Structure and Data Decay Evaluators (http://www-nds.iaea.org/nsdd/).}. It should be noted that while measured values of $Q_{tot}$
have typical errors below 0.1\%, the precision of evaluation of $Q_{H}$
is generally less certain as it depends on the shapes of $\beta$-decay spectra and precision of the measurement
of the branching ratios for complex decays with $\beta$-decay into excited
levels of daughter nuclei. 

The details of the $\beta$-spectra shape calculations are presented
below in section~\ref{subsec:Beta-decay-shapes}. The universal $\beta$-decay
shape is used in calculations with the exception of $^{210}$Bi in
the $^{238}$U chain, of $^{40}$K $\beta$-decay (3rd unique forbidden),
and of $^{87}$Rb (3rd forbidden non-unique). Note that the transition $^{212}$Bi$\rightarrow^{212}$Po in Table~\ref{tab:betaTh},
and all the transitions in~Table~\ref{tab:betaU}, are forbidden non-unique ones. 
As an illustration of possible bias due to the spectral shape, two values could be compared: the value $Q_{H}$=0.122 MeV obtained with assumption of the allowed shape and used in
earlier review~\cite{Fiorentini2007} for $^{87}$Rb evaluation is
twice as large as the value in Tab.~\ref{tab:Isotopes} calculated using the corrected shape. The
total error on energy release is not significantly influenced by the
corrections to the energy shapes, since only a small fraction of the
total energy is released in $\beta$-decays, typically below 5\% of
the total: 2.0 out of 47.7 MeV in the $^{238}$U chain, 1.5 out of 44.4
MeV in the $^{235}$U chain, and 1.5 out of 40.4 MeV in the $^{232}$Th chain.
Moreover, as will be discussed below, experimental spectra
for forbidden decays of high-Z nuclei are very similar to the allowed
shapes, and only minor contributions to the uncertainty of the total
energy release in the U and Th chains are expected. 

We can conclude that geoneutrinos are predominantly born as electron
antineutrinos in $\beta$-decays with an admixture of electron neutrinos
from electron captures. Neutrinos are produced only in the electron capture
of $^{40}$K. In contrast to the Sun powered by nuclear fusion and emitting neutrinos
at the rate of 
$L_{\nu}^{\odot}\simeq1.8\times10^{38}$ $\nu$/s, the
Earth ``shines'' essentially in antineutrinos produced in radioactive
$\beta^{-}$-decays, emitting $L_{\overline{\nu}}^{\oplus}\simeq3\times10^{25}$ $\overline{\nu}$/s, many orders of magnitude below the rate of neutrino production in the Sun. For comparison, a typical nuclear reactor produces  $L_{\nu}^{R}\sim\times 10^{20}$ $\overline{\nu}$/s. Because of the Sun-to-Earth distance the flux of the solar neutrinos at the Earth is strongly reduced and the difference in solar and geoneutrino fluxes is less pronounced, though still significant, $\Phi_{\nu}^{\odot}\sim6\times10^{10}$ cm$^{-2}$s$^{-1}$ neutrinos from the Sun versus $\Phi_{\overline{\nu}}^{\oplus}\sim3\times10^{7}$ cm$^{-2}$s$^{-1}$ of all antineutrinos from the Earth. 

The estimates above consider total flux of neutrinos of all flavours. In the process of propagation the original electronic neutrino flavour for both the solar neutrino and earth antineutrino is not conserved because of the neutrino oscillation mechanism, and
other neutrino flavours can be detected or missed depending on the detection method used.

\subsection{Radiogenic heat and geoneutrino luminosity of the Earth}

For a given amount of radioactive isotopes, the energy released in
radioactive decays is strictly connected to the amount of antineutrinos
produced in these decays. The total antineutrino production rate (or luminosity)
L$_{\nu}$ can be related to the mass of elements contained in the Earth using the data from Table~\ref{tab:Isotopes}
\begin{equation}
L_{\nu}=[(7.64\pm0.1)\cdot M(\text{U})+(1.63\pm0.1)\cdot M(\text{Th})+(2.765\pm0.007)\cdot10^{-3}\cdot M(\text{K})+...]\cdot10^{24}\quad\text{s}^{-1}.
\label{eq:Lum_nu}
\end{equation}
A similar relation between radiogenic heat production $\text{H}_{\text{R}}$
and the mass of the corresponding elements can be expressed as
\begin{equation}
H_{R}=[(9.849\pm0.013)\cdot M(\text{U})+(2.64\pm0.02)\cdot M(\text{Th})+(3.356\pm0.008)\cdot10^{-4}\cdot M(\text{K})+...]\quad\text{TW},\label{eq:HR}
\end{equation}
where masses of the corresponding elements present in the Earth are expressed in units of $10^{17}$~kg, and 1~TW is $10^{12}$~W\footnote{The errors in equations~\ref{eq:Lum_nu} and \ref{eq:HR} can be partially or fully correlated, especially for the U and Th contributions. The equations adapted from~\cite{Fiorentini2007}}. The estimates for the global mass of
U varies from 0.5 to 1.2$\times10^{17}$ kg (see Table~\ref{tab:Models-Masses}
and related discussion of the corresponding models in section~\ref{subsec:Models}), the Th and K mass are related to the U mass, and the ratio M(Th)/M(U) and M(K)/M(U) can
differ in various models. The chondritic value of M(Th)/M(U)=3.9
and the ratio M(K)/M(U)$\sim$10$^{4}$ are commonly used, following McDonough
and Sun's estimates~\cite{McDonough1995}.

The Earth's interior is hotter than its surface, this difference of temperatures feeds the heat flux through the Earth's surface. The present day total energy loss of the Earth is well established by the heat
flow measurements in the continents and well-tested physical models
for cooling of the sea floor. 
The most recent estimates are those
by Jaupart et al. providing the 46$\pm$3~TW heat flow~\cite{Jaupart}.
The authors revised the analysis of $\sim$25000 heat flow measurements
around the globe used in previous estimates, reassessing the required
corrections and errors. Another estimate, based on the data from $>38000$ heat flow measurements, gives the 47$\pm$2~TW heat flow~\cite{Davies2010}.
These estimates of the heat are in good agreement and are less
than 0.1\% of the incoming solar energy, 

\begin{figure}
\begin{centering}
\includegraphics[scale=0.25]{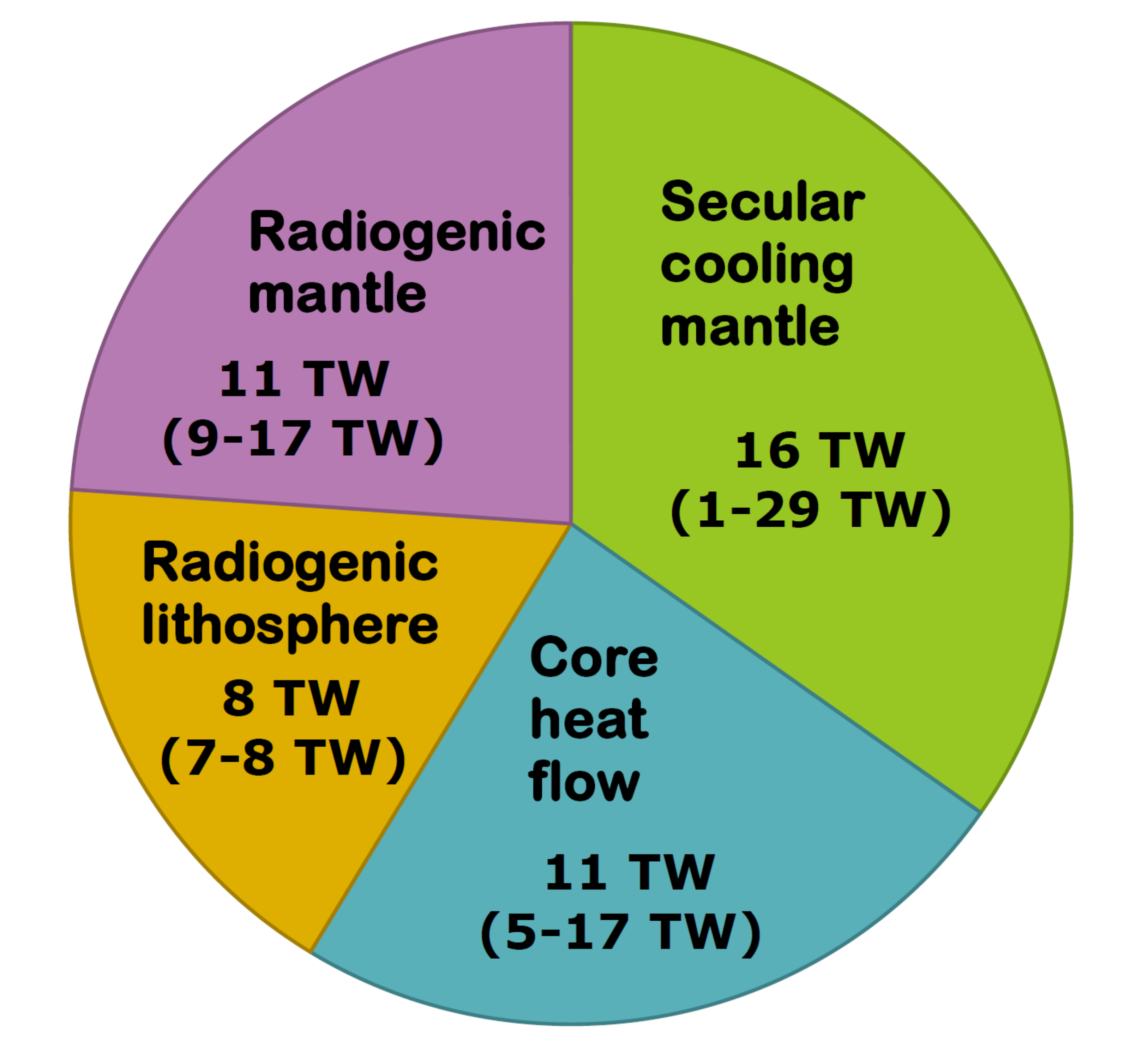}\includegraphics[scale=0.28]{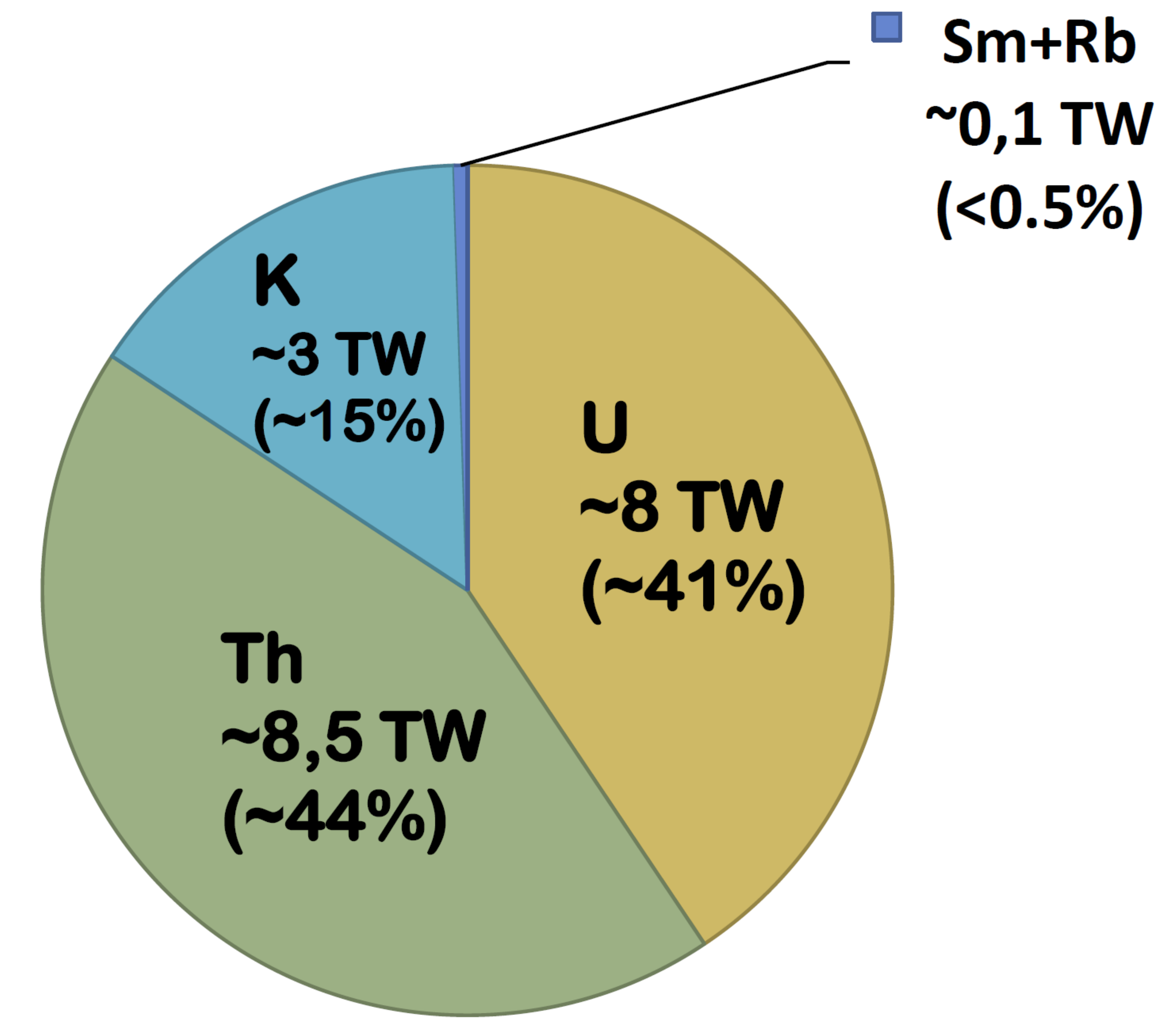}
\par\end{centering}
\centering{}\caption{\label{fig:EarthBudget}{\bf Left:} he energy budget of the Earth according to
\cite{Jaupart}, the radiogenic heat production splits between the
crust (lithosphere) and the mantle, the rest of the heat is provided
by secular cooling of the mantle and core heat flow.
{\bf Right:}Heat producing elements providing major contributions to the radiogenic heat production. The heat is calculated assuming abundances of the corresponding elements from McDonough and Sun's model~\cite{McDonough1995}. Other models can result in different absolute values of heat production, though the relative contributions are less sensitive to the model (see, e.g., Table~\ref{fig:Heat})}
\end{figure}

Radiogenic heat and decrease in the primordial heat content (secular cooling) constitutes the two primarily contributions to the total energy loss of the Earth. An insignificant
amount of heat is produced due to tidal effects, chemical differentiation,
crystallization in the D'' layer, etc. These minor sources produce
no more than one percent of the total heat. The ratio of the radiogenic
heat production to the surface heat flux is called the Urey ratio and is an important
quantity characterizing heat production in the Earth~\footnote{The Urey ratio is named after 
Harold Urey, the 1934 Nobel Prize laureate in Chemistry for the discovery of deuterium, as recognition of his contribution to cosmochemistry. As noted by Korenaga~\cite{Korenaga2008}, this key parameter for a planetary thermal budget could be a source of confusion because geophysicists and geochemists have defined the ratio in different ways.
Two different Urey ratios are being used: the bulk Earth Urey ratio and the convective Urey ratio. The bulk Earth Urey ratio in the geochemical literature usually denotes the contribution of internal heat generation in the entire Earth to the total surface heat flux. In contrast,
the convective Urey ratio is the ratio of internal heat generation in the mantle over the mantle heat flux. The latter definition is the original one introduced by Christensen~\cite{Christensen1985} for his parameterized convection models, and is commonly used by geophysicists while discussing the thermal history of the Earth. The contribution of continental crust does not directly participate in convective heat transfer and is excluded from both the internal heating budget and the surface heat flux.}. The energy budget
of the Earth is schematically presented in Fig.~\ref{fig:EarthBudget}: the radiogenic
heat production splits between the crust and the mantle, and the rest
of the heat is provided by secular cooling that can be determined
directly from ancient lava samples. The radiogenic heat production
in the continental crust responsible for up to 50\% of the Earth's
total radiogenic heat can be determined within 10\% precision. At
the same time, radiogenic heat generation in the Earth's
mantle remains uncertain, but its knowledge is important for geophysics
in order to understand the Earth's convective engine,
plates tectonics, and growth of continents. The latter process, in turn,
tends to deplete radioactive
elements in the Earth's mantle. The measurement of the geoneutrino fluxes that are directly
bound to the amount of the heat producing elements (HPEs) can provide better constrains
to the Earth's energy budgets. Because of the variations of the crust
thickness, the geoneutrino flux measurements in different geographical locations will
help separating a much less position-dependent mantle contribution. 

The total luminosity $L_{\nu}$ and hence the related heat production
$H_{R}$ can be estimated using the geoneutrino flux measured
by the antineutrino detector. The geoneutrino flux measured at a
certain position depends on the distribution of radioactive isotopes
in the Earth bulk and varies within an order of magnitude for different
locations on the Earth's surface~\cite{Huang2013}. One should also take into account
neutrino oscillations when comparing the expected antineutrino flux with
the measured one.

In stable continental regions, the main component of the heat flux
is the crustal radioactivity. In oceanic areas the higher surface
heat flux is due to the cooling of oceanic crust and bears no relation
to radioactivity. In thermal equilibrium, the geoneutrino flux is
expected to decrease over low heat flux regions and to increase over
high heat flux regions~\cite{Mareschal2012}.

\section{Geoneutrino flux calculation}

\begin{figure}
\begin{centering}
\includegraphics[scale=0.45]{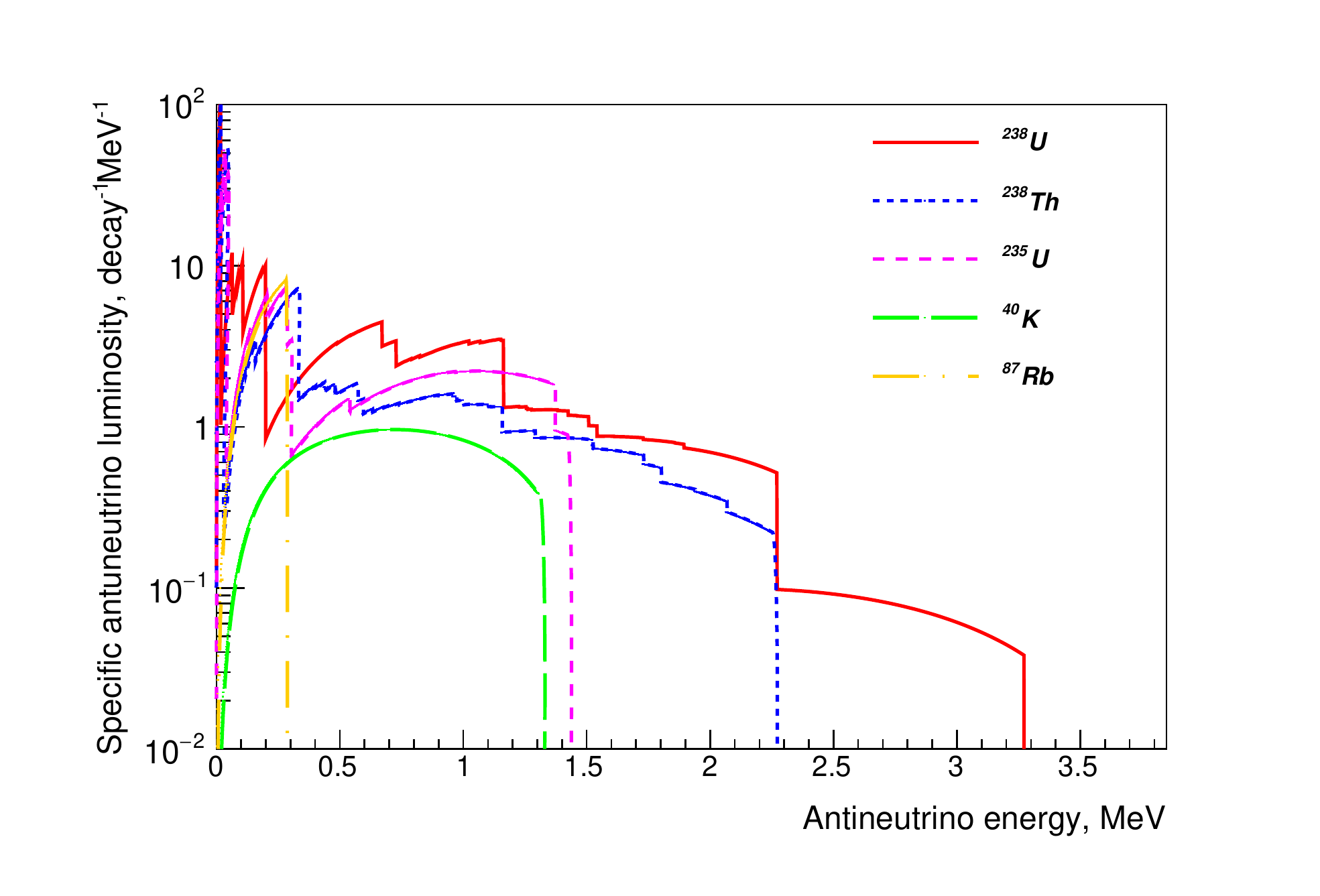}\includegraphics[scale=0.45]{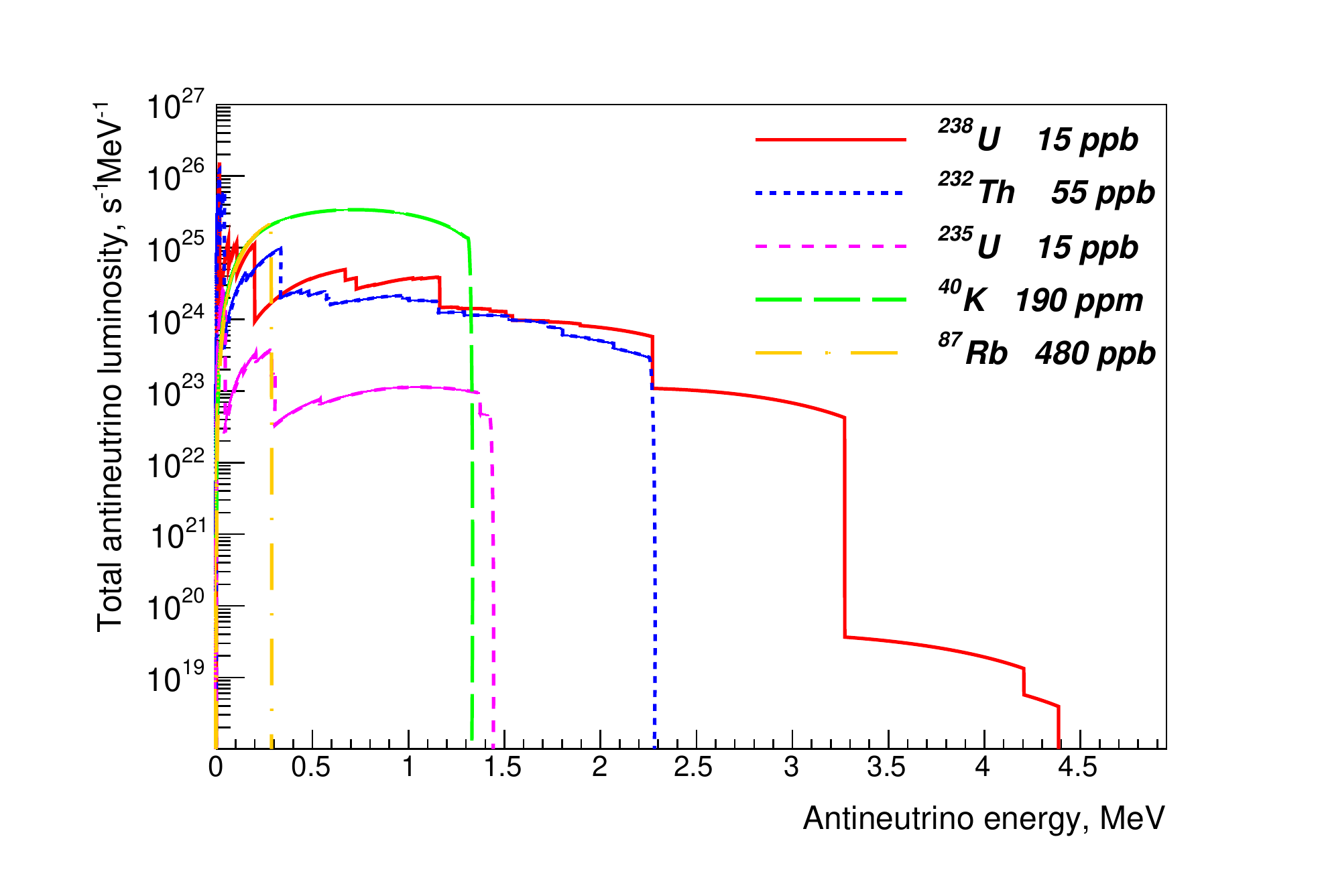}
\par\end{centering}
\protect\caption{\label{fig:Specific-antineutrino-spectra}The antineutrino energy
spectra normalized to one decay of the parent nucleus (left plot), and
total antineutrino differential luminosities (right plot). The abundances,
used in calculation, are shown after each isotope notation \cite{McDonough1995}. Figure adapted from~\cite{Fiorentini2007}. 
For interpretation of the references to colour in this figure legend, the reader is referred to the web version of this article.}
\end{figure}

The antineutrino energy spectra normalized to one decay of the parent
nucleus are shown in left plot of Fig.~\ref{fig:Specific-antineutrino-spectra}. The differential total antineutrino luminosities of the Earth $(\frac{dL_{\nu}(E)}{dE})$
calculated using abundances of natural isotopes in the Earth from~\cite{McDonough1995} are shown in the right plot of Fig.~\ref{fig:Specific-antineutrino-spectra}.

Geoneutrino flux at the detector's location can be calculated by
integrating the antineutrino fluxes from all possible points of origin
in the Earth and summing over all contributing $\beta$-decays. We assume
that non-electronic antineutrino species are lost for the detection and take into account the electron antineutrino survival
probability $P_{ee}$

\begin{equation}
\phi_{I}(E_{\nu})=n_{I}(E_{\nu})\cdot\intop_{V_{Earth}}\frac{\rho(\overrightarrow{r})}{4\pi|\overrightarrow{R_{D}}-\overrightarrow{r}|^{2}}\cdot A_{I}(\overrightarrow{r})\cdot P_{ee}(E_{\nu},|\overrightarrow{R_{D}}-\overrightarrow{r}|)d\overrightarrow{r},\label{eq:Flux}
\end{equation}

where the index $I$=\{\text{U,Th}\} denotes the U or Th chain (the summation should
be extended to include $^{40}$K and/or other species if needed),
$\rho(\overrightarrow{r})$ is the Earth's density at position $\overrightarrow{r}$,
$n_{I}(E_{\nu})$ is the antineutrino energy spectrum in the corresponding
chain normalized to the number of antineutrinos emitted in the chain ($\int n_{Th}(E)dE=$4,
$\int n_{U}(E)dE=$6), $\overrightarrow{R_{D}}$ is the position of
the detector, integration is performed over the Earth's volume $V_{Earth}$,
and $P_{ee}(E,L)$ is the electron antineutrino survival
probability for the antineutrino energy $E$ at distance $L=|\overrightarrow{R_{D}}-\overrightarrow{r}|$ from the antineutrino
source. $A_{I}(\overrightarrow{r})$ is the specific activity of the
chain $I$ at the position with coordinates $\overrightarrow{r}$

\begin{equation}
A_{I}(\overrightarrow{r})=\frac{a_{I}(\overrightarrow{r})\cdot C_{I}}{\tau_{I}m_{I}},\label{eq:AI}
\end{equation}

where $a_{I}(\overrightarrow{r})$ is the mass abundance of the corresponding
parent element, and $C_{I}$, $\tau_{I}$, and $m_{I}$ are the
isotopic abundance, life-time, and atomic mass of the isotope, respectively.

\subsection{\label{sec:oscillations}Neutrino oscillations}

The electron neutrino survival probability appears in equation (\ref{eq:Flux})
because the cross section of the neutrino interaction with the active
media of the detector depends on the neutrino flavour. All existing
neutrino oscillation data (with a few exceptions indicating the presence
of the fourth hypothetical sterile neutrino, a case that goes beyond
the scope of the review) can be described assuming a 3-flavour neutrino
mixing in vacuum. Mass eigenstates correspond to mass states of
neutrinos, called $\nu_{1}$, $\nu_{2}$ and $\nu_{3}$. It follows
from experimental data that neutrinos must be light with $m_{1},m_{2}$, and $m_{3}$ below
1~eV, and all the three ones must have different masses to allow for oscillations, while
the lightest neutrino still can be massless.
The "usual" flavour states produced in nuclear reactions are the mixture
of the mass eigenstates, the composition of which can change as a neutrino
propagates. We can state that at present all properties of the mixing
matrix describing neutrino mixing are well established experimentally
with respect to the precision needed for the geoneutrino flux estimations.
The electron antineutrino, being born in $\beta$-decay as a mass
state mixture corresponding to the electron flavour, after traveling
a distance $L=|\overrightarrow{R_{D}}-\overrightarrow{r}|$ could be
detected as another flavour with a certain probability. The distant
observer will still detect these neutrinos in electron flavour with
a (survival) probability $P_{ee}$, while the complementary
probability 1-$P_{ee}$ remains for observing other neutrino
flavours. If a detector is unable to register non-electron neutrino
flavours, the factor $P_{ee}$ should be applied to obtain the electron
flavour component. In the general case (e.g. in the case of a hypothetical
detector sensitive to the neutrino elastic scattering) one should
also take into account the presence of other flavour components.

The three-flavour electron neutrino survival probability is (see e.g.~\cite{Petcov2002}):
\begin{equation}
{P}_{ee}^{3\nu}(E_{\nu},L)=1-\left(\cos^{4}\theta_{13}\sin^{2}2\theta_{12}\cdot\Delta{}_{21}+\sin^{2}2\theta_{13}\left[\cos^{2}\theta_{12}\cdot\Delta_{31}+\sin^{2}\theta_{12}\cdot\Delta{}_{32}\right]\right),\label{eq:Pee3nu}
\end{equation}
where $\theta_{12}$ and $\theta_{23}$ are the mixing angles, and $\Delta_{ij}$
is defined as
\begin{equation}
\Delta_{ij}=\sin^{2}\left(1.27|\Delta m_{ij}^{2}|[\text{eV}^{2}]\frac{L\text{[m]}}{E_{\nu}\text{[MeV]}}\right),
\label{eq:Deltaij}
\end{equation}
with $\Delta m_{ij}^{2}=m_{i}^{2}-m_{j}^{2}$, the neutrino mass squared
difference. Note that by definition we have only two independent mass
squared differences because of $\Delta m_{31}^{2}=\Delta m_{32}^{2}+\Delta m_{21}^{2}$.
The values of $\Delta m_{21}^{2}$ and $\Delta m_{31}^{2}$ are measured
in oscillation experiments, and one of the two independent
mass squared differences, say $\Delta m_{21}^{2}$, turns out to be much smaller
in absolute value than the second one $|\Delta m_{21}^{2}|\ll|\Delta m_{32}^{2}|$,
providing that $\Delta m_{31}^{2}\approx\Delta m_{32}^{2}$. Neutrino experiments
data imply $\Delta m_{21}^{2}=(7.37\pm0.17)\cdot10^{-5}$~eV$^{2}$
and $\Delta m_{31}^{2}\simeq(2.525\pm0.030)\cdot10^{-3}$~eV$^{2}$
\cite{Capozzi}. Formula (\ref{eq:Pee3nu}) can be used directly in precise calculations.

It is convenient to introduce the length of oscillations $L_{osc}$. The phase in eq. (\ref{eq:Deltaij}) that is responsible for oscillations can be written as $\pi\frac{L}{L_{osc}}$  the length of oscillation $L_{osc}$ is then defined as

\begin{equation}
L_{osc}=\frac{\pi}{1.27}\frac{E_{\nu}\text{[MeV]}}{|\Delta m^{2}|[\text{eV}^{2}]}\text{[m]}.
\label{eq:Losc}
\end{equation}

There are two distinct length scales for oscillation, defined by two very different $\Delta m^{2}$, as noted above. For a given $\Delta m^{2}$, the length of oscillation depends only on energy.

The survival probability can be approximated by averaging over fast-oscillating
terms for energies in the range of interest (below the maximum energy
of 3.27 MeV in the geoneutrino spectra). Indeed, the length of oscillations
for $\Delta m_{12}^{2}$ is of the order of 100 km, and for $\Delta m_{32}^{2}(\Delta m_{31}^{2})$
it is even less. Both lengths are small compared to the Earth radius
of 6400 km, and the average survival probability is:

\begin{equation}
\left\langle {P}_{ee}^{3\nu}(E_{\nu},L)\right\rangle \simeq1-0.5\left(\cos^{4}\theta_{13}\sin^{2}2\theta_{12}+\sin^{2}2\theta_{13}\right).\label{eq:Pee3nu-1}
\end{equation}

Using $\sin^{2}\theta_{13}=0.0215\pm0.007$ and $\sin^{2}\theta_{12}=0.297\pm0.017$~\cite{Capozzi}, 
we obtain the electron antineutrino survival probability

\begin{equation}
\left\langle {P}_{ee}\right\rangle =0.558\pm0.015,\label{eq:Pee}
\end{equation}

which can be used in estimations instead of (\ref{eq:Pee3nu}). The
uncertainty quoted for $\left\langle{P}_{ee}\right\rangle $
is evaluated using the uncertainties of the parameters involved in the
calculations, and the error on substitution of $P_{ee}$ by
its average value is omitted here, since it depends on the method of
detection and the detector's geographical position. The substitution
of ${P}_{ee}(E_{\nu},L)$ by its average value can also lead
to distortion of the geoneutrino spectrum, which can in principle
induce additional systematic errors in far future high statistics
geoneutrino spectral fits, possibly larger than the error on the absolute
value of $\left\langle {P}_{ee}\right\rangle $. For example,
more detailed evaluations performed for the Jinping site~\cite{Wan2017}
give $P_{ee}$ values growing linearly from 0.55 for $E\sim$0
to 0.56 at $E$=3 MeV, with the average over the geoneutrino spectrum $\left\langle {P}_{ee}\right\rangle =0.553$
and the effective average $\left\langle {P}_{ee}\right\rangle =0.562$,
calculated by weighting over the cross section of the detection reaction, which is inverse
$\beta$-decay of the proton in this case~\footnote{These calculations are based on the earlier compilation of the oscillation
parameters from~\cite{PDG2016}.}. Similar estimates are provided by Dye in~\cite{Dye2012},
where he reports an upward shift and distortion of the energy spectrum introduced
by using the average oscillation probability. Uranium and thorium
within one oscillation length to the detector cause an increase
of the survival probability, making the effect more pronounced on
continental crust than on oceanic crust. The $P_{ee}/\left\langle P_{ee}\right\rangle $
ratio for the continental crust is growing almost linearly from 1.015
at 1.8 MeV to 1.04 at 3.5 MeV. The estimated deviations are in average
within the precision of (\ref{eq:Pee}), well beyond the sensitivity of
modern and near future geoneutrino detection facilities.

With the average electron neutrino survival probability, the formula
for the antineutrino spectrum reduces to

\begin{equation}
\phi_{I}(E_{\nu})=n_{I}(E_{\nu})\cdot\left\langle P_{ee}\right\rangle \cdot\intop_{V_{Earth}}\frac{\rho(\overrightarrow{r})}{4\pi|\overrightarrow{R_{D}}-\overrightarrow{r}|^{2}}\cdot A_{I}(\overrightarrow{r})\cdot d\overrightarrow{r},\label{eq:FluxAvg}
\end{equation}

where the term defined by the integral is the full flux of the antineutrino $\Phi_{I}$
from the isotope $I$ expressed in usual flux units of cm$^{-2}$s$^{-1}$, and the antineutrino spectrum at the detector from the isotope $I$ is

\begin{equation}
\phi_{I}(E_{\nu})=n_{I}(E_{\nu})\cdot\left\langle P_{ee}\right\rangle \cdot\Phi_{I},\label{eq:FluxAvgSimple}
\end{equation}

All calculations above are performed for the case of vacuum oscillations
neglecting the matter effect. While neutrinos travel through dense matter
(in our case through the Earth) their mixing angles acquire a density-dependent
matter term. In fact, the electron neutrino interacts through charged
and neutral currents with electrons in the medium, while those of
other flavours interact through neutral currents only. As a result,
a difference in the potential of interaction for electronic and other
flavours appears, $-\sqrt{2}G_{F}n_{e}$, where $n_{e}$ is the electron
density and $G_{F}$ is the Fermi constant. In the case of constant
density, the matter effect changes the mixing angles and masses, and
the basic oscillation formula can be used with the corresponding substitutions.
In the case of varying matter density, the calculations are more complex.
The electron density in the Earth indeed varies with depth. The changes in
density are especially noticeable at the boundaries of different layers
of Earth. The steep changes of electron density complicate the prediction
of survival probability, because changes in the wave function depend on
the state of the wave function and the speed of the variation. The survival
probabilities were calculated by solving numerically equations
of motion in \cite{Sanshiro},\cite{Giunti1998}. It was concluded
that the matter effect increases the survival probability by $\sim$2\%,
and the spectral distortions (deviations from the average for different
point of production) are below 1\%. The corresponding systematics should
be taken into account when using the formula for the vacuum oscillations.

\subsection{Modelling the distribution of heat producing elements}

\begin{figure}[!ht]
\begin{centering}
\includegraphics[scale=0.7]{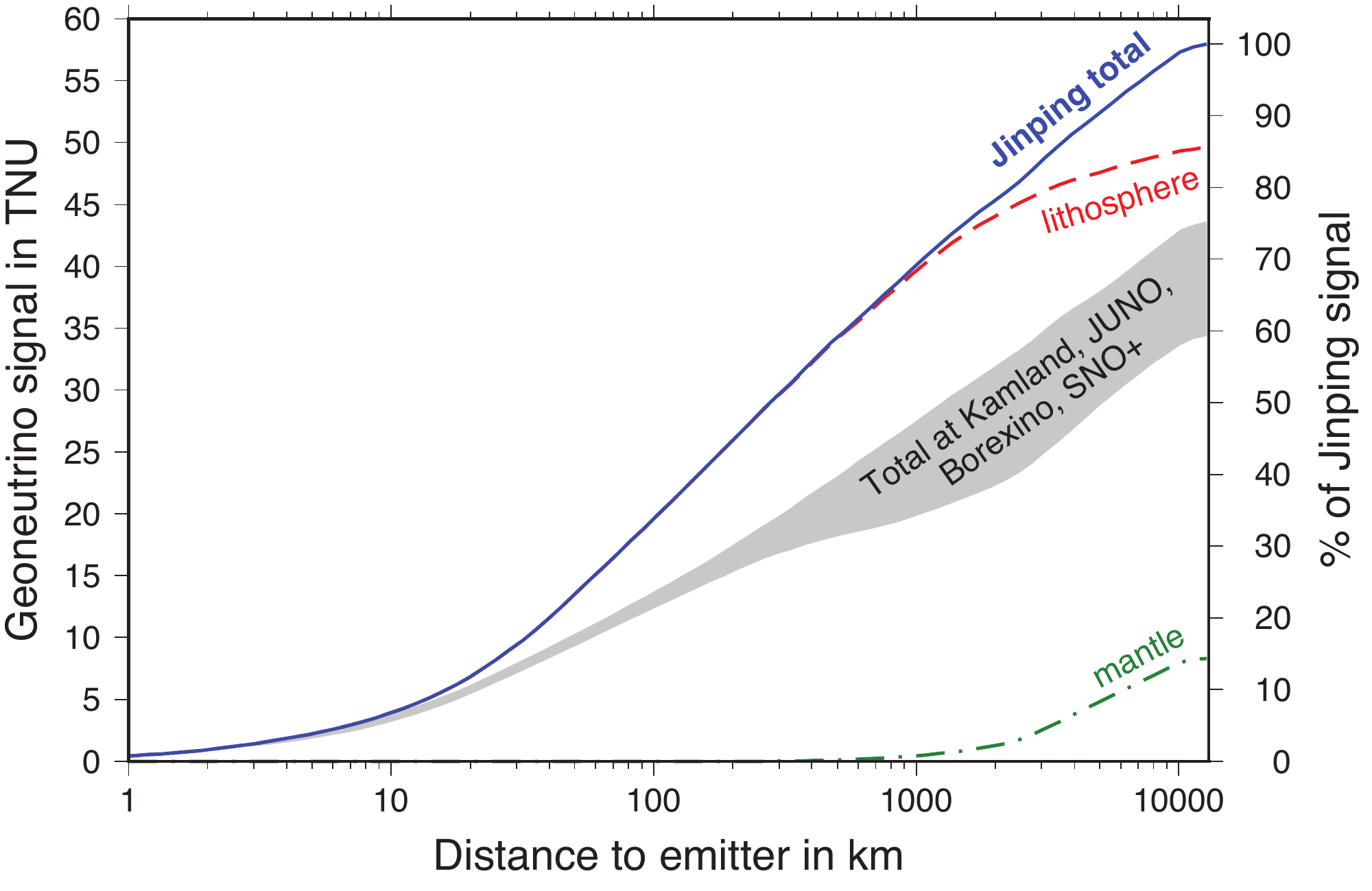}
\par\end{centering}
\caption{\label{fig:CumulativeSignal}Cumulative geoneutrino signal versus
distance to the fixed detector location. Geoneutrino signal is expressed
in absolute units (left vertical axis, see section \ref{subsec:detection}
for TNU definition) and as a percentage of the total signal expected
at the Jinping location (right vertical axis). Contributions
from lithosphere and mantle to the total Jinping signal are shown separately (denoted as "lithosphere" and "mantle" in the plot). Grey shaded area envelops
signals at other detectors. (Figure from \v{S}r\'{a}mek et al.~\cite{Sramek2016},
reproduced under Creative Commons Attribution 4.0 International License~\cite{CCL4.0})}
\end{figure}

The geoneutrino flux at a certain position depends on the distribution
of radioactive isotopes in the Earth bulk; in basic formula (\ref{eq:Flux})
it is the $\rho(\overrightarrow{r})A_{I}(\overrightarrow{r})$ term.
Calculations show that the geoneutrino flux varies within an order
of magnitude for different locations over the Earth's surface because
of the non-uniformities in isotope distribution. Half of the total
antineutrino flux arrives from distances of no more than 500~km, and
the closest 50~km contribute a quarter of the total. Thus, the local
area around the laboratory is the most important contributor to the
total signal, see Fig.~\ref{fig:CumulativeSignal}. The mantle contribution
at these distances is still negligible as the typical HPEs
abundances are three orders of magnitudes lower compared to the crustal
abundances. 

The information on density $\rho(\overrightarrow{r})$ comes from
seismic studies. The simplified structure of the Earth is well described
by the Preliminary Reference Earth Model (PREM)~\cite{PREM1981},
a one-dimensional model representing the average Earth properties
as a function of the planetary radius. It provides, in particular, a table
of the Earth density as a function of the radius that can be used in calculation.
The Earth interior is structured in the following way: from surface
to centre, a thin crust (thicker continental, $\sim$35~km in average,
and thinner oceanic, $\sim$8 km) and a mantle ($\sim$2900~km), both of
which are made up of silicates, and a Fe-Ni alloy core ($\sim$3500~km). The PREM tables can be used for the mantle densities. The mantle
volume is generally further subdivided in two parts: the main spherical
shell extending from the core-mantle boundary (CMB) to about 80~km
beneath the surface and a smaller mantle volume in between the main
shell and beneath the oceanic crust~\footnote{this is a geophysical mantle subdivision; the geochemical separation into Enriched Mantle and Depleted Mantle has different discontinuities and volumes, see also next comment.}. The latter breaks the spherical
symmetry of the geophysical response at the observation sites. 

The contribution from the mantle is model dependent. The so-called
Bulk Silicate Earth models (BSE, for their review see~\cite{Sramek2013})
predict abundances of the heat producing elements (HPEs) in the primordial mantle and assume that the total mass of the corresponding element is
at present distributed between the crust and the mantle. 
Hence, the mass of HPEs in the mantle can be estimated withing the BSE model by subtracting
the mass of the HPEs in the crust from the total one. Then the geoneutrino
signal from the mantle can be calculated considering different possible distributions 
of the remaining mass of the HPEs across the mantle.

A map of the density of the Earth's crust is provided by more detailed
three-dimensional models. The most recent are CRUST 1.0 \cite{CRUST1.0}
and LITHO 1.0 \cite{LITHO1.0}. CRUST 1.0 provides 8 structural layers
defined in 1\textdegree{}x1\textdegree{} cells: water,
ice, upper, middle, and lower sediments, and upper, middle, and lower crystalline
crust. LITHO 1.0 is a 1\textdegree{} tessellated model of the crust
and the uppermost mantle of the Earth extending into the upper mantle
to include the lithospheric lid and the underlying asthenosphere. The
model is parameterized laterally by tessellated nodes and vertically
as a series of geophysically identified layers, similar to CRUST 1.0
with the additional lithospheric lid and the asthenosphere\footnote{Lithospere is the rigid outer conductive part of the Earth, consisting of the crust and the mechanically coupled upper layer of the mantle beneath it; asthenosphere is the upper layer of the Earth's mantle below the lithosphere, in which convection is thought to occur. The both terms belong to the mechanical (or rheological) Earth layer subdivision, which includes the lithosphere, asthenosphere, mesospheric mantle, outer core, and inner core. More popular is chemical subdivision
of the Earth regions into the crust, the mantle (further subdivided into the upper and lower mantle), and the core, which can also be subdivided into the outer core and inner core.}. Details can be found in \cite{LITHO1.0}. 

The values of the HPEs concentration $A_{I}(\overrightarrow{r})$
are assigned to various crustal layers of the model. Earlier calculations
were based on CRUST~5.1 (a global crustal model with 5$\text{\textdegree}\times$5\textdegree{}
cells)~\cite{CRUST5.1} and its update to 2$\text{\textdegree}\times$2\textdegree{}
cells (CRUST~2.0)~\cite{CRUST2.0}. In particular, the reference model
by Fiorentini et al.~\cite{Mantovani2004}, further developed by Huang
et al.~\cite{Huang2013}, is based on the global CRUST~2.0 model. CRUST~2.0 provides a detailed crustal column for 2$\text{\textdegree}\times$2\textdegree{}
cells, which is not constrained by geophysical measurements but only
inferred from the geological type. A \textquotedblleft characteristic\textquotedblright{}
crustal column consists of 7 layers with different composition and
physical properties for each geological type (based on age and tectonic
setting) and is assigned to each cell depending on its geology. In
accordance with Mareschal et al.~\cite{Mareschal}, it is an oversimplification,
because it does not account for the heterogeneity of the continental
crust. These reference models provide a useful starting point for
interpreting geoneutrino observations. However, its practical
use is questioned by Mareschal et al.~\cite{Mareschal}, who note
that though CRUST~2.0 (and similarly its refined 1$\text{\textdegree}\times$1\textdegree{} version)
uses a compilation of seismic data, it is still a model. Neither CRUST~1.0 nor LITHO~1.0 provides uncertainties on the thickness of crustal
layers or rock density. \v{S}r\'{a}mek et al. in \cite{Sramek2016} expect
them to be of the order of 5\textendash10\%, but arguments for this
estimate are not provided. Huang et al.~\cite{Huang2013} note that
uncertainties in Moho depths\footnote{Mohorovi\v{c}i\'{c} discontinuity (Moho) corresponds to the boundary between
the Earth's crust and its mantle. The Moho lies at a depth of about
35~km below the continents and about 7 km beneath the oceanic crust and
is contained almost entirely within the lithosphere.} are a major source of uncertainty in the global crustal model. The previous 3SMAC topographic model~\cite{Nataf1996} included the
analysis of crust-mantle boundary developed by \v{C}adek and Martinec~\cite{Cadek1991}, in which the average uncertainties of continental
and oceanic crustal thickness are 5~km and 3~km (1$\sigma$) respectively
in agreement with the expectations of~\cite{Sramek2016}. Mareschal et
al. in~\cite{Mareschal} report that crustal thickness of CRUST~1.0
differs by up to 30\% from the real data of the seismic studies in Ontario
and Quebec. Dye~\cite{Dye2010} notes that a potentially significant
source of systematic uncertainty is lateral heterogeneity of the terrestrial
U and Th distributions. Whereas oceanic crust is quite uniform in
composition, lateral heterogeneity certainly exists in the continental
crust, as evidenced by uranium mining at specific locations. 

\v{S}r\'{a}mek et al in~\cite{Sramek2016} tested predictions of CRUST~1.0
against LITHO~1.0 for Jinping location. The results differ
by $\sim$5\%. The LITHO~1.0-based model prediction is higher, consistent
with the LITHO~1.0 Continental Crust (CC) around Jinping being thicker
compared to the CC of CRUST 1.0; globally the CC in LITHO~1.0
is 13\% more massive than in CRUST~1.0. 

\subsection*{Local area contribution}

The use of a simplified global model could lead to a significant systematic
error in antineutrino flux evaluation. The most sensitive parts are the
regions closest to the detector. The systematics induced by the crustal
model can be reduced by supplementing  models with detailed topographic information. In fact, $\sim$25\% of all geoneutrinos arrive
from distances below 50 km, which is within a 1\textdegree{}x1\textdegree{} cell, and $\sim$50\% of all antineutrinos arrive from a
distance below 500~km; thus, the detailed description of the
local area reduces the error of the most sensitive part. For example,
after the evaluation of the local contribution to the geoneutrino
signal at the Gran Sasso National Laboratory (LNGS) on the basis of a detailed
geological, geochemical, and geophysical study of the region, the total
predicted flux was reduced by 10\%~\cite{Coltorni2011}, compared
to the previous calculation based on general worldwide average assumptions.
The reason is the presence of a relatively thick layer of sediments
significantly depleted in U and Th, which was not considered in more general
studies. Thus, the need for detailed integrated geological study is
underlined by this work. The rest of the crust does not need precise
mapping of the geological regions and depends mainly on the total
quantity of HPEs. The estimates of the local contributions to the geoneutrino signal are available for the laboratories hosting or planning to host the geoneutrino experiments. These estimates will be discussed in detail in the sections devoted to the related experiments. 

\subsection*{Large mantle structures}

A homogeneous composition of seismically differentiated regions
in the mantle is usually assumed in models. However, there is experimental evidence
of inhomogeneities in the deep mantle. The global seismic tomography
reveals two large low shear velocity provinces (LLSVP) at the base
of the mantle off the west coast of Africa and near Tahiti in the
South Pacific~\cite{Garnero2008} with indication of a composition different
from the surrounding mantle~\cite{Sramek2013}. \v{S}r\'{a}mek
et al.~\cite{Sramek2013} examined the influence of the chemical composition
changes on the mantle signal. The studies demonstrate that the mantle
geoneutrino flux at the Earth's surface will have two distinct maxima
at about +20\% of the average mantle flux above the LLSVPs, assuming
the mantle is depleted in U and Th with respect to the LLSVPs. 

Roskovec et al.~\cite{Roskovec2018} considered the search for the so-called
"second continents", hypothesized
gravitationally stabilized regions enriched with upper crustal concentration
levels of radiogenic elements at a depth of 600 to 700 km. The geoneutrino
flux in these second continents can be detected with land- and/or ocean-based
detectors. 

\subsection{\label{subsec:Models}Models of the Earth}

The models discussed above provide the density at the given position of the mantle (PREM) and constrain the crust properties assigning the concentrations of HPEs to each of the crustal/lithospheric types. To complete the model and obtain the global distribution of HPEs, one needs a set of abundances of these elements in the mantle. The common major assumptions are:
\begin{itemize}
\item the core of the Earth does not contain HPEs, since U, Th, and
K are lithophile elements having affinity to the silicate minerals
and melts, and should not be present in the Earth's core;
\item the BSE concept is used to calculate partitioning
of the elements. The BSE model provides chemical composition considers of the primordial mantle, namely the mantle at the moment just after the core separation but prior to the start of the crust formation\footnote{According to McDonough and Sun~\cite{McDonough1995} the Silicate Earth and Primitive Mantle are synonymous};
\item during the formation of the Earth's crust, the mantle was depleted
in U and Th, while the continental crust was highly enriched in these HPEs. The crustal
abundances can be derived from the direct measurements, and the mantle
abundances can be calculated using the BSE constraint;
\item the homogeneous composition of seismically differentiated mantle regions
is usually assumed in the model. 
\end{itemize}
The BSE model is in good agreement with the majority of experimental
observations concerning the core and the upper mantle. The composition
of the primitive mantle is derived from the composition of the chondritic
meteorites, representing the majority of stone meteorites, with some
corrections for the observed abundances in the Earth. The composition
of chondrite meteorites is practically identical to the chemical composition
of the Sun, excluding the lightest elements, such as hydrogen and
helium. It is assumed that chondrites originate from protoplanet matter
surrounding the Sun by condensation and accretion of the dust with
intermediate heating. Chondrites does not show traces of melting and
correspond to non-differentiated planets (planets without crust),
i.e., to the composition of the primitive mantle. Compositional models
of the Earth can be grouped into three major classes in accordance with
the total predicted radiogenic heat of the Earth \cite{Dye2015,McDonough2016}:
low-Q models (10\textendash 15 TW), medium-Q models (17\textendash 22
TW), and high-Q models ($>$25 TW):
\begin{itemize}
\item \textbf{Low-Q models} (\textbf{cosmochemical}): these models state either 
that the Earth accreted with low quantities of Th and U (Enstatite
Earth model, \cite{Javoy1995,Javoy2010,Kaminski2013,Javoy2014}) or
that in the early stages of accretion the planet lost up to half of
its crust enriched in Th and U in collisions with massive external
bodies (the Collisional Erosion model \cite{ONeill2008,Caro2010,Warren2011,Jackson2013}).
The latter is one of the variations of the Non-Chondritic model \cite{Campbell2012}).
It is based on the composition of enstatite chondrites having the
closest isotopic similarity to the mantle samples and possessing high
enough iron content to explain the metallic core. The studies supporting the enstatite chondrite-derived composition are provided by Warren \cite{Warren2011} 
and Zhang et al.~\cite{Zhang2012}.
Models predict U concentration of 12$\pm$2 ppb~\footnote{ppb, parts per billion, is used to describe low concentration of substances, 1 ppb corresponds to $10^{-9}$ g/g and equals 1 ng of substance per g of solid (ng/g).}. 
\item \textbf{Medium-Q models (geochemical):} these models use samples of
the Earth's mantle, constraining the abundance ratios
to those of the carbonaceous chondrites\footnote{The class of carbonaceous chondrites meteorites is usually denoted CI, here ``C'' is for carbon and ``I'' is the first letter of the Ivuna location, where this class of meteorites was first found.} similar
to the solar photosphere in content of refractory lithophile elements
(having affinity to silicate minerals and melts), siderophile elements
(affinity to iron), and volatile elements \cite{McDonough1995,Ringwood1975,Jagoutz1979,Wanke1981,Hart1986,Allegre1995,Palme2003,Lyubetskaya2007}.
A typical chondritic ratio of Th/U masses is 3.9\footnote{Wipperfurth et al. recently reported the results of their high precision
study of the mass ratio in~\cite{Wipperfurth2018}, which they found to be M(Th)/M(U)=3.776$_{-0.075}^{+0.122}$.
We use a commonly accepted value of 3.9 throughout the text, as
it has been used in all experimental publications until now, and it is still
compatible with the value of Wipperfurth et al. within the 1$\sigma$
uncertainty.} and K/U masses $\sim$14000. Typical U concentration in these models
is 20$\pm$4 ppb. The underlying strong assumption is that by sampling
the top of the Earth one can restore the bulk composition of the mantle,
i.e., the mantle is roughly homogeneous in composition.
\item \textbf{High-Q models (geodynamical):} these models are based on a
physical approach with a boundary condition given by the surface heat
flux, and on the simple relationship between the heat output from the
convecting mantle and the vigor of convection described as a balance
between thermal buoyancy driving the dynamics and thermal and momentum
diffusion hindering the flow \cite{Schubert1980,Turcotte2001,Turcotte2002,Davies2010b}.
Trade-offs between buoyancy and viscosity versus thermal and momentum
diffusion rates allow for a range of solutions that can include Low-
to High-Q models. Typical U concentration in these models is 35$\pm$4
ppb. 
\end{itemize}

The compositional predictions of these three models for the BSE are shown in Tab.~\ref{tab:BSE}. The more detailed compilation of estimates for the abundances of HPEs for different Earth reservoirs: bulk continental crust, bulk oceanic crust, bulk mantle, and depleted mantle, can be found in \v{S}r\'{a}mek et al.~\cite{Sramek2013} together with estimates of power in TW and the mantle Urey ratio.

Global masses of U, Th/U mass ratio, and radiogenic heat production H(U+Th)
in the Earth according to different classes of BSE models are shown in Table~\ref{tab:Models-Masses}. 

\begin{table}
\caption{\label{tab:BSE}Estimates of HPEs abundances and corresponding radiogenic
power for BSE in cosmochemical, geochemical, and geodynamical models,
compiled by \v{S}r\'{a}mek et al.~\cite{Sramek2013}. The BSE mass is $\sim4\times10^{24}$ kg, and the corresponding total mass of HPE can be obtained multiplying its abundance by the BSE mass}
\centering{}%
\begin{tabular}{cccc}
\hline 
{Parameter} & {Cosmochemical} & {Geochemical} & {Geodynamical}\tabularnewline
 & {(Low-Q) } & {(Medium-Q) } & {(High-Q)}\tabularnewline
\hline 
{$a_{U}$, ppb} & {12$\pm$2} & {20$\pm$4} & {35$\pm$4}\tabularnewline
\hline 
{$a_{Th}$, ppb} & {43$\pm$4} & {80$\pm$13} & {140$\pm$14}\tabularnewline
\hline 
{$a_{K}$, ppm} & {146$\pm$29} & {280$\pm$60} & {350$\pm$35}\tabularnewline
\hline 
{Th/U} & {3.5} & {4.0} & {4.0}\tabularnewline
\hline 
{K/U} & {12,000} & {14,000} & {10,000}\tabularnewline
\hline 
{$Q_{H}$, TW} & {11$\pm$2} & {20$\pm$4} & {\footnotesize33$\pm$3}\tabularnewline
\hline 
\end{tabular}
\end{table}

\begin{table}
\caption{\label{tab:Models-Masses}Global masses of U, M(Th)/M(U) ratio, and the corresponding
radiogenic heat production H(U+Th) in the Earth according to different
BSE models (compiled by Huang et al. \cite{Huang2012})}
\vspace{2mm}
\centering{}%
\begin{tabular}{cccc}
\hline 
{Reference} & {M(U), 10$^{17}$ kg} & {M(Th)/M(U)} & {H, TW}\tabularnewline
\hline 
{Javoy et al. \cite{Javoy2010}} & {0.5} & {3.5} & {9.2}\tabularnewline
{Lyubetskaya and Korenaga \cite{Lyubetskaya2007}} & {0.7} & {3.7} & {13.4}\tabularnewline
{McDonough and Sun \cite{McDonough1995}} & {0.8} & {3.9} & {16.2}\tabularnewline
{Allegre et al. \cite{Allegre1995}} & {0.8} & {3.9} & {16.2}\tabularnewline
{Palme and O'Neil \cite{Palme2003}} & {0.9} & {3.8} & {17.6}\tabularnewline
{Anderson \cite{Anderson2007}} & {1.1} & {4.0} & {23.0}\tabularnewline
{Turcotte and Schubert \cite{Turcotte2002}} & {1.2} & {4.0} & {25.4}\tabularnewline
\hline 
\end{tabular}
\end{table}

The Earth's crust is highly enriched in U and Th, while
its mass is much less than that of the mantle. Therefore, the crustal
geoneutrino contribution dominates in continental observatories.
The prediction for the crust contribution is bounded more strongly than
the mantle contribution, the latter can be obtained extracting the crust contribution from the total. The difference in heat output of the models is in fact 
defined by the mantle contribution and the variability of predictions of the mantle heat is more pronounced as compared to the total heat, namely : (3$\pm2$)~TW for cosmochemical,
(12$\pm$4)~TW for geochemical and (25$\pm$3)~TW for
geodynamical models. Therefore, the measurement of the geoneutrino
signal from the mantle is a clue to distinguishing between the existing models.
Moreover, geoneutrinos are a tool for studying the identified large-scale
mantle structures, as discussed in the previous section.

\section{\label{subsec:detection}Detection of antineutrinos}

Antineutrino spectra expected from the $^{238}$U and $^{232}$Th chains,
as well as from the $^{40}$K decay, are presented in Fig.~\ref{fig:Specific-antineutrino-spectra}.
A fraction of antineutrino spectra from the $^{238}$U and $^{232}$Th
chains exceeds the threshold of the inverse beta-decay (IBD) of the free proton\footnote{Inverse beta-decay of the free proton is commonly abbreviated as IBD without mentioning the free proton. We will follow this convention if not stated otherwise explicitly}.
This makes possible registration of antineutrino by liquid scintillation detectors in the IBD of the proton
$\overline{\nu}_{e}+p\longrightarrow e^{+}+n$ with a threshold of $E_{\nu}\simeq$1.8~MeV. Today, all the available data on geoneutrinos are obtained with
this detection technique, and the near future detectors will also use the
IBD for the geoneutrino detection. It should
be noted that at the moment there are no dedicated projects aimed
at the geoneutrino studies. The projects have different primary goals, such as solar neutrino studies
in Borexino or reactor antineutrino studies in KamLAND, and the geoneutrino
studies are merely an extension of the mainstream programme. The need
for the detectors with lower threshold for $^{40}$K detection and
potential advantages of geoneutrino directional studies can trigger
development of specialized detectors based on other principles.
These possibilities will be discussed later.

A convenient unit to describe the rate of interactions in the IBD detectors 
is a Terrestrial Neutrino Unit (TNU) first introduced by Mantovani et al. in~\cite{Mantovani2004}. One TNU corresponds to 1 interaction
for $10^{32}$ target protons in one year; $10^{32}$ target protons
are contained in roughly 1 kt of typical liquid organic scintillator,
and few years is a typical expected exposure time for the geoneutrino
detection. Thus, the counting rate in TNU roughly corresponds to the expected
event rate per year in a 1 kt scale IBD detector. Let us note that the TNU 
defined in this way is suitable only for the reaction of the IBD of the proton,
in the case of other IBD reactions, the definition should be properly
changed. 

The characteristic features of the typical geoneutrino detector are
its size and strong requirements on the backgrounds. Shielding against
cosmogenic backgrounds needs a huge quantity of material and is easily
achieved by placing the detector deep underground, in a mine or
a tunnel under the mountain massif. There are few laboratories in
the world suited to host large detectors. A list of the most relevant
ones is presented in Table~\ref{tab:Locations}, and a more detailed list
can be found in~\cite{Ricci2015}. 

All the data on the geoneutrino
obtained until now are provided by two experiments, in LNGS (Borexino)
and at Kamioka (KamLAND). The SNO lab is ready to start the SNO+ experiment,
and there are plans to use the future data from the JUNO detector
for geoneutrino studies. The note on the primary goals also applies to these two upcoming experiments: the main goal of SNO+ will
be neutrinoless double beta decay search, and JUNO challenges the
problem of the mass hierarchy of neutrinos. The laboratory at Jinping plans to host two detectors able to search for geoneutrinos, but currently the final decision for the detector choice is missing. There are no approved experiments at other sites.

\begin{table}
\caption{\label{tab:Locations}Locations and rock overburden in meters of water equivalent (m.w.e.) of
existing labs hosting or having plans to build a geoneutrino detector,
existing or approved facilities are highlighted by bold. The last 4
rows present correspondingly: the total expected reactor antineutrino
signal (FER stays for full energy range), the expected reactor antineutrino
signal in the geoneutrino energy window (GEW, $E_{\overline{\nu_{e}}}<3.27$~MeV),  the predicted geoneutrino
signal and the reactor/geoneutrino ratio for the relevant energy
range. The calculations for KamLAND are performed for the actual situation
with Japanese reactors off. 
The data on the predicted reactor signals are compiled using the
estimates from~\cite{Han2016, Beacom2016, Ricci2015}.
The geoneutrino signals were calculated by Huang et al. in~\cite{Huang2013} using their Reference EARTH Model}
\vskip 2mm
\centering{}%
\begin{tabular}{ccccccccc}
\hline 
{\scriptsize{}Laboratory} & {\scriptsize{}LNGS} & {\scriptsize{}Kamioka} & {\scriptsize{}Sudbury} & {\scriptsize{}Jinping } & {\scriptsize{}Jiangmen} & {\scriptsize{}Baksan} & {\scriptsize{}Pyh\"{a}salmi} & {\scriptsize{}Hawaii}\tabularnewline
\hline 
{\scriptsize{}longitude} & {\scriptsize{}42.55$^{\circ}$N} & {\scriptsize{}36.42$^{\circ}$N} & {\scriptsize{}46.475 $^{\circ}$N} & {\scriptsize{}28.15$^{\circ}$N} & {\scriptsize{}22.12$^{\circ}$N} & {\scriptsize{}43.24$^{\circ}$N} & {\scriptsize{}63.66$^{\circ}$N} & {\scriptsize{}19.72$^{\circ}$N}\tabularnewline
{\scriptsize{}latitude} & {\scriptsize{}13.57$^{\circ}$W} & {\scriptsize{}137.31$^{\circ}$W} & {\scriptsize{}81.20$^{\circ}$W} & {\scriptsize{}101.71$^{\circ}$E} & {\scriptsize{}112.51$^{\circ}$E} & {\scriptsize{}42.70$^{\circ}$E} & {\scriptsize{}26.04$^{\circ}$E} & {\scriptsize{}156.32$^{\circ}$W}\tabularnewline
{\scriptsize{}depth, m.w.e.} & {\scriptsize{}3000} & {\scriptsize{}2700} & {\scriptsize{}6000} & {\scriptsize{}6720} & {\scriptsize{}2100} & {\scriptsize{}4900} & {\scriptsize{}4000} & {\scriptsize{}4-5000}\tabularnewline
\hline 
{\scriptsize{}Detector/proposal} & \textbf{\scriptsize{}Borexino} & \textbf{\scriptsize{}KamLAND} & \textbf{\scriptsize{}SNO+} & {\scriptsize{}Jinping } & \textbf{\scriptsize{}JUNO} & {\scriptsize{}Baksan} & {\scriptsize{}LENA} & {\scriptsize{}Hanohano}\tabularnewline
\hline 
{\scriptsize{}S(R)$_{\text{FER}}$, TNU} & {\scriptsize{}$85.2_{-1.8}^{+2.0}$} & {\scriptsize{}$27.4_{-0.6}^{+0.6}$} & {\scriptsize{}$193.9_{-4.5}^{+4.7}$} & {\scriptsize{}27.8} & {\scriptsize{}$1569\pm88$} & {\scriptsize{}$37.4_{-0.8}^{+0.9}$} & {\scriptsize{}$69.5_{-1.6}^{+1.7}$} & {\scriptsize{}$3.4_{-0.07}^{+0.08}$}\tabularnewline
{\scriptsize{}S(R)$_{\text{GEW}}$, TNU} & {\scriptsize{}$22.9_{-0.5}^{+0.6}$} & {\scriptsize{}$7.4_{-0.2}^{+0.2}$} & {\scriptsize{}$48.8_{-1.5}^{+1.7}$} & {\scriptsize{}6.8} & {\scriptsize{}$351\pm27$} & {\scriptsize{}$9.96_{-0.27}^{+0.28}$} & {\scriptsize{}$18.0_{-0.5}^{+0.5}$} & {\scriptsize{}$0.9_{-0.02}^{+0.02}$}\tabularnewline
{\scriptsize{}S(U+Th), TNU} & {\scriptsize{}$40.3_{-5.8}^{+7.3}$} & {\scriptsize{}$31.5_{-4.1}^{+4.9}$} & {\scriptsize{}$45.4_{-6.3}^{+7.5}$} & {\scriptsize{}$59.4\pm6.8$} & {\scriptsize{}$39.7_{-5.2}^{+6.5}$} & {\scriptsize{}$47.2_{-6.4}^{+7.7}$} & {\scriptsize{}$45.5_{-5.9}^{+6.9}$} & {\scriptsize{}$12.0_{-0.6}^{+0.7}$}\tabularnewline
{\scriptsize{}S(R$_{\text{GEW}}$)/S(U+Th)} & {\scriptsize{}0.6} & {\scriptsize{}0.2} & {\scriptsize{}1.1} & {\scriptsize{}0.1} & {\scriptsize{}8.8} & {\scriptsize{}0.2} & {\scriptsize{}0.4} & {\scriptsize{}0.1}\tabularnewline
\hline 
\end{tabular}
\end{table}

\subsection{Antineutrino detection in IBD of the proton}

The antineutrino detection reaction is the IBD of the proton:

\begin{equation}
\overline{\nu_{e}}+p\rightarrow n+e^{+}.\label{eq:inverse_beta}
\end{equation}
It was first applied in the pioneering research by F. Reines and C.L.
Cowan who conducted the first experiment on reactor antineutrino detection.
Liquid organic scintillators can be used as a detection medium, as
it is rich in free protons and cheap, permitting construction of large
detectors. At present, geoneutrino observations using IBD have been reported by two collaborations with the Borexino and
KamLAND organic liquid scintillation detectors respectively.

The reaction has a relatively high kinematic threshold due to a difference in mass between the products and the reactants: $E_{th}=\frac{(m_{n}+m_{e})^{2}-m_{p}^{2}}{2m_{p}}$
= 1.806 MeV, 2 keV higher than in the case where recoil energy
is neglected. The kinematic threshold of the reaction makes it impossible to detect 
lower-energy antineutrinos in IBD detectors. 

The energy of the incoming antineutrino $E_{\nu}$
is strongly correlated with the energy of the positron created as
a result of the interaction. The positron takes away the energy $E_{kin}=$
$E_{\nu}-1.806$ MeV in the form of the kinetic energy with a negligible
fraction of this excess energy left for the kinetic energy of the
neutron. The kinetic energy of the positron is partially transformed into scintillation
light detected by photosensitive elements of the detector - photomultiplier
tubes (PMTs). The annihilation of the positron produces two gamma-quanta
of 0.511 MeV energy each, also producing light flashes in the scintillator
through Compton scattering on electrons. Thus, the total amount of
scintillation light corresponds to the visible energy

\begin{equation}
E_{vis}=E_{kin}+2m_{e}=E_{\nu}-0.784\;\text{MeV}.\label{eq:Evis}
\end{equation}

Together with the positron, a neutron is released in reaction (\ref{eq:inverse_beta}).
This gives an opportunity to "tag" the reaction, looking for pairs
of events correlated in space and time. Indeed, the neutron is thermalized
very effectively in proton-rich media (in 10-20 collisions with protons),
and the thermal neutron has a relatively high cross section of interaction
with protons. It is captured on the proton with a typical time of
200-250 $\mu$s for liquid organic scintillators, releasing an easily-detectable monoenergetic gamma quantum

\begin{equation}
n+p\rightarrow D+\gamma(2.223\;\text{MeV}).\label{eq:np}
\end{equation}

In large-volume detectors all 2.2 MeV of gamma energy are lost in Compton
scattering on electrons and thus converted into light, and the gamma
is effectively detected. Therefore, the ``tag'' is provided by a 2.2
MeV ``delayed'' event observed in coincidence with a ``prompt''
one. In a liquid organic scintillator another neutron capture reaction
is possible: $^{12}\text{C(n,\ensuremath{\gamma})}{}^{13}\text{C}$.
In this case, a more energetic gamma quantum with $E_{\gamma}=4.945$
MeV is emitted. The relative probability of the neutron capture is
only about 1\% of that of the capture on the proton due to the large
difference in cross sections, and the contribution of this reaction
is insignificant. 

The time window defining the "true" coincidence can be chosen
on the basis of the expected accidental coincidence rate and the life-time
of the neutron with respect to the capture on the proton, the $5\tau$ window is typically used in the low background environment. The tagging technique
allows the effective reduction of the random coincidence background.
Further improvement of the technique can be achieved by using the vertex
reconstruction of prompt and delayed events offered by modern large-volume liquid scintillation detectors equipped with many photosensors.
The position is reconstructed using arrival times of photons at
each photosensor (for more detail and examples of the technique see
e.g. \cite{Sanshiro,Smirnov2003,Borex2014}). The position reconstruction
is also essential for the reduction of the background from the construction
materials in the active volume of the scintillator. The liquid scintillator (LS) should
be contained in a transparent vessel separating LS from the non-scintillating
buffer, both the thin nylon or thicker acryl
are used in modern detectors. The material of the vessel is unavoidably less clean with respect to radioactive impurities than the LS. The
inner section of the LS volume, called the fiducial volume (FV), provides the best
signal-to-background ratio and is selected off-line using the position reconstruction
code. In such a way, the remaining outer section serves as active
shielding for the FV. 

The typical drift distance of a neutron is few centimeters
compared to ten centimeters of the typical position reconstruction
precision, and two events in reactions
(\ref{eq:inverse_beta}) and (\ref{eq:np}) occur practically at the
same position. Hence, another condition can be imposed on the pair of events:
their positions should be reconstructed as close to each other as
possible. The selection criteria should be based on the balance
of the efficiency of the spatial cut and on the rate of accidental coincidences
satisfying the applied cut. The efficiency of the spatial cut
generally depends on the precision of the reconstruction algorithm.
The two basic selection criteria, time and
distance between events, can be combined with other specific features
of two correlated events, e.g., by rejecting pairs containing $\alpha$-like
events.

Compared to low energy neutrino detection, 
antineutrino detection in few MeV energy range is relatively easier in the LS due to the possibility
of using the delayed coincidence tag, and due to the fact that antineutrino detection reaction (\ref{eq:inverse_beta}) for these
energies has a higher cross section than the reaction of elastic neutrino-electron
scattering used in neutrino detectors. The elastic geoneutrino (antineutrino)
scattering off electrons will still contribute to the total untagged events
counting rate, but it is a small effect compared to the natural background
from elastic Solar neutrino scattering off electrons. 

\subsection{Cross section of IBD of the free proton}

The total cross section for reaction (\ref{eq:inverse_beta})
can be expressed in terms of the neutron lifetime $\tau_{n}$ and
the neutron decay phase-space factor $f=1.7152$ (includes the Coulomb,
weak magnetism, recoil, and outer radiative corrections, but not the
inner radiative corrections) neglecting terms of the order of $E_{\nu}/m_{p}$
as \cite{Wilkinson82}

\begin{equation}
\sigma_{n}(E_{\nu})\simeq\frac{2\pi^{2}}{m_{e}^{5}\cdot f\cdot\tau_{n}}E_{e}p_{e},\label{eq:cs_formula}
\end{equation}

where $E_{e}=E_{\nu}-(m_{n}-m_{p})=E_{\nu}-1.293$~MeV is the positron
energy neglecting the small neutron recoil energy, and $p_{e}$ is
the corresponding momentum, both measured in MeV. The numerical factor
before $E_{e}p_{e}$ in (\ref{eq:cs_anu}) depends inversely on the
neutron decay time; using a recent value $\tau_{n}=880.3\pm1.1$~s~\cite{PDG}, we obtain

\begin{equation}
\sigma_{n}(E_{\nu})\simeq9.57\cdot10^{-44}p_{e}E_{e}\;\text{cm}^{2}.\label{eq:cs_anu}
\end{equation}

The (small) energy-dependent outer radiative corrections to $\sigma_{tot}$ are given in~\cite{Vogel1984a,Fayans1985}. The corrections 
to the cross section of the order of $E_{\nu}/m_{p}$,
not negligible even at low energies, and the angular distribution
of the positrons are described by Vogel and Beacom~\cite{Vogel1999}.
The relatively small outer radiative corrections of the order of $\alpha/\pi$
can be accurately evaluated if necessary~\cite{Kurylov2003},
and the cross section of the inverse $\beta$-decay can be evaluated
with an accuracy of up to $\sim$0.2\%. The uncertainties of the IBD reaction 
cross section obtained in reactor experiments are
an order of magnitude larger (1.4-3.0\%~\cite{Zacek1986,Vidyakin1994,Declais1994}),
dominated by the uncertainties of the number of target protons estimation, and the reactor-related
uncertainties of the reactor power and 
time-dependent fuel composition, e.g., fission fractions.

A more general calculation of the neutrino-nucleon cross section
in the energy range up to several hundred MeV was discussed by Strumia
and Vissani in~\cite{Strumia2003}. They provided a simple approximation
of the result

\begin{equation}
\sigma_{n}(E_{\nu})\simeq10^{-43}p_{e}E_{e}\cdot E_{\nu}^{-0.07056+0.02018\ln E_{\nu}-0.001953\ln^{3}E_{\nu}}\;\text{cm}^{2}\label{eq:cs_strumia-vissani}
\end{equation}

that agrees within a few per-mille with the precise calculations for energies
up to 300~MeV, and provides the comparably good total accuracy, the authors
report 0.4\% total precision at low energy.

The differential geoneutrino event rate at the detector is

\begin{equation}
\frac{dN}{dE_{\nu}}=n_{p}\sigma_{n}(E_{\nu})\epsilon(E_{\nu})\phi(E_{\nu}),\label{eq:DiffRate}
\end{equation}

where $n_{p}$ is the number of target protons within the FV of the detector, $\epsilon(E_{\nu})$ is the detector efficiency
at energy $E_{\nu}$, and $\phi(E_{\nu})$ is the differential electron
antineutrino flux at the detector's location defined by (\ref{eq:Flux}).

\subsection{\label{subsec:geoneutrino-signal-IBD}Geoneutrino signal in the IBD
antineutrino detector}

The expected total geoneutrino signal $S_{I}$ for the isotope I
can be obtained by integrating (\ref{eq:DiffRate}) over energy from
the reaction threshold to the maximum energy of the corresponding
antineutrio flux

\begin{equation}
S_{I}=\intop_{E}dE_{\nu}\cdot n_{p}\cdot\epsilon(E_{\nu})\cdot\phi(E_{\nu})\cdot\sigma_{n}(E_{\nu})\simeq\left\langle {P}_{ee}\right\rangle \cdot \Phi_{I}\cdot n_{p}\cdot\intop_{E}dE_{\nu}\cdot\epsilon(E_{\nu})\cdot\sigma(E_{\nu})\cdot n_{I}(E_{\nu}).\label{eq:Signal}
\end{equation}

All the notations were introduced above, and the substitution of $P_{ee}(E_{\nu},L)$
by its average value is performed in accordance with (\ref{eq:FluxAvgSimple}).
Assuming the energy-independent detector efficiency, the term defined by the
integral can be replaced with an average cross section

\begin{equation}
S_{I}=\left\langle P_{ee}\right\rangle \cdot \Phi_{I}\cdot n_{p}\cdot\epsilon\cdot\sigma_{I},\label{eq:Signal-1}
\end{equation}

where the average cross section $\sigma_{I}$ is defined over the
corresponding neutrino spectrum

\begin{equation}
\left\langle \sigma_{I}\right\rangle =\frac{\intop_{E}dE_{\nu}\cdot n_{I}(E_{\nu})\cdot\sigma_{n}(E_{\nu})}{\intop_{E}dE_{\nu}\cdot n_{I}(E_{\nu})}.\label{eq:CSavg}
\end{equation}

Performing calculations for $^{238}$U and $^{232}$Th, one obtains
\begin{equation}
\left\langle \sigma_{^{238}\text{U}}\right\rangle =(4.08\pm0.04)\times10^{-45}\;\text{cm}^{-2},
\end{equation}
\begin{equation}
\left\langle \sigma_{^{232}\text{Th}}\right\rangle =(1.25\pm0.05)\times10^{-45}\;\text{cm}^{-2}.
\end{equation}

The main contribution to the uncertainties are induced by the uncertainties of intensities and end-point energies of the corresponding transitions quoted in Tables~\ref{tab:betaTh} and~\ref{tab:betaU}. Since the IBD cross section increases fast, the small uncertainties of the end-point energy in the $^{238}$U chain are inducing errors comparable to the ones due to the relatively larger uncertainties of the intensities of the transitions. 
Possible errors due to the deviations of $\beta$-spectra from the "universal" shape are not considered because of the lack of experimental measurements. 

With the average cross sections, the geoneutrino signal for 1 year can be calculated as

\begin{equation}
N(I)=\left\langle \P_{ee}\right\rangle \cdot\epsilon\cdot
\left\langle \sigma_{I}\right\rangle\cdot10^{45}
\cdot\frac{\Phi_{I}}{10^{6}}\cdot\frac{n_{p}}{10^{32}}\cdot
\frac{3.156\times10^{7}}{10^{7}},\label{eq:SignalTNU}
\end{equation}

or

\begin{equation}
N(\text{U})=(12.87\pm0.15)\cdot\left\langle {P}_{\text{ee}}\right\rangle \cdot\epsilon\cdot\frac{\text{\ensuremath{\Phi}}_{\text{U}}}{10^{6}}\cdot\frac{n_{p}}{10^{32}}\;\text{yr}^{-1},
\end{equation}

\begin{equation}
N(\text{Th})=(3.95\pm0.14)\cdot\left\langle {P}_{ee}\right\rangle \cdot\epsilon\cdot\frac{{{\Phi}}_{\text{Th}}}{10^{6}}\cdot\frac{n_{p}}{10^{32}}\;\text{yr}^{-1}\cdot
\end{equation}

The geoneutrino signal expressed in TNU is usually quoted
for 100\% detector efficiency and depends on the corresponding geoneutrino
flux as
\begin{equation}
S(\text{U})=(12.87\pm0.15)\cdot\left\langle {P}_{ee}\right\rangle \cdot\frac{{{\Phi}}_{\text{U}}}{10^{6}}\;\text{TNU}\simeq(7.2\pm0.2)\cdot\frac{\Phi_{\text{U}}}{10^{6}}\;\text{TNU},
\end{equation}

\begin{equation}
S(\text{Th})=(3.95\pm0.14)\cdot\left\langle {P}_{ee}\right\rangle \cdot\frac{{\ensuremath{\Phi}}_{\text{Th}}}{10^{6}}\;\text{TNU}\simeq(2.2\pm0.1)\cdot\frac{\Phi_{\text{Th}}}{10^{6}}\;\text{TNU}.
\end{equation}

The errors on the average cross section and on the average oscillation probability are propagated quadratically. Only the error on the $\left\langle {P}_{ee}\right\rangle$ from~\ref{eq:Pee3nu-1} is considered, though further contributions to the total signal uncertainty are possible as discussed in~\ref{sec:oscillations}.

In order to obtain the number of detected events, one should multiply the geoneutrino
signal expressed in TNU by the detector efficiency and scale the exposure
to $10^{32}$~protons$\cdot$yr. For the target mass of $10^{32}$~protons, 1~year exposure with the detection efficiency $\epsilon=1$,
and $\left\langle P_{ee}\right\rangle $=0.558, the
detector would count (7.2$\pm$0.2) events from the U chain and (2.2$\pm$0.1) events from the Th
chain for the incident flux of $10^{6}$~cm$^{-2}$s$^{-1}$. Since
the Th/U mass ratio of the Earth is M(Th)/M(U)$\simeq$3.9, and thorium
activity for the unit mass is 3.1 times lower, one expects $S$(Th)/$S$(U)$\simeq$1/4,
and though the global thorium mass is four times higher than that
of uranium, it contributes only $\sim$1/5 to the total geoneutrino
signal.

\begin{figure}
\begin{centering}
\includegraphics[scale=0.45]{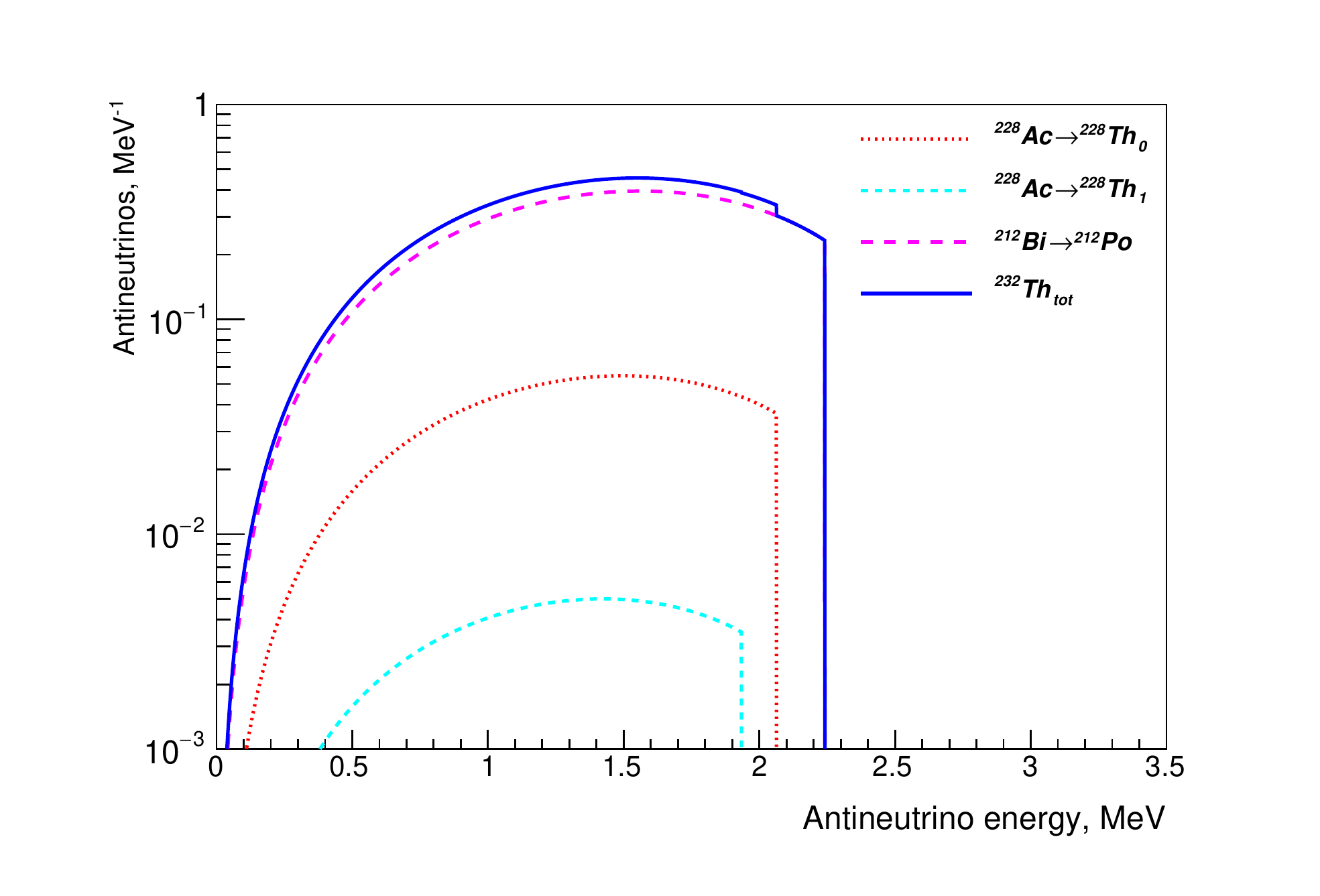}\includegraphics[scale=0.45]{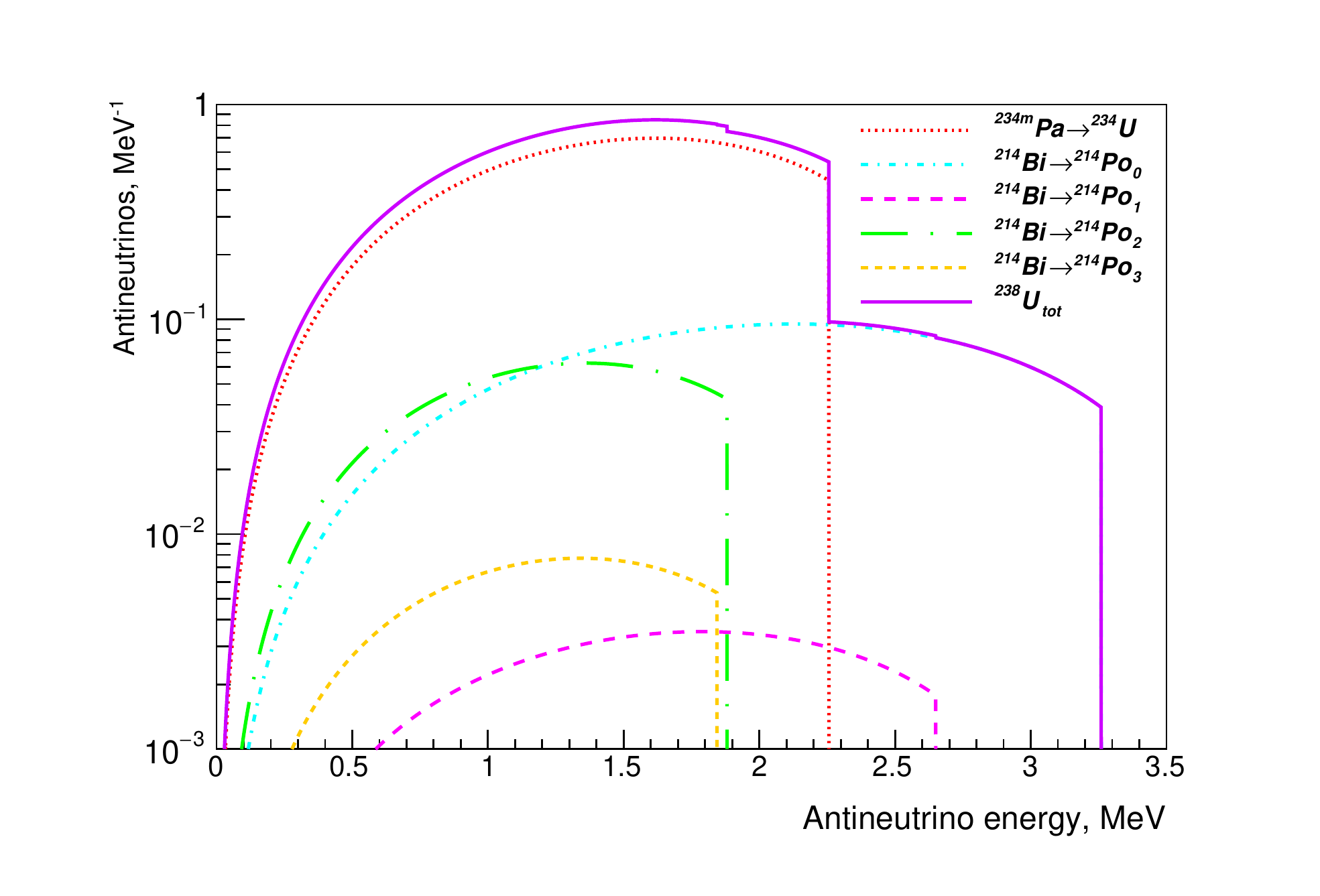}
\par\end{centering}
\protect\caption{\label{fig:UTh_spectra}Antineutrino spectra for $\beta$-decaying isotopes
in the $^{232}$Th (left) and $^{238}$U (right) chains with the end-point energy $Q>1.806$
MeV. The resulting total spectra are normalized to one parent decay. The end-point energies $Q$ used in calculations are reported in Tables~\ref{tab:betaTh} and~\ref{tab:betaU}. (Figure adapted from~\cite{Fiorentini2007}. For
interpretation of the references to colour in this figure legend,
the reader is referred to the web version of this article.)}
\end{figure}

\begin{figure}
\begin{centering}
\includegraphics[scale=0.45]{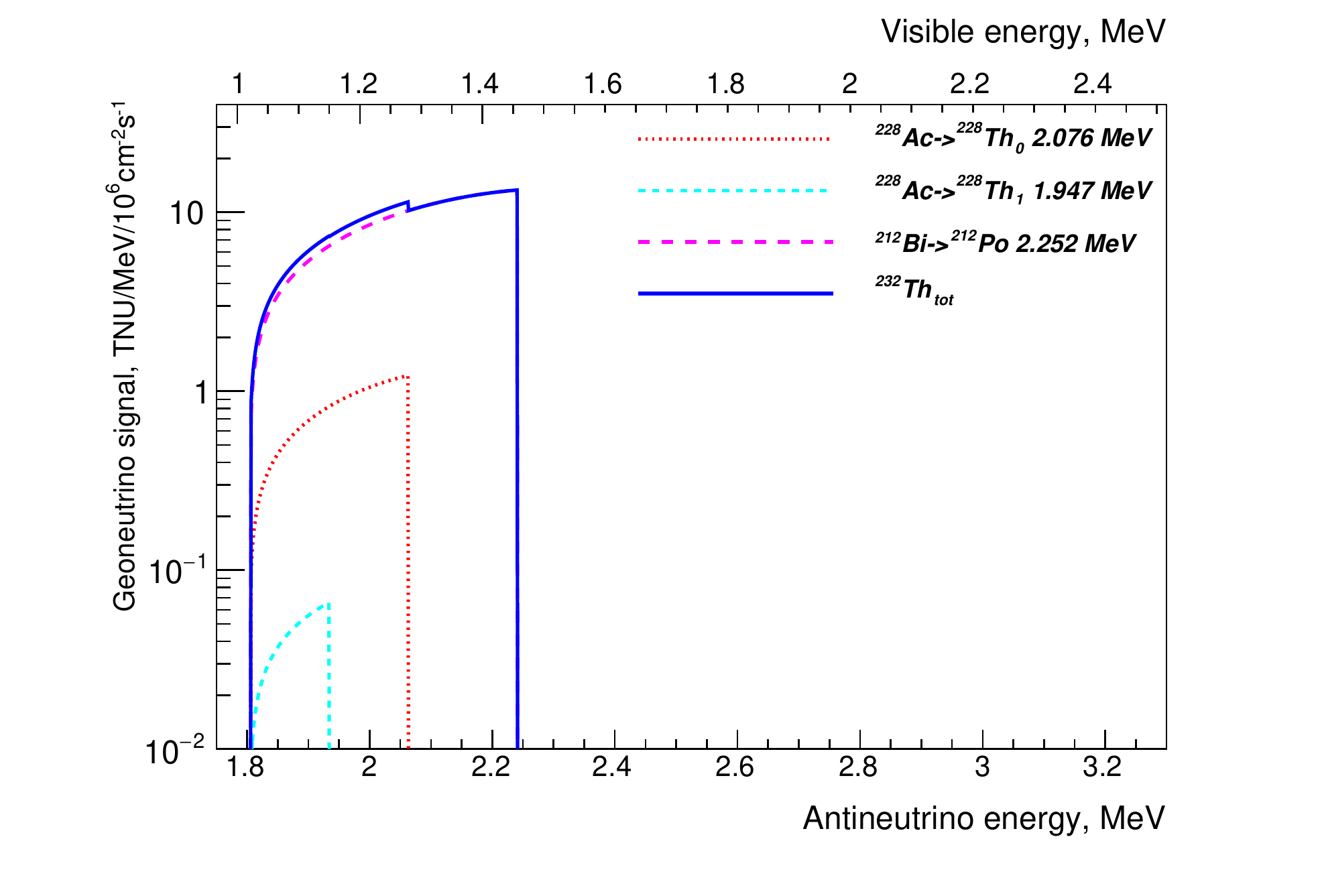}\includegraphics[scale=0.45]{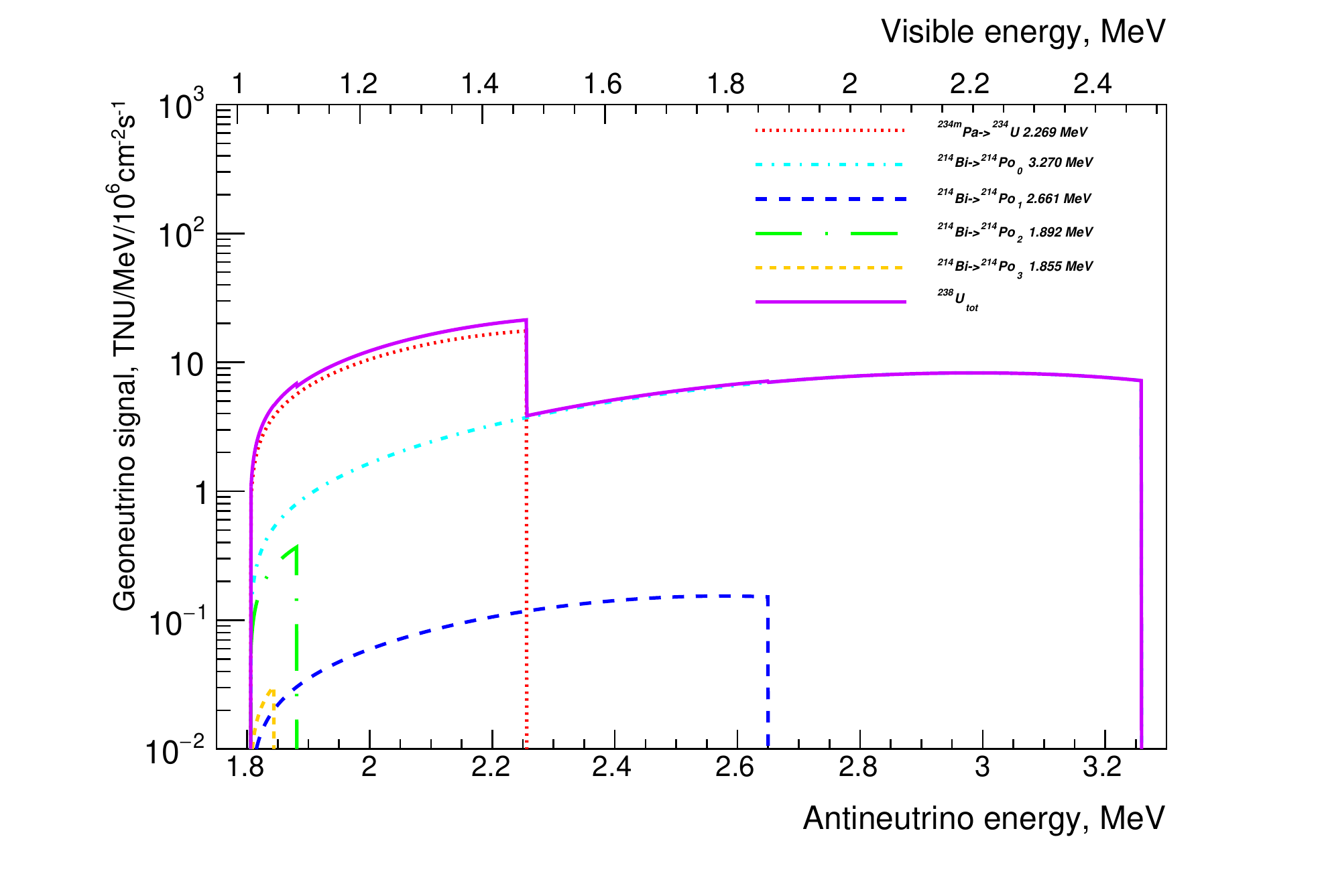}
\par\end{centering}
\protect\caption{\label{fig:UandTh_signal}Expected IBD antineutrino signal from decays
in the $^{232}$Th (left) and $^{238}$U (right) chains. Indices after the decay denotes the corresponding energy level of the daughter nucleus. An incident antineutrino
flux of $10^{6}$ cm$^{-2}$s$^{-1}$ is assumed in calculations. (Figure adapted from~\cite{Fiorentini2007}. For interpretation of the references to colour in this figure legend,
the reader is referred to the web version of this article.)}
\end{figure}

Because of the reaction threshold, the detection of antineutrinos
from decays of $^{40}$K and $^{87}$Rb is impossible in IBD reaction.
The same applies to all $\beta$-decays in the $^{235}$U chain
and most $\beta$-decays in the $^{238}$U and $^{232}$Th
chains. Only two isotopes in each chain have energetic enough $\beta$-decays
to provide a contribution to the IBD signal. In the $^{232}$Th decay
chain these are two branches of the $^{228}$Ac decay (to ground state
and to the first excited state), and the decay of the $^{212}$Bi to the ground
state. In the $^{238}$U chain, these are the decays of $^{234\text{m}}$Pa
and 4 branches of the $^{214}$Bi decay. 

The properties of $\beta$-decaying isotopes
in the $^{232}$Th and $^{238}$U chains contributing to the geoneutrino
signal are presented in Tables \ref{tab:betaTh} and \ref{tab:betaU},
respectively. The second column contains activity of the corresponding
reaction with respect to the activity of parent $^{232}$Th (or $^{238}$U).
The $I\pm \delta I$ column contains the relative intensity of the corresponding
decay (with respect to one parent decay) with the associated experimental
error. $Q$ is the $\beta$-decay end-point. $f_{Th}$, $f_{U}$ and $f_{tot}$ denotes the fraction of the signal
in the $^{232}$Th, $^{238}$U) chain and in total geoneutrino signal
respectively.  The neutrino spectra for U and Th are shown in Fig.~\ref{fig:UTh_spectra}. The corresponding neutrino signals expected in the IBD detector for the U and Th chains are shown in~Fig.~\ref{fig:UandTh_signal}.

\begin{table}
\caption{\label{tab:betaTh}Properties of $\beta$-decays in the $^{232}$Th
chain with $Q>$1.806 MeV contributing to the geoneutrino signal in IBD detectors. The intensity of the corresponding decay with respect to one parent decay of $^{232}$Th is presented in the second column. The branching ratio $I$ is shown together with its error $\delta I$. The end-point energy $Q$ of the corresponding decay is presented with its uncertainty $\delta Q$ too. The numeric uncertainties are shown in parentheses, denoting an uncertainty in the last significant figure(s).
Next column provides the type of the decay. The last two columns show the contribution $f_{Th}$ of the corresponding branch to the full geoneutrino signal from the $^{232}$Th chain, and the contribution $f_{tot}$ to the total signal from the both U and Th chains in assumption of the allowed $\beta$-decays shape and chondritic ratio of the parent isotope masses M(Th)/M(U)=3.9}
\vskip 2mm
\centering{}%
\begin{tabular}{ccccccc}
\hline 
{Decay}  & {Bq/Bq{[}Th{]}}  & {$I(\delta I)$}  & {$Q(\delta Q$), MeV}  & {Type}  & {$f_{\text{Th}}$,\%}  & {$f_{\text{tot}}$,\%}\tabularnewline
\hline 
{$^{228}$Ac$\rightarrow{}^{228}$Th}  & {1.000}  & {0.07(5)}  & {2.076(3)}  & {Allowed ($3^{+}\rightarrow2^{+}$)}  & {5.1}  & {1.1}\tabularnewline
  &   & {0.006(5)}  & {1.947(3)}  & {Allowed ($3^{+}\rightarrow4^{+}$)}  & {0.1}  & {0.0}\tabularnewline
\hline 
{$^{212}$Bi$\rightarrow^{212}$Po}  & {0.6406(6)}  & {0.5537(11)}  & {2.2521(17)}  & {1st forbidden ($1^{-}\rightarrow0^{+}$)}  & {94.7}  & {19.7}\tabularnewline
\hline 
\end{tabular}
\end{table}

\begin{table}
\caption{\label{tab:betaU}Properties of $\beta$-decays in the $^{238}$U chain
contributing to the geoneutrino signal in IBD detectors. The notations
are the same as in Table \ref{tab:betaTh}}
\vskip 2mm
\centering{}%
\begin{tabular}{ccccccc}
\hline 
{Isotope}  & {Bq/Bq{[}U{]}}  & {$I(\delta I)$}  & {$Q(\delta Q)$, MeV}  & {Type} & {$f_{\text{U}},$\%}  & {$f_{\text{tot}}$,\%}\tabularnewline
\hline 
{$^{234}$Pa$\rightarrow^{234}$U}  & {0.9984(4)} & {0.9757(4)}  & {2.269(4)}  & {1st forbidden ($0^{-}\rightarrow0^{+}$)}  & {38.6}  & {30.6}\tabularnewline
\hline 
{$^{214}$Bi$\rightarrow^{214}$Po}  & {0.99979(10)}  & {0.1910(17)}  & {3.270(11)}  & {1st forbidden ($1^{-}\rightarrow0^{+}$)}  & {60.6}  & {47.9}\tabularnewline
 &  & {0.0058(18)}  & {2.661(11)}  & {1st forbidden ($1^{-}\rightarrow2^{+}$)}  & {0.7}  & {0.5}\tabularnewline
 &  & {0.0735(5)}  & {1.892(11)}  & {1st forbidden ($1^{-}\rightarrow2^{+}$)}  & {0.1}  & {0.1}\tabularnewline
 &  & {0.0089(3)}  & {1.855(11)}  & {1st forbidden ($1^{-}\rightarrow0^{+}$)}  & {0.0}  & {0.0}\tabularnewline
\hline 
\end{tabular}
\end{table}

As one can see from Table~\ref{tab:betaTh}, the contribution from the
$^{228}$Ac is quite uncertain. It contributes 0.8$\%$ to the
total antineutrino signal uncertainty because of the poorly measured
relative intensity of the decay. Uncertainty of the  $^{214}$Bi decay intensity to the ground
state contributes another 0.4$\%$ to the total error of the signal.
The uncertainties of the end-point of the more energetic transitions 
in the U chain give another 0.8\% to the uncertainty of the total signal.
Summed quadratically, these contributions constitute 1.2$\%$,
and other transitions in both decay chains do not contribute significantly to the total signal uncertainty.
The uncertainties induced by the approximations
used in $\beta$-decay
shape calculation can, in principle, contribute to the total signal uncertainty, and these errors cannot be, in general, propagated quadratically because of the unknown correlations. 

\begin{figure}
\begin{centering}
\includegraphics[scale=0.45]{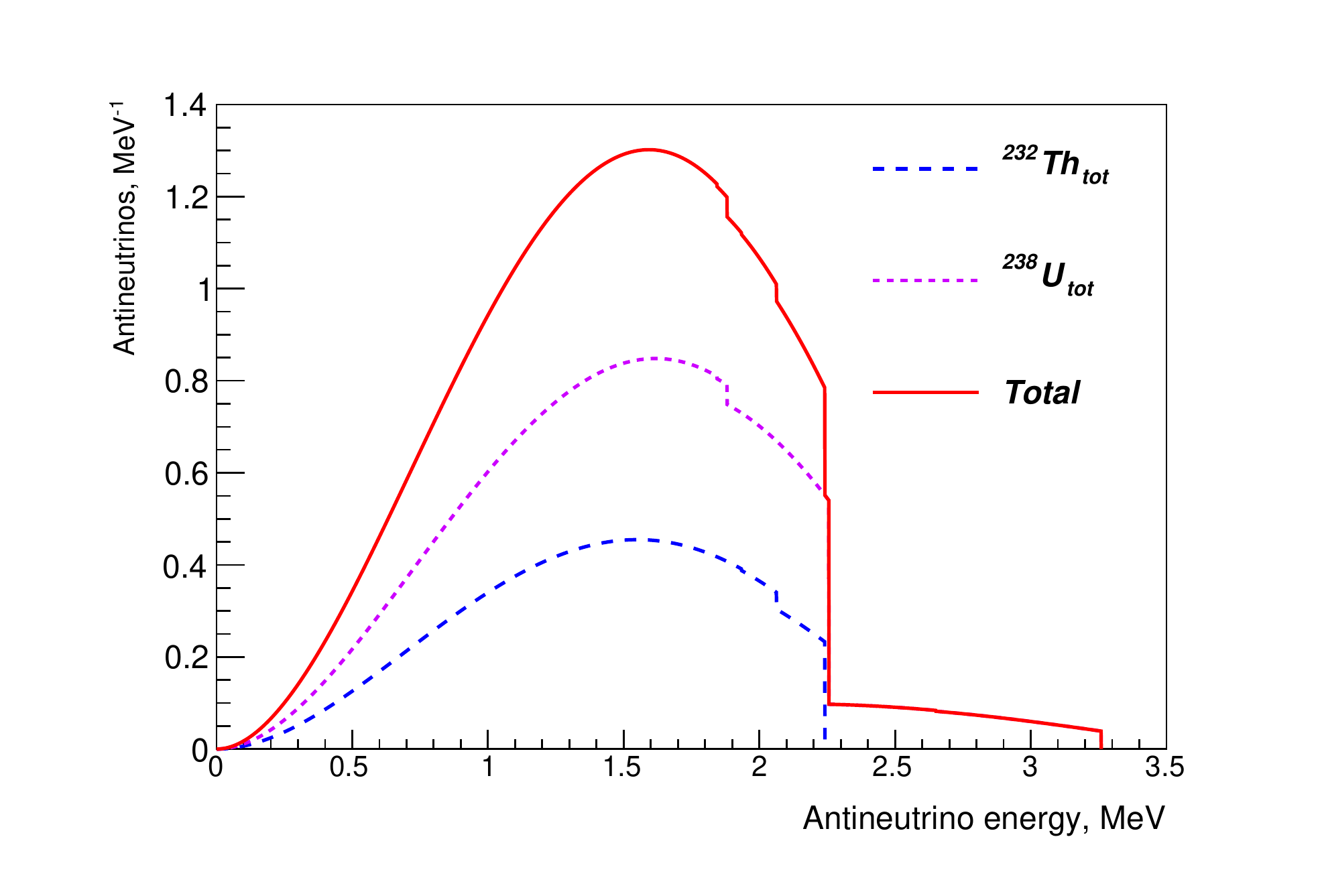}\includegraphics[scale=0.45]{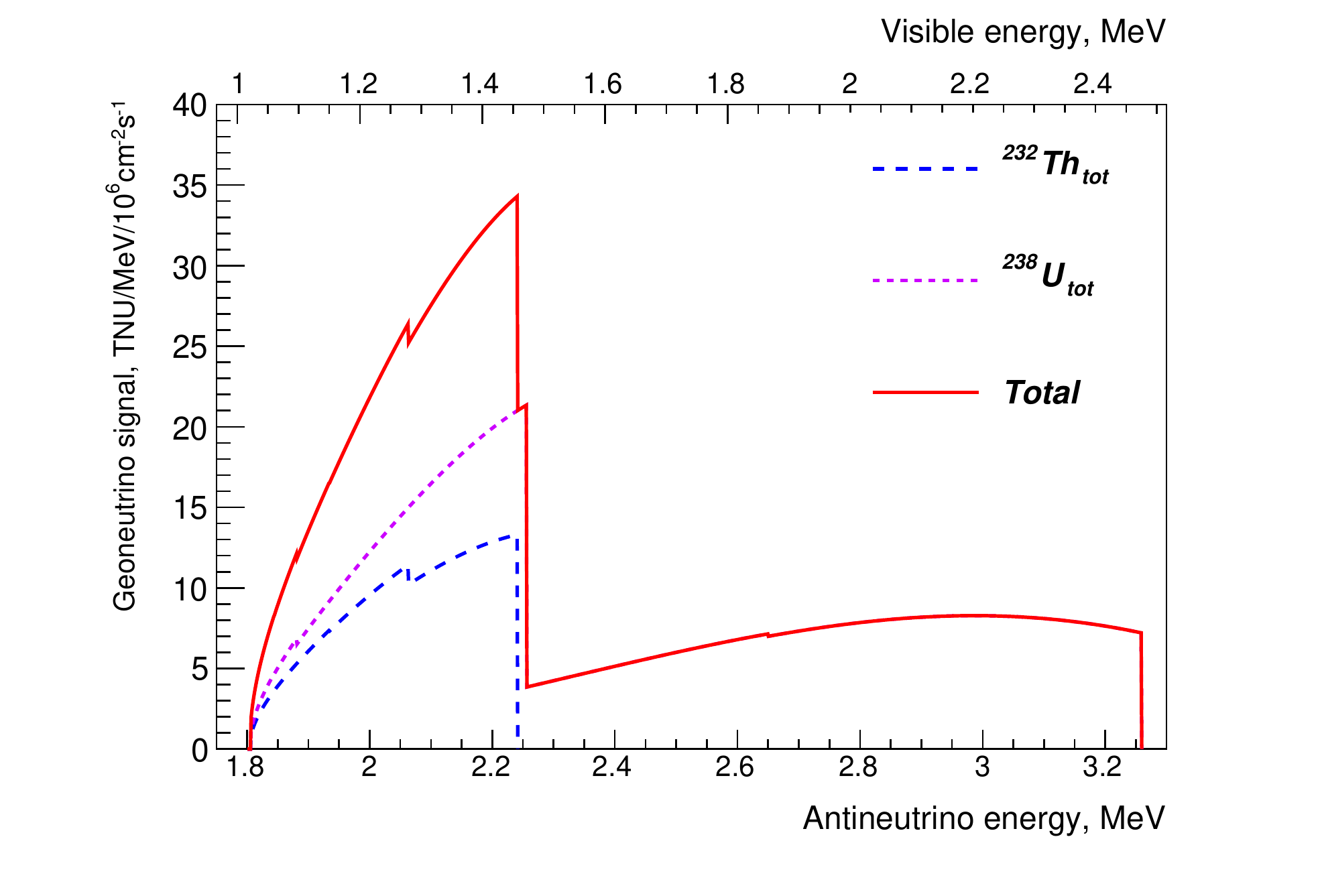}
\par\end{centering}
\protect\caption{\label{fig:UandTh_nu}\textbf{left:} resulting antineutrino spectra
for $^{232}$Th and $^{238}$U chains from isotopes with end-point
energy $E_{0}>1.806$ MeV and their sum for chondritic Th/U mass ratio. Total spectrum is normalized to unity. \textbf{}\protect \linebreak{}
\textbf{right:} expected IBD antineutrino signal from decays in $^{232}$Th and $^{238}$U chains and total signal. Chondritic ratio of Th and U masses and incident antineutrino flux of $10^{6}$~cm$^{-2}$s$^{-1}$ are assumed in calculations. (For interpretation of the references to colour in this figure legend, the reader is referred to the web version of this article)}
\end{figure}

The total neutrino spectrum calculated for the chondritic ratio of
Th/U masses M(Th)/M(U)=3.9 is shown in the left plot of Fig.~\ref{fig:UandTh_nu}.
Theoretical IBD spectra 
from the U and Th chains on the assumption of the secular equilibrium in the
chains and at the chondritic ratio of Th/U masses 
are shown in the right plot of Fig.~\ref{fig:UandTh_nu}. The 
spectra are calculated for an ideal detector with an infinite resolution
The visible difference in the
spectra makes contributions from the U and Th potentially distinguishable,
though the low-energy part of the U spectrum has a noticeable contribution
very similar to the Th spectrum, weakening the discrimination power
of the U and Th signals. The separation of the U and Th contributions will be discussed in section~\ref{sec:UThratio}.

\subsection{\label{subsec:Beta-decay-shapes}Beta-decay shapes}

For the allowed $\beta$-decays with massless neutrinos, the energy
spectrum of electrons has the following theoretical (``universal'')
shape \cite{Morita1963}

\begin{equation}
dN_{\beta}(w,w_{0})=\frac{1}{N_{0}}F(w,Z_{D})C(w)pw(w_{0}-w)^{2}dw,\label{eq:beta_universal}
\end{equation}

where $w$ and $p$ are the full energy of the electron and its momentum
($p=\sqrt{w^{2}-1}$), respectively, $w_{0}$ is the $\beta$-decay
end-point, all measured in units $m=h=c=1$, and $F(w,Z_{D})$ is the Fermi function describing the correction for screening
by atomic electrons in a nuclear Coulomb field, where $Z_{D}$ is the
charge of the daughter nucleus, and $C(w)$ describes departures from
the allowed shape. With energy-independent contributions omitted the relativistic
Fermi function $F(w,Z_{D})$ could be expressed as

\begin{equation}
F(w,Z_{D})=p^{2(\gamma-1)}e^{\pi y}|\Gamma(\gamma+iy)|^{2},\label{eq:FwZD}
\end{equation}

where $\Gamma$ is the complex gamma-function, $\gamma=\sqrt{1-(\alpha Z_{D})^{2}}$,
$y=2\pi\alpha Z_{D}\frac{w}{p}$, and $\alpha$ is the fine structure
constant.

The normalization factor $N_{0}$ provides the normalization to unity

\begin{equation}
\int_{1}^{w_{0}}N_{\beta}(w,w_{0})dw=1.\label{eq:Norm}
\end{equation}

The correction of the $\beta$-decay spectrum due to the electron
screening can be easily taken into account by replacing $F(w,Z_{D})$
with (Rose correction)

\begin{equation}
F(w-v_{0},Z_{D})\sqrt{\frac{(w-v_{0})^{2}-1}{w^{2}-1}}\frac{w-v_{0}}{w},\label{eq:FwZ}
\end{equation}

where $v_{0}\simeq1.13\alpha^{2}Z^{4/3}$. The correction is important
for the low energy part of the $\beta$-decay spectrum, i.e., relatively important
for the neutrino spectra calculation. Another correction is due to
the finite nuclear size, negligible for allowed transitions, and
a radiative correction is due to internal bremsstrahlung (negligible
for the low $w_{0}$ $\beta$-decays). Both were neglected in our
calculations of the energy release in the $\beta$-decays of the isotopes
from the U and Th decay chains and $^{40}$K.

The $\beta$-decays can involve transitions to excited nuclear states
``i'' with different branching ratios $b_{i}$ (with $\sum_{i}b_{i}=1$).
Transitions to excited states are typically accompanied by energy
relaxation involving a single gamma or a gamma cascade and/or less abundant
conversion electrons.

The $\beta$-decays are classified according to the spin change $\Delta J=\left|J_{i}-J_{f}\right|$ and
the parity change $\Delta\pi=\pi_{i}\pi_{f}$, where indices $i$ and $f$ denote the
initial and the final spin $J$ and parity $\pi$ of the nuclear states,
respectively. Decays with $\Delta J^{\Delta\pi}=0^{+},1^{+}$ are
allowed, decays with $\Delta J^{\Delta\pi}=0^{-},1^{-},2^{+},3^{-},4^{+}...$
are forbidden non-unique with the degree of forbiddenness defined by $\Delta J$,
and decays with $\Delta J^{\Delta\pi}=2^{-},3^{+},4^{-}...$ are forbidden
unique with the degree of forbiddenness defined by $\Delta J-1$. From
the theoretical point of view, the forbidden unique $\beta$-decays
are simpler, since the rate of the decay and the shape of the spectrum are defined
by only one nuclear matrix element (and thus they are ``unique'').
For allowed and forbidden unique transitions, Behrens and B\"{u}hring~\cite{Behrens1982} demonstrated that in a first approximation,
but with excellent precision, the nuclear current component can be
factored out, and only the dynamical factor from the coupling with
the leptons remains. However, all forbidden transitions involved
in formation of the geoneutrino signal are nonunique ones. The calculation
of the $\beta$-spectra for the nonunique transitions is far more complicated
because the structures of the initial and final nuclear states have
to be taken into account. There are good qualitative reasons to assume
the allowed shape for several first forbidden nonunique $\beta$-decays
if the $\xi$ approximation is fulfilled~\cite{Schopper1966}, i.e.,
when the Coulomb energy of the $\beta$-particle emitted at the nuclear
surface is much higher than the maximum energy of the transition,
$2\xi=\frac{\alpha Z}{R}\gg E_{o}$. As noted by Mougeot~\cite{Mougeot2018},
there are no studies evaluating the applicability of the $\xi$ approximation
or setting a limit on the ratio $2\xi/E_{0}$ for which it would
definitely provide good results.

Hayen et al.~\cite{Hayen2018} discussed the precision calculation
of the allowed $\beta$-decay spectra for low to medium Z nuclei
considering all the corrections
up to $\sim10^{-4}$ for energies down to 1~keV.
They point out the poorly described problem of the $\beta$-spectra
calculations at low energies. In particular, the screening correction,
usually introduced as an energy shift $v_{0}$, does not allow
calculations for energies lower than $v_{0}$, which results in discontinuities
for energies typically below 10 keV. Another low-energy
effect is the exchange effect corresponding to the case where a beta
electron is not directly emitted into the continuum but into a bound
orbital of the daughter atom, accompanied by the simultaneous emission
of a bound electron. It leads to enhancement of the beta emission
intensity, in particular at low energy. First discussed by Bahcall
\cite{Bahcall1963}, it was quantitatively evaluated for several $\beta$-decaying nuclides in \cite{Harston1992}. Its implementation for allowed
transitions can be found in the BetaShape code~\cite{Mougeot2012}.
Measurements of the $\beta$-spectrum from the forbidden transition of $^{241}$Pu
shows the evidence of the effect~\cite{Loidl2014}. These low-energy
effects tend to increase the $\beta$-spectrum, and the mirror neutrino
spectrum will be correspondingly enhanced at its high-energy part.
The effect in the observed signal will be amplified because of the
fast increase of the IBD cross section with energy, and the enhancement of the
$\beta$-spectrum by 0.1\% of the total one at low energies will result in the 0.6\% increase
of the U signal and the 0.9\% increase of the Th signal.

The shapes of the $\beta$-decay spectra have only little been studied experimentally
since the late 1970s. These spectra were thought to be known accurately
enough. Theoretical studies stopped in the 1980s. However, at present,
it became obvious that there are not enough data available, in particular
for high forbidding orders and at low energy~\cite{Bisch2013}. Precise knowledge 
of the shape of energy spectra and their uncertainties,
especially in the low-energy region of the $\beta$-spectrum, is
needed for accurate geoneutrino flux estimations. The same applies to
the feeding probabilities in complex decay schemes. 

An attempt to measure the shape of the $^{214}$Bi ($\beta+\gamma$) decay was reported in~\cite{Fiorentini2010}. The experimental spectrum obtained with the CTF detector~\cite{CTF98a,CTF98b}. 
was fitted with a model function. First, the feeding probability $p_{0}$
of the ground state was defined from the fit assuming the universal $\beta$-decay shape. The feeding probability $p_{1}$ of the first excited state in this fit was fixed at the 
central value from~\cite{TOI-1995}. Other transitions result in $\beta$-decays with
energies below the geoneutrino detection threshold, and do not contribute to the geoneutrino signal. Their probabilities were fixed at the central values from~\cite{TOI-1995} too. 
The $p_{0}$ was  found  to  be $0.177\pm0.004$, consistent with
the indirect estimate of the table of isotopes\footnote{The consistency of the experimental result in~\cite{Fiorentini2010} was checked against the value $p_{0}=$0.182$\pm$0.006
from~\cite{TOI-1995}. The updated value of $p_{0}$ is $0.1910\pm0.0017$~\cite{ENSDF},
which differs by 3.2$\sigma$ from the measured in CTF. Therefore, the result of~\cite{Fiorentini2010} could indicate
the presence of deviations from the allowed universal shape, and more data are needed to draw the statistically significant conclusions.}.  

At the next step of the analysis
the shape of the $\beta$-spectrum with end-point $E_0$=3.27~MeV corresponding to the transition to the ground state was allowed to vary applying an adjustable form-factor parameter. The values of $p_{0}$ and $p_{1}$ in this fit were constrained at the values from~\cite{TOI-1995}. 
It was found that the effect of uncertainties of the $\beta$-decay shape on the geoneutrino signal is mostly negligible. The conclusion reflects the fact that the effect of
shape variations and that of the values of $p_{0}$ and $p_{1}$ on the signal are anticorrelated in the fit of the $^{214}$Bi spectrum with free shape parameter and free 
$p_{0}$ and $p_{1}$. Therefore, if the spectrum is deformed in a way that there are more (fewer) low-energy
electrons, the corresponding best-fit value for $p_{0}$ is lower
(higher). If $p_{0}$ ranges from 0.13 to 0.20, the
signal would change only by about $\pm2\%$, i.e., the resulting signal weakly
depends on the spectral shape~\cite{Fiorentini2010}. Larger statistics and reduction of systematics should allow 
the testing of possible deviations of the neutrino spectrum from
the one predicted using the universal shape~\cite{Fiorentini2010}.

\subsection{Uncertainties of the predicted geoneutrino signal}

The principal factors responsible for uncertainty of the expected
geoneutrino signal are related to  masses and spatial distribution of radioactive isotopes in the Earth, in contract to the radiogenic heat that is related only to the total masses of the HPEs.  As noted above, the measurement of the geoneutrino signal from the mantle is a clue to distinguishing between the existing models. The mantle contribution can be obtained by subtracting the crust contribution from the total measured signal, and, thus,  the minimal possible mantle signal uncertainty is defined by absolute uncertainty of the crust contribution. Since, the accurate prediction of the crustal signal is of primary importance for extraction of the mantle signal. The total uncertainty of the predicted geoneutrino signal from the crust includes geological, geophysical and geochemical uncertainties, and, to the less extent the uncertainties related to the antineutrino physics. The chemical properties of the Earth are less understood compared to its physical structure. The typical uncertainties of the U and Th content in the BSE model are 15-20\% or worse~\cite{McDonough1995}. The uncertainties of the distribution of HPEs, or geophysical uncertainties, contributes 5\textendash10\% to the total uncertainty of the crustal signal prediction~\cite{Sramek2016}. Other uncertainties of the expected geoneutrino signal are less important (including uncertainty in the oscillation parameters, uncertainties of the $\beta$-decay shapes, uncertainties in the branching ratios, etc.). These antineutrino physics uncertainties constitutes $\sim3$\%, still beyond the sensitivity of the experiments.

It should be noted that since the U, Th, and K, the main contributors to the geoneutrino
signal, are all lithophile elements with similar geochemical properties, their content in different Earth reservoirs is positively correlated. In contrast, the total mass constraint of
the Bulk Silicate Earth model anti-correlates these element abundances
in complementary reservoirs. The problem of the uncertainties and
correlations in geoneutrino flux calculation was investigated in~\cite{Fogli2006}. 
The authors constructed a "geoneutrino source model"
(GNSM) for the U, Th, and K abundances in the main Earth reservoirs
and then used it to predict geoneutrino signal. They showed
the possibility for the future data to constrain the error matrix of the model itself.
It should be noted, however, that present experimental uncertainties
(connected mainly with low statistics) appear to be significantly
larger than those in the GNSM.

The bias in the estimation of the geoneutrino signal related to the
specific properties of the probability distributions used for the
description of the element abundances are discussed by Takeuchi et
al.~\cite{Takeuchi2019}. The log-normal distribution is widely
used for rock composition models~\cite{Huang2013,Sramek2016,Huang2014},
as was suggested by Ahrens \cite{Ahrens1954}, based on the observation
that igneous rocks commonly show skewed compositional distributions.
The use of log-normal probability density functions (PDFs) can lead
to bias in mean values, if the maximum likelihood method is used to
fit the sample distributions. The geometric mean is unbiased in log-normal
models, but it is generally different from the arithmetic mean. The gamma distribution was found to be a better choice than the normal or log-normal distributions for the purpose of unbiased flux calculation~\cite{Takeuchi2019}.

Another source of bias could arise from the use of median values for
description of rock compositions. For skewed distributions, the median
differs from the mean, and the result will have an unpredictable mean
or median if median values are used in estimations or in mass-balance
calculations. Also, after rejecting "non-typical"
samples in distribution tails, the mean of the filtered samples depends
on the filter cut-off and hence the resulting mean becomes arbitrary.
All these systematic uncertainties can be avoided by using the proper
distribution shape.

Based on the discussed points, a method of 3D crustal modeling was developed
by Takeuchi et al. \cite{Takeuchi2019} with uncertainty estimation that is truly probabilistic and reproducible,
as opposed to \textquotedblleft uncertainties by comparison\textquotedblright{}
or \textquotedblleft subjective/arbitrary estimation\textquotedblright .

\section{Sources of background in IBD experiments}

Experimental errors in geoneutrino experiments are associated mainly
with low statistics and with the presence of backgrounds, as it is
very difficult to ensure a background-free environment for counting
such rare events. Principal contributions to the background come either
from antineutrino sources other than geoneutrinos (reactor antineutrinos)
or from events mimicking antineutrino interactions. The latter can
be produced by cosmic muons (cosmogenic background) or by intrinsic
residual radioactive contamination of LS (internal background). Other
natural antineutrino sources such as a diffuse flux of neutrinos from
past supernovae (supernova relic neutrinos with, the flux is $\sim10$~cm$^{-2}$s$^{-1}$~\cite{Ando2004}) and atmospheric neutrinos
($\sim1$~cm$^{-2}$s$^{-1}$~\cite{Gaisser2002}) are
negligible compared to the geoneutrino flux ($\sim10^{6}$~cm$^{-2}$s$^{-1}$)
and do not contribute to the background.

\subsection{Antineutrino candidates selection}

Selection criteria for the pairs of correlated events (antineutrino
candidate events) are built using the reconstructed distance $\Delta R$
between two events, the time difference between the events $\Delta T$, the
cut on events energy, and the cuts on the interaction types in the case where the detector
provides a possibility of distinguishing events caused by $\beta$'s or
$\gamma$'s from those caused by heavier particles, alphas and/or
protons. The set of events selected using the cuts can still have
an admixture of "false" events due to the random coincidence of single
background events passing all the selection criteria, and/or due to
the $(\beta,n)$ decays of the spallation products, events caused
by fast neutrons, etc. These backgrounds will be considered in more
detail in this section.

The antineutrino candidate selections in general include:
\begin{enumerate}
\item A cut on the prompt event energy, which should accept events with
energies from the kinematic threshold of the IBD to the highest energy
expected for the considered process. Energy resolution of the detector should
be taken into account.
\item A cut on the delayed event energy should be tuned to accept the peak
of 2.2 MeV $\gamma$ from the neutron capture. The cut should take into
account the energy resolution of the detector and the possible partial
escape of $\gamma$'s to the outer part of the detector.
\item A cut on the time difference between the prompt and the delayed events,
$\Delta T$, usually ranges from the minimum possible time defined
by the detector and its electronics properties to the upper limit defined by optimizing the ratio of efficiencies for accepting true
events and the leakage of the random background.
\item A cut on the relative distance between the prompt and the delayed
events, $\Delta R$. The upper limit for $\Delta R$ is to be chosen
by optimizing the signal-to-background ratio.
\end{enumerate}
The efficiency of the basic cuts depends on the detector's characteristics
(energy resolution, vertex reconstruction precision) and internal
background in the chosen FV and can depend on the energy
of the prompt event. The efficiency is lower at the sensitive volume
borders due to the possible escape of the delayed $\gamma$ from the
detector. In practice, it reaches 70-85\% in large volume liquid scintillation detectors.

The set of the basic cuts for selecting the correlated events can
be complemented with $\alpha/\beta$ selection cuts, since both prompt
and delayed events are caused by light ionizing particles (electron-like
events). A cut on the FV was applied in both the Borexino
and KamLAND analyses to remove excessive external background.
The false events caused by passing muons are removed by excluding
the time range after the identified muon; the duration of the excluded
time is defined by the type of expected background. For neutron-related backgrounds it is enough to exclude $\sim$2 ms, and exclusion
of long-lived ($\beta$+n) isotopes needs veto times of $\sim$2 s.

The physical background from nuclear reactors is irreducible and can be separated
from geoneutrinos by applying a spectral cut. The spectral cut should also
allow for the shape of the residual backgrounds, given that they
are still present after all the cuts applied. 

\subsection{\label{sec:Reactor}Reactor antineutrinos}

Until now the geoneutrino flux measurement was a side-product of
projects having other main goals, namely solar neutrino fluxes measurement
in the case of Borexino, or oscillations of reactor antineutrinos
in the case of KamLAND. The reduction of background from reactors was
not an option in the projects. The search for the neutrino mass
hierarchy with the next generation of antineutrino detectors, e.g., 20~kt LS JUNO experiment~\cite{JUNO}, requires a location even closer to the reactors, which would provide the maximal manifestation of the effect searched for, but at the same time making more difficult the extraction of the geoneutrino signal
because of the stronger reactor antineutrino background.
On the other hand, these detectors will provide higher statistics
of events, in principle allowing for the use of statistical separation
of geoneutrios and reactor antineutrinos on the basis of directionality.
The reactor antineutrino flux depends on the power of the nuclear reactors
and on the distance between the reactors and the detector. Keeping
track of power history of near reactors helps in statistical extraction
of the expected signal. 

Nuclear reactors generate heat by controlled fission of uranium and
plutonium isotopes. The main components of the nuclear fuel are $^{235}$U,
$^{238}$U, $^{239}$Pu, and $^{241}$Pu providing more than 99.9\%
of the total power, and thus only these actinides are considered. Their
relative contributions to the total thermal power of the plant depend
on the reactor type and on the burn-up stage of the individual core.
Antineutrinos are emitted in $\beta^{-}$
decays of fission products. The energy spectrum of reactor antineutrinos
extends up to $\sim$10 MeV, covering the full range of the geoneutrino
spectrum. The total antineutrino spectrum from the reactor for a single fission event can be expressed as a sum of individual contributions

\begin{equation}
\Phi_{R}(E)=\sum_{I=1}^{4}f_{I}\phi_{I}(E),\label{eq:PhiR}
\end{equation}

where the fission fraction $f_{I}$ is the contribution of the isotope $I$,
$\sum f_{I}=1$, and $\phi_{I}(E)$ is its antineutrino energy spectrum.
The thermal power of the reactor $W_{th}$ is expressed through the
number of fission rates $R$ and the thermal energy release $Q_{I}$
per one fission of the isotope $I$ or through the thermal power $P_{I}=R \cdot f_I \cdot Q_I$ generated by the isotope $I$ 

\begin{equation}
W_{th}=R\sum_{I=1}^{4}f_{I}Q_{I}=\sum_{I=1}^{4}P_{I}.\label{eq:Power}
\end{equation}

The corresponding effective energies $Q_{I}$ are determined from
the energy released in fission with account taken of the energy loss
due to the antineutrino emission and extra energy produced by neutron
captures on the reactor materials. The values of  $Q_{I}$ were calculated 
by Kopeikin et al.~\cite{Kopeikin2004}. 

Similarly to the fission fractions, the power fractions $p_{i}$ can be defined as

\begin{equation}
p_{I}=\frac{P_{I}}{W_{th}}=\frac{f_{I}Q_{I}}{\sum_{I=1}^{4}f_{I}Q_{I}}.\label{eq:pI}
\end{equation}

The power fraction $p_{I}$ is the fraction of the total thermal power produced by the fission of the isotope $I$.  If the power fractions $p_{I}$ are known, equations (\ref{eq:PhiR})
and (\ref{eq:Power}) can be used to obtain the antineutrino spectra
from the reactor's thermal power using isotopic antineutrino spectra $\phi_{I}(E)$

\begin{equation}
S(E_{\nu})=W_{th}\cdot\sum_{I=1}^{4}\frac{p_{I}}{Q_{I}}\phi_{I}(E_{\nu}),\label{eq:SE}
\end{equation}

The antineutrino spectrum $\phi_{I}(E)$ of the isotope $I$ is a weighted
sum of antineutrino spectra from all fission fragments for the corresponding
isotope. Typically, the number of involved isotopes is over 800. In
practice, the antineutrino spectrum of a given fission fragment is
not measured, but obtained from the corresponding $\beta$-spectrum.
The latter in many cases is not known with sufficient precision, or its computation
is complicated by many possible branches. Consequently, a summation
of the antineutrino spectra will result in a limited accuracy of the
final antineutrino spectrum \cite{Fallot2012,Mueller2011}, but in
many cases it is the only available option.

Another method is based on the reconstruction of the reactor antineutrino
spectrum from the measured integral $beta$-spectrum of all fission fragments
\cite{VonFeilitzsch1982,Schreckenbach1985,Hahn1985,Haag2014,Huber2011}.
In order to convert a measured electron spectrum into the antineutrino
one, the total spectrum is fitted with a set of allowed $\beta$-spectra
over the preliminary defined set of virtual end-point energies that can also
be the fit parameters. The antineutrino spectrum is then
obtained by replacing the electron energy $E_{e}$ in each branch
by its "mirror" reflection $E_{0}$ \textminus{} $E_{\overline{\nu}}$. The procedure guarantees good reproduction of the experimental
electron spectrum, but the converted spectrum depends on the uncertainties
related to effects of nuclear structure \cite{Huber2011}. In particular,
Hayes et al. \cite{Hayes2014} pointed out that uncertainties could
arise due to the forbidden decays and could make up to 30\% of all
the spectra. Correspondingly, a highly debated "reactor anomaly",
consisting in a 6\% deficit of antineutrinos detected in short-baseline
reactor antineutrino experiments compared to the prediction based on the
converted aggregate $\beta$-spectrum and presumably constituting 
the 3$\sigma$ effect~\cite{Mention2011}, can be a result of underestimation
of the theoretical errors. 

In the last several years, a significant deviation from the
prediction in the 4-6 MeV energy range is observed by a number of reactor antineutrino experiments, see~\cite{DayaBay2017, DayaBay2017c, DoubleChooz2014, RENO2018, NEOS2018} and review~\cite{Hayes2016}. More specifically, an excess of events observed in the measured spectrum at these energies has 4.4$\sigma$ local significance~\cite{DayaBay2017c}. With respect to the geoneutrino energies, the feature appears at higher energies and shouldn't significantly influence the geoneutrino analysis. At present, the physical reasons for the neutrino spectral bump are unclear, but the spectral distortion can be accounted for in the analysis by using the experimental measurements.

In geoneutrino studies the antineutrino spectra are predicted using
available information on the reactors' thermal power, their geographical
position, mode of operation, types, etc. The non-oscillating
antineutrino spectra are taken from parameterizations, with the
most recent one provided in~\cite{Mueller2011}. The spectral shapes
are very similar to the older parameterization given in~\cite{Huber2004}
with a 3.5\% higher total flux. Phenomenological parameterization
of fission antineutrino spectra for a given isotope has a polynomial
form

\begin{equation}
\phi_{I}(E_{\nu})=\exp(\sum_{p=1}^{6}\alpha_{p,I}E_{\nu}^{p-1}).\label{eq:PhenomenologicalReactor}
\end{equation}

\begin{figure}
\begin{centering}
\includegraphics[scale=0.5]{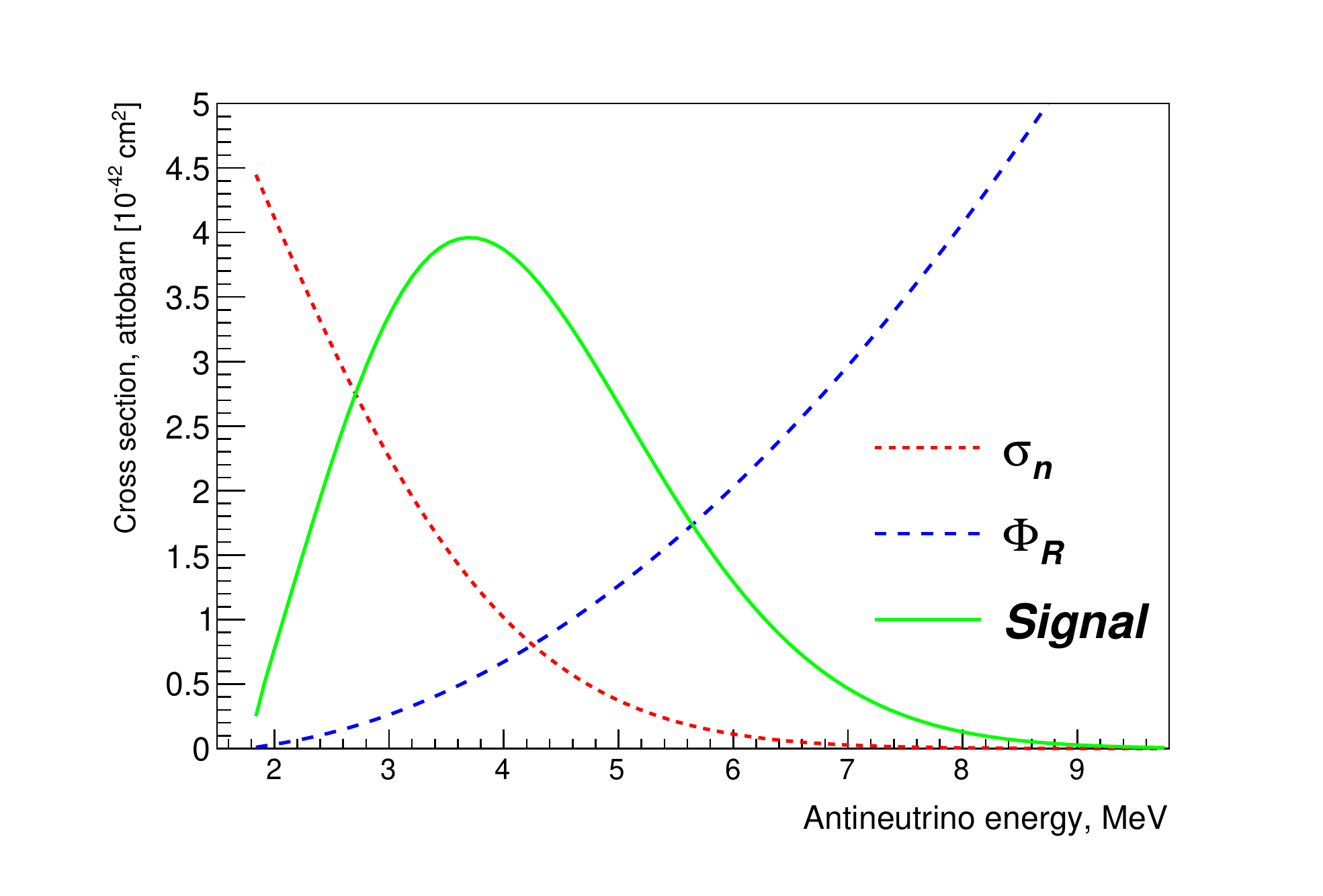}
\par\end{centering}
\caption{\label{fig:ReactorSignal}IBD cross section (increasing function, blue), the non-oscillating
reactor antineutrino flux (decreasing distribution, red), and the resulting reactor signal (bell-like shape, green). The vertical
scale shows the IBD cross section absolute values measured in attobarn ($10^{-42}$~cm$^{2}$), while other two shapes are arbitrary normalized.}

\end{figure}

The coefficients $\alpha_{p}$ for each of the involved isotopes $I$
are provided in~\cite{Mueller2011}. The product of the fast-decreasing
reactor spectrum and the steeply increasing IBD cross section gives
a bell-like shaped spectrum of the detected signal with a maximum around
4~MeV, schematically shown in Fig.~\ref{fig:ReactorSignal}.

A reactor operation schedule is required for estimating its contribution
to the detected antineutrino spectrum in a given time interval. The load factor (LF) is used as a measure of
the reactor time profile. It is defined as the ratio between the net
electrical energy produced during the reference period and the net
electrical energy that would be produced if the reactor operated in the
steady mode. The load factor data are published by the International
Atomic Energy Agency (IAEA) \cite{IAEA}, providing the data on a
monthly timescale and as an annual average. It is assumed in the calculation
that the published values of electrical LFs are equal to the corresponding thermal LFs.

Summarizing, the signal from the reactor antineutrino can be evaluated
as~\cite{Baldoncini2015}

\begin{equation}
\frac{dn_{R}(E_{\nu})}{dE_{\nu}}=\epsilon(E_{\nu})\cdot N_{p}\cdot T\cdot\sum_{R=1}^{N_{Reactors}}\frac{W_{th}^{R}}{4\pi L_{R}^{2}}LF_{R}\cdot\sum_{I=1}^{4}\frac{p_{I}}{Q_{I}}\phi_{I}(E_{\nu})P_{ee}(E_{\nu},L_{R})\sigma_{n}(E_{\nu}),\label{eq:ReactorSpectrum}
\end{equation}

where the index $R$ runs over all world reactors $N_{reactors}$ and $I$ runs
over 4 isotopes, $E_{\nu}$ is the antineutrino energy, $\epsilon(E_{\nu})$ is the detector
efficiency, $N_{p}$ is the number of target protons, $T$ is the data taking
time, $W_{th}^{R}$ is the thermal power of the $R$th reactor, $LF_{R}$
is the load factor of the $R$th reactor, $\phi_{I}(E_{\nu})$ is
the antineutrino spectrum of the isotope $I$, $P_{ee}(E_{\nu},L_{R}$) is the electron
antineutrino survival probability at the distance $L_{R}$ to the reactor. In general the signal also depends on the neutrino oscillations parameters,
and in the 3-flavour scenario of oscillations the dependence is expressed by formula
(\ref{eq:Pee3nu}) if matter effects are neglected. The effective  energies
$Q_{I}$ and the thermal power fractions $p_{I}$ are defined above, and $\sigma_{n}(E_{\nu})$ is the IBD reaction cross section.

The information on LFs and thermal powers is provided on the monthly basis, and, hence, for the practical usage the further month-by-month summation of the spectra provided by formula (\ref{eq:ReactorSpectrum}) is needed, with the values of $T$, $W_{th}^{R}$ and $LF_{R}$ evaluated for each month.

\begin{figure}[!h]
\begin{centering}
\includegraphics[width=0.30\textwidth,angle=90]{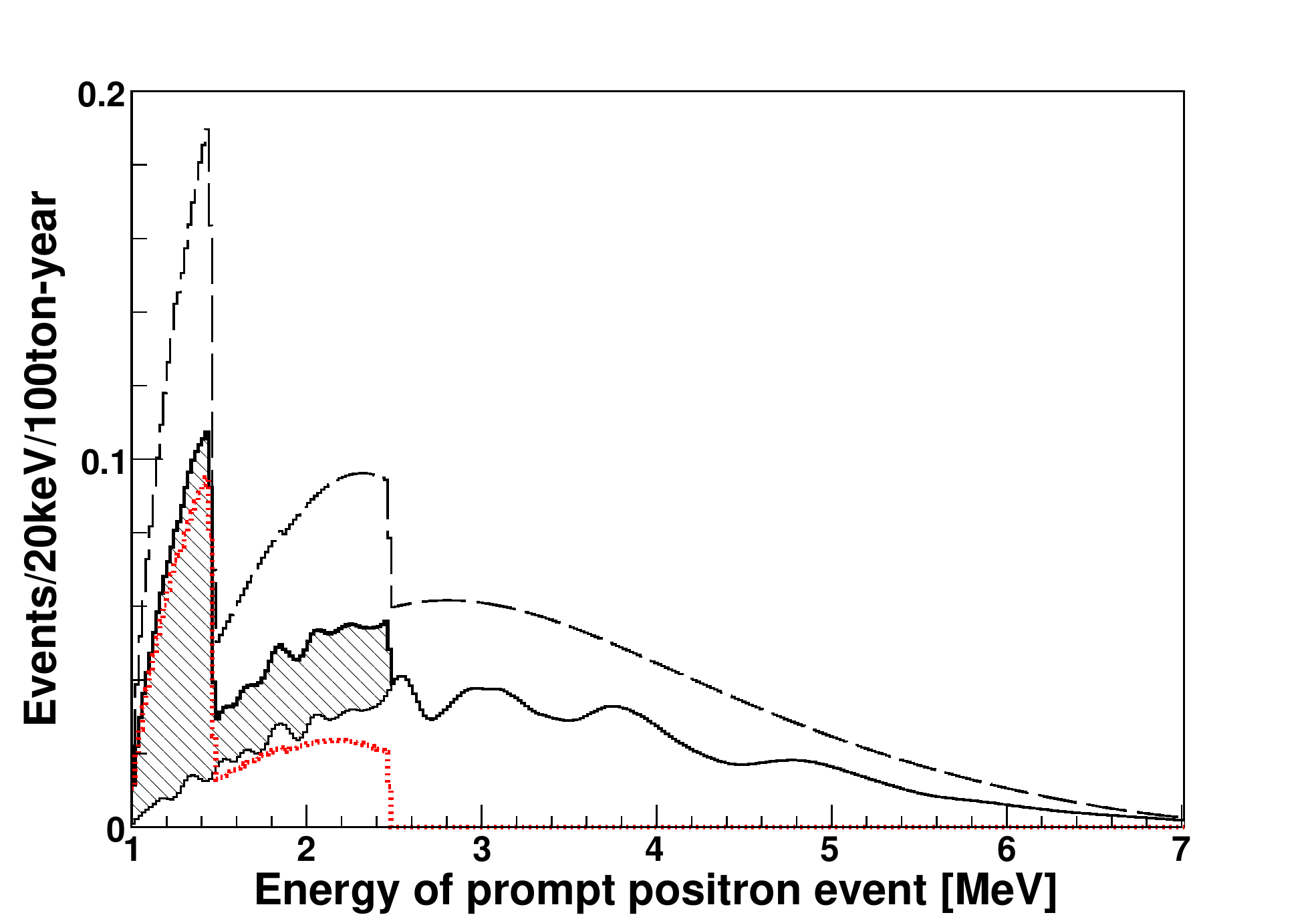}\includegraphics[width=0.45\textwidth]{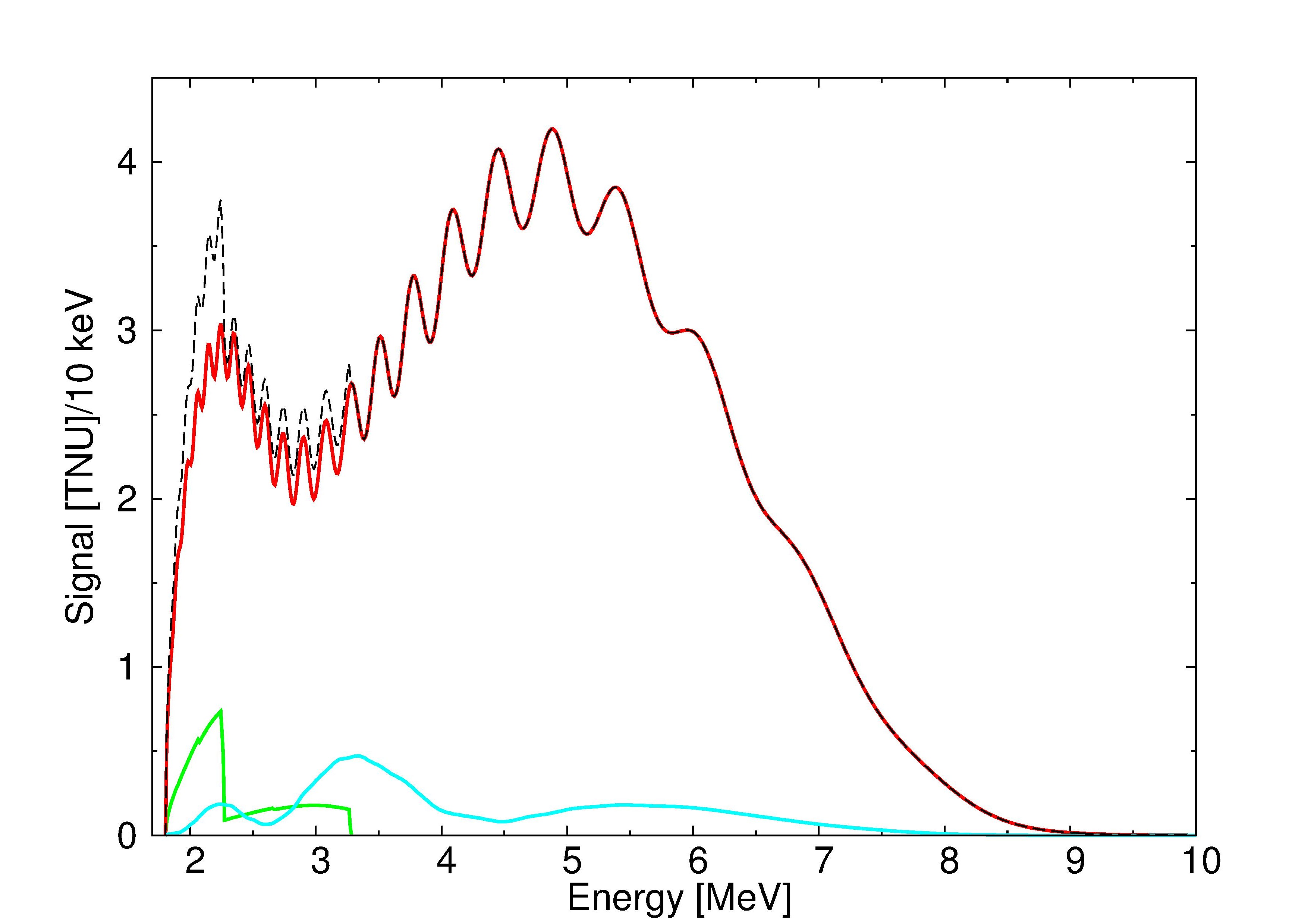}
\par\end{centering}
\caption{\label{fig:brx-expected}\textbf{left:} total antineutrino
spectra expected at Borexino's location, no energy resolution is applied.
Non-oscillating (or total antineutrino flux) and oscillating (electron
antineutrino) spectra are shown with dashed and solid lines, respectively.
The geoneutrino contribution is shown by the red line. Its contribution to the oscillated spectrum corresponds to the shaded area (Fig. from \cite{Bellini2013}).
\textbf{}\protect \linebreak{}
\textbf{right:} antineutrino energy spectra expected at JUNO's
location. The geoneutrino energy spectrum is visible at the beginning of the energy scale (green). The reactor
antineutrino spectrum is shown before (the lower curve expanding over all energy scale, cyan) and after adding the
contribution of the Yangjiang and Taishan nuclear power plants (the upper solid line, red), reactor ON scenario.
The reactor antineutrino spectra are computed assuming normal neutrino mass hierarchy
and taking into account neutrino oscillation. The total spectrum (black
dashed lines) is obtained assuming the reactors ON scenario (Figure
from~\cite{Strati2015}. For interpretation of the references to
colour in this figure legend, the reader is referred to the web version
of this article.)}
\end{figure}

There are about 450 nuclear power plants over the world, concentrated
mainly in Europe, North America, Japan, and Korea. The oscillation
length for antineutrinos with energies of a few MeV is at a level
of 100 km, and the expected signal spectra demonstrate a characteristic
oscillation pattern. The geographic location of the reactors is important
for the calculations. An example of the expected reactor spectrum
compared to the geoneutrino signal for two extreme cases for Reactor/Geo
antineutrino signals ratio is shown in Fig.~\ref{fig:brx-expected} for the Borexino (with a
moderate Reactor/Geo signals ratio) and JUNO (with the highest value of the Reactor/Geo
signals ratio) detectors; another example is presented in Fig.~\ref{fig:SNO}, where
the reactor spectrum for the closest reactor in the Sudbury laboratory
shows pronounced oscillation features.

The total expected reactor antineutrino signal, the expected reactor
antineutrino signal in the geoneutrino energy window ($E_{\overline{\nu_{e}}}<3.27$
MeV), the predicted geoneutrino signal, and the reactor-to-geoneutrino
ratio for the relevant energy range are presented in Table~\ref{tab:Locations}.
The data on the predicted reactor signals are compiled using the
estimates from~\cite{Han2016} for JUNO, that
include 2 future reactors, contributing more than 90\% to the total, from~\cite{Beacom2016} for Jingping site, and from~\cite{Ricci2015} for the rest of the locations.
The geoneutrino signals were calculated by Huang et al. in~\cite{Huang2013} using their Reference EARTH Model. The ratio between the predicted
reactor signal in the Low Energy Region (LER) and the expected geoneutrino
signal can be considered as a factor characterizing the sensitivity
of the detector placed at the corresponding location with respect
to the geoneutrino fluxes.

In addition to short-lived fission products discussed above, the exhausted
fuel contains isotopes with decay time of the order of 1 year that
can contribute to the reactor antineutrino signal. The pressurized light
water power reactor (PWR) operates for 11 months, followed by shutdown
of 1 month to replace 1/3 of the spent nuclear fuel (SNF), which is generally
kept for several years in a water spent fuel pool not far from the reactor,
forming an additional small but potentially non-negligible source
of antineutrinos. The KamLAND collaboration estimated an additional
contribution of 2.4\% to the reactor antineutrino signal from the spent
fuel \cite{KamLAND2005}. For the Chooz experiment, an increase of
1.0-1.5\% in the LER of the reactor antineutrino spectrum
was predicted by Kopeikin et al. \cite{Kopeikin2006}. Similarly,
it was found that the spent nuclear fuel spectrum influenced
the softer part of the Daya Bay experiment antineutrino spectrum, with
the maximum contribution from the SNF of about 3.0\% \cite{Ma2017}.

Since the background from reactors is caused by antineutrinos indistinguishable
from the geoneutrinos, it can be separated from the geoneutrino spectrum
only by applying a spectral fit. The model spectra should account for
the shape distortions due to antineutrino oscillations. In practice,
both the reactor antineutrino spectrum and the number of events
are calculated, but the normalization is left as a free parameter
in the fit over the entire energy scale. 

The antineutrino flux from the reactors is affected by planned upgrades
in which different reactor cores are turned off, which results in expected
variations in the total flux. The known time evolution of the reactors power
provides additional information for separating this background. This
possibility was exploited by the KamLAND collaboration before the
almost total shutdown of the Japanese reactors in 2011.

\subsection{Cosmogenic backgrounds}

Muons produced in primary cosmic ray interactions in the atmosphere penetrate
deep underground. Though the cosmic muon flux is reduced in the underground
labs by orders of magnitude, it is still high enough
to induce significant background. The site depth is the main factor
influencing the cosmic muon flux, which decreases exponentially with
depth. Among existing underground sites the best shielded against
cosmic muons is SNO laboratory (Canada) placed in a 2
km deep mine and providing shielding equivalent to 6010~meters of
water. The Baksan laboratory in the Caucasus mountains (Russia) provides
4800 m.w.e. shielding. The LNGS laboratory
in Italy, where the Borexino detector is placed, provides 3400~m.w.e.
shielding; the KamLAND site is less protected (2700~m.w.e.), and the
future JUNO site will be even less protected (2000~m.w.e.) against
cosmic muons. See Table~\ref{tab:Locations} for more detail.

Hagner et al.~\cite{Hagner2000} studied production of radioactive
isotopes by the muon spallation reaction in an organic LS
using the Super  Proton  Synchrotron muon beam at CERN. The  production  yields were  studied  with  100  and  190  GeV  incident muons and extrapolated at other  energies assuming  a  power-law dependence  on  the incident muon  energy. Direct  measurements of the  production  yield of cosmogenic isotopes  were performed with  large underground  detectors: the Mont Blanc liquid scintillation detector (LSD)~\cite{LSD1989}, the large-volume  detector  (LVD)~\cite{LVD2003}, the Borexino~\cite{Borex2014} in LNGS, and the KamLAND~\cite{KamLAND2010}. The direct measurements were compared with calculations on the base of MUSIC~\cite{MUSIC},  FLUKA~\cite{FLUKA},  and GEANT4~\cite{GEANT4} programs.  
Some of these isotopes are long-lived
and a false candidate event could be produced by a random coincidence
of two decay, following the muon. But the most dangerous cosmogenic
background from the point of view of its similarity to the IBD events
is associated with neutron-rich isotopes produced by spallation with
a $\beta$-decay branch followed by the emission of a neutron and thus perfectly
matching the antineutrino-induced signal. These are long-lived $^{8}$He
($T_{1/2}=119.1$ ms, with $\beta+n$ branching ratio of 16\% and
Q=10.65 MeV \cite{ENSDF}) and $^{9}$Li ($T_{1/2}=178.3$ ms, with
$\beta+n$ branching ratio of 50.8\% and Q=13.6 MeV \cite{ENSDF}).
The end-points of both $\beta$-spectra in $\beta+n$ branches
are big enough to cover completely both the geoneutrino and the reactor spectra. Because
$\beta+n$ decays are indistinguishable from antineutrino events,
the best way to suppress the corresponding background is complete
or partial exclusion of events occurring after an identified muon passing through
the detector. In the first case all events in the detector volume
are excluded from the analysis in the fixed time window after the passing muon. The veto time should be optimized to provide the best balance
between the cosmogenic background reduction efficiency and the loss of
the live time. The typical veto time is 2 seconds. In the case of a partial
veto, only the cylindrical volume around the muon track, where spallation
is possible, is excluded from the analysis for the fixed time. This
helps to gain the live time, but the method is more complicated as
it requires reconstruction of the muon track and introduces additional uncertainties to the exposure calculations. 

Thus, the IBD detector must be equipped with an efficient muon
detector and/or muon tracker. The criteria for the choice of the
muon veto depend on the site depth.

Another background related to cosmic muons is caused by fast neutrons
(single or multiple) produced outside the detector, i.e., by muons
not identified by the muon detector. Fast neutrons have a long free
path, and hence can reach the active volume of the detector entering from outside.
The fast neutron elastically scattered off the proton can produce a high
energy prompt event, and then thermalize and be captured onto a proton,
producing an antineutrino-like false event. The amount of these events
depends on the site depth and detector geometry, and is
typically comparable to the residual background from the 
isotopes with the $\beta+n$-decay mode. The prompt signal due to the proton scattered off the neutron should, in principle, be distinguishable from the signal induced by a positron
in LS due to different time profiles of the scintillation flash.
A heavy particle produces flashes with a more pronounced "slow"
component, i.e., its time distribution has longer tails compared to ones in
time distribution of scintillations caused by electrons and gammas, allowing
partial discrimination of this type of background.

\subsection{Intrinsic backgrounds and external $\gamma$ background}

Because of the presence of residual radioactivity in the LS some false
events can be selected after applying the cuts, forming the so-called
random coincidences background. The possible low-level contamination
of the LS with radioactive impurities can be due to the cosmogenic $^{14}$C,
naturally present in the organic LS, or due to the radioactive noble gases 
transferred to the LS when in  the contact with air (normally containing cosmogenic
$^{39}$Ar, technogenic $^{85}$Kr from nuclear power plants exhaust,
and $^{222}$Rn produced in the $^{238}$U decay chain), or it could
be residual intrinsic contamination with $^{40}$K or isotopes
from the decay chains of $^{232}$Th and $^{238}$U. Some of the
daughter isotopes can be present not in secular equilibrium with progenitors
due to different efficiencies of purification and other processes. In order to provide an acceptably low level of random coincidences count,
the LS of the detectors should be purified and kept in conditions
that prevents its contamination. The choice of the proper LS purification
strategies can reduce these contaminations to very low levels, as
was demonstrated by the Borexino collaboration. 

The random coincidence spectrum and its counting rate can be easily reproduced
on the basis of single (uncorrelated) event spectra by picking pairs of
events in an off-time coincidence window (one random event and another
one from the time window far away from the one used to search for
the candidates) and applying the same selection criteria as used to select
candidate events. The overall rate of random coincidences depends
quadratically on the single events count, i.e., on the residual radioactivity.
The prompt signal energy spectrum coincides with a single events spectrum
if the events are uniformly distributed over the detector volume,
otherwise the non-uniform distribution of sources can result in the deformation of the spectrum shape.  

Another source of background tightly linked with internal LS contamination
is the $(\alpha,n)$ reaction on $^{13}$C, which is naturally present
in organic scintillators with isotopic abundance of 1.3\%. Monoenergetic
$\alpha$-particles with the 5.3 MeV kinetic energy are produced in decays
of $^{210}$Po from the $^{238}$U decay chain \cite{Harissopulos2005}.
The $^{210}$Po have 138.4 days life-time.
In both the Borexino and the KamLAND detectors the activity of $^{210}$Po
was found to be higher than those expected in the case of secular equilibrium with
$^{238}$U or its most probable source of $^{210}$Pb deposited from
the decays of $^{222}$Rn. More detail can be found in sections dedicated
to the Borexino and KamLAND experiments.

An antineutrino candidate event can be mimicked by the $^{214}$Bi decay
from the $^{222}$Rn decay chain followed by a rare branch in the $^{214}$Po
$\alpha$-decay. The life-time of the $^{214}$Po isotope is close to
the neutron capture time in the LS. If the energy conditions for the
prompt and delayed signals are satisfied, then this delayed coincidence
would pass the simple selections cuts. The energy of $\alpha$ is
7.69 MeV in 99.99\% of decays, and in the remaining $10^{-4}$ cases the energy is released through 6.9~MeV $\alpha$ accompanied by a 0.8 MeV $\gamma$. The visible energy
in the latter case is much higher, since light yield of $\alpha$-particles
is highly suppressed compared to the light yield of light ionizing
particles and can fall into the energy range chosen for the delayed
event selection. The $\alpha/\beta$ discrimination technique can
be applied to distinguish pure 2.2 MeV $\gamma$ from the $\alpha+\gamma$
decay in the considered case.

The LS is contained in a transparent vessel. Despite very
thin walls of the vessel, it is still a relatively strong source of $\gamma$'s
because of many orders of magnitude higher content of radioactive
impurities compared to the intrinsic contamination of the LS. Also,
the energetic $\gamma$'s from the decay of radioactive isotopes in
other construction materials can penetrate the passive shielding
(buffer). To reduce these backgrouds, the inner part of the whole
volume, or the FV, can be used in the analysis, and the outer part
of the LS volume will serve as an active veto against the external
backgrounds. The software-defined dimensions of the FV depend on
the trade-off between the acceptable count of random coincidences and
the mass of the antineutrino target. 

Background from the radioactive contamination of LS can be further reduced
by applying the $e^{+}/e^{-}$ discrimination method \cite{Franco2011},
successfully used by the Borexino collaboration to reduce background
from $\beta^{+}$-decay of cosmogenic $^{11}$C \cite{Borexino2012b}.
The idea of the method is based on the difference in the photon emission
time distribution. In 50\% of the cases the decay of the positron occurs through
a bound orthopositronium state, which delays the positron decay in the LS
by few nanoseconds. This turns out to be enough to be used in statistical
positron tagging. In a simple case of other backgrounds being pure
electronic ($\beta$-decays) the amount of background events should correspond
to the difference between the amount of electron-like and positron-like
events, giving additional information for background suppression.

\section{Current experiments and their results}

The first indication of a non-zero geoneutrino signal was reported
by KamLAND in 2005 \cite{KamLAND2005}, and the 90\% confidence interval for
the total number of detected geoneutrinos of 4.5 to 54.2 was found
with the Th/U mass ratio fixed in the fit at the chondritic value
of 3.9. In April 2010, the first high significance confirmation of
the geoneutrino signal detection came from the Borexino \cite{Borexino2010}.
The collaboration reported the presence of a non-zero geoneutrino signal
in the acquired antineutrino data set at a 99.997\% confidence level.
Both experiments provided several updates of their results with the
most recent ones released in 2015 by Borexino \cite{Borexino2015} and
in 2016 by KamLAND \cite{Watanabe2016}, the latter still has a preliminary
status.

Both experiments exploit large volume liquid scintillation detectors
with 300 and 1000 tonnes of the target mass respectively, placed
deep underground to protect the setup against cosmic muons. The
antineutrino detection in both detectors is performed via the IBD of the proton.

\subsection{Borexino experiment}

Borexino is a large volume, unsegmented organic liquid scintillation detector located underground at the Laboratori Nazionali del Gran
Sasso, Italy. The complete up-to-date technical description of the
Borexino detector is reported in \cite{Borexino2002,Borexino2009}.
Primarily designed for high precision real-time measurement of
the flux of monoenergetic (862 keV) $^{7}$Be solar neutrino via $\nu-e$
elastic scattering interactions in the scintillator, Borexino is also
an extremely sensitive antineutrino detector with practically zero
non-reactor backgrounds. Because both $\beta$ and $\gamma$ interactions
are essentially indistinguishable from the sought-after solar neutrino
induced events, the measurement was made possible by the deep purification of the LS
from radioactive contaminants, well below the 10$^{-16}$ g of U and Th per g
of the LS~\cite{Borexino2008}. The feasibility of reaching the level
of radiopurity required by Borexino was first proven in the tests
performed in 1996 with Borexino's counting test facility (CTF),
a first detector sensitive to radioactive contaminations at subtrace
levels below $10^{-16}$ g/g in U and Th, out of the reach of other techniques~\cite{CTF98a,CTF98b}.

\begin{figure}
\begin{centering}
\includegraphics[scale=0.4]{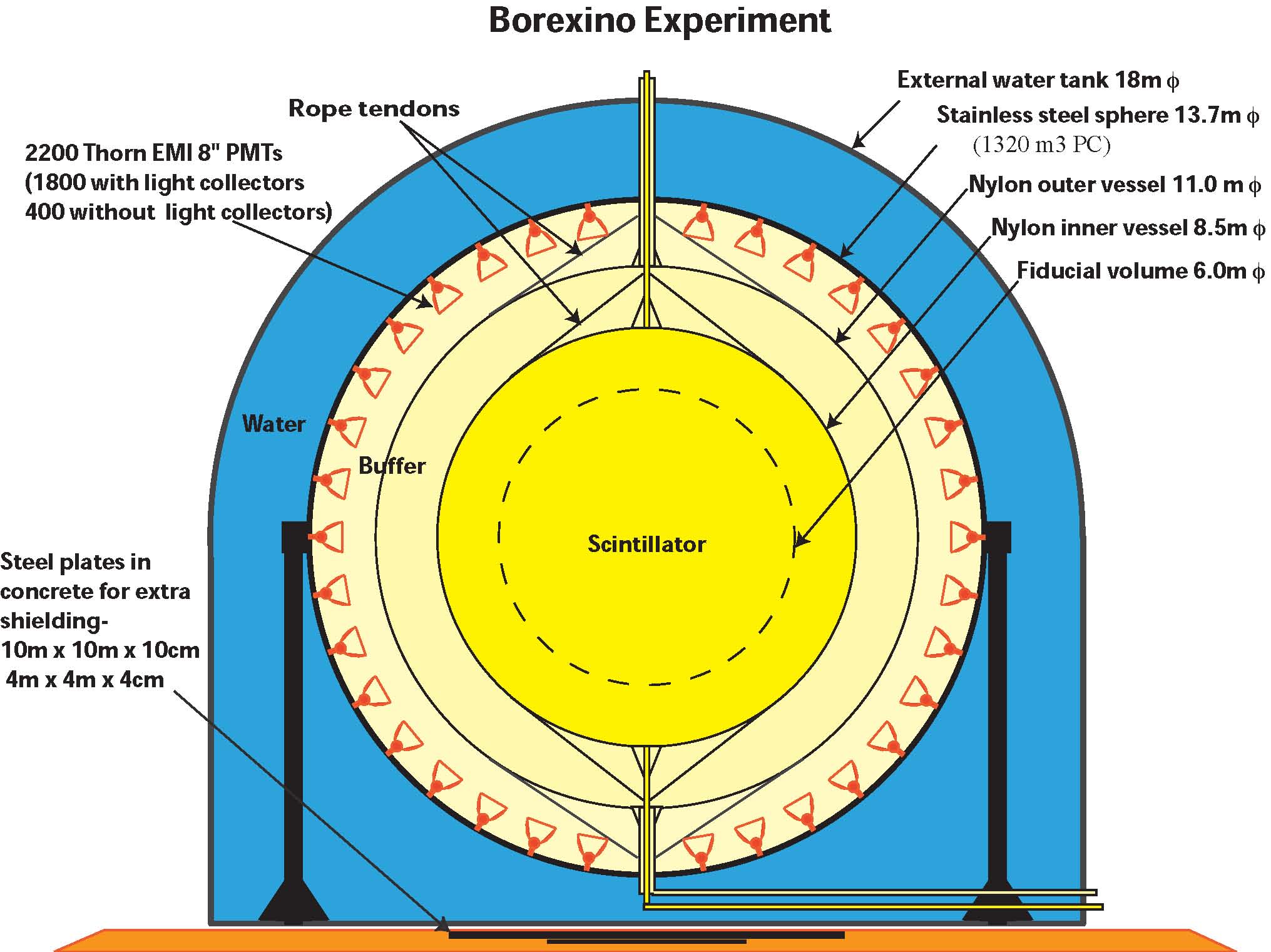}
\par\end{centering}
\caption{\label{Detector}Sketch of the Borexino detector (courtesy of Borexino collaboration).}
\end{figure}

The Borexino detector is schematically depicted in Fig.~\ref{Detector}.
The central spherical core of 278~t design value contains organic
LS solution, composed of pseudocumene
(1,2,4-trimethylbenzene C$_{6}$H$_{3}$(CH$_{3}$)$_{3}$) doped
with 1.5 g/l (0.17\% by weight) PPO fluor (2,5-diphenyloxazole, C$_{15}$H$_{11}$NO).
The LS is contained within an 8.5-m-diameter Inner Vessel (IV), made
of a thin (125 $\mu$m) transparent nylon carefully selected and handled
in order to achieve maximum radiopurity~\cite{Benziger2007}.
The Borexino is equipped with 2212 large area photocathode PMTs. The inward looking 8" PMTs are uniformly distributed over the inner surface of 13.7-m-diameter stainless steel sphere (SSS). All
but 384 PMTs are equipped with light concentrators that
are designed to reject photons not coming from the active scintillator
volume, thus reducing the background caused by the radioactive decays
in the buffer liquid or by penetrating $\gamma$'s from the PMTs.
The 384 PMTs without concentrators are used to study this background
and help identify muons that cross the buffer but not the IV. 

The space between the IV and the SSS is filled with 1000~t of
quenched pseudocumene (PC) buffer fluid, composed of PC with an addition of 5 g/l of dimethylphthalate (DMP, C$_{6}$H$_{4}$(COOCH$_{3}$)$_{2}$). A small amount of DMP quenches the undesired scintillation yield of the buffer fluid by a factor of 20. The amount of light produced by external $\gamma$-rays originating from the PMTs and surrounding materials is quenched below the trigger level, hence avoiding the processing of these events and preventing the trigger electronics overloading.

The buffer is further subdivided into two regions
by a second transparent nylon vessel 11.5~m in diameter, or outer
vessel (OV), made of the same material as IV, and preventing $^{222}$Rn
emanated by the external materials (steel, glass, PMT
materials) from being transported close to the IV with convective fluxes.
Since the PC/PPO solution is slightly lighter (about 0.4\%) than the
PC/DMP solution, the IV is anchored to the bottom ("south pole" of the SSS) with a set of nylon strings. The buffer between the IV and the
SSS provides shielding against external backgrounds. The choice
of PC as a buffer material is based on the fact that it matches both
the density and the refractive index of the scintillator, thus reducing
the buoyancy force for the nylon vessel and avoiding optical mismatch
that otherwise would complicate the vertex reconstruction of events. 

The SSS is placed inside a large dome-like water tank (WT) filled
with ultra-pure water that provides shielding from high energy $\gamma$
rays and neutrons emerging from the surrounding rock. The tank has
a cylindrical base with a diameter of 18~m and a hemispherical
top with a maximum height of 16.9~m. Since the densities of the buffer and LS are $\sim$12\% below the water density, the SSS is anchored to
the ground by 20 steel legs to compensate for the mechanical stress
caused by $\sim$150 t of buoyancy force. The muon flux, although
reduced by a factor of $10^{6}$ by the 3800 m.w.e. depth of the Gran
Sasso Laboratory, is of the order of 1 m$^{-2}$h$^{-1}$, corresponding
to about 4300 muons per day crossing the detector. This flux causes the background well
above the Borexino requirements, and muons identificaton system is needed with efficiency providing a strong reduction factor of about $10^{4}$. Therefore, the WT is designed
as a \v{C}erenkov muon detector and muon tracker (outer detector,
OD) with additional 208 PMTs, The OD is particularly important for detecting
muons skimming the central detector and inducing signals in the energy
region of interest for neutrino physics. In order to maximize the
light collection efficiency, the SSS and the interior of the WT surface
are covered with a layer of Tyvek, a white paper-like material made
of polyethylene fibers.

Pure PC is a scintillator itself, but an addition of a small quantity
of PPO greatly improves the time response and shifts the emission
wavelength spectrum to higher values, thus better matching the PMT photocathode quantum efficiency maximum. The PC/PPO solution adopted as a liquid scintillator
satisfies specific requirements: high scintillation yield ($\sim10^{4}$
photons/MeV), high light transparency (the mean free path is typically
8 m) and fast decay time ($\sim$3 ns), all essential for good energy
resolution, precise spatial reconstruction, and good discrimination
between $\beta$-like events and events due to $\alpha$-particles. Furthermore,
several conventional petrochemical techniques are available to purify hundreds of tonne of fluids needed by Borexino. 

The LS based on PC provides an opportunity for distinguishing
heavy ionizing particles from the light ones. The methos is based on the difference in the time profiles of the corresponding
signals. Lighter ionizing particles produce faster signals, distinct
from the ones produced by the heavier particles with longer tails in the time distribution of the
light emission. Several techniques were studied in Borexino for the $\alpha/\beta$ discrimination, including the optimal Gatti filter
\cite{CTF2008,Borexino2014}. The Gatti filter weights bin per bin the time profile of a signal using two reference distributions. Constructed in this way Gatti parameter $g_{\alpha\beta}$ gives a numeric estimate of the similarity of the time profile of the signal to one of the two reference profiles. The $\alpha/\beta$ discrimination is
used for the antineutrino candidate selection to distinguish the
false signal due to the $^{214}$Po($\alpha+\gamma$) decay with positive $g_{\alpha\beta}$ values from the true one induced by
2.2 MeV $\gamma$ from the neutron capture on the proton (negative $g_{\alpha\beta}$).

The position of each scintillation
event in Borexino is reconstructed off-line with a spatial reconstruction
algorithm using the timing pattern of hit PMTs. The code is
tuned using the data from several calibration campaigns with radioactive
sources inserted in different positions inside the detector~\cite{Borexino2012}. A maximum
deviation of 5 cm between the reference and reconstructed source positions
is observed at a radius of $\sim4$ m, close to the bottom of the
IV.

\begin{table}
\caption{\label{Table:technical}Principal technical characteristics of the
Borexino and KamLAND detectors. The light yield is characterized by the amount of the detected photoelectrons (PEs)}
\vskip 2mm
\centering{}%
\begin{tabular}{ccc}
\hline 
{Characteristic} & {Borexino} & {KamLAND}\tabularnewline
\hline 
{LS} & {PC+PPO(1.5 g/l)} & {PC(20\%) in C$_{12}$H$_{26}$+PPO(1.36 g/l)}\tabularnewline
{LS density, g/cm$^{3}$} & {0,880 (at 15 C)} & {0.780 (at 11.5 C)}\tabularnewline
{Geometrical coverage, \% of 4$\pi$} & {31} & {34}\tabularnewline
{Number of PMTs} & {2212(8'')} & {1325 (17'')+554(20'')}\tabularnewline
{Attenuation length} & {$>$10 m at 430 nm} & {10 m at 400 nm}\tabularnewline
{Light yield, PEs/1 MeV} & {$\sim$500 for 2000 PMTs} & {$\sim$500}\tabularnewline
 & ($\sim$11,500 photons) & ($\sim$8,300 photons) \tabularnewline
{Energy resolution (1$\sigma$), \%} & {$\sim 5 \cdot \sqrt{E\text{(MeV)}}$} & {$\sim 6.4 \cdot \sqrt{E\text{(MeV)}}$}\tabularnewline
{IV size (radius), m} & {4.25 } & {6.5}\tabularnewline
{Full mass, t} & {278 } & {1000}\tabularnewline
{Energy threshold for e$^{-}$, keV} & {180} & {\ensuremath{\sim}800}\tabularnewline
{Muon detection efficiency } & {$10^{-4}$ inefficiency} & \tabularnewline
{Trigger rate, cps} & {11} & \tabularnewline
{Spatial resolution, cm} & {$\sim$14 @ 1 MeV} & {$\sim 12 \cdot \sqrt{E(MeV)}$}\tabularnewline
\hline 
\end{tabular}
\end{table}

Technical characteristics of the Borexino and its achieved performances
are summarized in Table~\ref{Table:technical}. 
The lower threshold on electrons energy used in the analyses is quoted, the 50\% efficiency of the trigger corresponds to significantly lower energy, $\sim$50~keV. The number of PMTs corresponds to the nominal amount of the installed PMTs. The detector was never operating the full set of the PMTs since significant amount of the PMTs ($\sim$10\%) failed after the filling of the detector with water.

Borexino has
an energy resolution of $\sim$5\% at 1~MeV, very good one for a liquid scintillation detector of its size. This
is the result of the high light yield and high transparency of the LS. The latter ensures the excellent uniformity of the detector's response throughout the volume, and, hence, low non-statistical contribution to the energy resolution.
The number of working PMTs slowly decreases with time with a
loss of about 50 per year~\cite{LNGS2017}. In the last period used in solar
neutrino analysis~\cite{Borexino2018n}, the number of working channels
was 1769 at the beginning of the data taking period (December 2011)
and only 1383 at its end (May 2016). The energy resolution correspondingly decreased 
from the initial 5\% at 1 MeV observed at the start
of the data taking with $\sim$1900 operating channels to $\sim$6\%
at 1 MeV in 2016. 

\subsubsection{Radiopurity}

The LS radiopurity was essential for the Borexino solar neutrino programme.
The components of the 278~t of the LS and 889~t of the buffer liquid
were purified in a devoted purification system that combined distillation, water
extraction, gas stripping, and filtration. The PPO was purified in
a concentrated PPO solution in PC (master solution) by filtration
and water extraction and distillation. All the storage and transportation
tanks and in general all the surfaces in contact with LS were thoroughly
cleaned and kept sealed after being filled with ultrapure nitrogen. The PC was prepared
from crude oil extracted from very old Libyan layers, which guaranteed
the low level of the contamination in $^{14}$C that could not be removed
by cleaning. The principles of operation, design, and construction
of the purification system and the review of the requirements and methods
to achieve system cleanliness and leak-tightness can be found in \cite{Benziger2008}. 

The levels of radiopurity achieved in Borexino and presented in Table~\ref{Radiopurity}, provide an important reference
point for future experiments.  
The rate of $^{210}$Po events is given in counts per day (cpd)
for 1 t of the LS. For other isotopes the counting rates are much lower, and either
the corresponding units per 100 t of the LS are used, or the contamination
is expressed in grams of the contaminant contained in 1 g of the LS.
The $^{14}$C abundance is calculated from the evaluated counting
rate of $40\pm1$ Bq/100t~\cite{Smirnov2016}. The contents of $^{238}$U
and $^{232}$Th are obtained from the detected counting rates of
$^{222}$Rn (by counting $^{214}$Bi-Po coincidence events) and that of $^{220}$Rn
(by $^{212}$Bi-Po coincidence events), assuming secular equilibrium
in the corresponding decay chain. The $^{210}$Po decay is not in
equilibrium with the parent nuclei, as could be concluded from the
observed decrease in the $^{210}$Po counting rate with the characteristic
time constant \cite{Borexino2014} corresponding to the $^{210}$Po
life-time T$_{1/2}$=138.4~d. The observed long-term time variations of the $^{210}$Po
counting rate were more complex compared to a simple exponential
decay. Significant instantaneous increases of the counting rate were
associated with operations and changes in the environment, see~\cite{Borexino2014}
for detail. The $^{210}$Po rate was as low as $\sim$2.6 cpd/t in 2017~\cite{Borexino2018n}.
Since the only possible source of technogenic $^{85}$Kr, contained in the gaseous
exhaust of the nuclear power plants, is the atmospheric air, it is
expected that the count of $^{39}$Ar will be proportional to
the ratio of $^{85}$Kr/$^{39}$Ar in the air. 

\begin{table}
\caption{\label{Radiopurity}Radiopurity levels achieved in two phases of
Borexino compared to the typical abundances and the Borexino solar
programme requirements.}
\vskip 2mm
\centering{}%
\begin{tabular}{ccccc}
\hline 
{\footnotesize{}Background} & {\footnotesize{}Typical abundance (source)} & {\footnotesize{}Borexino goals} & {\footnotesize{}Borexino-I \cite{Borexino2014}} & {\footnotesize{}Borexino-II \cite{Borexino2018n}}\tabularnewline
\hline 
{\footnotesize{}$^{14}$C/$^{12}$C, g/g} & {\footnotesize{}$\sim10^{-12}$ (cosmogenic)} & {\footnotesize{}$\sim10^{-18}$} & {\footnotesize{}$(2.7\pm0.1)\cdot10^{-18}$} & {\footnotesize{}$2.7\cdot10^{-18}$}\tabularnewline
{\footnotesize{}$^{11}$C, cpd/100t} & {\footnotesize{}(cosmogenic)} & {\footnotesize{}not specified} & {\footnotesize{}27.4$\pm0.3$} & {\footnotesize{}26.8$\pm0.2$}\tabularnewline
{\footnotesize{}$^{238}$U, g/g} & {\footnotesize{}$10^{-6}-10^{-5}$ (dust)} & {\footnotesize{}$\sim10^{-16}$ (1 $\mu$Bq/t)} & {\footnotesize{}$(1.6\pm0.1)\cdot10^{-17}$} & {\footnotesize{}$<9.4\cdot10^{-20}$(95\% C.L.)}\tabularnewline
{\footnotesize{}$^{232}$Th, g/g} & {\footnotesize{}$10^{-6}-10^{-5}$ (dust)} & {\footnotesize{}$\sim10^{-16}$} & {\footnotesize{}$(6.8\pm1.5)\cdot10^{-18}$} & {\footnotesize{}$<5.7\cdot10^{-19}$(95\% C.L.)}\tabularnewline
{\footnotesize{}$^{222}$Rn, cpd/100t} & {\footnotesize{}100 cm$^{-3}$ (air)} & {\footnotesize{}10} & {\footnotesize{}$\sim$1} & {\footnotesize{}0.1}\tabularnewline
{\footnotesize{}$^{40}$K, g{[}K$_{nat}${]}/g} & {\footnotesize{}$2\cdot10^{-6}$ (dust)} & {\footnotesize{}$\sim10^{-15}$} & {\footnotesize{}$<1.7\cdot10^{-15}$95\% C.L.} & {\footnotesize{}n.a.}\tabularnewline
{\footnotesize{}$^{210}$Po, cpd/t} & {\footnotesize{}surfaces ($^{222}$Rn, $^{210}$Pb)} & {\footnotesize{}$\sim10^{-2}$} & {\footnotesize{}80 (max. initial)} & {\footnotesize{}2 (initial)}\tabularnewline
{\footnotesize{}$^{210}$Bi, cpd/100t} & {\footnotesize{}high ($^{222}$Rn, $^{210}$Pb)} & {\footnotesize{}not specified} & {\footnotesize{}20-70} & {\footnotesize{}$\sim20$}\tabularnewline
{\footnotesize{}$^{85}$Kr, cpd/100t} & {\footnotesize{}1 Bq/m$^{3}$ (air, \cite{Zuzel2004})} & {\footnotesize{}$\sim1$} & {\footnotesize{}$30.4\pm5.5$} & {\footnotesize{}$6.8\pm1.8$}\tabularnewline
{\footnotesize{}$^{39}$Ar } & {\footnotesize{}13 mBq/m$^{3}$ (air, \cite{Zuzel2004})} & {\footnotesize{}$\sim1$} & {\footnotesize{}$\sim$0.4} & {\footnotesize{}$\sim1.3\times10^{-2}\cdot\text{R}(^{85}\text{Kr})$}\tabularnewline
\hline 
\end{tabular}
\end{table}

It should be noted that for some contaminants the achieved levels
exceed the demands of the solar neutrino programme. As a result,
a very low random coincidence count caused
by intrinsic contamination is expected in Borexino. Nevertheless, external sources can
produce $\gamma$'s energetic enough to penetrate into the sensitive volume. All
construction materials were carefully selected in order to reduce their
amount. The closest to the sensitive volume material is nylon
of the containment vessel and its supporting strings, and their contamination
would be most critical for the Borexino programme. Nylon was carefully
selected to guarantee low contamination with $^{238}$U (1.7 ppt
measured) and the absence of potassium in view of the fact that the nylon ropes are frequently
treated with potassium salts to improve their mechanical properties~\cite{Benziger2007}. However, the residual contamination of the IV restricts the use of the whole IV
volume, and the FV cut is used in the analysis, to exclud the most external layer of the LS. 

\subsubsection{Selection of antineutrino candidates}

The antineutrino candidate selection criteria in Borexino are presented
in Table \ref{candidates-selection}. 

The cosmogenic backgrounds are removed by using the muon veto tag.
If a muon passing through the OD without entering the ID is identified,
a 2 ms time window is excluded from the analysis in order to avoid
false antineutrino candidate events caused by neutrons. Additionally,
a 2 s veto is applied after each muon passing through the scintillator
volume in order to suppress the $^{9}$Li\textendash{} $^{8}$He
background. The total loss of the live-time due to these vetoes is
about 11\%. The low energy cut on the prompt event is set slightly below
the kinematic threshold of the IBD allowing to take into account
the energy resolution broadening. The low energy threshold of the
neutron energy window was increased in the latest analysis, since in this
analysis the data set with increased $^{222}$Rn was included. This
contamination was observed after the LS purification tests.
No high energy cut was applied on the prompt event. The cut on the delayed event
energy was tuned to accept the peak of the 2.2~MeV $\gamma$ from the
neutron capture in the whole FV used in the geoneutrino analysis.
The $\alpha/\beta$ selection cut based on a Gatti filter~\cite{Borexino2014}
is applied on the delayed signal to prevent false coincidence due to the $^{214}$Po($\alpha+\gamma$) events
leakage; the cut was absent in the first analysis
of 2010. No pulse shape cut is applied on the prompt candidate. A
cut on the time difference between the prompt and the delayed events
is applied, requiring 20$<\Delta T<$1280~$\mu$s. The upper limit corresponds to 5~lifetimes
of the neutron in the LS of 259.7$\pm$2.4~$\mu$s~\cite{Borexino2013a}.
The lower bound of the time cut is related to the specific features
of the Borexino electronics, where the hardware trigger opens a window
13.5~$\mu$s long, and thus two correlated events with $\Delta T<$~13.5~$\mu$s will belong to one trigger, and a special algorithm would
be needed to extract the antineutrino candidate. To avoid this unnecessary complication, $\Delta T>$~20~$\mu$s was chosen slightly exceeding the length of a single event window. A cut on the relative distance between the prompt
and the delayed events is applied, demanding $\Delta R<$1~m. In the latest
analysis two cuts were added: the ``multiplicity'' cut and the fast ADC (FADC) one. The multiplicity cut rejects events either preceded or followed
by neutron-like events within a 2 ms window, as these are most likely
the events caused by two fast neutrons. The FADC cut consists in the complementary
check of candidate events using an independent 400 MHz digitizer acquisition
system installed with the aim of increasing the dynamic range of
the Borexino electronics for better supernova event processing. 

The total detection efficiency of these basic candidate selection criteria within the FV
(all but the muon veto cut) was determined through
 Monte Carlo (MC) simulations to be $\epsilon=(84.2\pm1.5$)\%, not depending
on energy. External background from ($\alpha,n$) reactions in the
buffer was removed by excluding the events occurring in the
vicinity of the IV wall, namely $\Delta d<$30~cm (or $\Delta d<$25~cm in earlier
analyses). The Borexino spatial selection does not assume spherical
symmetry of the IV. The position of the IV walls is reconstructed on a weekly basis using events originating from the 
radioactive contaminants of its material. The overall uncertainty of the total
exposure includes systematic errors of the reconstruction
of the vessel shape (1.6\%) and that of the position of the prompt event
(3.8\%), other selection criteria contribute no more than 1\%.

The time and radial distributions of the candidate events were checked against the
expected ones. The distribution of the time difference between
the delayed and the prompt events should be compatible with that of the
neutron capture time. All prompt events should have a negative Gatti parameter,
confirming that they are not caused by $\alpha$'s or protons. 

\begin{table}
\caption{\label{candidates-selection}Antineutrino candidate
selection in the KamLAND and Borexino analyses. Indices ``p'' and ``d''
stand for the prompt and delayed events. Borexino performs selections
in the ``light yield scale'' (measured in photoelectrons);
$\Delta R$ is the reconstructed distance between the prompt and delayed
events; $\Delta T$ is the time difference between the prompt and delayed
events; $R_{IV}(\theta,\phi)$, in Borexino's case, is the radial
position of the IV wall in the direction $(\theta,\phi)$;
$T_{\mu}$ is the time passed after the preceding muon. The radius
of the KamLAND inner vessel is R$_{det}=6.5$m, and thus the external
50 cm are cut out. The selection criteria for both KamLAND and Borexino
have been changed since the first analysis. We show here the
ones from the latest published papers.}
\vskip 2mm
\centering{}{}%
\begin{tabular}{ccc}
\hline 
{Cut} & {KamLAND-2013} & {Borexino-2015}\tabularnewline
\hline 
{Prompt event energy} & {$0.9<E_{p}<8.5$ MeV} & {$Q_{p}>$408~PEs}\tabularnewline
{Delayed event energy} & {$1.8<E_{d}<2.6$ MeV } & {$860<Q_{d}<1300$~PEs}\tabularnewline
{} & {or $4.4<E_{d}<5.6$ MeV } & \tabularnewline
{$\Delta R$} & {$\Delta R<2$ m} & {$\Delta R<1$ m}\tabularnewline
{$\Delta T$} & {$0.5<\Delta T<1000$ $\mu$s } & {$20<\Delta T<1280$ $\mu$s }\tabularnewline
{Spatial} & {$R_{p}<6$ m and $R_{d}<6$ m } & {$R_{IV}(\Theta,\phi)-R_{p}(\Theta,\phi)>0.30$ m}\tabularnewline
{} & {$R_{d}>2.5$ m, $\rho_{d}>2.5$ m and $Z_{d}>2.5$ m} & \tabularnewline
{$T_{\mu}$} & {$T_{\mu}>2$ s (showering muons)} & {$T_{\mu}>2$ s (every muon in ID)}\tabularnewline
{} & {+ $T_{\mu}>2$ ms (all muons)} & {+ $T_{\mu}>2$ ms (all muons)}\tabularnewline
{Likelihood} & {\footnotesize{}$L>L_{cut}$($E_{p}$)} & {none}\tabularnewline
{Pulse shape} & {none} & {$g_{\alpha\beta}^{d}<0.015$}\tabularnewline
{Multiplicity} & {none} & {multiplicity=1}\tabularnewline
{FADC} & {none} & {FADC}\tabularnewline
\hline 
\end{tabular}{\footnotesize \par}
\end{table}

\subsubsection{Geoneutrino analysis}

Borexino reported the first geoneutrino observation at more than
4$\sigma$ C.L. in 2010, providing updates of the analysis in 2013 and
2015. The progress of the analysis is presented in Table~\ref{Geo-Table}.

\begin{table}
\caption{\label{Geo-Table}Borexino results on the geoneutrino flux measurements. The results are presented for the electron antineutrino flux since only antineutrinos of this flavour are detected. The total
flux of all antineutrino flavours is approximately 1/$\left\langle P_{ee}\right\rangle\simeq$1.8, times higher.
Exposure of 1~t$\cdot$yr in KamLAND corresponds to $8.48\times10^{28}$
protons$\cdot$yr, and 1~TNU corresponds to the $1.1\times10^{5}$ cm$^{-2}$s$^{-1}$
electron antineutrino flux. The last column presents the probability of the of
the absence of the geoneutrino signal in the corresponding data set
(null-hypothesis)}
\vskip 2mm
\centering{}%
\begin{tabular}{cccccccc}
\hline 
{\small Year} & {\small Live-time,} & {\small Exposure,} & {\small number of} & {\small number of} & {\small geo $\overline{\nu_{e}}$} & {\small geo $\overline{\nu_{e}}$} & {\small P(H$_{0}$)}\tabularnewline
& {\small days} & {\small protons$\cdot$yr } & {\small IBD} & {\small geo $\overline{\nu_{e}}$s} & {\small signal,} & {\small flux} & \tabularnewline
&  & {\small $\times10^{31}$} & candidates & & TNU &   {\small  $\times10^{6}$cm$^{-2}$s$^{-1}$} & \tabularnewline
\hline 
{\small 2010 \cite{Borexino2010}} & {\small 537.2} & {\small 1.52} & {\small 21} & {\small $9.9_{-3.4}^{+4.1}$} & {\small $65_{-22}^{+27}$} & 
{\small $7.4_{-2.5}^{+3.1}$} &
{\small $3\cdot10^{-5}$ ($4.2\sigma$)}\tabularnewline
{\small 2013 \cite{Borexino2013}} & {\small 1363} & {\small $3.69\pm0.16$} & {\small 46} & {\small $14.3\pm4.4$} & {\small $38.8\pm12.0$} & 
{\small $4.4\pm1.4$} &
{\small $6\cdot10^{-6}$ ($4.9\sigma$)}\tabularnewline
{\small 2015 \cite{Borexino2015}} & {\small 2056} & {\small $5.46\pm0.27$} & {\small 77} & {\small $23.7_{-5.7}^{+6.5}$} & 
{\small $43.5_{-10.7}^{+12.1}$} &
{\small $5.0^{+1.4}_{-1.2}$} & {\small $3.6\cdot10^{-9}$ ($5.9\sigma$)}\tabularnewline
\hline 
\end{tabular}
\end{table}

The antineutrino spectra expected in Borexino are shown in Fig.~\ref{fig:brx-expected}.
The geoneutrno contribution was calculated assuming a chondritic Th/U
mass ratio. To make the effect of neutrino oscillations evident,
the spectra are shown with and without neutrino oscillations.
For the geoneutrinos, the oscillations result roughly in the scaling of
the absolute normalization of the spectrum. In contrast, the spectral
shape of reactor antineutrinos demonstrates visible oscillation patterns
in the absence of the smearing by the detector's energy resolution. The
geoneutrino and antineutrino spectra as well as those for other
backgrounds, are simulated using a Geant-4-based MC model of
the detector~\cite{Borexino2018}. The code produces input spectra by
converting the initial energy randomly picked from the corresponding
spectrum to the chosen detector's observable. The non-uniformity of the detector response 
function and non-linearities of the energy scale are automatically accounted with this approach. 
The MC code was tuned using the data of the extensive
calibration campaign with a set of gamma sources. The energy of the
gamma rays ranged between a few hundred keV and 9~MeV~\cite{Borexino2012}.
These calibration data were used to reduce the systematic error
associated with all Borexino results and to optimize the MC
simulation of the detector response. 

The spectra obtained in this way are used to fit the experimental data.
In the geoneutrino analysis, the most suitable observable
(energy estimator) is the total collected charge from PMTs expressed in the amount
of detected photoelectrons (PEs). It offers the best energy resolution above 1 MeV, in 
particular due to the fact that its spatial variation across the FV is the smallest among the energy estimators used in the Borexino analyses~\cite{Borexino2014}. The number of PMTs in operation is defined on-line and stored for each event. Using this information, the energy estimator
of the event is scaled (normalized) to 2000 working PMTs in order to exclude the non-statistical
contribution to the energy resolution, that otherwise will appear because of the variations of the number of PMTs in operation. Another correction
is applied to take into account the non-uniformity of the light collection for
events of the same energy at different positions. The energy of each
event is reconstructed using the total amount of light registered
by all PMTs of the detector (measured in PEs) and
corrected with a position-dependent light collection function $f(x,y,z)$,
which relates the light yield at a point $(x,y,z)$ to that of the event
with the same energy deposited at the centre of the detector, $Q(x,y,z)=f(x,y,z)\;Q_{0}$.
For events at the centre of the detector with energies above 1~MeV the total amount of 
collected light is linear with respect to energy, as shown by
calibrations~\cite{Borexino2010b}. 

The expected event rate and the spectral shape of the reactor antineutrinos
were calculated by considering all reactors around the world. The
variation of the thermal power of individual cores was accounted for
by using the monthly mean load factor provided by the IAEA \cite{IAEA}.
The power fractions (PFs) $^{235}$U : $^{238}$U : $^{239}$Pu : $^{241}$Pu
of a typical reactor used in the calculations were 0.56:0.08:0.30:0.06
with a systematic error of 3.2\% due to possible differences among
the fuels of different cores and the unknown stage of burn-up in each
reactor \cite{Borexino2010}. For 35 European reactors using the Mixed
Oxide (MOX) technology 30\% of thermal power was considered
to have 0.000 : 0.080 : 0.708 : 0.212 PFs \cite{Borexino2010}.
In the latest analysis the more precise PFs of 0.542 : 0.411 : 0.022 : 0.0243 were used for the 46 cores using heavy water moderator
\cite{Borexino2015,Borexino2013}; the associated correction is very
small, $\sim0.1$\%, as only two reactors of this type are located
in Europe. The matter effect on the neutrino propagation increases
the observed signal by 0.6\%, and 1\% increase is due to the contribution
of the long-lived fission products in the spent fuel. The predicted signal
from the nuclear reactors corresponds to (87$\pm$4) TNU. 

The non-reactor background sources in Borexino are at negligible
levels. In the most recent analysis the set of 77 candidate events contained
$0.78_{-0.10}^{+0.13}$ event from the major part of the non-reactor backgrounds,
and less than 0.65 event at 90\% C.L. can be associated with the remaining
possible sources of background, namely from ($\alpha,n$) reactions
in the buffer and from the fast neutrons generated by muons crossing
the rocks \cite{Borexino2015}. The upper limits in the latter case
are set in a conservative approach, since the absense of precise information
does not allow more accurate calculations. 

The rate and shape of the accidental coincidences spectrum was
obtained by shifting the delayed time window to 2\textendash 20
seconds, while keeping all other cuts unchanged. The energy spectrum
of these events is below 3 MeV, overlapping the geoneutrino
spectrum. Time correlated events have been searched for in the (2~ms,~2~s) time window. A negligible amount of correlated events with
a $\sim$1~s time constant were identified, and their contribution
in the antineutrino time window was determined. The average rate of $^{210}$Po decays
in the data set is determined to be 14.1$\pm$0.2 cpd/t, this value is used to
constrain the $(\alpha,n)$ reacton contribution.

\begin{figure}
\begin{centering}
\includegraphics[scale=0.5]{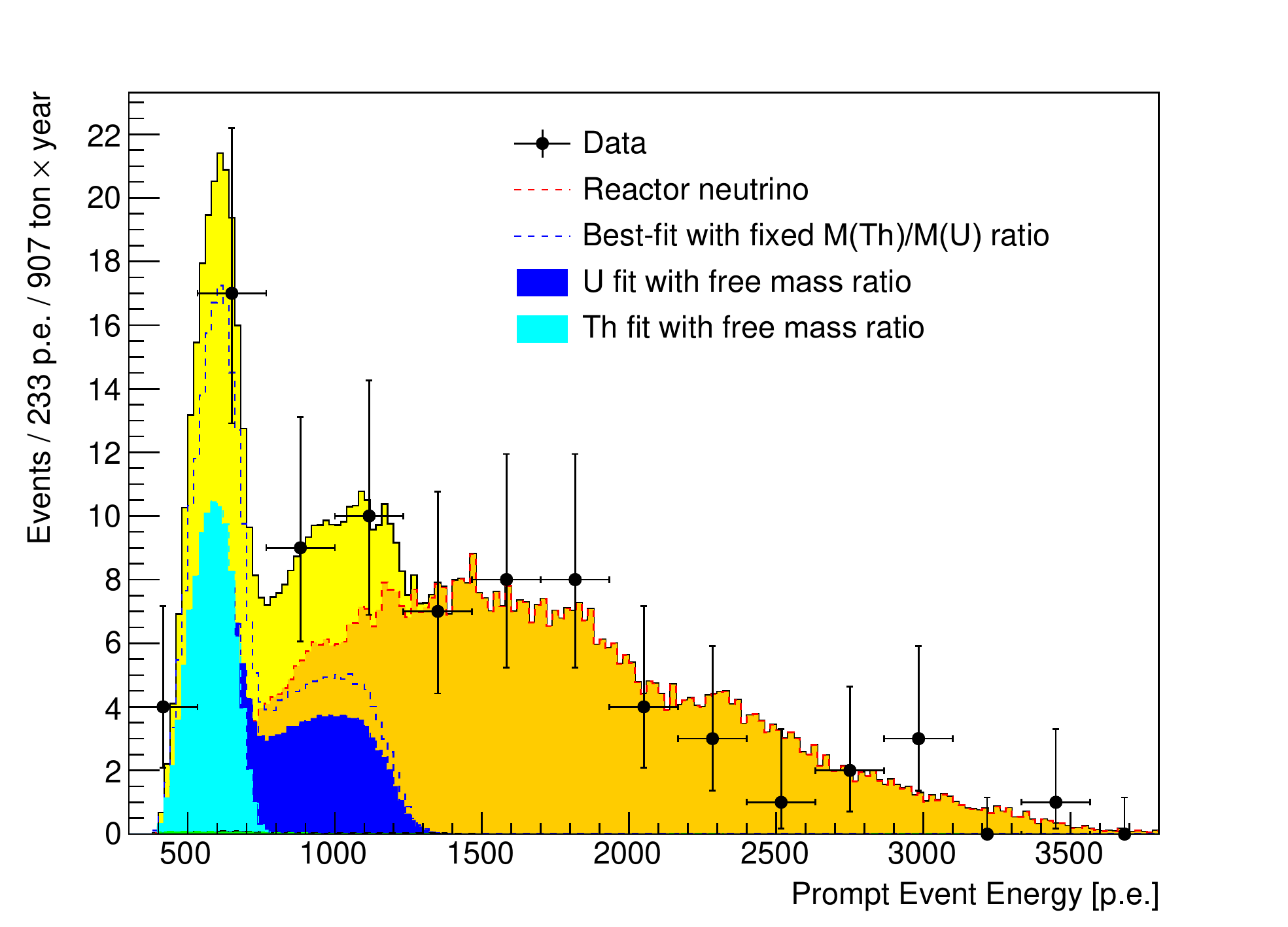}
\par\end{centering}
\caption{\label{fig:BorexFit}The light yield
spectrum of prompt events of antineutrino candidates in Borexino, the energy scale is in units of PEs. The 
best-fit in assumption of the chondritic Th/U mass ratio
corresponds to the upper curve fitting the data (yellow coloured area). The best-fit reactor antineutrino spectrum coincides with the best fit curve above the end-point of the geoneutrino spectrum (orange
colored area), while the best-fit spectrum of geoneutrino
is shown with a dashed line (characteristic curve at low energies, blue). The contributions of U and Th corresponding to the best-fit with free Th/U mass ratio are shown too. The Th contribution has bell-like shape visible at the beginning of the scale (light-blue coloured area), the U contribution expands to $\sim$1300~PEs (blue colored area)
(Fig.~from~\cite{Borexino2015}. For interpretation
of the references to colour in this figure legend, the reader is referred
to the web version of this article)}

\end{figure}

An unbinned likelihood fit of the energy spectrum of 77 selected prompt
$\overline{\nu_{e}}$ candidate events shown in Fig.~\ref{fig:BorexFit}
was performed with the reactor and geoneutrino spectra obtained
by MC simulation. The Th/U mass ratio was fixed to a chondritic value
of 3.9 in the geoneutrino spectrum. Either the measured background energy spectrum (accidental coincidences)
or the ones obtained through the MC simulations ( $^{9}$Li, $^{8}$He, ( $\alpha$
, n) backgrounds) were used in the fit. The other components were not included in the fit because of the uncertainty in their energy spectrum and negligible ($\sim$1\%) contribution. The normalizations of the geoneutrino
spectrum and the reactor antineutrino spectrum were both left free fit
parameters without implying any constraints. In contrast, the normalizations of the background components were constrained as pull parameters around the expected
values. 

The fit returns 
\begin{equation}
N_{geo}=23.7_{-5.7}^{+6.5}\text{(stat)}_{-0.6}^{+0.9}\text{(syst)}
\end{equation}
geoneutrino events and
\begin{equation}
N_{react}=52.7_{-7.7}^{+8.5}\text{(stat)}_{-0.9}^{+0.7}\text{(syst)}
\end{equation}
events caused by the reactor antineutrinos, or, expressed in TNU units
\begin{equation}
S_{geo}=43.5_{-10.4}^{+11.8}\text{(stat)}_{-2.4}^{+2.7}\text{(syst)}\quad\text{TNU}
\end{equation}
and
\begin{equation}
S_{react}=96.6_{-14.2}^{+15.6}\text{(stat)}_{-5.2}^{+4.9}\text{(syst)}\quad\text{TNU}.
\end{equation}

The results include systematic uncertainties from both the exposure
uncertainty of 4.8\% and the uncertainty of 1\% arising due to the
energy calibration. The observed geoneutrino signal corresponds to
$\overline{\nu_{e}}$ fluxes at the detector of $\phi(\text{U})=(2.7\pm0.7)\times10^{6}$~cm$^{-2}$s$^{-1}$ for the U chain and $\phi(\text{Th})=(2.3\pm0.6)\times10^{6}$~cm$^{-2}$s$^{-1}$ for the Th chain, respectively. Here statistical
and systematic uncertainties are summed in quadrature. Borexino observes a
non-zero geoneutrino contribution in the experimental data with the 5.9$\sigma$
significance. The null hypothesis for geoneutrino observation has
a probability of $3.6\times10^{-9}$.

\subsubsection{Future}

The Borexino detector will continue to accumulate data for at least
another year, until the end of 2020. By the end of 2018 available
statistics was increased by $\sim$40\%, compared to the 2015 analysis. The collaboration
plans to improve the performance of the muon-veto cut following KamLAND's
approach. This should save about 9\% of livetime, and 13\% of exposure
can be recovered by increasing the FV (by reducing the excluded
outer shell to 10 cm) and decreasing the lower energy cut for the delayed event.
About $4$\% of exposure can be recovered by including pairs of events
occurring below the lower time cut of $20$ $\mu$s excluded in previous
analyses. Increasing the statistics and using the optimized selection
cuts will make it possible to reduce the uncertainty of the total geoneutrino
flux measurement from the actual 26\% down to $\sim$20\% \cite{Kumaran2019}. 

\subsection{KamLAND experiment}

KamLAND is a reactor antineutrino experiment with a 180 km baseline, perfectly
suited to the search for neutrino oscillations in the so-called large
mixing angle (LMA) parameter region; at these parameters the expected oscillation length roughly corresponds to 100 km.

\begin{figure}
\begin{centering}
\includegraphics[scale=0.5]{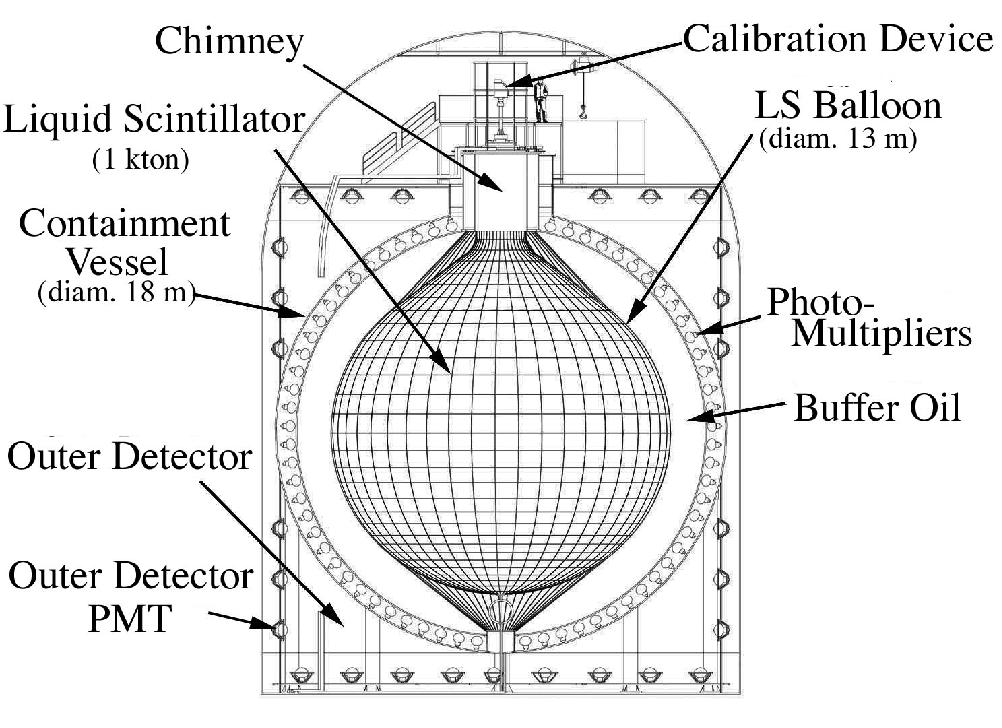}
\par\end{centering}
\caption{\label{DetectorKamLAND}Schematic view of the KamLAND detector (courtesy of KamLAND collaboration)}
\end{figure}

The KamLAND detector is located in the Kamioka mine in Japan. Its design
is similar to that of the Borexino detector (see Fig.~\ref{DetectorKamLAND}).
A dome-like stainless steel structure with a cylinder base 18 m in diameter
is filled with 3.2 kt of ultrapure water. A 9-m-radius stainless
steel sphere placed inside the cylinder divides the detector into
two volumes: external, serving as a water \v{C}erenkov detector of cosmic
muons, and internal, containing the neutrino target within a transparent
vessel with 135-$\mu$m-thick walls with the total volume of 1171$\pm$25
m$^{3}$. The Inner Vessel (IV) supported by a network of Kevlar
ropes contains 1000 tonnes of 20\% solution of pseudocumene in dodecane (C$_{12}$H$_{26}$), with addition of
1.36$\pm$0.03 g/l of PPO fluor, the latter serves to improve
the scintillating properties. The mixture density is 0.780 g$\cdot$cm$^{-3}$
at 11.5\textdegree{}~C. The PC increases light emission but decreases transparency.
The composition of the scintillator was adjusted to obtain the maximum
light yield for the events at the centre, suffering more from the LS
transparency compared to the non-central ones. Scintillation flashes
within the inner detector (ID) are observed by 1879 PMTs of two types:
1325 specially developed 20-inch PMTs masked to 17-inch diameter in order to
improve timing characteristics and 554 older type
20-inch ones, reused from the Kamiokande experiment. The PMTs
provide a total geometric coverage of 34\%. The light yield before
the second purification was 300 PEs/MeV with 17-inch PMTs only and
500 PEs/MeV for the complete set of 17+20-inch PMTs. The attenuation length was measured by a dye-laser, it is about 10~m at the 400 nm~wavelength.
The light output of the LS is 8300 photons/MeV \cite{Gando2012}.

Mineral oil (a mixture of 53\% normalparaffin, C$_{12}$H$_{26}$,
and 47\% of isoparaffin, C$_{n}$H$_{2n+2}$, n$\sim$14), serving
as a protection against external $\gamma$'s and neutrons, fills
the gap between the 9-m-radius SSS and the
IV. The buffer liquid is 0.04\% less dense than the LS. The detector
has an additional transparent sphere with the radius of 8.3 m
dividing the buffer into two subvolumes. It serves as a barrier
against possible convective transfer of radon emanating from the
SSS and the PMTs. The \v{C}erenkov light in the Water Tank (WT) is detected
by 225 PMTs mounted on the internal WT wall and forming the outer
detector (OD).

The KamLAND front-end electronics system based on the analog
transient waveform digitizer (ATWD) captures PMT signals in 128 10-bit
digital samples at intervals of 1.5 ns. Each ATWD captures three gain
levels of a PMT signal providing a wide dynamic range from 1 to 1000~PEs covering the energy deposits of up to 30~MeV~\cite{Suzuki2014}. To reduce the dead time, two ATWDs are attached to each PMT.
The trigger electronics receives 125~ns logical signals from
front-end discriminators set at the 0.15~PE ($\sim$0.3~mV) threshold and
counts the number of triggered ID and OD PMTs with a sampling rate
of 40~MHz. The trigger level is set at the number corresponding to
the deposited energy of $\sim$0.8 MeV. Only fast 17" PMTs are
used for the trigger. The OD readout is triggered in a similar way.

The energy of events is reconstructed off-line using the total collected
charge corrected for the variation of the number of PMTs in operation, dark noise, solid angle, shadowing
by suspension ropes, optical transparencies, and scattering properties
in the LS. The relation between the total collected charge (or visible
energy $E_{vis}$) and the deposited energy $E_{dep}$ is non-linear
and is modelled for each type of particle, taking into account ionization
quenching (Birks' mechanism) and \v{C}erenkov light emission. The
scale is adjusted using 2.2~MeV $\gamma$ from the neutron
capture on the proton as a reference (i.e., by setting $E_{vis}$=$E_{dep}$). 
From 5 March 2002 to 27 February 2003 the geometric coverage was only 22\%, and 
the observed energy resolution was $\sim$7.4\% at 1~MeV approximately scaled as
1/$\sqrt{E_{vis}}$. For the rest of the data, with the 20-inch PMTs installed, the energy resolution improved to $\sim$6.4\% at 1~MeV.

The energy scale was calibrated using a number of radioactive sources
deployed in various positions within the 5.5~m distance from the detector centre
and covering a range of energies between 0.28 and 6.1~MeV \cite{Berger2009}.
The sources used for calibrations included $\gamma$-sources: $^{203}$Hg,
$^{68}$Ge, $^{65}$Zn, and $^{60}$Co; and n + $\gamma$ sources:
$^{241}$Am + $^{9}$Be and $^{210}$Po + $^{13}$C (the latter
was proposed by McKee et al. \cite{McKee2008}). The energy scale
was also calibrated using internal radioactive contaminants distributed
in the IV, such as $^{40}$K and $^{208}$Tl, the tagged delayed fast
coincidence of $^{212}$Bi \textminus{} $^{212}$Po decays from the $^{232}$Th chain,
and the delayed coincidence of $^{214}$Bi \textminus{} $^{214}$Po decays from the
$^{238}$U chain, cosmogenic $^{12}$B and $^{12}$N, and $\gamma$'s
from neutron captures on the proton and the $^{12}$C.

Vertices of low-energy ($<30$~MeV) events are reconstructed using
times of arrival of photons to the PMTs. In order to compensate
for the possible difference in the PMTs time response, the timing calibration
is performed using short light pulses from the dye laser delivered
through the optical fiber to the detector centre. Calibration
sources were used to verify the vertex reconstruction code by comparing the reconstructed position
against the known position of the source. An average position
reconstruction uncertainty of less than 3~cm was found for events
with energies in the range from 0.28 to 6.1~MeV. The vertex reconstruction
is essential as it is used in the selection cuts to estimate the distance between
the prompt and delayed events, and it provides aso the input for the active veto against
the external backgrounds by removing events in the vicinity of
the more radioactive walls of the IV. In the spherical detector the
latter selection criterion in its simplest form will be the radial
cut. The absolute precision of the vertex reconstruction is responsible
for the uncertainty of the FV mass, i.e., the number of target
protons. The statistical precision of the point-like event reconstruction
is $\sim$14~cm at 1~MeV. The position resolution for the 2.506~MeV $\gamma$ from the $^{60}$Co source is 19~cm, scaled as 30/$\sqrt(E\text{[MeV]})$~cm with energy~\cite{Suzuki2014}. More detail can be found in~\cite{KamLAND2010,KamLAND2008}.

The dominating contribution to the antineutrino signal detected
by KamLAND before the Fukusima event originated from the $\simeq$70 Japanese
reactors with total thermal power of 130~GW, located at
an average distance of 180~km from the detector, with an expected reactor
antineutrino flux of $\sim10^{7}$~$\text{cm}^{-2}s^{-1}$. Antineutrinos
are detected in the IBD of the proton. The thermalized neutron life-time is
207.5$\pm$2.8~$\mu\text{s}$ before being captured on the proton. 

KamLAND presented its first results on reactor antineutrino oscillations with the 162 t$\cdot$yr exposure in 2003~\cite{KamLAND2003}. The next analyses
used exposure of 766 t$\cdot$yr~\cite{KamLAND2005a} and 2.881 kt$\cdot$yr~\cite{KamLAND2008}
respectively. In the latter analysis the statistical significance of spectral
distortions exceeded the 5$\sigma$ level, and the analysis of the reactor
spectra was performed from the IBD threshold including among others the geoneutrino
contribution. The first geoneutrino study appeared in 2005 on the
data set from 7.09$\times$10$^{31}$ target proton years~\cite{KamLAND2005}.
The total number of detected geoneutrinos
of 4.5 to 54.2 for a 90\% confidence interval 
was reported with the Th/U mass ratio fixed at 3.9.
The upper limit on the antineutrino flux was set to 1.62$\times10^{7}$
cm$^{-2}$s$^{-1}$ at 99\% C.L., corresponded to a limit of 60 TW for radiogenic heat production from the U and Th chains.

Together with the accumulation of new data, the selection cuts were tuned,
including the decrease of the energy threshold and increase of the
FV by one third, due to the increase of the radial cut from 5 to 5.5 m. The straightforward
increase of the statistics with time would not result in significant 
improvement of the uncertainties because of the presence of the strong background
from the $(\alpha,n)$ reaction
on $^{13}$C in the low-energy part of the spectrum, caused by the high content of $^{210}$Pb. The isotope have life-time of 22.7~yr, supporting the daughter $^{210}$Po with a shorter life-time, which is the
main source of $\alpha$'s in the LS\footnote{In contrast, in Borexino "unsupported" $^{210}$Po was observed, 
not in equilibrium with its parent $^{210}$Pb.})~\cite{KamLAND2005b}. In 2007 a purification campaign began
with the aim of achieving purity levels suitable for the solar neutrino programme.
As a result, the contribution of the $^{13}\text{C(\ensuremath{\alpha},n)}^{16}\text{O}$
reaction decreased by a factor of $\sim$20, and the signal-to-background
ratio was significantly improved in the energy range below 2.6 MeV.
In the next round of analysis~\cite{KamLAND2011a}, the KamLAND collaboration
used the data from March 9, 2002 to November 4, 2009 (2135 days).
The FV was further increased by 0.5 m, up to 6 meters,
and the total exposure reached 4.116 kt$\cdot$yr. 

\begin{figure}
\begin{centering}
\includegraphics[scale=0.4]{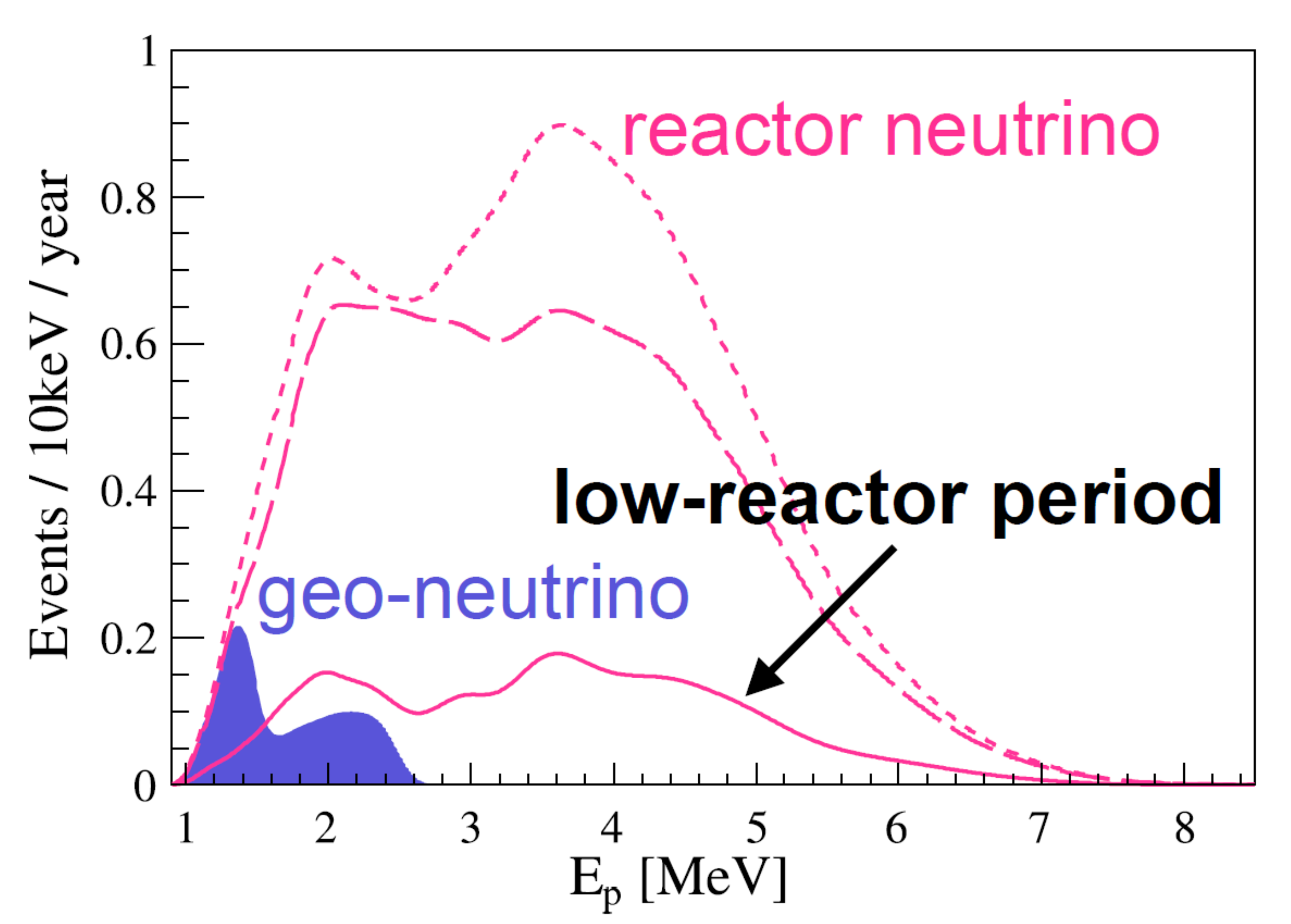}
\par\end{centering}
\caption{\label{fig:KamLAND-reactors}Expected reactors and geoneutrino energy spectra before and after
shutdown of the Japanese reactors. The reactors energy spectrum after the partial shutdown is marked as low-reactor period in the plot. $E_p$ is visible energy of the prompt event (Fig. from \cite{Watanabe2016}). }

\end{figure}

\begin{figure}
\begin{centering}
\includegraphics[scale=0.3]{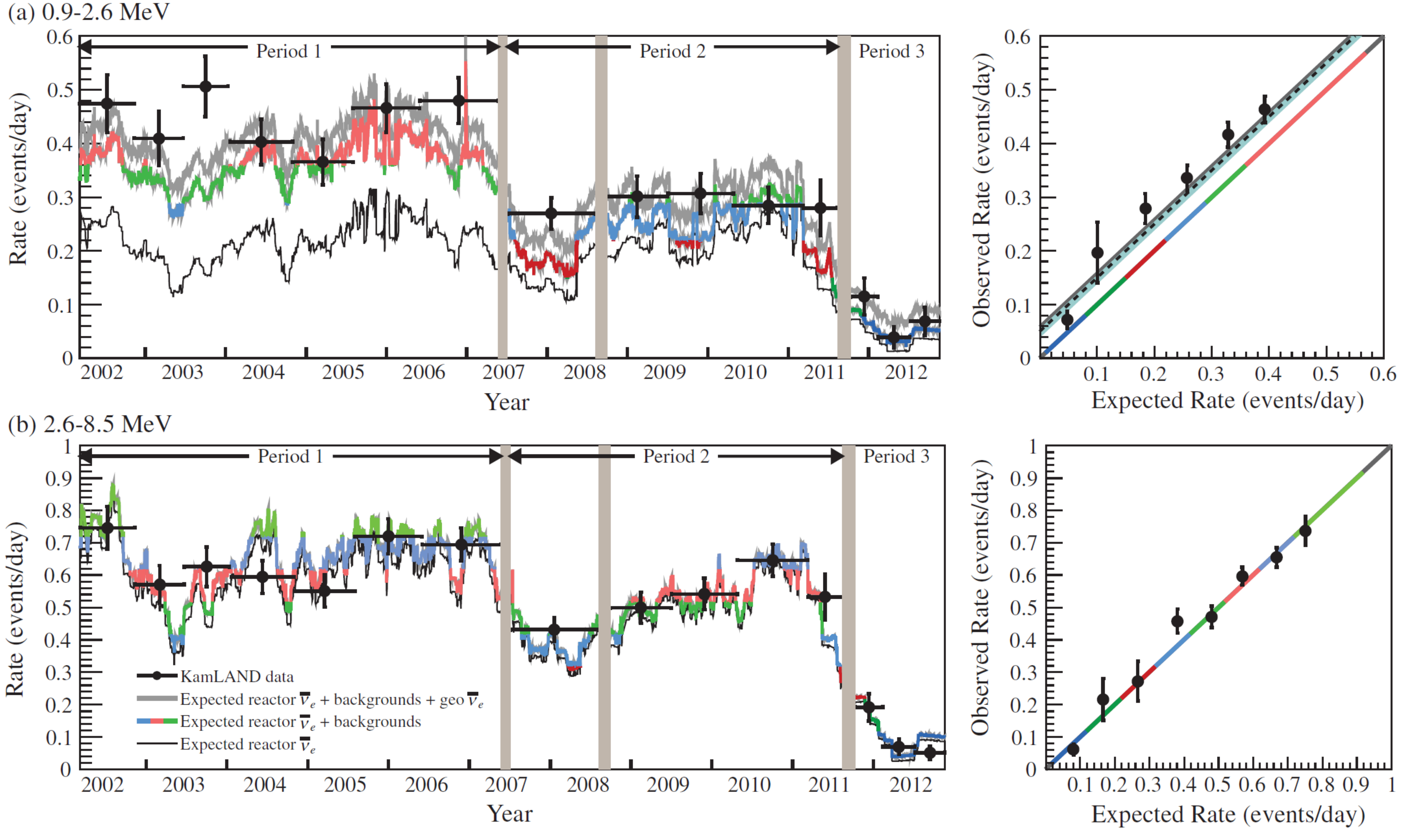}
\par\end{centering}
\caption{\label{fig:KamLAND-periods}Time evolution of expected and observed
rates at KamLAND for antineutrinos with prompt event visible energies (a) in the geoneutrino energy
window, 0.9-2.6 MeV in this analysis, and (b) above the geoneutrino energy window, 2.6-8.5
MeV. The points indicate the measured rates, while the curves show
the expected rate variation for different combinations of signals
and backgrounds and are almost indistinguishable in the lower plot.
In the upper plot the lines indicate reactor antineutrinos (lower
thin line), reactor antineutrinos plus backgrounds (coloured middle line),
and reactor antineutrinos+backgrounds+geoneutrinos (gray upper line). The
data periods excluded from the analysis are shaded by thick bars.
In the right panel the observed event rate for each group is plotted
at the exposure-weighted versus expected event rate. The efficiency-corrected
best-fit value of the geoneutrino rate from the full spectral analysis
(dashed line), its 1$\sigma$ error (shaded region), and the model
expectation (gray line) are drawn for comparison. (Figure from \cite{KamLAND2013}).}

\end{figure}

In September 2011, KamLAND started to take data with the updated
experimental facility, KamLAND-Zen, with the main goal of the search for neutrinoless
double beta decay. The target isotope $^{136}\text{Xe}$ is dissolved in 13 t
of the LS contained in a small balloon with radius 1.54 m placed at the
detector centre. The presence of the balloon
does not interfere much with the geoneutrino studies. Only a very small
fraction of the total volume is affected, which can be removed by applying
the corresponding spatial cut. An important point for the geoneutrino
study in the KamLAND-Zen experiment was in 2012, when, following
the Fukusima event, the Japanese nuclear power plants were switched
off for checks and maintenance. The idea to use the information on
the thermal power of nuclear plants in view of the geoneutrino
signal extraction was discussed before these dramatic events \cite{KamLAND2011b}.
After May 2012, when the last reactor was switched off, the reactor-to-geoneutrino
signal ratio is practically the same as in the Borexino experiment. The
decrease of the reactor signal is illustrated in Fig.~\ref{fig:KamLAND-reactors}.

The total data set can be divided into three periods. Period I with 1486
days of live-time refers to the data taken from 2002 to May 2007,
before the LS purification campaign that continued until 2009. Period
II with 1154 days of live-time refers to the data taken until the autumn of 2011
during and after the LS purification campaign. Period III started
since September 2011 after the installation of the internal Xe balloon
for KamLAND-Zen experiment. Time evolution of expected and observed
antineutrino rates in KamLAND for two energy ranges (below and above
the energy geoneutrino edge) is shown in Fig.~\ref{fig:KamLAND-periods}.
The contribution of geoneutrinos to the second region is negligible.
The oscillation parameters used to calculate the expected reactor
antineutrino rate were the best-fit values from the global oscillation
analysis.

\subsubsection{Radiopurity}

Initially, the LS and the buffer oil were purified
by a water extraction process during the detector filling in March
2002. The levels of achieved radiopurity were $(3.5\pm0.5)\cdot10^{-18}$~g/g for $^{238}$U, $(5.2\pm0.8)\cdot10^{-17}$~g/g for $^{232}$Th,
and $<2.7\cdot10^{-16}$~g/g for potassium. Nevertheless, contamination with $^{85}$Kr at the level of 883$\pm$20~$\mu$Bq/kg and
with daughter isotopes of $^{222}$Rn was observed. The decay rate of $^{210}$Pb was 58.4$\pm$1.1~$\mu$Bq/kg,
as inferred from the decay rates of its unstable daughters $^{210}$Bi and $^{210}$Po~\cite{KamLAND2015}.
Though the resulting single event
counting rate of 8.1$\times$10$^{7}$~(kt$\cdot$d)$^{-1}$ was acceptable
for the antineutrino programme, the solar neutrino one would be
impossible without further purification. The presence of a significant amount of $^{210}$Po estimated
at 95~Bq in the whole volume~\cite{Sanshiro} was also a source of
noticeable $(\alpha,n)$ background for geoneutrino study.

The KamLAND collaboration developed methods for efficient removal
of Kr and Pb from the LS aiming to achieve the radiopurity levels for low-energy
physics programme~\cite{Keefer2015}. The second purification campaign was performed
in 2007-2009 by distillation and nitrogen purging of each LS component.
The distillation is expected to reduce high-boiling point metal elements
such as $^{210}$Pb and $^{40}$K, while nitrogen carries away the
radioactive noble gases $^{222}$Rn, $^{85}$Kr and $^{39}$Ar. After
the second purification, the radioactive impurities were further reduced 
to $(1.5\pm1.8)\cdot10^{-19}$~g/g in $^{238}$U, to $(1.9\pm0.2)\cdot10^{-17}$~g/g in $^{232}$Th,
and in to a limit $<2.7\cdot10^{-16}$~g/g in natural K.
The content of the $^{222}$Rn daughters was reduced to 2~mBq/m$^{3}$
for $^{210}$Po, below $1$~mBq/m$^{3}$ for $^{210}$Bi, and to $\sim0.1$~mBq/m$^{3}$ for $^{85}$Kr. The overall reduction factors for counting rates
of $^{85}$Kr, $^{210}$Bi, and $^{210}$Po are about 6$\times$10$^{-6}$,
8$\times$10$^{-4}$, and 5$\times$10$^{-2}$, respectively. After
the second purification, the main source of accidental coincidences
are events from the vessel walls, possibly due to dust contamination
during the IB film assembly and air leakage during the LS purification.

\subsubsection{Selection of antineutrino candidates}

The antineutrino candidate selection criteria of the more elaborated KamLAND analysis of 2013 are presented
in Table~\ref{candidates-selection}. KamLAND performs energy selections
using reconstructed energy. Note the difference in the radial cuts: 
Borexino's is based on the position of the prompt event, and in KamLAND
both prompt and delayed events should be within the FV radius. This
leads to a relative depletion of the outer layer with detected candidates in the KamLAND analysis.
The selection of the FV decreases the amount of target protons: in KamLAND a
50 cm layer is removed, while in the first analysis a much thicker layer of
150 cm was removed. For the crossing muons KamLAND used more
advanced selections compared to Borexino, which were aimed to tag showering muons only. This results in a
moderate loss of 4\% live-time in comparison to 11\% for Borexino; 
the difference per crossing muon is even bigger if the higher muon
flux at the Kamioka site is taken into account.

The low energy cut on the prompt event is set at the kinematic threshold
of the IBD. The high energy cut corresponds to the maximum energy
of the reactor antineutrino spectrum. The cut on the delayed event
energy accepts the peak of the 2.2~MeV $\gamma$ from the neutron capture
on the proton, and the 4.95~MeV $\gamma$ from the neutron capture on $^{12}$C.
No pulse shape cuts are applied in the KamLAND analysis. A cut on the
time difference between the prompt and the delayed events requires
0.5$<\Delta T<$~1000~$\mu$s. The upper limit corresponds to
5~life-times of the neutron in the KamLAND LS (207.5$\pm$2.8 $\mu$s).
A cut on the relative distance between the prompt and the delayed
events, $\Delta R<$2~m, is less stringent compared to Borexino. The
software-defined FV in KamLAND is a sphere with a radius of 6~m. In the last period, an additional FV cut in the antineutrino analysis is applied
to exclude the contamination from the Xe balloon. In the lower hemisphere
an inner spherical volume with a radius of 2.5~m is excluded, and the
cylindrical volume with a radius of 2.5~m is excluded in the upper
hemisphere.

An additional event selection is applied to suppress accidental coincidence
backgrounds while maintaining high efficiency for antineutrinos. A
figure-of-merit is constructed on the basis of the PDFs for antineutrinos rate, $f_{\nu}$,
and accidental coincidences rate, $f_{acc}$, obtained by a combination of
MC, data-driven, and analytical methods. The PDF is based
on six cut parameters: $E_{p}$, $E_{d}$, $\Delta R$, $\Delta T$, $R_{p}$, $R_{d}$. For each candidate event the ratio $L=\frac{f_{\nu}}{f_{\nu}+f_{acc}}$
is calculated and tested against the value maximizing the figure-of-merit
$\frac{S}{\sqrt{S+B_{acc}}}$ for prompt energy intervals of 0.1 MeV.
$S$ is the expected antineutrino signal assuming an oscillation-free
reactor spectrum and predicted geoneutrino fluxes. $B_{acc}$ corresponds
to the number of accidental background events and is measured using
an out-of-time delayed coincidence window (10~ms~$<\Delta T<$~20~s). The efficiency of the selection cuts defined in this way has strong
energy dependence and is calculated from the ratio of the MC-generated
antineutrino events that passed the selections to the total number of the
generated events. The systematic uncertainty is evaluated using calibrations
with the $^{68}$Ge and $^{241}$Am-$^{9}$Be sources~\cite{KamLAND2011a}. 

\subsubsection{Geoneutrino analysis}

The KamLAND geoneutrino analyses are presented in Table~\ref{Geo-Table-KL}. In this section we mainly discuss the KamLAND analysis from~\cite{KamLAND2013}, if not otherwise explicitly stated. The more recent analysis from 2016 still has preliminary status.

Reactor fluxes are predicted from reactor operation records that
include the thermal power, fuel exchange, and reshuffling data for
all Japanese commercial reactors. The information is provided by
a consortium of Japanese electric power companies. The thermal power
generation is measured within 2\% precision. The relative fission yields
averaged over the live-time period for the 2013 publication were
(0.567:0.078:0.298:0.057) for ($^{235}$U: $^{238}$U: $^{239}$ Pu: $^{241}$Pu), respectively. A calculation of the geoneutrino flux
at KamLAND is based on the reference Earth model of Enomoto et al.~\cite{Enomoto2007}; 109 and 27 geoneutrino events from U and Th,
respectively, were expected in the 2013 dataset. 

\begin{figure}
\begin{centering}
\includegraphics[scale=0.28]{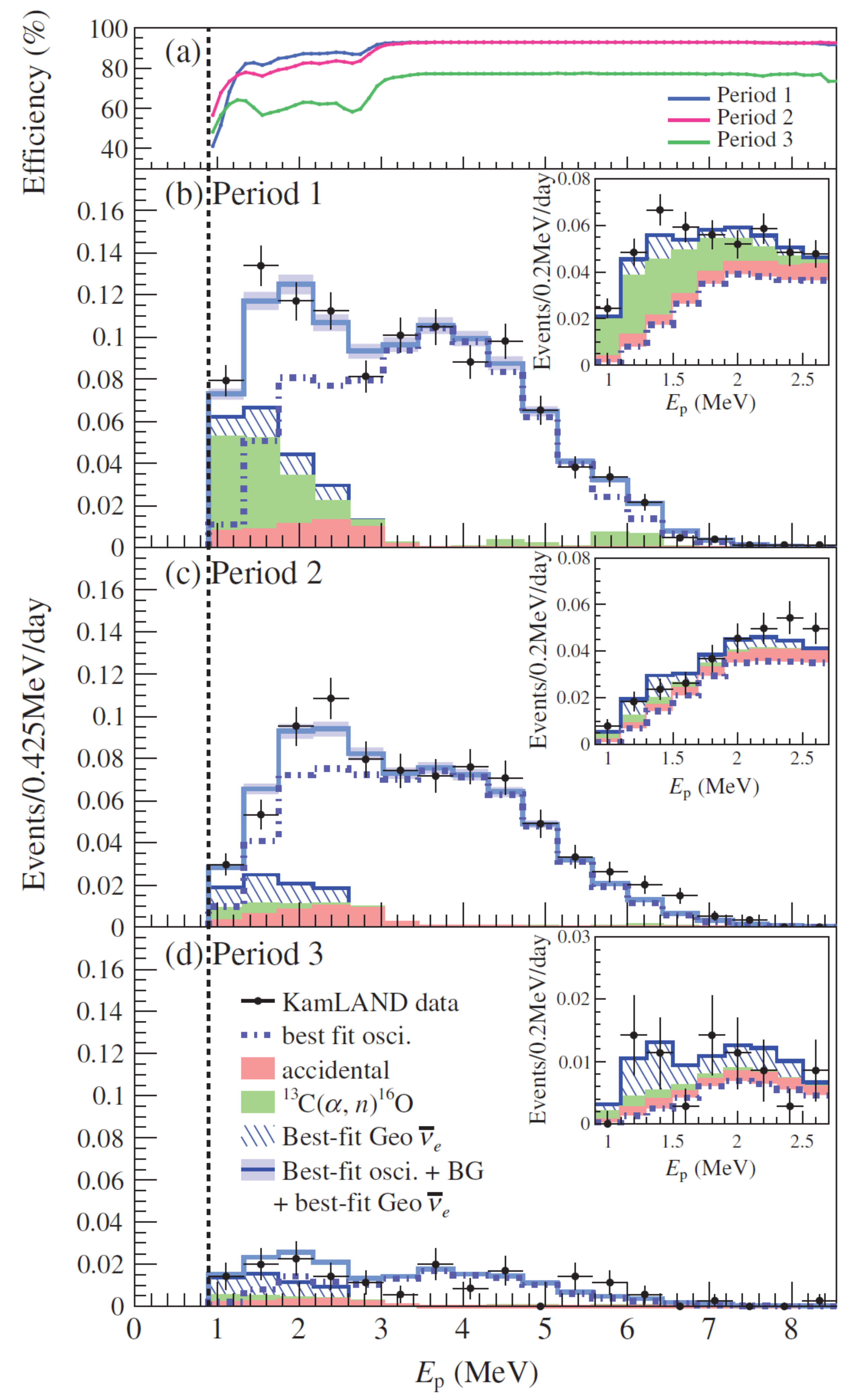}\includegraphics[scale=0.4]{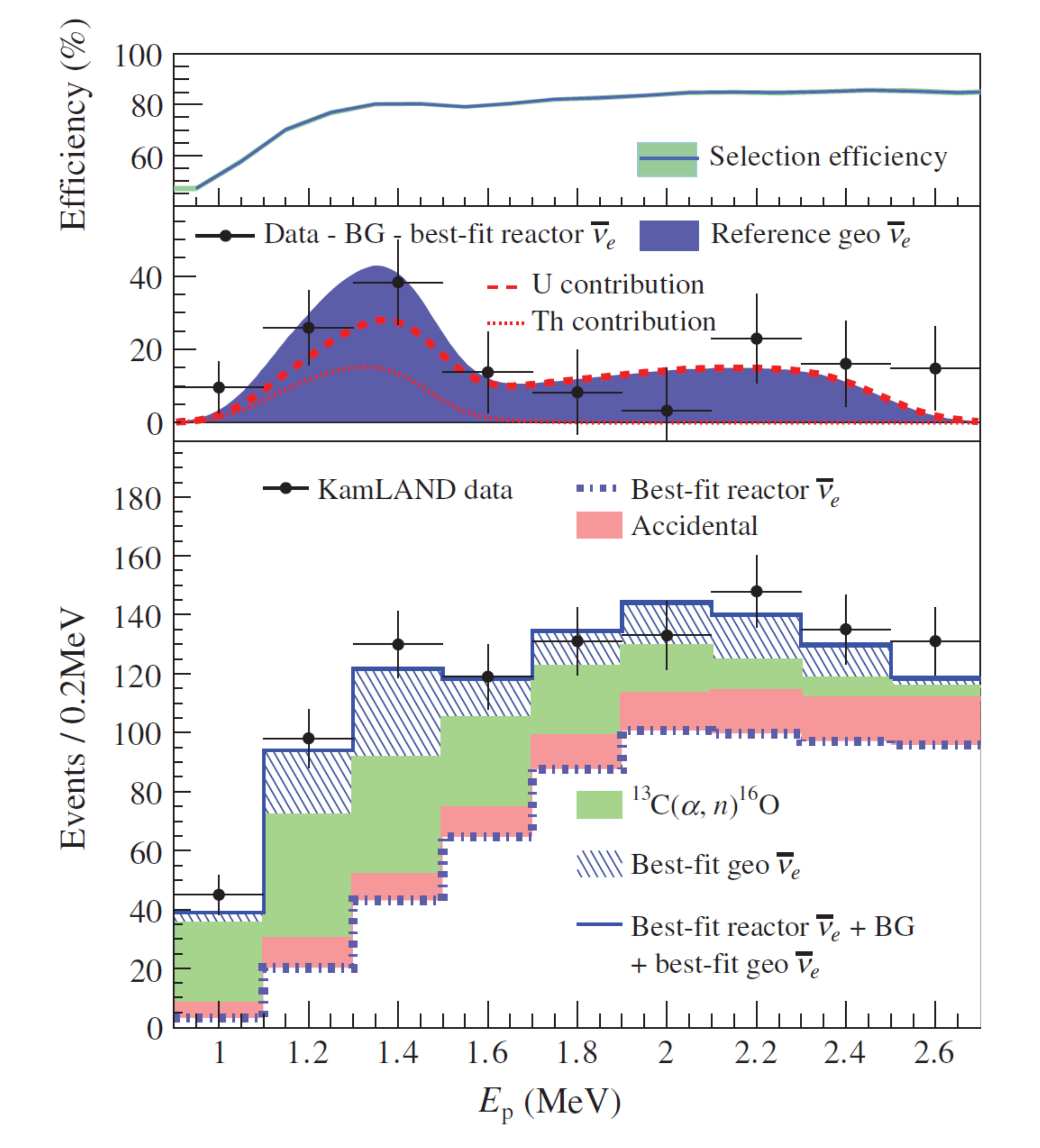}
\par\end{centering}
\caption{\label{fig:KamLAND-expSpectra}\textbf{\footnotesize{}Left:} visible energy spectrum of $\overline{\nu_{e}}$ candidates prompt event shown separately
for 3 data-taking periods, signed as Period I, Period II and Period III from top to bottom. The visible energy spectra of $\overline{\nu_{e}}$
candidate events in the geoneutrino energy range are shown in the
insets with a finer binning. The very top panel shows the energy
dependence of the $\overline{\nu_{e}}$ selection efficiency for each
period.\protect \linebreak{}
\textbf{\footnotesize{}Right:} energy spectrum of the $\overline{\nu_{e}}$
events in the geoneutrino region for the total data set up to
2013. Bottom panel: data together with the best-fit background and
geoneutrino contributions. The shaded area corresponds to the background
and geoneutrino contributions. The middle panel presents the observed
geoneutrino spectrum after subtraction of other contributions. The
dashed and dotted lines show the best-fit U and Th spectral contributions,
respectively, and the blue shaded area corresponds to the estimation
using the geological model~\cite{Enomoto2007}. The upper inset shows the selection efficiency curve. (Figures from~\cite{KamLAND2013}).}

\end{figure}

The energy spectrum of $\overline{\nu_{e}}$ candidate events with visible energy above the
0.9 MeV energy threshold are shown in Fig.~\ref{fig:KamLAND-expSpectra}
separately for all 3 data taking periods analyzed in~\cite{KamLAND2013}.
The background, reactor, and geoneutrino contributions are the best-fit
values from the KamLAND-only analysis. The decrease of the $(\alpha,n)$
background after the purifications is clearly seen in the second period.
In the third period also the reactor antineutrino background decreased.
In the right plot of Fig.~\ref{fig:KamLAND-expSpectra} the total
spectrum for all data taking periods up to 2013 is shown. The
fit in the bottom panel includes all available constraints on the
oscillation parameters.

The geoneutrino fluxes are extracted together with the neutrino oscillation
parameters by using an unbinned maximum-likelihood method incorporating
the event rate and the prompt energy spectrum shape, and accounting for their time variation. The fit is performed in the full energy range from 0.9 to 8.5 MeV. The
expression for $\chi^{2}$ is more complicated compared to the Borexino analysis,
and includes two contributions for the time-varying event rate,
the time-varying energy spectrum shape, five penalty terms for backgrounds
constrained with accordance of their estimates, a penalty term for
systematic uncertainties, and a penalty term for the oscillation parameters
(see~\cite{KamLAND2013} for more detail). The normalizations for
the uranium and thorium signals can be treated independently and allow
the analysis without assuming any particular Earth model. The $\chi^{2}$
includes period-dependent parameters describing the uncertainties
of the reactor $\overline{\nu_{e}}$ spectrum, the energy scale, the
event rate, and the energy-dependent detection efficiency; all these
parameters are constrained. The overall reactor rate uncertainties
for Period 1, and Periods 2 and 3 are 3.5\% and 4.0\%, respectively.
Systematic uncertainties are conservatively treated as being fully
correlated across all data-taking periods. Finally, a penalty term
provides an optional constraint on the neutrino oscillation parameters
by using solar, accelerator, and short-baseline reactor antineutrino
experiments.

The best-fit to the unbinned data with free N(U) and N(Th) yields
N(U)=116 and N(Th)=8 geoneutrino events from U and Th, respectively.
The corresponding fluxes are $\Phi$(U)=N(U)$\times$2.01$\times$10$^{4}$=2.33$\times10^{6}$
and $\Phi$(Th)=N(Th)$\times$6.88$\times10^{4}$=0.55$\times10^{6}$,
both measured in cm$^{-2}$s$^{-1}$. The joint confidence intervals
for the sum N(U)+N(Th) and the asymmetry factor $\frac{N(U)-N(Th)}{N(U)+N(Th)}$
agrees with the expectation from the geological reference model
of Enomoto et al. \cite{Enomoto2007}. Assuming the chondritic Th/U mass
ratio of 3.9, the total number of U and Th geoneutrino events is N(U+Th)=$116_{-27}^{+28}$,
corresponding to the electron antineutrino flux $\Phi$(U+Th)=(3.4$\pm0.8)\times10^{6}$
cm$^{-2}$s$^{-1}$ at the KamLAND location or to the all-flavours total antineutrino
flux of $(6.2\pm1.5)\times10^{6}$ cm$^{-2}$s$^{-1}$. 

The latest results of KamLAND were presented in 2016~\cite{Watanabe2016}.
The analyzed data set included a $\sim$3.5-year-period with few reactors operated, and $\sim$2-year live-time with all Japanese reactors switched off.
The number of detected geoneutrinos, $164_{-25}^{+28}$
events, is measured with the 17\% uncertainty in agreement with expectation.
The evidence of the geoneutrino signal achieves $7.9\sigma$. 
It is confirmed that an excess of events observed by some reactor experiments
in the 4-6 MeV range (see section~\ref{sec:Reactor}) has no impact on the geoneutrino
results. The effect of the reactor spectrum uncertainty is much smaller
than the statistical uncertainty of the geoneutrino signal.

\begin{table}
\caption{\label{Geo-Table-KL}KamLAND results on the geoneutrino flux measurements.
The results are presented for the electron antineutrino flux since only antineutrinos of this flavour are detected. The total
flux of all antineutrino flavours is approximately 1/$\left\langle P_{ee}\right\rangle\simeq$1.8, times higher.
Exposure of 1~t$\cdot$yr in KamLAND corresponds to $8.48\times10^{28}$
protons$\cdot$yr, and 1~TNU corresponds to the $1.1\times10^{5}$ cm$^{-2}$s$^{-1}$
electron antineutrino flux. The last column presents the probability of the of
the absence of the geoneutrino signal in the corresponding data set
(null-hypothesis)}
\vskip 2mm
\centering{}%
\begin{tabular}{cccccccc}
\hline 
{\small Year} & {\small Live-time,} & {\small Exposure,} & {\small number of} & {\small number of} & {\small geo $\overline{\nu_{e}}$} & {\small geo $\overline{\nu_{e}}$} & {\small P(H$_{0}$)}\tabularnewline
& {\small days} & {\small protons$\cdot$yr } & {\small IBD} & {\small geo $\overline{\nu_{e}}$s} & {\small signal,} & {\small flux} & \tabularnewline
&  & {\small $\times10^{32}$} & candidates & & TNU &   {\small  $\times10^{6}$cm$^{-2}$s$^{-1}$} & \tabularnewline
\hline 

{\small 2005 \cite{KamLAND2005}} & {\small 749.1} & {\small $0.709\pm0.035$} & {\small 152} & {\small $25_{-18}^{+19}$} &
{\small $43_{-32}^{+35}$} & {\small $5.1_{-3.6}^{+3.9}$} &
{\small $4.6\%$}\tabularnewline
{\small 2008 \cite{KamLAND2008}} & {\small 1486} & {\small $2.44$} & {\small \textendash{}} & {\small $73\pm27$} 
& {\small $39.4\pm14.3$} & {\small $4.4\pm1.6$} & {\small 0.45$\%$}\tabularnewline
{\small 2011 \cite{KamLAND2011b}} & {\small 2135} & {\small $3.49\pm0.07$} & {\small 841} & {\small $106_{-28}^{+29}$} 
& {\small $38.5_{-9.8}^{+10.7}$}
& {\small $4.3_{-1.1}^{+1.2}$} & {\small $3\cdot10^{-5}$}\tabularnewline
{\small 2013 \cite{KamLAND2013}} & {\small 2991} & {\small $4.90\pm0.1$} & {\small \textendash{}} & {\small $116_{-27}^{+28}$} 
& {\small $30.4\pm7.2$}
& {\small $3.4\pm0.8$} & {\footnotesize{}$2\cdot10^{-6}$}\tabularnewline
{\small 2016 \cite{Watanabe2016}} & {\small 3901} & {\small $6.39$} & {\small 1130} & {\small $164_{-25}^{+28}$} 
& {\small $34.9_{-5.4}^{+6.3}$}
& {\small $3.9_{-0.6}^{+0.7}$} & {\small 7.92$\sigma$}\tabularnewline
\hline 
\end{tabular}
\end{table}

\subsubsection{Future}

KamLAND continues data acquisition in its KamLAND-Zen phase with the Japanese
nuclear reactors switched off, detecting about 14 geoneutrino events
per year. The latest results were presented at the end of 2016,
and the data from at least three more years are available for the analysis.
The construction of a new, cleaner inner balloon is envisaged to meet
the needs of the neutrinoless double beta decay search. For the same
purpose, techniques to increase the light yield are studied. The
possible improvements include: increasing the geometric coverage using
Winston cones for light collection,
installing higher quantum efficiency 20" PMTs instead of the older ones masked
to 17\textquotedblright{}, and using new LAB-based LS with better transparency~\cite{Inoue2018}. The estimated increase
of the light yield is factor 1.8, 1.9 and 1.4 for each of the improvements, respectively, or factor 4.5 in total.
These measures should
further improve the sensitivity of KamLAND to geoneutrinos, helping to discriminate
the backgrounds and to increase the IBD detection efficiency.

\subsection{\label{sec:UThratio}Extracting Th/U global mass ratio}

The Th/U mass ratio plays an important role in the Earth geoscience. It is important for the  understanding of the early Earth evolution. The distributions of U and Th could be also used as benchmark for estimating the distribution of other refractory lithophile elements. Refractory elements are condensed from the nebula at high temperatures, and are assumed to be accreted in chondritic proportions, with limited variability in their ratio. The decay sequences of the both elements stops at different lead isotopes, and the Th/U mass ratio in a sample can be evaluated based on measured $^{232}$Th/$^{238}$U molar values and their time integrated Pb isotopic values. The detailed analysis of the available data was recently performed by Wipperfurth et al.~\cite{Wipperfurth2018}, the authors obtained high precision planetary Th/U mass ratio of 3.876$\pm$0.016. The geoneutrinos can provide the independent check of the validity of the underlaying assumptions, since the proportion of the Th and U contribution to the total spectrum reflects the global Th/U mass ratio, given the local contribution is known.

The theoretical (for ideal detector with infinite resolution) IBD
spectra from the U and Th chains with the assumption of secular equilibrium
in the chains and at the chondritic ratio of Th/U masses M(Th)/(U)=3.9
are shown in Fig.~\ref{fig:UandTh_nu}. The visible difference in
the spectra makes contributions from U and Th potentially distinguishable,
but the low-energy part of the U spectrum has a contribution very
similar to the Th spectrum. In fact, the end-point for the Th spectrum
(E$_{0}$=2.252 MeV from $^{212}$Bi $\beta$-decay with 64\% branching
ratio) is practically the same as the end-point of the $^{234}$Pa$^{\text{m}}$
$\beta$-decay (E$_{0}$=2.267~MeV with branching $\sim$100\%). The
difference above 2.267~MeV is due to $\beta-$decay of $^{214}$Bi
to the ground state with a branching of $\sim$19\%. The fraction of
events in the $^{238}$U chain spectrum above the E=2.267~MeV point is about
0.5 of the total U signal. Considereng that the expected signal
from the $^{238}$U is a factor of $\sim$3.5 higher for chondritic M(Th)/M(U)=3.9, a simple
statistical estimate show that in the absence of other backgrounds,
the uncertainty of the S(Th)/S(U) ratio will be a factor of 5 higher
compared to the statistical uncertainty of the number of the detected geoneutrino events. 

Dye in~\cite{Dye2016} estimated number of events needed to
distinguish between different Th/U mass ratios. The estimated number
of events needed to obtain absolute uncertainty of 0.1 (this would translate to 2.5\% uncertainty for the chondritic M(Th)/M(U) ratio) on the M(Th)/M(U) exceeds $10^{5}$, the non-realistic statistics for the modern detectors. With more realistic statistics of $10^{3}$ geoneutrino events
one can expect the absolute uncertainty of the order of~1 (or 25\% of the chondritic ratio) on the Th/U mass ratio.

\begin{figure}
\begin{centering}
\includegraphics[scale=0.7]{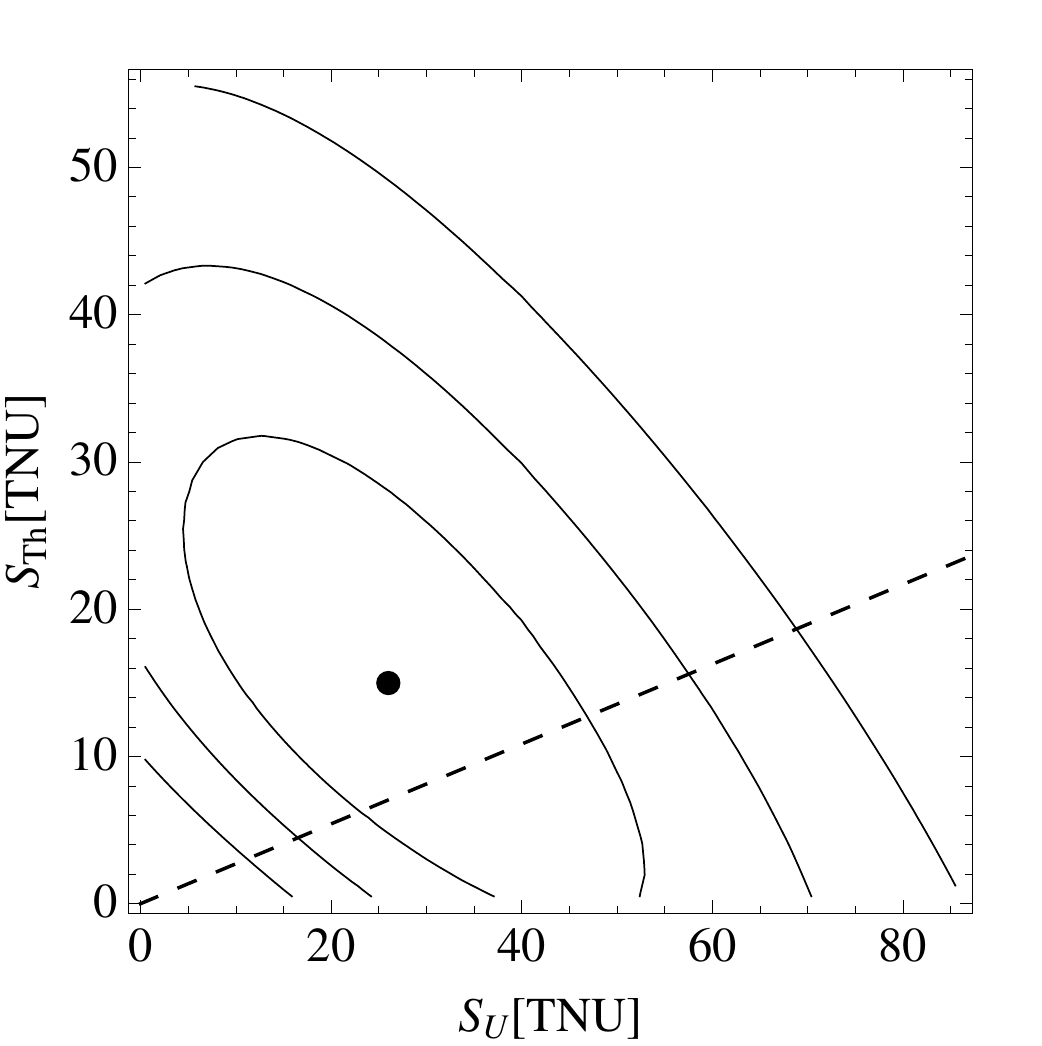}
\par\end{centering}
\caption{\label{fig:BorexContour}Best-fit contours corresponding to 1, 2 and
3$\sigma$ obtained by the unbinned likelihood fit with both the U and
Th signals left free parameters. The dashed line corresponds to the chondritic
Th/U mass ratio (Figure from \cite{Borexino2015})}
 
\end{figure}

A fit with the U and Th normalizations left as free parameters was performed by the Borexino collaboration in~\cite{Borexino2015}.
The U and Th best-fit contributions are shown in Fig.~\ref{fig:BorexContour}.
The obtained contours indicate a non-zero U signal at a confidence
level slightly above 1$\sigma$, and the Th contribution is still
compatible with the zero signal. The amount of geoneutrino events observed
in Borexino until now ($\sim$25) does not leave a chance to obtain a
reasonable estimate for the M(Th)/M(U), as its expected uncertainty
is approaching 100\% even in the absence of reactor antineutrinos.

KamLAND obtained an upper limit on the Th/U mass ratio $<19$ (90\% C.L.)
in~\cite{KamLAND2013}, indicating the separation of the U and Th antineutrino
signals. The analysis was updated using $\sim$2.5 years more data acquired
after all Japanese reactors were switched off~\cite{Watanabe2016}.
For the first time the positive indication of the U
signal at $3.44\sigma$ C.L. in the fit with a free Th/U mass ratio was reported. Also, the first "global" Th/U mass ratio measurement M(Th)/M(U)=$4.1_{-3.3}^{+5.3}$ was obtained
with the upper limit for the mass ratio of Th/U improved to $<17.0$ at 90\% C.L. All the values are consistent with the chondrite data and the BSE
models. It is the first Th/U mass ratio measurement which is not compatible
with zero at the 1$\sigma$ level. The M(Th)/M(U) ratio obtained by KamLAND
with $\sim$160 geoneutrino events is weaker compared to the simple
estimate presented above because of the presence of backgrounds
in the experimental data. japanese nuclear plants

\subsection{Obtaining radiogenic heat from the measured geoneutrino signal}

The total radiogenic heat is proportional to the total mass of U and
Th in the Earth. In contrast, as follows from formula (\ref{eq:Flux}),
the measured geoneutrino signal depends not only on the total mass
of U and Th but also on their distribution throughout the Earth.
The uncertainty in HPEs distribution translates into an uncertainty in the distribution of energy sources in the Earth. 
Therefore, the measured geoneutrino fluxes do not provide a unique measure of the heat production and can correspond to a range of heat power depending on the distribution of HPEs. 

Due to the fact
that the distribution of HPEs in the crust is relatively
well-studied, the variety of models is mainly defined by the
distribution of HPEs in the mantle. The extreme values
for the same amount of heating material can be obtained assuming either
a uniform distribution of U and Th throughout the mantle or a layered
one. 

Fof fixed amount of uranium in the mantle, the minimal
geoneutrino signal from the mantle would be produced by a thin uranium
layer at the bottom of the mantle (sunken-layer hypothesis). The
maximum geoneutrino signal corresponds to the uniform distribution
of the same amount of uranium in the mantle (homogeneous hypothesis).
These two extreme cases give respectively the mantle signal from uranium $S_{M}^{min}=11.3\times M(U)$
and $S_{M}^{max}=16.2\times M(U)$ TNU~\cite{Bellini2013}. 

The geoneutrino contributions from the crust and the mantle can be combined to obtain predictions of the surface geoneutrino flux for the 
known amount of HPEs, or, equivalently, for a fixed heat flow.
The U and Th in the crust should correspond to the observational values, and the rest of the U and Th within the values given by the BSE model is assigned to the mantle. The highest geoneutrino signal is modeled by assigning the maximum acceptable value to the crust and distributing the remaining fraction uniformly through the mantle. The lowest geoneutrino signal is obtained in a similar way, assigning the minimum value compatible with observations to the crust and distributing the remaining fraction in a thin layer at the bottom of the mantle. The total amount of radioactive elements is constrained by the measured heat flow 47\textpm{}2~TW. Reaching this limit would correspond to
a fully radiogenic Earth model. On the other hand, the minimal geoneutrinos signal is obtained with the minimal possible mass of U and Th in the crust and a negligible amount in the mantle.

In this way, the two extreme total signals, $S_{high}$ and $S_{low}$, expected at the chosen site for a given heat flow H(U+Th) can be obtained. In the plot Signal vs Heat the corresponding lines will bound the range of corresponding allowed values for geoneutrino signal and heat. The plot of is site dependent, though the slope of lines is universal.

\begin{figure}
\begin{centering}
\includegraphics[scale=0.32,angle=270]{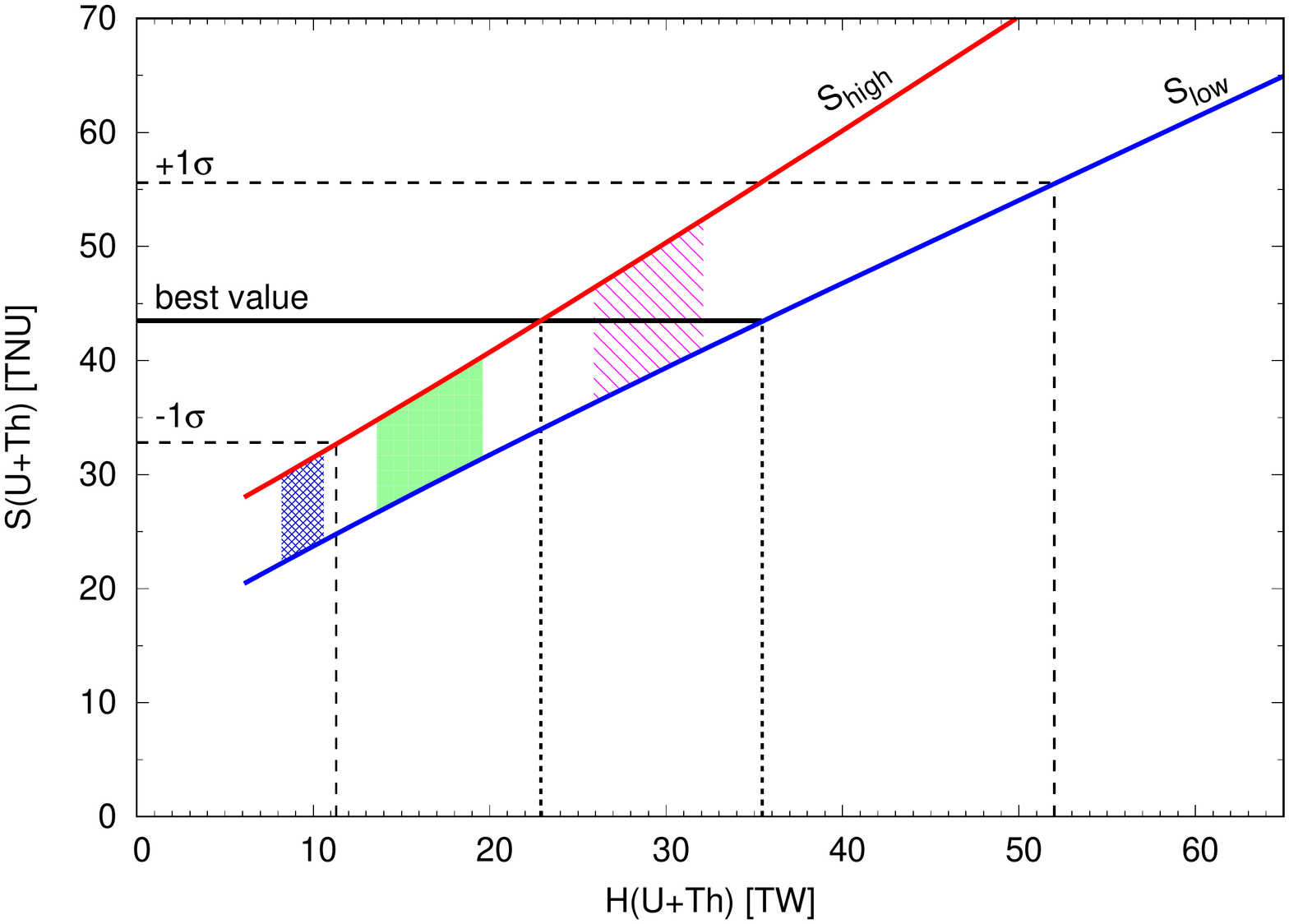}\includegraphics[scale=0.32, angle=270]{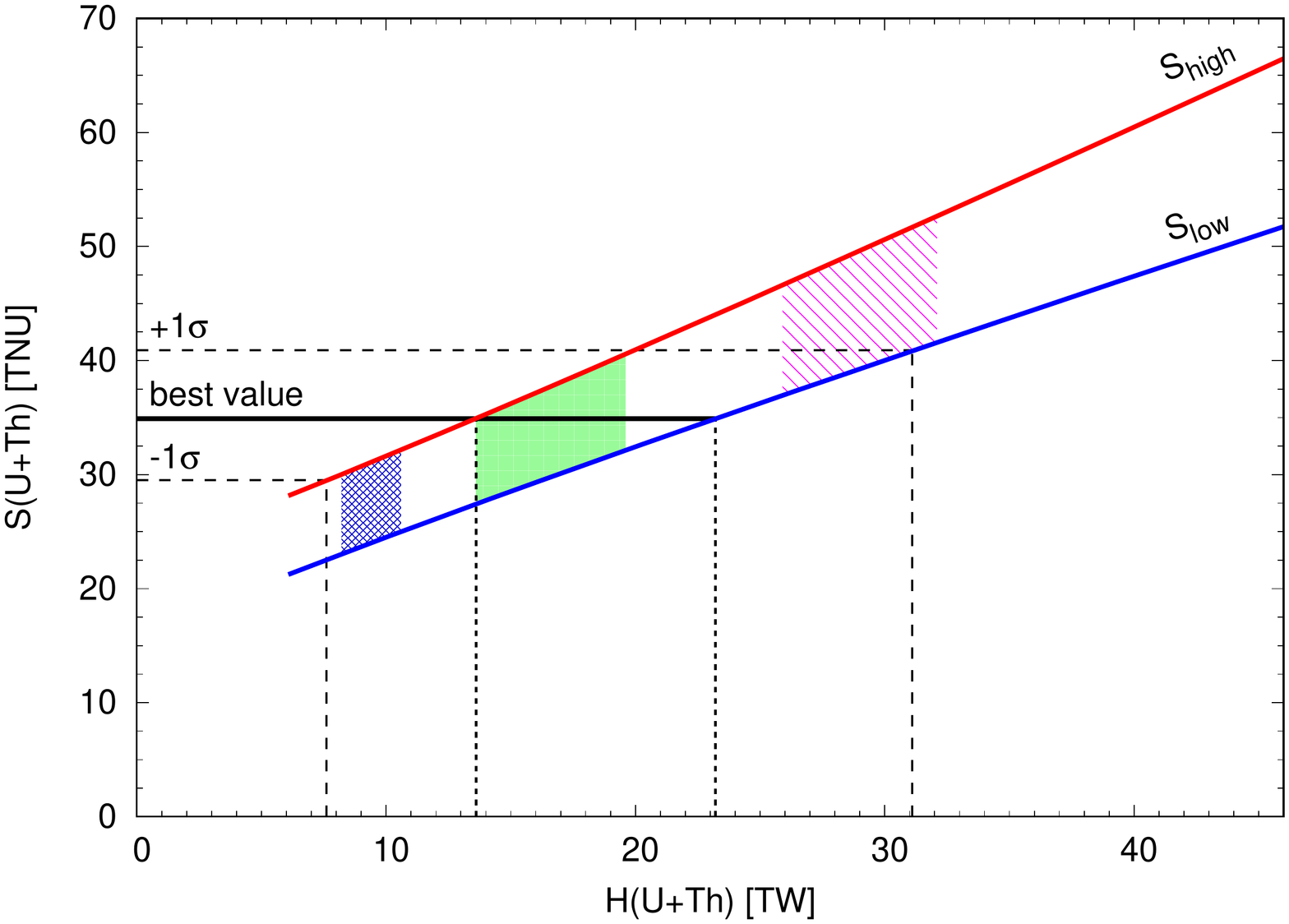} 
\par\end{centering}
\caption{\label{fig:Heat}The expected geoneutrino signal from U and Th in Borexino (left) and KamLAND (right)
as a function of radiogenic heat released in radioactive
decays of these chains~\cite{Borexino2015}. Three filled regions
correspond, from the left to the right, to the cosmochemical, geochemical,
and geodynamical BSE models. Best experimental values from Borexino
and KamLAND, together with 1$\sigma$ errors, are reported in corresponding
plots (Figures generously provided by Fabio Mantovani). }
\end{figure}

In Fig.~\ref{fig:Heat} the expected geoneutrino signal in Borexino
(left) and KamLAND (right) are shown as a function
of the produced radiogenic heat. The upper (red in the colour print) and lower (blue) lines correspond
to the high and low models, respectively. The error in the prediction
of the crustal signal is taken into account, and so are the different U
and Th distributions through the mantle, as described above. Three
filled areas in~Fig.~\ref{fig:Heat} represent three classes of
BSE models, based on the heat values from the classification by \v{S}r\'{a}mek
et al.~\cite{Sramek2013}: 11$\pm$2~TW, 20$\pm$4~TW, and 33$\pm$3~TW for the cosmochemical, geochemical and geodynamical estimates,
respectively. The horizontal lines represent the results
of Borexino of 2015~\cite{Borexino2015} and the latest result of KamLAND~\cite{Watanabe2016},
respectively. The Borexino results are compatible with all BSE models
within 1$\sigma$, while the KamLAND measurements are compatible with
all models within 2$\sigma$, slightly disfavouring the more energetic cosmochemical
ones. 

The radiogenic heat production for the sum of U and Th corresponding
to the central value of the Borexino data fit is in the range 23-36~TW, and 13-23~TW for the KamLAND. Considering the 1$\sigma$ interval and the minimum and maximum models, the 68\% C.L. interval for H(U+Th) radiogenic heat covers  11\textendash52~TW range for Borexino. Assuming the chondritic ratio for M(Th)/M(U) and M(K)/M(U)=10$^{4}$, the total measured terrestrial radiogenic
power is restricted from the Borexino data to H(U+Th+K)=33$_{-20}^{+28}$ TW at 68\% C.L., to be compared with the global terrestrial
power output of 47$\pm$2~TW. The KamLAND constraints for the radiogenic heat
are more stringent, H(U+Th) is in the interval 8\textendash31~TW at 68\%~C.L., see Fig.~\ref{fig:Heat}.

\subsection{\label{sec:Mantle-signal}Mantle signal}

The measurement of the geoneutrino signal from the mantle is a clue
to distinguish between the existing models (see \ref{subsec:Models}).
Both the Borexino and KamLAND collaboration provided estimations
of the mantle signal based on the total measured signal and the predictions
for the crust signal. Because of the dominating contribution from
the crust in both locations and the low available statistics, the
differential measurement has a high uncertainty. 

In Borexino the contribution to the total geoneutrino signal
from the local crust (LOC) $S_{geo}^{LOC}=(9.7\pm1.3)$~TNU was estimated by Fiorentini et al.~\cite{Fiorentini2012}\footnote{In this paper a first attempt to extract the mantle signal from the combined analysis of the first high significance Borexino and KamLAND data was performed}, while
the contribution from the rest of the crust (ROC) is $S_{geo}^{ROC}=(13.7\pm2.8)$~TNU according to Huang et al.~\cite{Huang2013}, and the total signal
from the crust is $S_{geo}^{Crust}=(23.4\pm2.8)$~TNU
\footnote{Fiorentini et al.~\cite{Fiorentini2012} reported separate values for
U and Th, $S$(U)=7.81$\pm$1.00~TNU and $S$(Th)=1.86$\pm$0.27~TNU,
obtained using $\left\langle{P}_{ee}\right\rangle $=0.55. The uncertainties are summed assuming fully
correlated U and Th contributions, and the total signal from the LOC
is $S_{LOC}$= 9.7$\pm$1.3~TNU. The best value of the
ROC is reported by Huang et al.~\cite{Huang2013}, in the row FFC
(for Far Field Crust, another term used for the ROC) of Table 2, $S$(U)=10.3$_{-.2.2}^{+2.6}$~TNU and
$S$(Th)=3.2$_{-0.7}^{+1.1}$~TNU. The total signal is $S$(U+Th)=13.7$_{-2.3}^{+2.8}$~TNU with the central value of $S$(U+Th) not equal to $S$(U) + $S$(Th) because
of the sum of the two non-symmetric distributions. In the Borexino
paper~\cite{Borexino2015} the highest value of the uncertainty was
considered, $S$(U+Th)=13.7$\pm$2.8~TNU. Since the two components
ROC and LOC are uncorrelated, the central values are summed and the uncertainties
are propagated in quadrature, resulting in $S_{Crust}$=23.4$\pm$2.8~TNU~\cite{Mantovani2019}.}, corresponding to the average predicted number of events from the
crust $N^{Crust}_{geo}$=12.75$\pm1.53$~events. Using the
experimental likelihood profile for the $N_{geo}$ and assuming the Gaussian
distribution of the $N^{Crust}_{geo}$ uncertainty and Poissonian
statistics for the number of detected events, one can obtain the 68\% C.L. interval for
the signal from the crust in Borexino, which corresponds
to 5.8-19.6 events with the maximum at 11.4 events. Recalculated in the
TNU, it is $S(Mantle)=20.9_{-10.3}^{+15.1}$~TNU.
The probability of observing at least one event from the mantle is
98\% \cite{Borexino2015}. 

\begin{figure}
\begin{centering}
\includegraphics[scale=0.27]{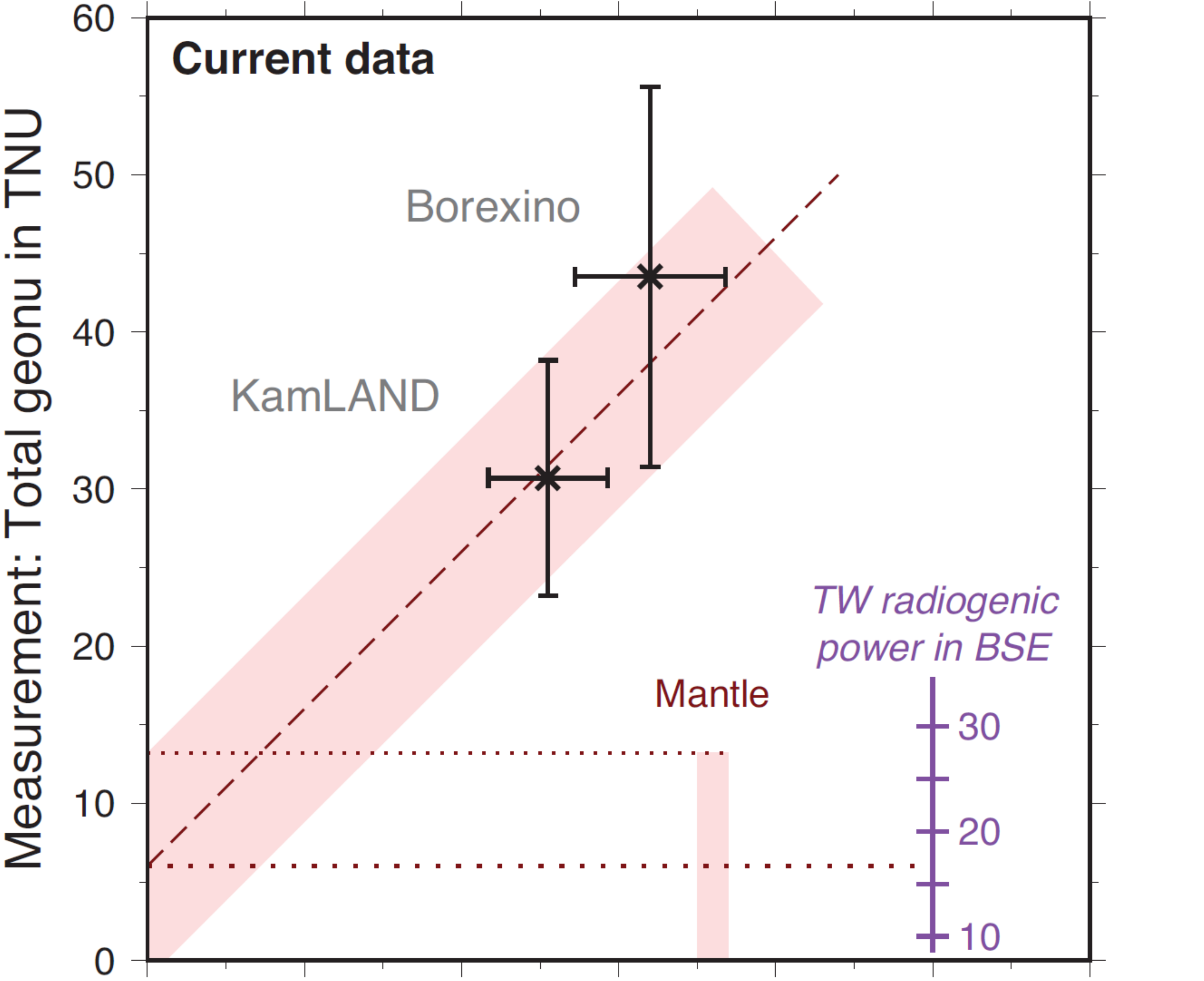}\includegraphics[scale=0.25]{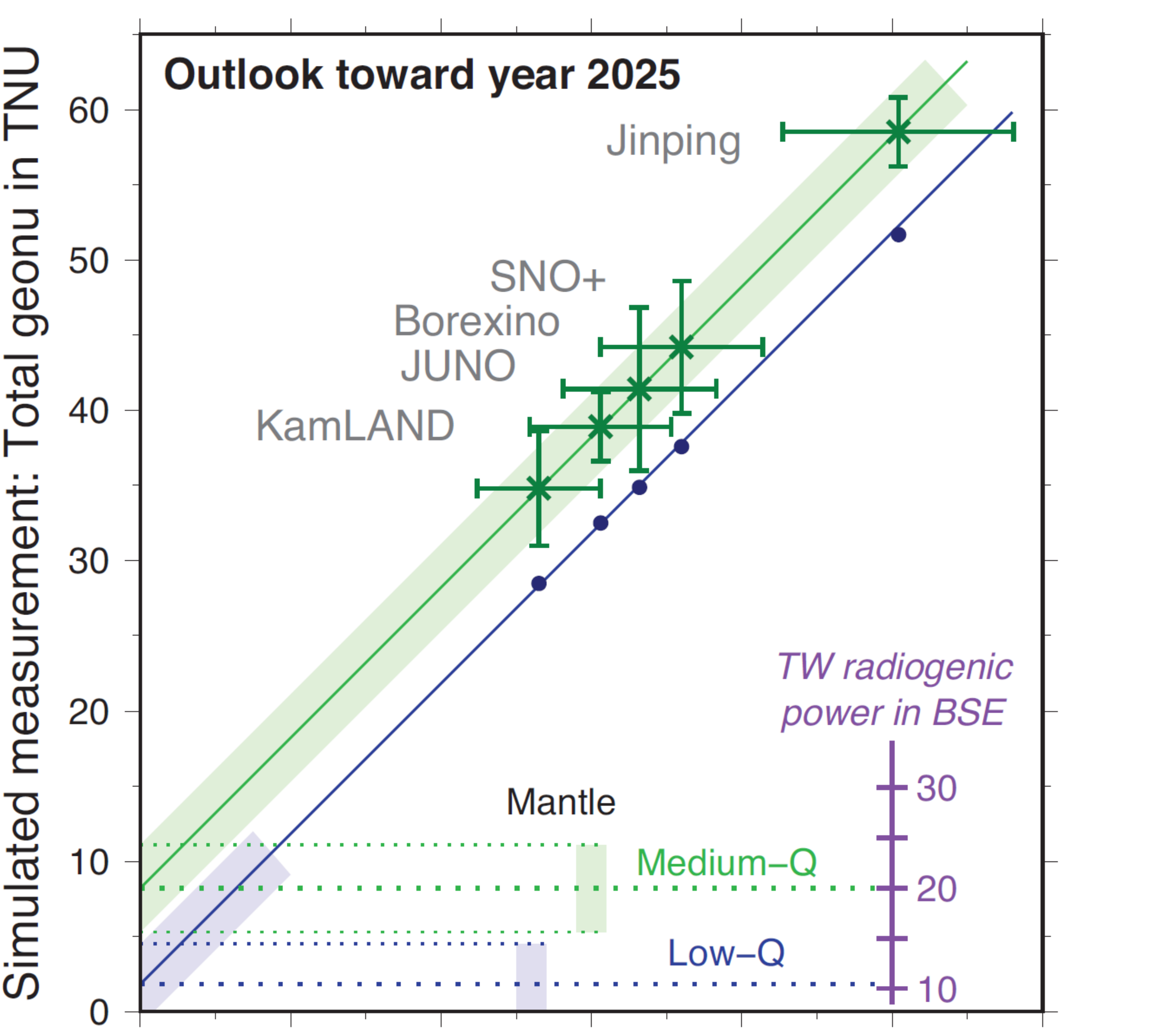}
\par\end{centering}
\begin{centering}
\includegraphics[scale=0.25]{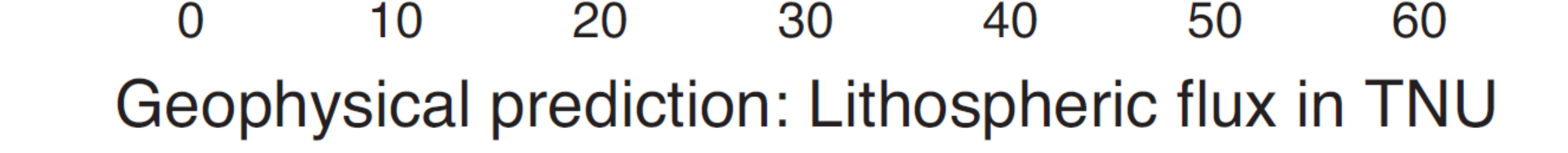}\includegraphics[scale=0.25]{Figures/Ondrej/4slopeone-axis}
\par\end{centering}
\centering{}\caption{\label{fig:ShramekMantle}\textbf{Left}: measured geoneutrino
signal in Borexino and KamLAND (2013 measurement) versus geological
predictions of the lithospheric flux. \protect \linebreak{}
\textbf{Right: }expected geoneutrino signals in existing (extrapolated to 2025) and future detectors versus geological predictions of the lithospheric
flux. (Figure from \v{S}r\'{a}mek et al. \cite{Sramek2016}, reproduced under
Creative Commons Attribution 4.0 International License \cite{CCL4.0}).}
\end{figure}

In Fig.~\ref{fig:ShramekMantle} the measured geoneutrino signal in Borexino
and KamLAND is plotted versus the geological predictions of the lithospheric
flux. In accordance with \v{S}r\'{a}mek et al.~\cite{Sramek2016}, analyses
of the existing data (KamLAND and Borexino combined) give a result with a
large uncertainty on the mantle flux, 6.0 \textpm{} 7.2 TNU\footnote{The analysis in \cite{Sramek2016} does not include the 2016 results of
KamLAND~\cite{Watanabe2016}.} (left plot in Fig.~\ref{fig:ShramekMantle}).

Future detectors will provide more opportunities for the mantle
signal separation. SNO+ in Sudbury (Canada) will soon start taking
data. Another detector sensitive to the geoneutrino, JUNO, is
under construction in the Jiangmen laboratory in Southern China. The mantle signal
at Jiangmen is a relatively small fraction ($\sim$14\%) of the total
geoneutrino signal, but a combination of the JUNO measurement with
those of other experiments will allow extracting the mantle contribution. Figure~\ref{fig:ShramekMantle} (from
\v{S}r\'{a}mek et al.~\cite{Sramek2016}) demonstrates the underlying principle.
The measured geoneutrino flux is plotted versus the geological estimate
of the flux from the lithosphere that does not include the contribution
from the convecting mantle. As a result, a best-fit line with a slope
of 1 will intersect the physics input axis at the point corresponding
to the mantle contribution. The mantle signal is reasonably assumed
to be the same for all the considered locations. The range of possible
interceptions provides the total uncertainty that includes
both the uncertainties of the measurements and the uncertainty of
the lithospheric flux predictions. 

In order to extract statistically significant information on
the mantle contribution from the measured total signal, a more accurate
estimation of the geoneutrino flux from the local area at each site is required. For this purpose a broad collaboration of particle physicists earth scientists is necessary, joining their efforts to provide surveys and descriptions of the geology, seismology, heat flow, and geochemistry of the regional lithosphere.

\subsection{Search for the hypothetical geo-reactor}

The nuclear reactor at the Earth's core was theoreticized
by Herndon \cite{Herndon1996}. Hollenbach and Herndon \cite{Hollenbach2001}
showed that a planetary-scale nuclear reactor could operate over the
lifetime of the Earth. The theory requires some U and Th sink to the core
instead of accumulating in the crust rising from the early molten
Earth as a slag. The proto-Earth geo-reactor hypothesis has been recently
reassessed by Degueldre and Fiorina \cite{Degueldre2016}. They found
its build-up and subsequent operations unlikely. Nevertheless, there is no experimental evidence against the hypothesis. The natural nuclear reactor would
explain the peculiarities in the isotopic content of the Earth, in particular
the observed high $^{3}$He/$^{4}$He ratio found in volcanic hotspot
lavas, like those erupted at Hawaii and Iceland, which is up to 50 times
the present atmospheric ratio. The geo-reactor would explain the ratio
in a natural way as $^{3}$He would be a product of tritium decay in reactor. As discussed by Raghavan \cite{Raghavan2002},
this hypothetical source in the thermal
power range of 1 to 10 TW could be the power source driving the deep Earth plumes and
the Earth's geo-dynamo mechanism responsible for the Earth's
magnetic field. Hypotheses for presently existing natural breeder
reactors propose deep-earth locations, including the centre of the
core~\cite{Herndon1996}, the inner core boundary~\cite{Rusov2007},
and the core-mantle boundary~\cite{deMeijer2008}. Herndon notes in
\cite{Herndon2014} that if a nuclear fission reactor should occur
in those places, then geo-reactors would anyway melt down to the Earth's
centre.

The KamLAND data were used to test the hypothesis of a natural
nuclear reactor in the Earth's core assuming its stable
power output over the data-taking period and $^{235}$U:$^{238}$U
= 0.75:0.25 composition. The modelled contribution from the reactor
was allowed to vary in the fit together with contributions from U
and Th geoneutrinos. The oscillation parameters were constrained
from the solar, accelerator, and reactor antineutrino data. The limits
on the geological reactor power were obtained from the fit, $<3.1$~TW
at 90\% C.L. (or $<3.7$~TW at 95\% C.L.)~\cite{KamLAND2013}, improving
the previous results. 

A geo-reactor with two dominant fission nuclides $^{235}$U (76\%) and $^{238}$U (23\%) composition (as discussed by Herndon and Edgerley~\cite{Herndon2005}) was
excluded by Borexino using the same likelihood fit as for the geoneutrino
analysis with the fixed mass ratio M(Th)/M(U)=3.9, adding the simulated geo-reactor
spectrum in the fit. The number of reactor events $N_{R}$ was constrained
to the expected value of $N_R=33\pm2.4$ events. The upper limit on the geo-reactor power is
4.5~TW at 95\%~C.L.~\cite{Borexino2013}, close to the one obtained by KamLAND.

Dye \cite{Dye2008} discussed a method for the geo-reactor detection by
analyzing the characteristic oscillation pattern in its energy spectrum.
Indeed, if the geometrical size of the natural reactor is small compared
to the length of oscillations and the distance to the detector is
fixed, the geo-reactor energy spectra should demonstrate oscillation
patterns. Characteristics of this distorted spectrum could reveal the
distance to the geo-reactor with an uncertainty of $\sim$150 km, and
using estimates from several Earth-surface sites locate a solitary
geo-reactor by triangulation, discriminate the models, and provide
precision power measurement. The method to locate an electron
antineutrino source by measuring spectral distortions utilizes the
Rayleigh test. This statistical test returns the power of spectral
distortions for an assumed length of the neutrino path. The amplitude
of distortions depends on the energy resolution; hence, the future
high resolution JUNO detector (see \ref{subsec:JUNO}) will have advantages
when applying the method. The exposure necessary to locate a geo-reactor
at the centre of the Earth (6371~km), close to the inner core boundary
(5149~km), and at the core-mantle boundary (2981~km) with the probability
of 95\% using the JUNO type detector (3\%/$\sqrt{E}$ energy resolution
and $\sim10^{33}$ target protons) is found to be 20, 6, and 0.4~
$10^{33}$ protons$\cdot$yr for 1~TW reactor, respectively. 

\section{New detection methods}

Antineutrino detection with IBD detectors has clear advantages:
the LS is cheap allowing construction of detectors with target media
large enough for detection of antineutrinos, the main antineutrino
background comes from reactors and can
be reduced by choosing locations far away from the operating nuclear reactors; moreover,
the reactor energy spectra are well-understood and predictable. Unfortunately,
the high energy threshold of the IBD on protons makes impossible the
detection of geoneutrinos from the third most important contributor
to the geoneuino spectra, $^{40}$K. That is why physicists are
looking for detection methods with a threshold below the $^{40}$K
end-point energy.

Some recent proposals are concerned with the directionality. The methods are
based either on the modification of the IBD detectors (sectioning
or doping with $^{6}$Li) with the purpose of improving the angular sensitivity
or on the application of other known detection techniques, e.g., gas-filled
TPC. The latter uses antineutrino- electron scattering and suffers
from the strong irreducible background from the solar neutrinos that
are both much more numerous and have a factor of $\sim$6 higher interaction cross
sections in the region of interest.

\subsection{Directionality of the geoneutrino signal}

\begin{figure}
\begin{centering}
\includegraphics[scale=0.54]{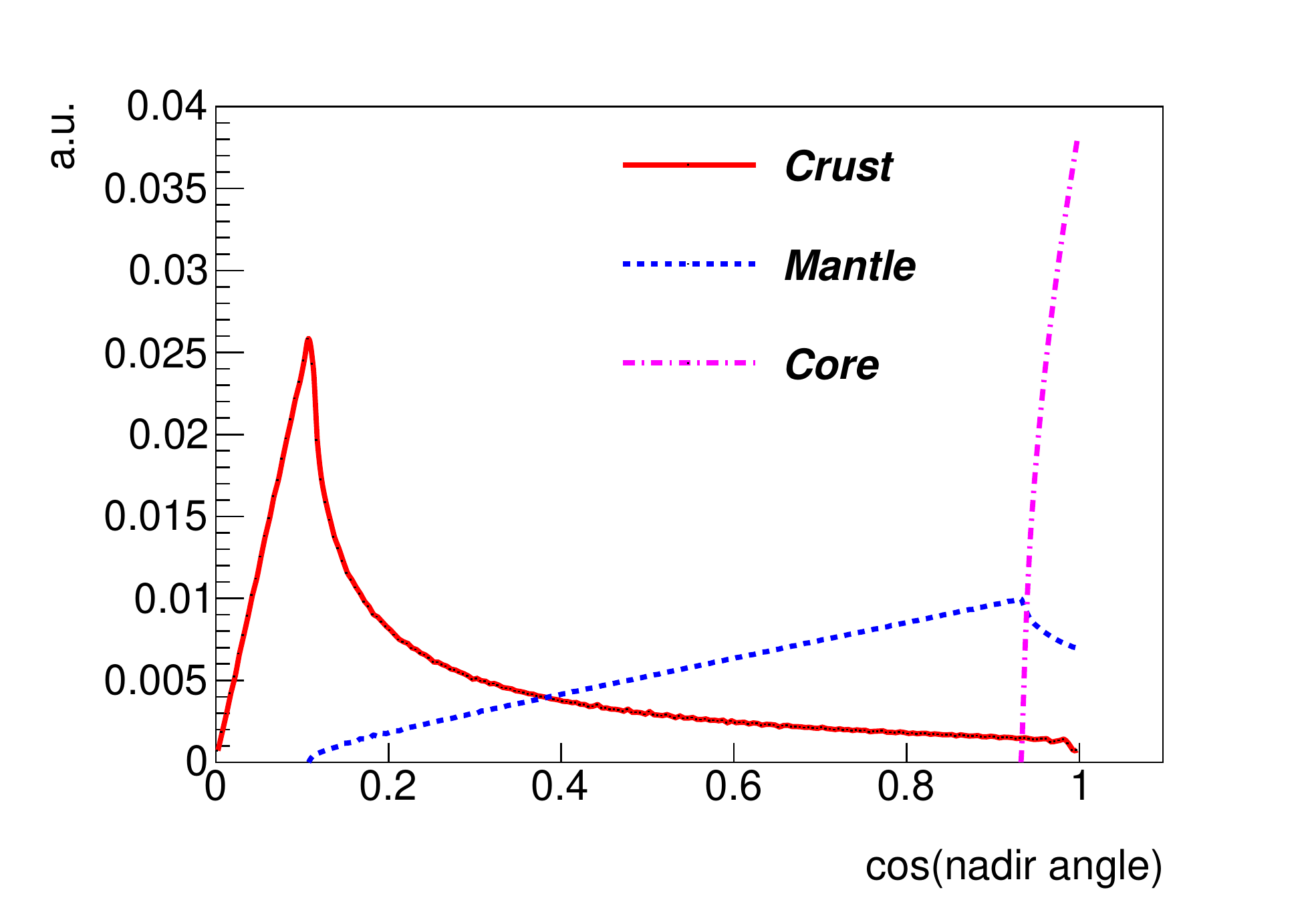}\includegraphics[scale=0.25]{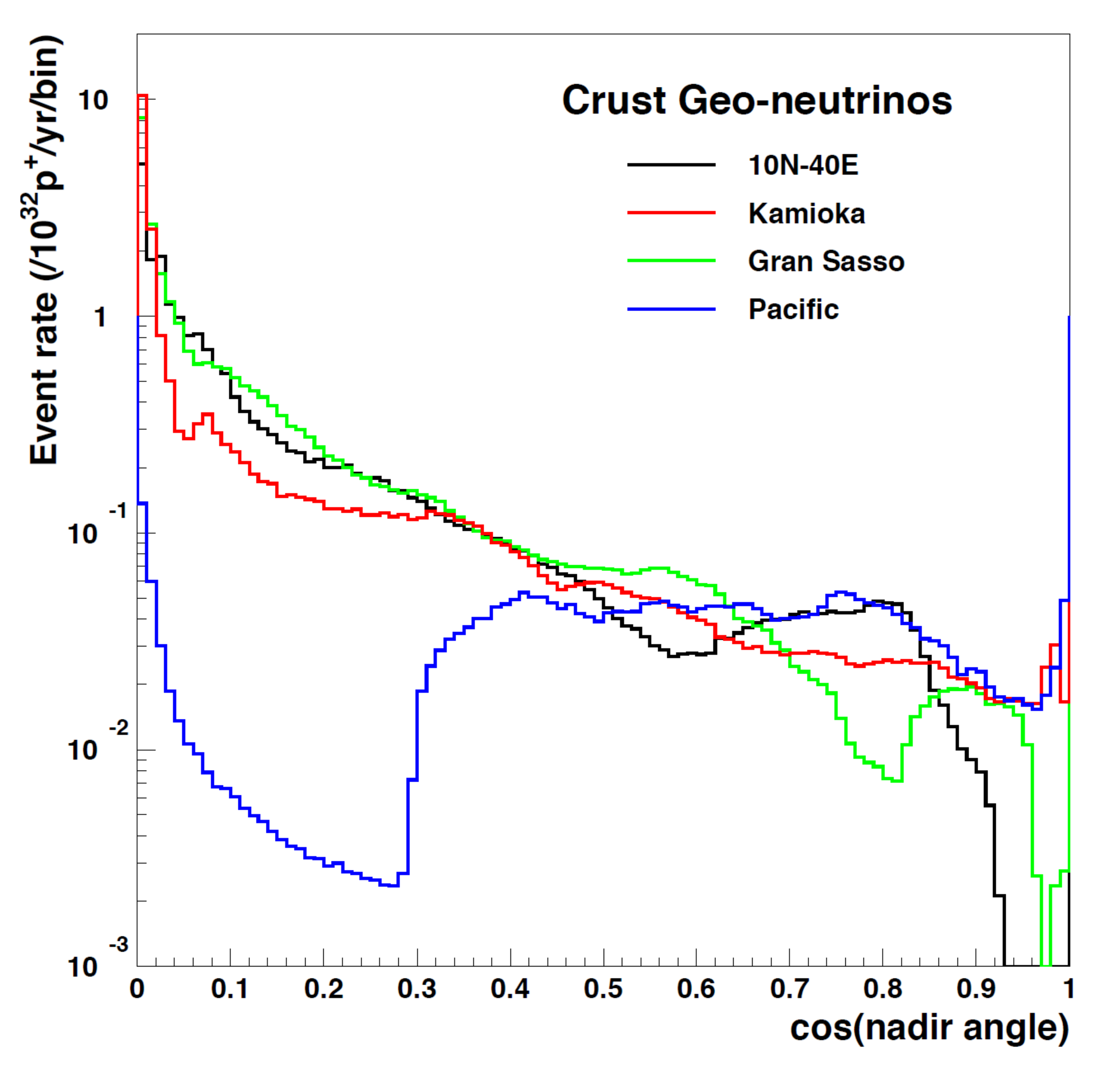}
\par
\end{centering}
\caption{\label{fig:Angles}\textbf{Left}: zenith angle distributions of geoneutrinos
produced in the uniform crust (red), mantle (blue) and core (magenta)
calculated in a simplified model with uniform distribution of heating
elements. 
\protect \linebreak{}
\textbf{Right}: predicted zenith angle distributions of crustal geoneutrinos
at some geographic positions. (Figure from~\cite{Dye2015}. (For interpretation
of the references to colour in this figure legend, the reader is referred
to the web version of this article)}
\end{figure}

Directional measurement of the antineutrino events would provide additional information that can be used to disentangle geoneutrinos from isotropic background and reactor antineutrinos whose points of origin are known. Moreover, it could help separating crust and mantle components in the total geoneutrino flux. Simple model calculations of the angular
distribution of the antineutrinos originating in the crust, mantle,
and hypothetically in the core of the Earth are shown in~Fig.~\ref{fig:Angles}. Uniform distribution of HPEs in the 40~km crust, mantle, and core is assumed in calculations. The area under first two curves is normalized to unity, and the core distribution is reduced by a factor 3 for the sake of better visual presentation. It can be seen that while the angular distribution of crustal geoneutrinos concentrates close to the horizon, the angular distribution of mantle geoneutrinos spreads over a range of angles below, and the hypothetical core contribution is expectedly concentrated close to vertical. In general, because about 25\% of the signal arrives from the distances comparable to the crust thickness, one should expect the same fraction of the signal to be distributed almost uniformly for the continental crust. The "ideal" crust distribution peaks at the direction of tangent to the mantle from the observation point, in this direction the line of sight is longest. For the thickness of the uniform crust of 40 km, this point corresponds to $\cos\theta\sim 0.09$.  As shown in~\cite{Fields2006} the rise of the angular distribution is linear before this point. For the uniform crust the peak would thus give the position of the inner edge~\cite{Fields2006}. The peak in the crust angular distribution corresponds to the maximum possible nadir angle from the mantle, the feature is clearly visible in the mantle angular distribution. The angular distribution for the core would concentrate at the vertical direction, with minimum $\cos\theta$ about 0.94. The corresponding peak in the mantle distribution is expected at this value, indicating the corresponding boundary.

The peak at small angles expected for the uniform approximation doesn't appear in more detailed calculations, see left plot in~Fig.~\ref{fig:Angles}. The crust is not uniform, the highest abundances of HPEs are expected in the upper crust and sediments, and the "horizontal" signals from the thin upper layer, especially in the near field, dominate in the distribution~\cite{Dye2019}.

The absence of U and Th in the core is generally assumed, but there exists experimental evidence that potassium can form alloys with iron~\cite{Lee2003,Rama2003}, and so the possibility that the Earth's core might contain significant amounts of $^{40}$K is not completely excluded.
Nimmo et al.~\cite{Nimmo2004} argued that along with other explanation the presence of the potassium in the core at the level $\sim$400~ppm would satisfy the present-day inner core size, surface heat flux, mantle temperature and cooling rate, and positive core entropy production.

Sensitivity to the mantle geoneutrino signal at the continental sites
is masked by a much stronger signal from the crust, as at least 75\% of
the flux come from the crust. Measurements with deep underwater detectors on the seafloor where the crustal contribution is minimal, as proposed by Dye et al.~\cite{Dye2006}, would be a straightforward solution,
but practical realization of the proposal is problematic. Exploiting
the geoneutrino directionality would be a good alternative to measuring
far away from continents. This could be achieved with a directional
detector capable of resolving the angular distribution of incoming
antineutrinos. Determining the direction of the incoming geoneutrino
was considered just after the first report of KamLAND on geoneutrinos
by Fields et al.~\cite{Fields2006} and Suzuki~\cite{Suzuki2006}. 

\begin{figure}
\begin{centering}
\includegraphics[scale=0.6]{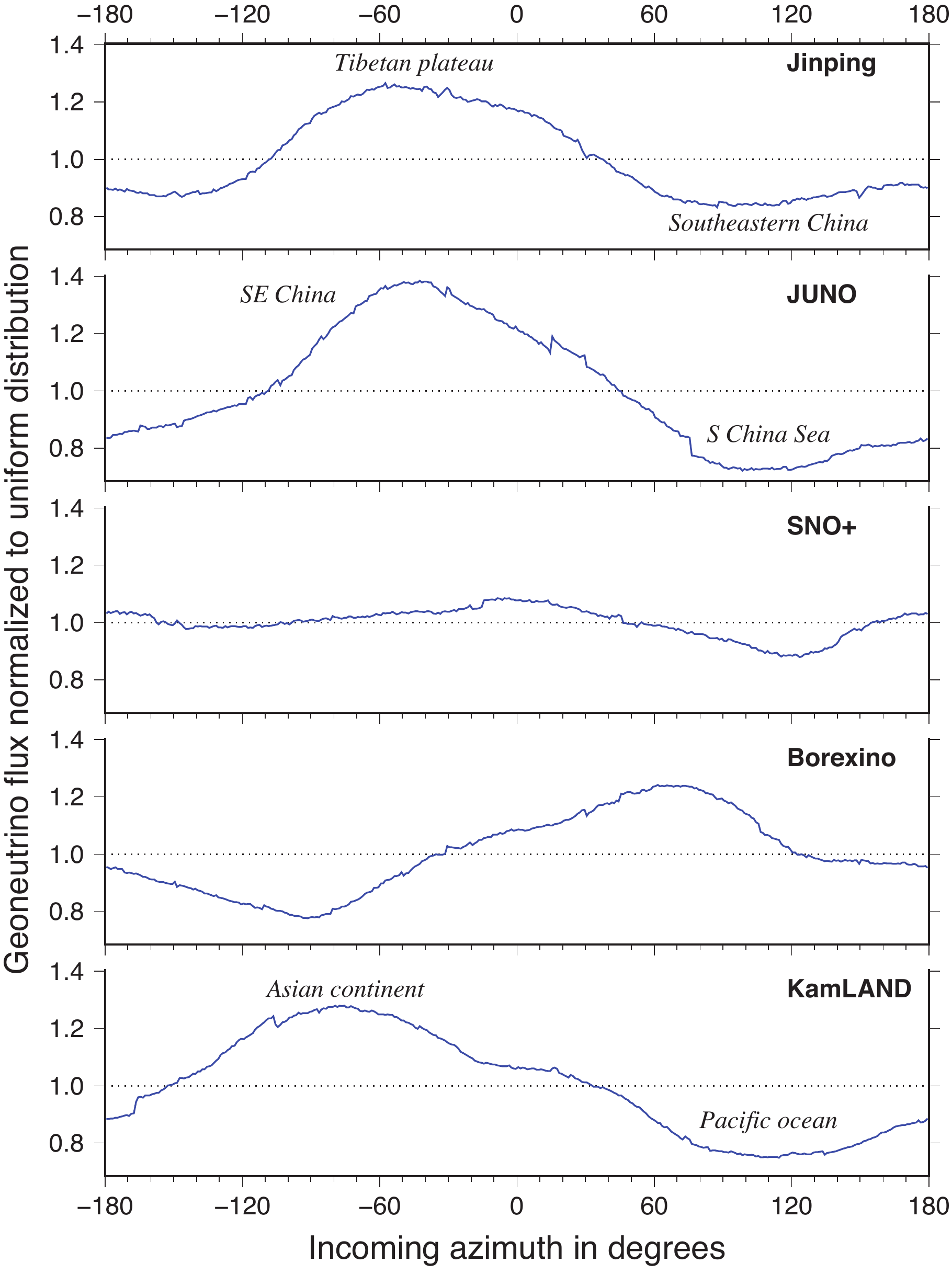}
\par\end{centering}
\caption{\label{fig:Shramek1}Predicted signal at 5 laboratories as a function
of azimuth of incoming geoneutrino. Signal in each bin is shown with respect to the one expected in a case of uniform distribution at the corresponding detector. (Figure from \v{S}r\'{a}mek et al. \cite{Sramek2016}, reproduced
under Creative Commons Attribution 4.0 International License \cite{CCL4.0}).}
\end{figure}

Calculations using the detailed topographic information show
that the distribution of geoneutrino signals demonstrate detectable
anisotropy not only in the nadir angle but also in the azimuthal one.
Calculations of azimuthal anisotropy of incoming geoneutrinos
were performed by \v{S}r\'{a}mek et al. \cite{Sramek2016}. As one can
see in Fig.~\ref{fig:Shramek1}, all existing sites but SNO demonstrate
significant azimuthal anisotropy due to the vicinity either to the
thicker crustal structures (Jinping) or to the thinner oceanic crust
(JUNO, KamLAND). 

If measured, the angular distribution of geoneutrinos can be inverted to give
tomography of the terrestrial radioisotope distribution \cite{Fields2006}.

\subsection{Directionality in IBD reaction in liquid scintillator}

Because the only experimental data on geoneutrinos acquired thus
far are collected with IBD detectors, it will be natural to examine
their angular sensitivity more closely. In fact, displacement of neutrons
in the direction of incoming antineutrinos was observed in reactor
experiments: the G\"{o}sgen, Bugey, and Palo Verde collaborations registered
neutron displacement for segmented scintillator detectors by detecting
the neutron in a segment farther away from the reactor with respect
to the segment where the positron was observed. The unsegmented CHOOZ
experiment reported measurement of the average neutron-positron
separation and determination of the incoming antineutrino direction
with an uncertainty of 18\textdegree{} (half-aperture of a cone) at
the 68\%~C.L. with 2700 events statistics~\cite{Apollonio2000}. 

Using the results of Vogel and Beacom \cite{Vogel1999,Kurylov2003},
one can evaluate the total cross section as well as the positron angular
distribution. As a rule, the positron angular distribution is not
detectable in IBD experiments. However, the angular distribution of
the recoil neutrons is accessible. Since in the laboratory system
the proton is at rest, the neutron is initially emitted at a forward
angle with a maximum at

\begin{equation}
\cos\theta_{n}=\frac{\sqrt{2E_{\nu}(m_{n}-m_{p})-((m_{n}-m_{p})^{2}-m_{e}^{2})}}{E_{\nu}}. \label{eq:Theta_n}
\end{equation}

At the IBD threshold the neutron emission direction is exactly forward, and at the
reactor antineutrino energies (below 10 MeV) it is still mainly in
the forward direction. The average angle $\overline{\cos\theta_{n}}$ is naturally
closer to unity than its maximum value defined by (\ref{eq:Theta_n}). 

The basic idea of the directional studies is to search for the statistical
displacement of the neutron with respect to the prompt event ($\overrightarrow{\Delta R}=\overrightarrow{R_{prompt}}-\overrightarrow{R_{delayed}}$).
The displacement of the emitted positron is negligible and can not provide
any reasonable information regarding the direction of the incoming antineutrino.
In contrast, the neutron is emitted in a relatively narrow range of angles
(within $\sim$55 degrees \cite{Batygov2006}) around the direction
of the antineutrino with energies up to 100 keV for reactor antineutrinos.
The emitted neutron is slowed down in the proton-rich media. The
average angle in a single collision is $\overline{\cos\theta_{n}}=\frac{2}{3A}$.
In scattering on protons (atomic number A=1) it is possible for the neutron 
to preserve directionality. The slowdown occurs very quickly, and the displacement
from the production point is determined by the first few scatterings,
still preserving the initial direction. The later isotropic diffusion
process does not change the average displacement, which, in the case
of CHOOZ, was observed to be 1.7~cm.

The position of the event can be reconstructed in liquid scintillation detectors with
a typical precision of 10-15 cm. The single 2.2 MeV $\gamma$ from the
neutron capture has a comparable mean free path, adding uncertainty
to the reconstructed point of origin, as it can travel up
to 1 meter away. Also, the prompt event is not a point-like energy
deposit, as it is a composition of the energy deposit of
the positron kinetic energy and the 511 keV gammas from the positron
annihilation. The 511 keV gammas are emitted in opposite directions, and
thus the centre of gravity of the deposited energy should, in general,
coincide with position of the event, though some additional uncertainties
to the position reconstruction of the parent event will be added. Simulations
show that the typical displacement of two reconstructed vertices
of the IBD pair of events is $\sim$1.5 cm in the organic scintillator,
much lower than the value that can be detected for a single event. Assuming large
sample of the events, a statistical analysis is possible. As a na\"{i}ve
estimation, with 100 events one can expect 1$\sigma$ sensitivity to this displacement with 15 cm vertex reconstruction precision. Due to the fact that much larger datasets are accumulated with antineutrino reactor experiments,
detection of the displacement does not look impossible.

Feasibility of the directionality technique was studied
by Batygov using Monte-Carlo simulations of a large volume IBD detector~\cite{Batygov2006}. 
MC predicts a statistically measurable delayed
signal displacement in the direction of the incident antineutrino,
making directionality studies with unloaded liquid scintillation detectors possible.
As expected, the effect is small compared to the natural spread of
the reconstructed signals. Obtaining a statistically significant result
would require thousands of antineutrino events even for the simplest
case of a point-like antineutrino source. For KamLAND-sized detectors
this implies unrealistic run times of hundreds of years. A larger detector
placed near the operating reactors may provide enough statistics
within a reasonable time. Similar conclusions were obtained in another study
by Domogatski et al.~\cite{Domogatski2006}. The authors state that
information obtained with one 30-kt target mass unloaded liquid scintillation detector
using directional separation of the incoming geoneutrino flux is useful
but limited. More definite information could be obtained only with
a much larger detector. 

To solve the problem of low statistics, two methods were discussed in~\cite{Batygov2006}. 
A bigger, as large as 100-kt, unloaded liquid scintillation detector
should collect enough events for simple terrestrial antineutrino directionality
studies in several years, provided the vertex reconstruction accuracy
is not worse than that in KamLAND and the background is low enough.
Another improvement can be achieved by minimizing the spatial spread
of the delayed event with specially chosen LS loading. Loading the
LS with $^{6}$Li will provide a $^{3}$H+$\alpha$ delayed signal instead
of a single gamma from the neutron capture on the proton. The loaded LS
should offer about two to four times better statistical efficiency
of incident angle determination
compared to unloaded scintillators. In order to take full advantage
of the method, high vertex reconstruction quality is needed, which is especially
important in this case. Simulations show that directional studies
are highly affected by the presence of backgrounds. Thus, special
care should be taken to avoid backgrounds, including the reactor
background and internal contamination. 

The idea of doping the LS with $^{6}$Li has been recently considered
more thoroughly by Tanaka and Watanabe \cite{Tanaka2014}. They studied the resolving power of a $^{6}$Li loaded liquid scintillation detector for the
detection of a hypothetical magma chamber under the Hida Mountains
located in the northern Japanese Alps. The estimated size of the magma
chamber is $\sim$16000 km$^{3}$. Assuming the average U and Th concentrations
equal to the surface values (5 ppm and 20 ppm), it can be estimated
that the expected contribution of the magma chamber to the antineutrino
flux is $\sim$10\% of the entire geoneutrino flux expected at the
KamLAND detector, namely 2.0 TNU and 1.6 TNU from the U and Th decays, respectively,
versus total of 38.5$\pm$7.7 TNU. Simulations show that a 3-kt
detector located at the KamLAND site can separate the magma chamber
contribution from the total geoneutrino flux at a $3\sigma$ confidence
level in 10 years. The technique is applicable to resolving crust
contributions from the mantle, as well as neutrinographic imaging
of the Earth's interior. \v{S}r\'{a}mek et al. \cite{Sramek2013}
estimated that the flux contribution from two large low shear velocity
provinces (LLSVPs) is comparable to the flux contribution from the
hypothetical magma chamber. Such excess fluxes can be imaged with
these directionally sensitive techniques.

A modification of existing antineutrino detection techniques based
on IBD and, according to the authors, better suited for directional studies
was proposed by Safdi and Suerfu\cite{Safdi2015}. They suggest using segmented detector with separate target and capture planes,
calling it SANTA (segmented antineutrino tomography apparatus). The
neutrons in this detector are captured in the thicker boron-loaded
layers. The location of the IBD event and the momentum of the positron
are determined by tracking the positron's trajectory
through the detector. The method's capability was demonstrated through
MC simulations. The use of the SANTA technique for the geoneutrino
detection will necessarily demand a lot of physical space. A kilotonne
scale SANTA for the geoneutrino detection consisting of a stack of 1000
modules with 10x10 m cross-section and 1 cm thick would stretch
for around 1~km, making the practical application of the technique for
the geoneutrino study questionable. 

While practical directionality measurement is impossible with existing
detectors, including KamLAND, the scientific significance of such
research would justify construction of detectors better suited
for this purpose.

\subsection{Water based IBD detectors}

Water is an inexpensive proton-rich medium that would be of a great benefit
in constructing extra-large geoneutrino detectors. Pure water
is used in large volume (50 kt) \v{C}erenkov detector
Super-Kamiokande~\cite{SuperK2011} to study the high-energy portion of the solar
neutrino spectrum. The detector registers electrons scattered by solar
neutrinos. Directionality of the \v{C}erenkov light is used to suppress
backgrounds. Due to the low light yield, the energy threshold in
water \v{C}erenkov detectors is typically too high (above 3.5 MeV)
to detect correlated pairs of events caused by geoneutrinos. An advanced
study of neutron tagging in a water \v{C}erenkov detector was reported
by the Super-Kamiokande collaboration \cite{Watanabe2009}. The tagging
efficiencies of thermal neutrons was evaluated in the 0.2\% GdCl$_{3}$-water
solution and in pure water. They were determined to be 66.7\% and
20\%, respectively, for events above 3 MeV with the corresponding background
probabilities of $2\times10^{-4}$ and $3\times10^{-2}$. This is
still not enough to challenge the detection of antineutrino events
in the energy range of interest for the geoneutrino studies.

The amount of light produced by \v{C}erenkov radiation is 50 times
lower than in liquid scintillators, and a significant fraction of light
is emitted in the UV region, where water loses transparency and
PMTs are not sensitive. Adding a wavelength shifter to water helps
to register photons otherwise invisible to PMTs, increasing the amount
of detected light by a factor of 2-3~\cite{Dai2008},\cite{Sweany2011}.
These studies were aimed at improving the sensitivity
of existing water \v{C}erenkov detectors, but not for the geoneutrino
studies. Adding chemicals to water can reduce transparency, which
influences the sensitivity.

A more promising approach to using water-based liquid scintillator
(WbLS) has recently been discussed by Alonso et al.~\cite{Alonso2014}. WbLS is
a novel scintillation medium for large liquid detectors, in which
scintillating organic molecules and water are co-mixed using surfactants
\cite{Yeh2011}. The possible mixture extends from pure water to pure
scintillator, offering a great possibility of tuning the detection
media to fit the physics programme. Compounds with a low content of LS
will work in a regime close to that of the water \v{C}erenkov detectors,
but offering higher light yield. In contrast, liquid scintillator
detectors with the addition of water will gain directionality and
metal-loading capability. A variety of solvents are available. In
addition, different wavelength shifters and other additives can be
used to adjust the timing properties of the final mixture.

The development of WbLS with a low attenuation length will allow the
construction of large-volume multipurpose detectors, suitable for the
geoneutrino studies.

\subsection{IBD and resonant electron capture on other nuclear targets}

IBD and resonant electron capture on other nuclear targets in view
of the geoneutrino detection were discussed in early papers~\cite{Krauss1984,Kobayashi1991}.
A list of possible nuclear targets can be found in Table~2 of the seminal article by Krauss
et al.~\cite{Kobayashi1991}. In general, the majority
of considered targets will need huge quantities of material, orders of magnitude larger than those needed for detectors exploiting IBD of
protons, to provide a comparable counting rate. The lowest energy
threshold of 1.041~MeV is provided by a $^{3}$He target. This would
allow the detection of antineutrinos from $^{40}$K. Unfortunately,
the necessary amount of $^{3}$He is beyond modern capabilities.

The use of a low-threshold target isotope $^{106}$Cd for antineutrino detection, not listed by Krauss et al. in~\cite{Krauss1984}, was proposed by Zuber~\cite{Zuber2003}\footnote{The use of the isotope for geoneutrino detection was first discussed by Chen~\cite{Chen2005}}. The natural abundance
of $^{106}$Cd is very low, 1.25\%, and enrichment will be needed.
The $^{106}$Cd isotope is one of the most suitable nuclei to search
for the double beta plus processes having a high-energy release in
the decay. It is also favoured for the possible resonant $0\nu2\epsilon$
transitions to excited levels of $^{106}$Pd. Serious efforts are
applied in order to develop large-scale set-ups for the neutrinoless
double-beta decay searches. A cadmium- tungstate (CdWO$_{4}$) solid state
215~g detector with an enrichment of 66\% in $^{106}$Cd is available
for tests~\cite{Belli2010}. It was developed within the framework of the collaboration
between the DAMA and INR-Kiev groups. CdWO$_{4}$ is a high density (7.9~g/cm$^{3}$), high atomic number scintillator. It emits light at a
wavelength of 475~nm and has a total light yield of about 15~photons/keV.
The decay time was found to be $\sim$15~$\mu$s for different source
energies. The possibility of using CdWO$_{4}$ crystals to construct both 
segmented and unsegmented detectors is being analyzed. The decay time
can be shifted to the nanosecond timescale using dopants during crystal
production. A fast scintillator can be used to reconstruct the vertex
of the events \cite{Szczerbinska2011}. There are currently two other
experiments in progress searching for the double beta decay of $^{106}$Cd:
the TGV-2 experiment in the Modane underground laboratory in France
utilizing 32 planar HPGe detectors with 16 thin foils of enriched $^{106}$Cd
(23.2~g) installed between the detectors~\cite{Briancon2015}, and
the COBRA experiment, using an array of CdZnTe semiconductors ($\sim$380~g of natural Cd) at the Gran Sasso underground laboratory~\cite{Ebert2013}.
COBRA is intended for a large-scale experiment with a total source
mass of about 400~kg of CdZnTe enriched in another "double-beta" isotope
$^{116}$Cd to about 90\%. 

The detection reaction is IBD on cadmium

\begin{equation}
\overline{\nu_{e}}+^{106}\text{Cd}\rightarrow^{106}\text{Ag}+e^{+}.\label{eq:inverse_beta-cd}
\end{equation}

The energy threshold of the reaction is 1.216~MeV, below the end-point
of the $^{40}$K $\beta$-spectrum (1.311~MeV). The full positron energy
is in the 1.02 \textendash{} 1.12~MeV energy range. Two transitions
are possible: the 0$^{+}\rightarrow1^{+}$ allowed transition to the ground
state or the 0$^{+}\rightarrow6^{+}$ transition to the first excited
state (89.66~keV).

\begin{figure}
\begin{centering}
\includegraphics[scale=0.5]{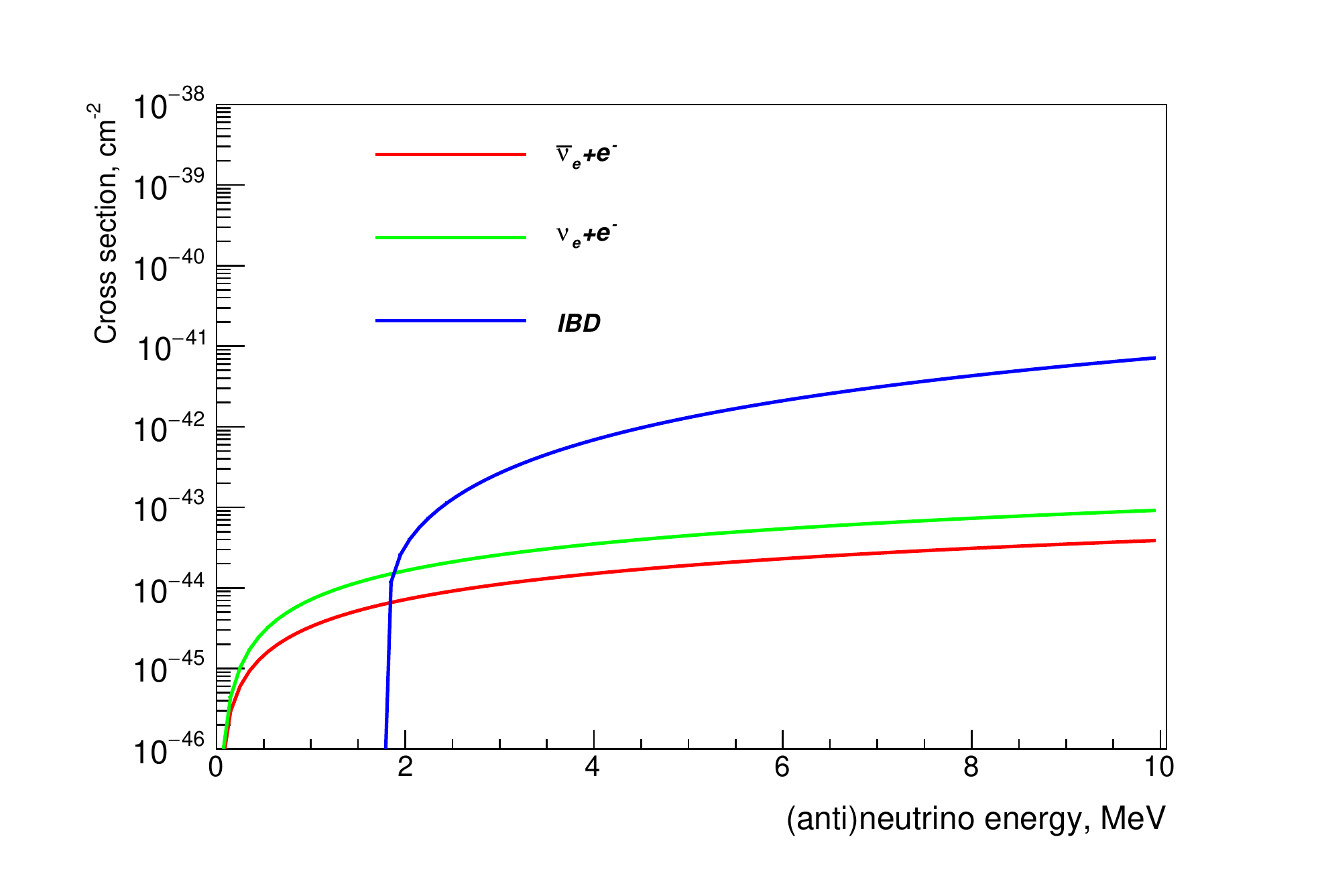}
\par\end{centering}
\protect\caption{\label{fig:Neutrino-cross-sections}Neutrino cross sections for three detection reactions: elastic scattering of neutrino and antineutrino off electrons with no energy threshold, and antineutrino interactions on protons (IBD reaction) with 1.8~MeV threshold}
\end{figure}

The unstable silver isotope $^{106}$Ag decays into the stable palladium
$^{106}$Pd with a half-life of 24~min through two concurrent processes.
In 41\% of cases it decays through electron capture (EC) ``invisible''
to the detector
\begin{equation}
^{106}\text{Ag}+e^{-}\rightarrow^{106}\text{Pd}+\nu_{e},
\end{equation}

and in 59\% of cases it undergoes $\beta^{+}$-decay
\begin{equation}
^{106}\text{Ag}\rightarrow^{106}\text{Pd}+e^{+}+\nu_{e}.
\end{equation}

The $\beta^{+}$ end-point energy is E$_{0}$=1.943 MeV in 82.9\%
of all transitions, and it is E$_{0}$=1.431 MeV accompanied by a 0.5119~MeV gamma from the first excited state (16.3\% of all transitions).
The pair of 511~keV gammas from the annihilated positron emitted in IBD of cadmium to silver and followed by another couple of
511~keV gammas produced by $^{106}$Ag $\beta^{+}$-decay provides
a tag for the antineutrino events. The clear disadvantage of this method
is a large delay between these
two correlated events corresponding to a 24~min half-life of $^{106}$Ag. But at the same time, the spatial separation
is extremely small. Therefore, the spatial coincidence can be used
to reject the background.

\subsection{Other detection methods}

There is no energy threshold for the elastic antineutrino scattering
process

\begin{equation}
\overline{\nu_{e}}+e^{-}\rightarrow\overline{\nu_{e}}+e^{-},
\end{equation}

potentially providing sensitivity to the low energy portion of the geoneutrino
spectra. The elastic scattering of neutrinos on electrons was successfully
used in water \v{C}erenkov (SuperKamiokaNDE, SNO) and liquid scintillator
detectors (Borexino) to detect solar neutrinos. The cross section
of the elastic antineutrino scattering is 2-3 times lower compared to the
neutrino scattering (see Fig.~\ref{fig:Neutrino-cross-sections})
and the flux of geoneutrinos is 2-4 orders of magnitude lower compared
to that of solar neutrinos. Thus, the use of this process for the
geoneutrino studies is challenging. The directionality of the scattered
electron offers a potential opportunity to suppress the neutrino background.
The method is successfully used in water \v{C}erenkov detectors
at higher energies. 

A technique with a certain potential for application to directionality studies
was considered by Aberle et al.~\cite{Aberle2012}. The authors
propose to use directional \v{C}erenkov light in LS to reconstruct the
direction of electrons. Light with a wavelength shorter than $\sim$370~nm 
is absorbed and re-emitted isotropically, making this fraction
indistinguishable from scintillation. \v{C}erenkov photons, on
average, arrive earlier with respect to the scintillation light, the
latter being affected by chromatic dispersion and the finite time
of the scintillation processes. A Geant4 simulation of a spherical
detector with the radius of 6.5 m with scintillator properties matching
a KamLAND-like scintillator showed that the early light could be effectively
separated, if photodetectors with the transit time spread of 0.1 ns are
used. This can be further improved if more red-sensitive photodetectors
or scintillators with narrower emission spectra are used. The algorithms
developed for water \v{C}erenkov detectors were able to converge on
reasonable reconstructed vertices and directions for the simulated
energies of electrons of 5~MeV, 2.1~MeV, and 1.4~MeV relevant for the
neutrinoless double beta decay search. At lower energies the reconstructed
vertex is wider by $\sim$1~cm. Though the authors do not consider the
geoneutrino detection as one of the potential tasks, the technique,
in principle, can be used in elastic scattering experiments with energies
above the \v{C}erenkov limit. The directionality in this case is needed
to separate the geoneutrino events from the much higher irreducible background induced by solar neutrinos.

\begin{figure}
\begin{centering}
\includegraphics[scale=0.3]{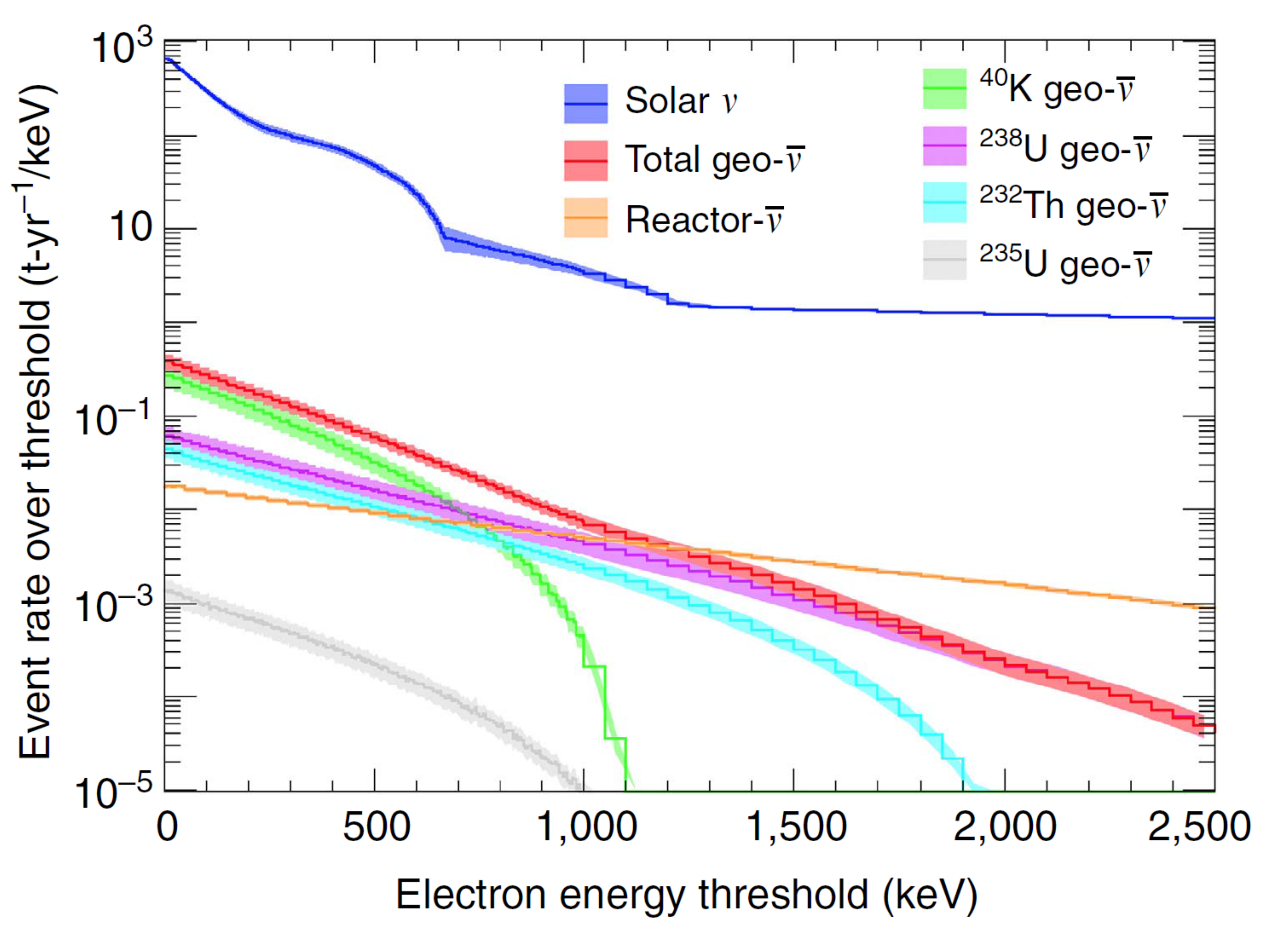}
\par\end{centering}
\protect\caption{\label{fig:Neutrino-Leyton} Event rates over threshold shown versus electron energy threshold for solar neutrino (upper curve, blue), geoneutrino (below the solar one at lower energies, red) and reactor antineutrino (below the solar one at higher energies, orange) events, assuming a CF$_{4}$ target and 55\% survival probability. Individual contributions to the total geo-neutrino flux are visible below the total geoneutrino signal: from $^{40}$K (ends at $\sim$1.1 MeV, green), $^{238}$U (coincides with total geoneutrino signal above 2.5 MeV, violet), $^{232}$Th (ends at $\sim$1.9 MeV teal), and $^{235}$U (ends at $\sim$1 MeV, grey) are also shown. (Figure from Leyton et al. \cite{Leyton2017}, reproduced
under Creative Commons Attribution 4.0 International License \cite{CCL4.0}).}
\end{figure}

A method for measuring the geoneutrino flux from $^{40}$K by using
antineutrino-electron elastic scattering was recently considered by Leyton
et al \cite{Leyton2017}. Gas-filled time projection chambers (TPCs)
are discussed in this work as a detector meeting the necessary requirements
on the backgrounds and directional sensitivity. The challenge of the method is shown in Figure~\ref{fig:Neutrino-Leyton}, the expected geoneutrino signal is two orders of magnitude below the irreducible background from the solar neutrino elastic scattering. Detector resolution,
efficiency, and energy thresholds assumed in the estimations are based
on the performance of other successful detectors \cite{MUNU,Dujmic2008}.
In particular, the MUNU collaboration \cite{MUNU} used a 1~m$^{3}$
TPC filled with CF$_{4}$ at 1~bar for searches of the antineutrino
magnetic moment exactly in the energy range relevant for the geoneutrino
studies. The energy resolution was determined to be 10\% at 200~keV
and 6.8\% at 478~keV. The TPC demostrated good angular resolution
for recoil electrons, namely 15\textdegree{} at 200~keV, 12\textdegree{}
at 400~keV, and 10\textdegree{} at 600~keV. The exposures needed to
access $^{40}$K, mantle, and core geoneutrinos at 95\%~C.L. are
140~tonne-years, 2.6~ktonne-years, and 200~ktonne-years, respectively.
Enhancing the angular resolution of the detector results in a significant
reduction in the required exposures. 

A gas-filled TPC large enough for antineutrino searches would have an enormous
size. The weight of 1~m$^{3}$ of the CF$_{4}$ gas at normal conditions
is about 4 kg. Leyton et al.~\cite{Leyton2017} consider high-pressure
time projection chambers operating at pressures up to 10\textendash 15
bar as a better solution, noting the trade-off between the necessary
mass and degradation of the angular sensitivity due to multiple
scattering at higher pressures. Assuming a CF$_{4}$ (or Xe) target
operating at 10~bar, a 15-tonne detector will have a FV
of 403 (260)~m$^{3}$ corresponding to a cube of 7.4 (6.4)~m on one
side.

\section{Upcoming experiments and new projects}

At present only two detectors, Borexino and KamLAND, are taking data.
Soon, the SNO+ experiment will join the pool of the geoneutrino detectors,
and JUNO is under construction. Below we discuss the potential of SNO+ and JUNO
and review other existing proposals.

\subsection{SNO+}

SNO+ is the successor to the SNO experiment at SNOLAB in Sudbury, Ontario,
Canada. The project was initiated in 2005. Significant efforts have been 
undertaken to transform the heavy water detector into a large volume liquid
scintillator detector. The main goal of this multipurpose neutrino
experiment is a search for neutrinoless double beta decay in $^{130}$Te,
but SNO+ also aims to study low-energy solar neutrinos, and geo and reactor antineutrinos.

SNO+ started filling with liquid scintillator on October 25, 2018~\cite{SNO2018}.  This experiment should be the third to provide geoneutrino results following the first studies by KamLAND and Borexino. 

SNO+ reuses the existing infrastructure of SNO, including a 6-m-radius
acrylic vessel (AV) surrounded by 9394 inward-facing PMTs mounted on a concentric stainless steel support structure with a
radius of 8~m. The AV will be filled
with the LS and surrounded with ultra-pure
water shielding. The depth of the SNO+ detector site is 6010~m.w.e.,
providing effective shielding against backgrounds induced by cosmic muons.
As few as 70~muons a day are expected. For comparison, in Borexino there are
4000 muons a day for a smaller detector. The centre of the detector
is 2039~m below the surface, and the surface is 309~m above sea level.
In contrast to the Borexino and KamLAND experiments, SNO+ does not
have an external tank. Water fills the space between the cave walls
and the AV.

SNO+ will use 780~tonnes of linear-alkylbenzene (LAB) based LS, well-suited for electron antineutrino detection. 
The LAB
scintillator formula can be roughly considered as C$_{18}$H$_{30}$,
with the 0.86~g/cm$^3$ density. MC simulations of the amount of
light collected in SNO+ yield 1200 PEs per MeV of energy
deposited, as a result of high geometric coverage (54\%) and high
light yield of undiluted LAB~\cite{Chen2006}. 780 t of LAB contain
7.34\texttimes{}10$^{31}$ target protons. The predicted geoneutrino
signal at the SNO+ location is around 40~TNU. The event rate in SNO+
would thus be around 29 events per year compared to $\sim$32 events
from the distant nuclear reactors in the same energy range. 

The SNO+ detector is located in a region of a thick continental crust.
The crust and the local geology in the vicinity of SNO+ were
extensively studied since this is a mining location. Huang et al.~\cite{Huang2014} constructed a detailed 3D model of the regional crust by using geological, geophysical, and geochemical information. The model includes six 2x2 degrees tiles centered at the location of the SNO lab, and spreading by 3 tiles in latitude and 2 longitude. Crustal cross sections obtained from seismic surveys were used to characterize the crust and assign
uncertainties to its structure. The average continental crust depth
in the studied area is 42.3$\pm$2.6~km. The geological structure of
the upper crust was reconstructed on the basis of regional geology.
The abundances of U and Th and their uncertainties in each upper crustal
lithologic unit were determined directly from analyses of representative
samples. The average chemical compositions of the middle and lower
crust were obtained using global compilation of the chemical compositions
of the corresponding rocks. The total regional crust contribution
to the geoneutrino signal at SNO+ is predicted to be $15.6_{-3.4}^{+5.3}$~TNU with the Huronian Supergroup near SNO+ dominantly contributing
to the uncertainty $7.3_{-3.0}^{+5.0}$~TNU. This is a glacial layer rich in uranium
and thorium deposited with variable thickness
throughout the region. The presence of the strongly contributing geological
feature nearby would require further study to decrease its contribution
to the uncertainty of the geoneutrino signal. The bulk crustal geoneutrino
signal at SNO+ was estimated by Huang et al~\cite{Huang2014} to be
$30.7_{-4.2}^{+6.0}$~TNU. The total geoneutrino signal at SNO+ is
predicted to be $40_{-4}^{+6}$~TNU without accounting for uncertainties
of the signal from the continental lithospheric mantle and the convecting
mantle. The geoneutrino signal from the crust predicted by the global-scale
reference model~\cite{Mantovani2004} based on the average composition
of the global upper continental crust was higher, $34.0_{-5.7}^{+4.0}$~TNU, as commented in~\cite{Huang2014}. This is due to the fact that
Archean to Proterozoic Canadian Shield has lower U and Th concentrations.
Local heat flow measurements revealed $\sim$25\% higher heat flow
than the average for the entire Canadian Shield~\cite{Perry2009}. The higher heat flux is very likely correlated
to higher local uranium and thorium abundances and thus can be related
to the conclusions by Huang et al.~\cite{Huang2014}. The surface
sediment layer may be a significant contributor. 

More recent estimates come from Strati et al. \cite{Strati2017}.
A combination of geological observations, extensive geochemical sampling,
and geophysical surveys was used to create a coherent 3D geological
model of a 50$\times$50~km upper crustal region surrounding SNOLAB.
A 3D numerical model of U and Th distribution predicts a contribution
of $7.7_{-3.0}^{+7.7}$~TNU out of the crustal geoneutrino signal of
$31.1_{-4.5}^{+8.0}$~TNU. The uncertainties of the local contributions
are even larger than those estimated by Huang et al.~\cite{Huang2014}.
The large uncertainty of the local crust geoneutrino signal restricts
SNO's capability of discriminating among different BSE compositional models
of the mantle, and a more detailed geophysical characterization of the
3D structure of the heterogeneous Huronian Supergroup is needed.

The SNO+ programme includes solar neutrino studies, which are very demanding from
the point of view of the internal backgrounds. This automatically
guarantees the radiopurity levels required for the geoneutrino studies.
Cosmic muon fluxes will be reduced by an order of magnitude compared
to Borexino. The total interaction rate for electron antineutrinos
by the world's nuclear power reactors is estimated
at 129 events per 1 year of exposure. Roughly 1/4 of this rate occurs
in the same energy range as the geoneutrinos, and the signal- to-background
ratio for geoneutrinos in SNO+ will be approximately 1:1. In summary,
the design characteristics of the SNO+ detector before the double-beta
phase (with the pure scintillator) are favourable from the point of view
of backgrounds for the geoneutrino detection. In the Te-loaded phase
the low energy backgrounds are expected to be about 50\textendash150
times higher than in the pure scintillator phase, which can make the
extraction of the geoneutrino signal more difficult~\cite{Andringa2016}.

\begin{figure}[!ht]
\begin{centering}
\includegraphics[scale=0.4]{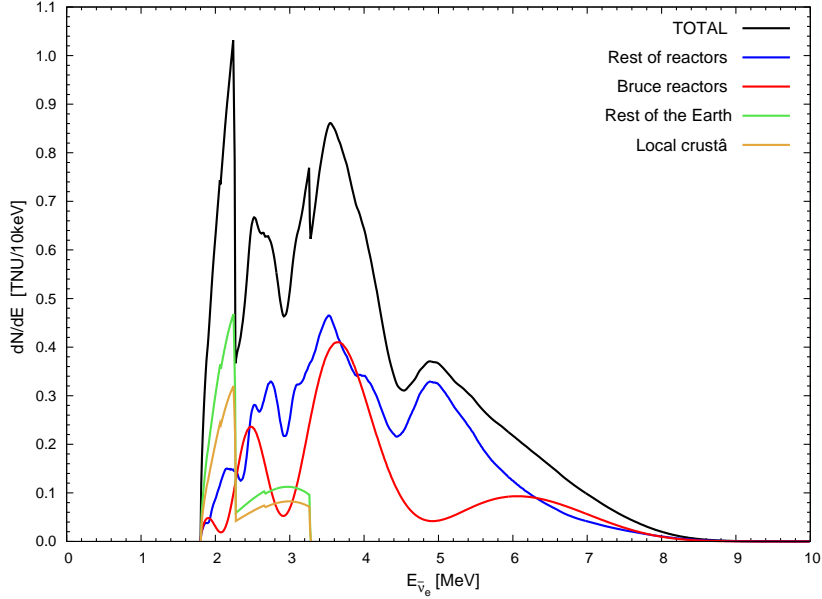}
\par\end{centering}
\caption{\label{fig:SNO}Antineutrino energy spectra expected at SNO+ as estimated
by Baldoncini et al. \cite{Baldoncini2016}. The geoneutrino spectrum
components are shown separately: the contribution of the local crust (smaller contribution, orange)
and the rest of the Earth (larger contribution, green). The reactor antineutrino spectrum
is separated in two parts to underline the contribution from the closest
(240 km) Bruce power station reactors (strongly oscillating lower spectrum, red), and the contribution from
the rest of the reactors (less oscillating middle spectrum, blue). The sum of reactor and the
geoneutrino components provides the total antineutrino spectrum (the resulting highest spectrum, black).
(For interpretation of the references to colour in this figure legend,
the reader is referred to the web version of this article.)}
\end{figure}

The SNO+ sensitivity to the total geoneutrino flux is expected to
be dominated by statistical uncertainties with an accuracy close
to that of Borexino for similar data-taking periods. The larger volume
of the SNO+ detector compensates for the higher reactor background rate~\cite{Andringa2016}. The antineutrino flux from the reactors will
be affected by the planned upgrades when different reactor cores will
be turned off, which will result in an expected total flux reduction below
10\% at each moment. The known time evolution of the reactor power
provides additional help in separating this background \cite{Andringa2016}.
The antineutrino spectrum from reactors at SNO+ was estimated
by Baldoncini et al.~\cite{Baldoncini2016}. The results of the caculations
are presented in Fig.~\ref{fig:SNO}.

At the first step, the analysis with the global Th/U mass ratio fixed
according to standard geological models is planned. The effect of
local variations of the ratio is being studied. In further analysis
SNO+ aims to separate both the uranium and thorium contributions and
the mantle and crust contributions in a global analysis of the geoneutrino
spectrum including the KamLAND and Borexino data~\cite{Andringa2016}.

\subsection{\label{subsec:JUNO}JUNO}

JUNO is a multipurpose neutrino oscillation experiment. The first
of the third generation of reactor antineutrino experiments, it is designed to determine
the neutrino mass hierarchy and to measure the neutrino oscillation
parameters with a greatly increased precision. One of the declared goals
of the experiment is the geoneutrino flux measurement. The start
of the data acquisition is planned for 2021. 

The Jiangmen Underground Neutrino Observatory is located in Kaiping,
Jiangmen, Guangdong province in Southern China. The position of the
detector, equally distant ($\sim53$ km) from two nuclear power plants
(Yangjiang and Taishan nuclear power plants, NPPs), was optimized in view
of the primary goal of the neutrino mass hierarchy determination. The large
mass of the detector (20 kt) and strong requirements on the LS radiopurity
and energy resolution (3\% at 1 MeV) are also dictated by this goal.

The JUNO project was approved by the Chinese Academy of Sciences in February
2013. Six reactor cores of Yangjiang NPP, each with the thermal power
(TP) of 2.9~GW, will run in 2020, before the start of the data acquisition in JUNO.
The Taishan NPP is planned to opertae four cores of 4.59~GW(TP) each. The
total TP of the Yangjiang and Taishan NPPs will be 35.73~GW. It is possible that the last two cores in Taishan will not
operate by 2021; in that case the total power would be 26.55~GW(TP)
at JUNO's start-up. A run with reduced power would be very useful for
the geoneutrino studies.

The JUNO detector will be located underground at a depth of 700~m, providing
protection against cosmic rays comparable to that of the KamLAND
site. The surrounding rock is granite with the average density of 2.66~g/cm$^{3}$ along a 650~m bore hole. The activities of U, Th,
and $^{40}$K in the rock around the experimental hall are measured
to be 130, 113, and 1062~Bq/kg, respectively. The JUNO detector consists
of a 20~kt central liquid scintillator detector contained in an acrylic
tank immersed in a water pool. 17000 20-inch PMTs installed on a
stainless steel sphere and facing the LS will provide optical coverage
larger than 75\%. A muon tracker will be installed on top of the detector.
The water pool is equipped with PMTs to detect the \v{C}erenkov light
from cosmogenic muons and thus also serves as an active muon veto
\cite{Lu2017}. 

A high energy resolution in the energy spectrum is extremely important
for resolving the neutrino mass hierarchy, which will also be an advantage
in the geoneutrino analysis. The designed energy resolution is the
highest among the state-of-the-art detectors, such as BOREXINO or KamLAND.
To prevent energy resolution degradation, the Earth's magnetic field
of about 0.5~Gs at the experimental site will be compensated by
a specially designed set of coils surrounding the water pool and ensuring
proper operation of the PMTs, which, due to their size, are extremely
sensitive to magnetic fields. The LAB is
chosen as a detection medium due to its excellent transparency, high
flash point, low chemical reactivity, and good light yield. The LS contains 3~g/l of PPO serving as the
fluor and 15~mg/l p-bis-(o-methylstyryl)-benzene (bis-MSB) as the wavelength shifter. An attenuation length larger than 25~m at 430~nm after the proper purification is reported for the mixure~\cite{Yang2017}.

\begin{figure}
\begin{centering}
\includegraphics[scale=0.25]{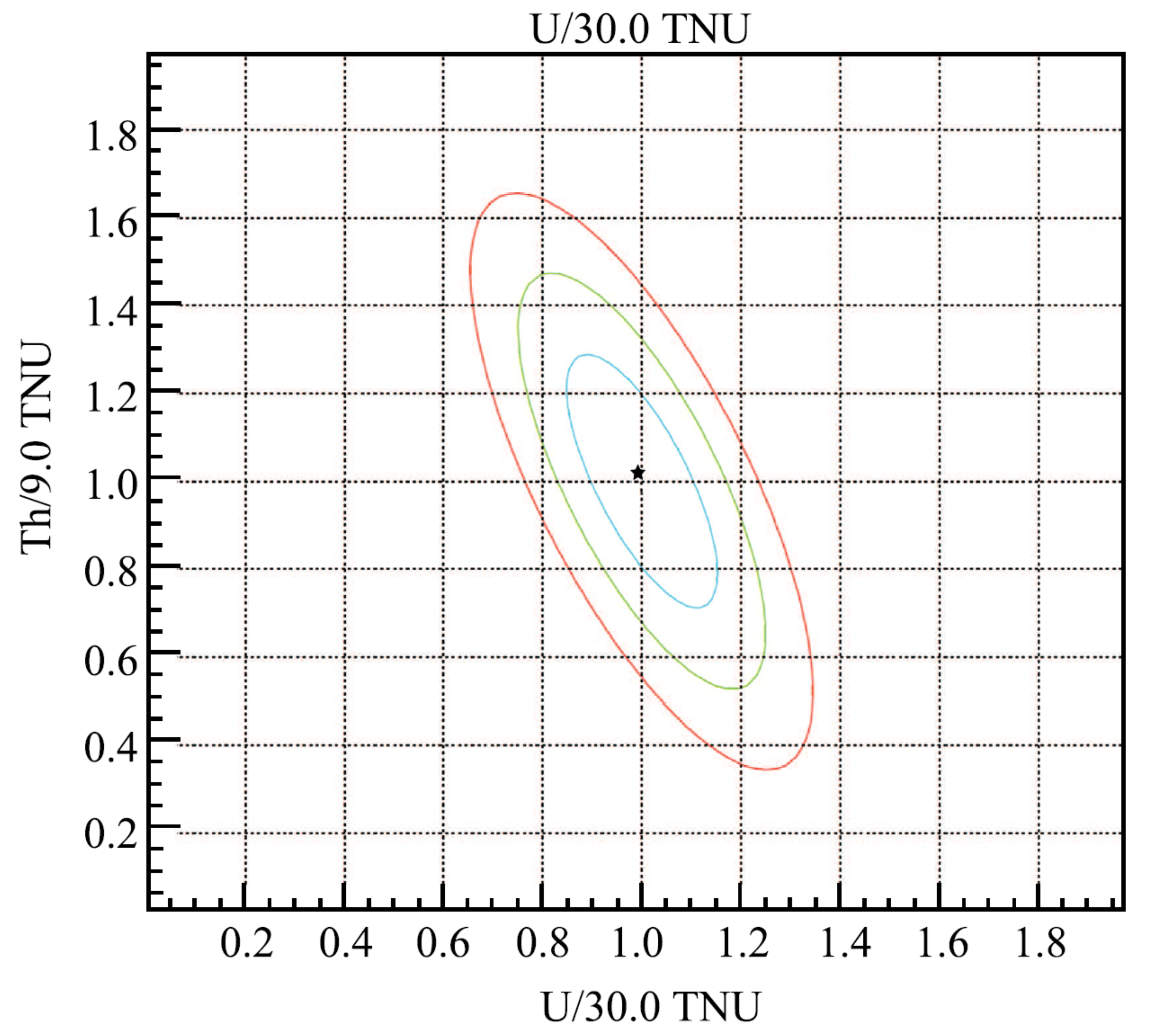}\includegraphics[scale=0.28]{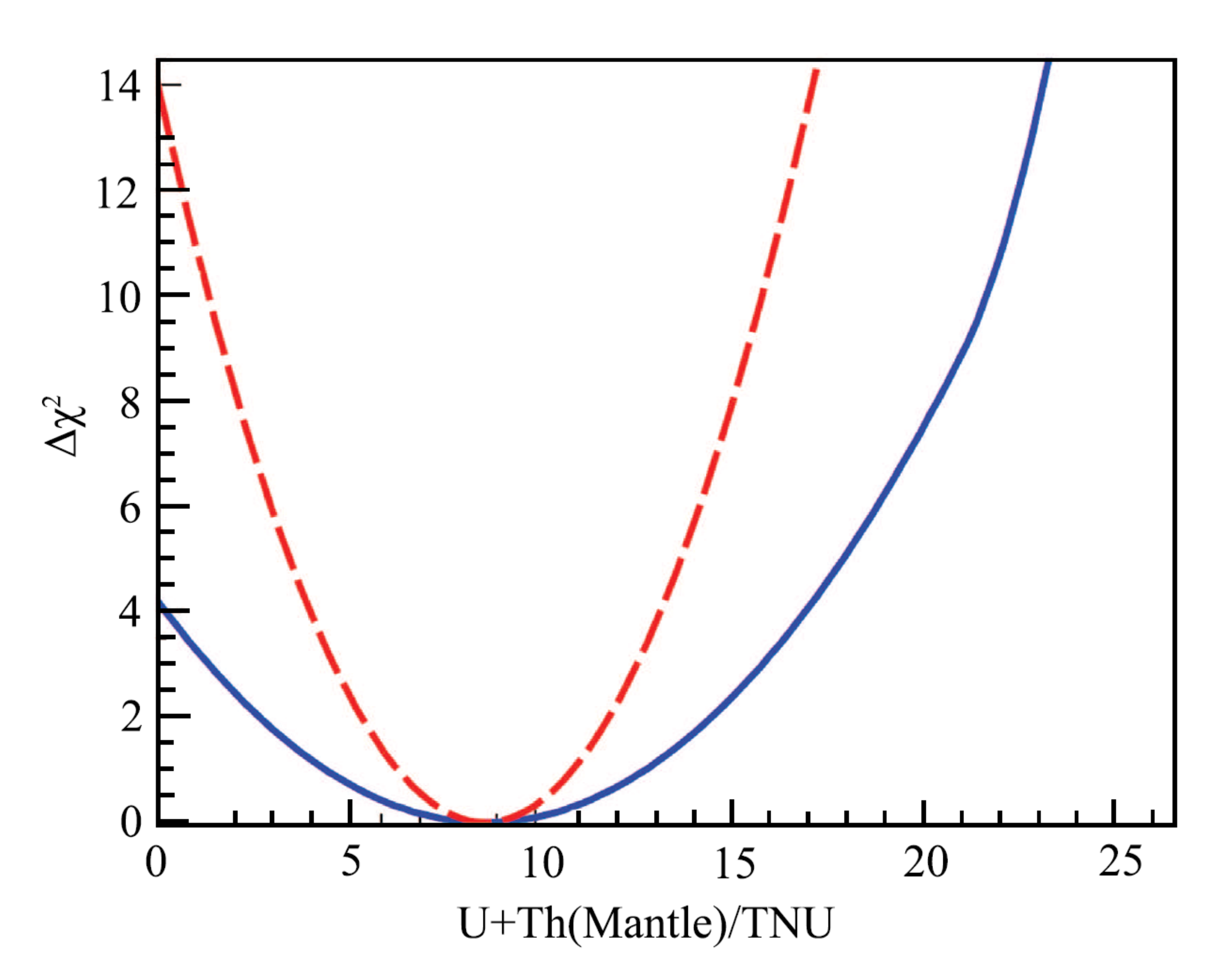}
\par\end{centering}
\centering{}\caption{\label{fig:CumJUNO}\textbf{left:} estimate of JUNO's sensitivity
to the Th/U mass ratio presented as 1,2 and 3$\sigma$ contours for the
free Th/U mass ratio fit of the model data corresponding to 10 years
of JUNO detector running \protect \linebreak{}
\textbf{right:} estimate of the possibility of separating the
mantle contribution from the total signal presented as a $\chi^{2}$ profile. The flatter
profile (blue) corresponds to the crust signal prediction with the 18\%
uncertainty, and the narrower profile corresponds to a better knowledge
of the crust contribution with the 8\% uncertainty (Figure from \cite{Han2016})}
\end{figure}

The same as at the other locations, at the JUNO site 50\% of the
total geoneutrino signal is provided by the closest $\sim$500~km of the Earth,
see Fig.~\ref{fig:CumulativeSignal}. The three-dimensional global reference
model of Huang et al. \cite{Huang2013} was used to estimate the signal
from lithospheric geoneutrinos at the JUNO site. Two main sources
of uncertainties are the physical structure of the lithosphere and
the abundances of HPEs in the reservoirs. The expected geoneutrino
signals at JUNO from the global Bulk Crust and from the Continental
Lithospheric Mantle are 28.2$_{-4.5}^{+5.2}$~TNU and 2.1$_{-1.3}^{+2.9}$~TNU respectively~\cite{Strati2015}. The distributions of U and
Th abundances are log-normal, the quoted central values are medians,
and the errors are asymmetric. The adopted mantle model divided into
two spherically symmetric domains (Depleted Mantle and Enriched Mantle),
refers to the McDonough and Sun's BSE model~\cite{McDonough1995} based
on a primitive mantle having U and Th masses M(U)=$8.1\times10^{16}$~kg and M(Th)=$33\times10^{16}$~kg, respectively. The mantle contribution
in JUNO is estimated to be \ensuremath{\sim}9~TNU~\cite{Strati2015}.
The total expected geoneutrino signal is $39.7_{-5.2}^{+6.5}$~TNU~\cite{Strati2015}. The present crustal models are affected by uncertainties, which are comparable to the mantle contribution~\cite{Strati2015}.

Similar to that of SNO+, the JUNO physics programme includes solar neutrino
studies. The demands on the LS radiopurity for the solar neutrino programme
are an order of magnitude stronger than for the antineutrino one~\cite{JUNO}.
The radiopurity levels necessary for the antineutrino programme (10~ppt in U and Th in acrylic and 10$^{-15}$~g/g in the LS) would also be
well suited for the geoneutrino studies. The cosmic muon fluxes
are comparable to the ones at the KamLAND site, and the general antineutrino
selection will provide a clear set of antineutrino data. The most
serious source of uncertainties in the geoneutrino studies will come
from the strong reactor background, and, in particular, from the uncertainties
of the reactor spectrum shapes at low eneries. 

All of the contributions from nuclear cores operating worldwide in
2013 were found to provide $95.3_{-2.4}^{+2.6}$~TNU by
Baldoncini et al.~\cite{Baldoncini2015}. The contributions of the
future Taishan and Yangjiang nuclear power plants are estimated on the basis
of the following parameters: 80\% antineutrino detection efficiency,
FV with 17.2~m radial cut (18.35~kt of LS
corresponding to 1.285$\times$10$^{33}$ free protons), 35.8~GW(TP)
and 53~km baseline for each core in the Taishan and Yangjiang power plants~\cite{Li2013}. The uncertainties on the predicted reactor antineutrino
signal are estimated to be $\sim$5.6\%, including the effect of uncertainties
on the neutrino oscillation parameters, which can be further reduced
after the JUNO measurement of the neutrino oscillation parameters.
The expected total reactor antineutrino rate is 16100$\pm$900 events
per year~\cite{Han2016} (1569$\pm$88 TNU), with the dominating contribution
from Taishan and Yangjiang ($>$90\%) to the total reactor antineutrino
signal. About 1/4 of the events occurs in the geoneutrino energy
window (351$\pm$27~TNU in $E_{\overline{\nu_{e}}}<3.27$~MeV), making it
difficult to extract the geoneutrino signal, which is almost
an order of magnitude lower. 

Non-antineutrino backgrounds were studied by Han et al.~\cite{Han2016}.
The expected rates of the accidental and cosmogenic ($^{7}$Li and
$^{8}$He) backgrounds are comparable with the expected geoneutrino
rate, but only a relatively small fraction of the corresponding background
spectra falls into the geoneutrino energy window. The test of JUNO geoneutrino
sensitivity in~\cite{Han2016} is based on the fit of a
large set of simulated spectra constructed by combining a random set
of the initial signal and backgrounds. The spectral fit is repeated 10000
times for the statistics corresponding to 1, 3, 5, and 10 years of
full lifetime after cuts. The precision of the geoneutrino measurement
with a Th/U mass ratio fixed at the chondritic value was found to be 13\%,
8\%, 6\%, and 5\%, respectively, if the uncertainty of the reactor
spectrum is $\sim$1\%. Indeed, the theoretical study by Hayes et
al. \cite{Hayes2014} shows that reactor antineutrino spectra have
at least a 4\% theoretical uncertainty, which would be difficult to reduce
within a purely theoretical framework. Thus, a near detector discussed
by Forero et al.~\cite{Forero2017} as a necessary tool for the neutrino mass hierarchy study\footnote{In a more recent study by Danielson et al.~\cite{Danielson2019}
it is argued that there is no need for the near detector from the point of view of the neutrino mass hierarchy measurement; nevertheless, the near detector
will still be needed for the robust geoneutrino signal measurement.} would also be extremely useful to solve the issue of the reactor spectrum precision.

Reguzzoni et al.~\cite{Reguzzoni2019} used gravimetric data
to improve the quality of the crustal geoneutrino signal prediction
in JUNO. The gravimetric data are from the Gravity field and the steady-state
Ocean Circulation Explorer (GOCE) mission~\cite{Drinkwater2003}.
The model is built by inverting gravimetric data over the 6\textdegree{}\texttimes{}4\textdegree{} area centered at the JUNO location. In
contrast to the existing global models, the GOCE Inversion for Geoneutrinos
at JUNO (GIGJ) provides a site-specific subdivision of the crustal
layer masses. The local rearrangement of the crustal layer thicknesses
results in a $\sim$21\% reduction and a $\sim$24\% increase of the
middle and lower crust expected geoneutrino signal, respectively.
Correspondingly, the uncertainties of the geoneutrino signal predictions
at JUNO induced by uncertainties in the uranium and thorium abundances
in the upper, middle, and lower crust are reduced by 77\%, 55\%, and
78\%. 

The $\sim$10\% precision of the geoneutrino signal measurement is expected
within 3 years of data taking, if the precise description of
the low energy portion of the reactor spectrum will be available, where a $\sim$1\%
precision level is needed. The same set of data will allow measurement of
the U contribution with a 20\% precision and Th contibution with a 30\% precision~\cite{JUNO}, i.e., for the first time a reasonable experimental
constraint of the global Th/U ratio can be provided (see left plot in Fig.~\ref{fig:CumJUNO}). 

The possibility of extracting the mantle signal from the JUNO data will strongly depend on the precision of the predictions of the local crust contribution. Two probable scenarios are shown in the right plot in Fig.~\ref{fig:CumJUNO}. As discussed in section~\ref{sec:Mantle-signal},
a combination of the JUNO measurement with those of other experiments
will allow the mantle contribution to be extracted from the global geoneutrino
measurements (see Fig.~\ref{fig:ShramekMantle}).

\subsection{Jinping Underground Laboratory}

The Jinping Neutrino Experiment is a proposed neutrino observatory
for low-energy neutrino physics in the China JinPing Laboratory (CJPL)~\cite{Li2015}, the most suited site for low background experiments
due to its 2400~m depth under the Jinping Mountain in Sichuan province.
The experimental site is located in a long tunnel at least 950 km
away from the NPPs in operation and under construction.
The Jinping Neutrino Experiment collaboration plans to build two detectors filled with LS
or slow (water-based) LS as a neutrino target and
detection material that will detect antineutrino events through the IBD of the protons. The design mass of each of the two detectors is
1.5~kt~\cite{Beacom2016}. The slow LS candidates for MeV-scale neutrino
experiments at Jinping were considered by Guo at al.~\cite{Guo2019}.
A 1-t-scale prototype detector is under development~\cite{Wang2017}.

Preliminary sensitivity studies for the Jinping detector based on
assessments of the site and potential detector designs were conducted~\cite{Wan2017}. The muon rate at Jinping is as low as $(2.0\pm0.4)\times10^{-10}$~cm$^{-2}$s$^{-1}$ with 6720~m.w.e. rock overburden greatly
suppressing the corresponding backgrounds.

The predicted geoneutrino signal is $59.4\pm6.8$~TNU with an expected background
from nuclear reactors of only $6.7\pm0.1$~TNU in the geoneutrino
energy range \cite{Beacom2016}. 

\subsection{Other projects}

LENA (Low Energy Neutrino Astronomy) was a project of 50~kt deep underground
multipurpose liquid scintillator detector~\cite{Wurm2012} with a
vast physics programme far beyond the one of its progenitor 50~kt
Borex detector. A cylindrical shape 30x100~m of the LS container
was considered in the project. Due to its size, LENA would have had
a moderate LY of 240~PEs/MeV. Two possible locations for the experiment
have been considered: at the Pyh\"{a}salmi
site the expected geoneutrino signal is 51.3$\pm$7.1~TNU, while at the Fr\'{e}jus site it is 41.4$\pm$5.6~TNU. In both cases, a detailed analysis of the contribution from the
local crust surrounding the detector would be important in order to
further constrain the expected signal and to better interpret the
measured signal. LENA would detect about 1000 geoneutrino
events per year. The main antineutrino background, namely that from
the nuclear power plants, would be several times lower in the Finland
location: this would make it a preferable location from the point
of view of the geoneutrino measurement. In the geoneutrino energy
window, the expected reactor antineutrino signal would be about 20
to 37~TNU depending on the construction of several new power plants in Finland, while it would be about 145~TNU at Fr\'{e}jus, as estimated on the basis of the
thermal power of NPPs as reported in 2009.

Another possible site for the geoneutrino exploration is the Baksan Neutrino
Observatory (BNO) in Russia. The geographic position of the observatory
far away from operating nuclear power plants provides for the geoneutrino studies the environment 
free of the reactor antineutrinos.
A multipurpose 10~kt liquid scintillation detector is being discussed~\cite{Barabanov2017}.
The predicted geoneutrino signal is $52.6$~TNU with expected background
from nuclear reactors of only $14.4$~TNU in the geoneutrino energy
range.

\begin{figure}
\begin{centering}
\includegraphics[scale=0.2]{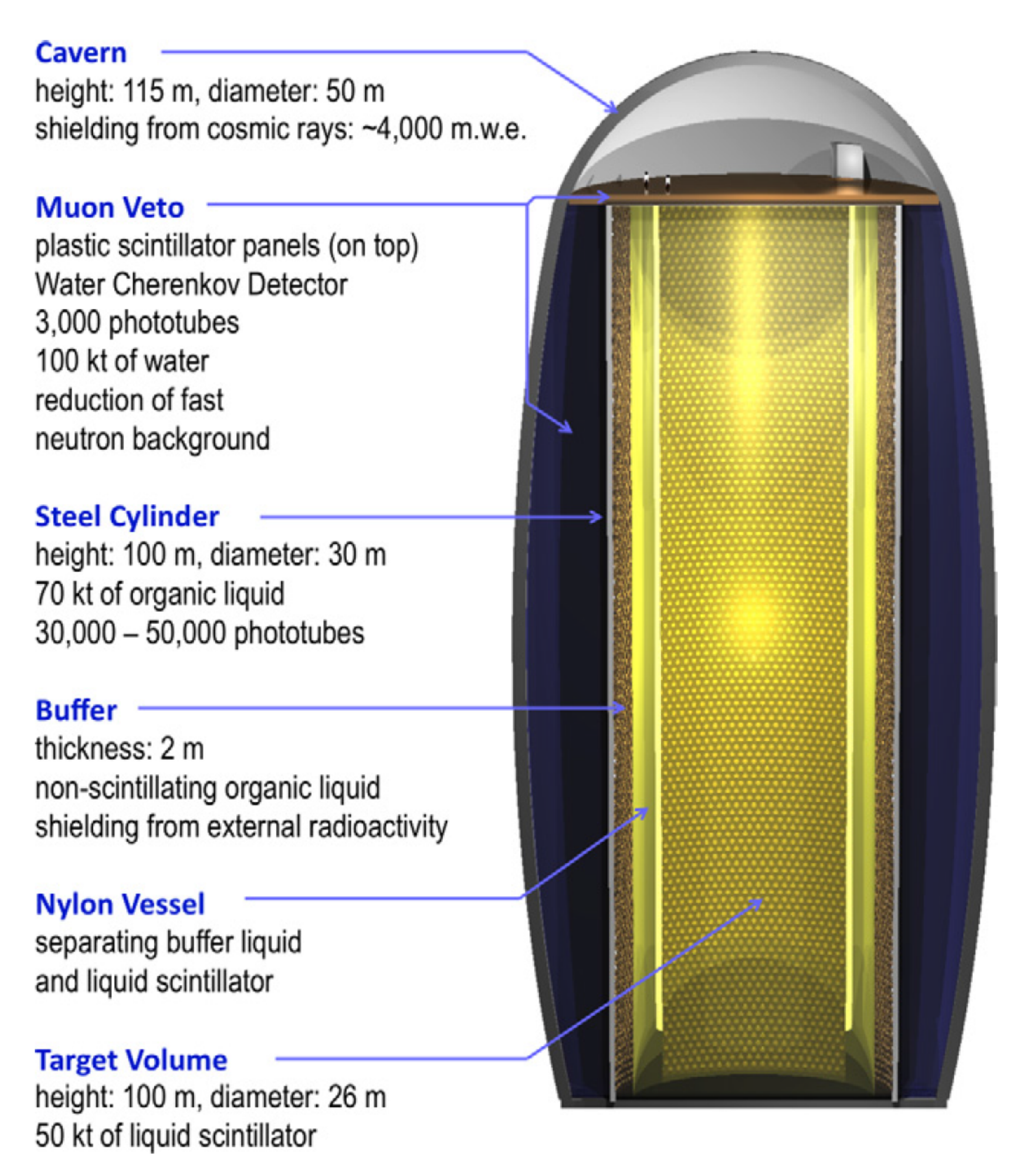}\includegraphics[scale=0.18]{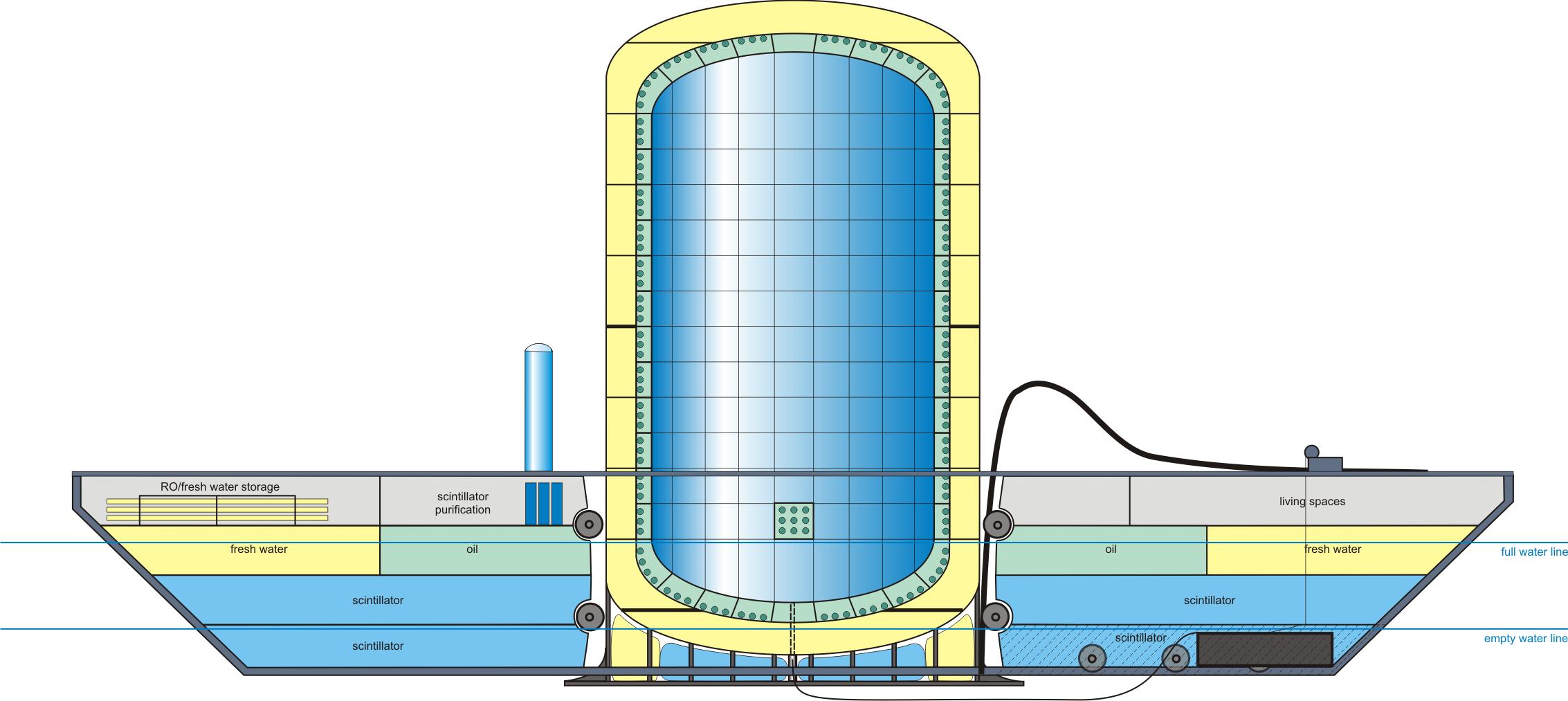}
\par\end{centering}
\caption{\label{fig:LENA}LENA (Fig. from \cite{Wurm2012}) and HANOHANO concepts
(Fig. from \cite{HANOHANO})}

\end{figure}

Hanohano is a project of a 10~kt-scale liquid scintillation detector to be placed 
3-5 km underwater~\cite{Learned2008},\cite{HANOHANO}. It should
be a portable device deployed from the barge, and it is intended for measuring
the mantle contribution to the total geoneutrino signal. Its location
far away from the continental crust will guarantee the overwhelming
contribution from the mantle. About 100 geoneutrino events per year
are expected. 
The combination of the data from multiple sites and the data from an oceanic
experiment would provide valuable information for measuring the mantle
contribution.

\section*{Conclusions}

The Borexino and KamLAND experiments acquired about 190 IBD events
from the geologically produced antineutrinos. These experiments demonstrated
the presence of the radiogenic component in the Earth's heat flux and
the capacity to directly assess the amount and distribution of Th
and U inside the Earth. The observations of geoneutrinos are consistent
with expectations, but the accumulated data set is still too small to draw conclusions
regarding the Earth's models. Additional exposure from existing and
upcoming experiments will provide further improvement. The mantle
signal separation could be enhanced by refinements of the predicted local crust flux. In addition to testing the models of the BSE, the geoneutrino can be used to assess
the compositional nature of large structures in the deep mantle. 

The detection of the $^{40}$K contribution is still beyond the capabilities of modern
detectors. New techniques with a lower threshold and/or directional
sensitivity are needed to observe this signal.

\section*{Acknowledgements }

The author is grateful to M.Baldoncini, S.Dye, R.Han, L.Ludhova,  O.
\v{S}r\'{a}mek, V.Strati, M.Wurm and to the Borexino and KamLAND collaborations,
who kindly granted the permission to use the figures from their publications,
and in particar to F.Mantovani who provided updated plots for Fig.~\ref{fig:Heat}. The author
warmly thanks M.Potapov, C.Kullenberg and R.Vasin for the careful reading of the manuscript. The author feels indebted towards two anonymous reviewers whose detailed comments helped to improve substantially the review.


\begin{thebibliography}{100}

\bibitem{Joly1909}J. Joly, {\it Radioactivity and geology, an account of the influence of radioactive energy on terrestrial history} (NY, D. van Nostrand) (1909)

\bibitem{Cowan1956}C.L. Cowan Jr., F. Reines, F.B. Harrison, H.W. Kruse, and A.D. McGuire, \Journal{Science} {124} {103} {1956}

\bibitem{Fiorentini2007}G. Fiorentini, M. Lissia, F. Mantovani, \Journal{Phys.Reports} {453} {117} {2007}

\bibitem{Marx1960}G. Marx, N. Menyh\'{a}rd, \Journal{\em Communications of the Konkoly Observatory} {48} {1} {1960}

\bibitem{Markov1964}M. Markov, {\it Neutrino} (Nauka, Moscow) (1964)

\bibitem{Eder1966}G. Eder, \Journal{\em Nucl.Phys.} {78} {657} {1966}

\bibitem{Marx1969}G. Marx, \Journal{\em Czech.J.Phys.} {19} {1471} {1969}

\bibitem{Krauss1984}L.M. Krauss, S.L. Glashow, and D.N. Shramm, \Journal{\NATURE} {310} {191} {1984}

\bibitem{Kobayashi1991}M. Kobayashi and Y. Fukao, \Journal{\em Geophysical Research Letters} {18} {633} {1991}

\bibitem{Raghavan1998}R.S. Raghavan, et al., \Journal{\PRL} {80} {635} {1998}

\bibitem{Raghavan1988}R.S. Raghavan and S. Pakvasa, \Journal{\PRD} {37} {849} {1988}

\bibitem{Rothschild1998}C.G. Rothschild, M. Chen, F.P. Calaprice, \Journal{Geophys.Res.Lett.} {25} {1083} {1998}

\bibitem{KamLAND2005}T. Araki, et al., KamLAND collaboration, \Journal{\NATURE} {436} {499} {2005}

\bibitem{Borexino2010}G. Bellini, et al., Borexino collaboration, \Journal{\PLB} {687} {299} {2010}

\bibitem{ENSDF}http://www-nds. iaea.org/nsdd/

\bibitem{Leutz1965}H. Leutz, G. Schulz, and H. Wenninger, \Journal{\em Z. Phys.} {187} {151} {1965}

\bibitem{Mougeot2018}X. Mougeot, \Journal{\PRC} {91} {055504} {2015}

\bibitem{McDonough1995}W.F. McDonough, S. Sun, \Journal{\em Chem. Geol.} {120} {223} {1995}

\bibitem{Enomoto2006}S.Enomoto, \Journal{\em Earth, Moon, and Planets} {99} {131} {2006}

\bibitem{Jaupart}C. Jupart, S. Labrosse, F. Lucazeau and J.-C. Mareschal {\it Treatise on Geophysics, 2-nd
edition, Vol 7} (Oxford, Elsevier) 223 (2015)

\bibitem{Davies2010}J.H. Davies, D.R. Davies, \Journal{\em Solid Earth} {1} {24} {2010}

\bibitem{Korenaga2008}J. Korenaga, \Journal{\em Reviews of Geophysics} {46} {RG2007} {2008}

\bibitem{Christensen1985}U.R.~Christensen, \Journal{\em J. Geophys. Res.} {90} {2995} {1985}

\bibitem{Huang2013}Y. Huang, V. Chubakov, F. Mantovani, R.L. Rudnick, and W.F. McDonough, \Journal{\em Geochem., Geophys., Geosyst.} {14} {2003} {2013}

\bibitem{Mareschal2012}J.-C.~Mareschal, C.~Jaupart, C.~Phaneuf, C.~Perry,  \Journal{Journal of Geodynamics} {54} {43} {2012}

\bibitem{Petcov2002}S.T. Petcov, M. Piai, \Journal {\PLB} {533} {94} {2002}

\bibitem{Capozzi}F.Capozzi, et al, \Journal {\PRD} {95} {096014} {2017}

\bibitem{Wan2017}L. Wan, G. Hussain, Z. Wang, and S. Chen, \Journal {\PRD} {95} {053001} {2017}

\bibitem{PDG2016}K.A. Olive, \Journal {\CPC} {40} {100001} {2016}

\bibitem{Dye2012}S. T. Dye,  \Journal {Rev. Geophys.} {50} {3007} {2012}

\bibitem{Sanshiro}S. Enomoto, {\it Neutrino Geophysics and Observation of Geo-Neutrinos at KamLAND} (Ph. D. Thesis, Tohoku University, Japan) (2005)

\bibitem{Giunti1998}C. Giunti, C. Kim, and M. Monteno, \Journal {\NPB} {521} {3} {1998}

\bibitem{Sramek2016}O. \v{S}r\'{a}mek, et al. \Journal{\em Sci. Rep.} {6} {33034} {2016}

\bibitem{CCL4.0}https://creativecommons.org/licenses/by/4.0/

\bibitem{PREM1981}A.M. Dziewonski, D.L. Anderson, \Journal{\em Phys. Earth and Planetary Interiors} {25} {297} {1981}

\bibitem{Sramek2013}O. \v{S}r\'{a}mek, W.F. McDonough, E.S. Kite, V. Leki\'{c}, S.T. Dye, S. Zhong, \Journal{\em Earth and Planetary Science Letters} {361} {356} {2013}

\bibitem{CRUST1.0}G. Laske, G. Masters, Z. Ma and M. Pasyanos, \Journal{\em Geophys.Res. Abstracts} {15} {Abstract EGU2013-2658} {2013}

\bibitem{LITHO1.0}M.E. Pasyanos, T.G. Masters, G. Laske, and Z. Ma, \Journal{\em J. Geophys. Res.} {119} {2153} {2014}

\bibitem{CRUST5.1}W.D. Mooney, G. Laske, and T.G. Masters, \Journal{\em J. Geophys.Res.} {103} {727} {1998}

\bibitem{CRUST2.0}C. Bassin, G. Laske and G. Masters, \Journal{\em EOS Trans AGU} {81} {F897} {2000}

\bibitem{Mantovani2004}F. Mantovani, L. Carmignani, G. Fiorentini and M. Lissia, \Journal {\PRD} {69} {1} {2004}

\bibitem{Mareschal}J.-C. Mareschal, C. Jaupart and L. Iarotsky, {\it Neutrino Geoscience. Chapter: 4} (Open Academic Press) (2016)

\bibitem{Nataf1996}H.-C. Nataf, and Y. Richard, \Journal{\em Phys. Earth Planet. Inter.} {95} {101} {1996}

\bibitem{Cadek1991}O. \v{C}adek, and Z. Martinec, \Journal{\em Stud. Geophys. Geod.} {35} {151} {1991}

\bibitem{Dye2010}S.T. Dye, \Journal{\em Sci. Lett.} {297} {1} {2010}

\bibitem{Coltorni2011}M. Coltorti, et al., \Journal{\em Geochimica et Cosmochimica Acta} {75} {2271} {2011}

\bibitem{Garnero2008}E. Garnero and A. McNamara, \Journal{\em Science} {320} {626} {2008}

\bibitem{Roskovec2018}B. Roskovec, O. \v{S}r\'{a}mek, W.F. McDonough, arXiv:1810.10914 (2018)
 
\bibitem{Dye2015}S.T. Dye, Y. Huang, V. Leki\'{c}, W.F. McDonough, W.F. and O. \v{S}r\'{a}mek, \Journal{\em Phys. Proc.} {61} {310} {2015}

\bibitem{McDonough2016}W.F. McDonough, W. F. {\it Deep Earth: Physics and Chemistry of the Lower Mantle and Core} (Wiley,American Geophysical Union) (2016) 145

\bibitem{Javoy1995}M. Javoy, \Journal{\em Geophys. Res. Lett.} {22} {2219} {1995}

\bibitem{Javoy2010}M. Javoy, et al., \Journal{Earth Planet. Sci. Lett.} {293} {259} {2010}

\bibitem{Kaminski2013}E. Kaminski, and M. Javoy, \Journal{\em Earth Planet. Sci. Lett.} {365} {97} {2013}

\bibitem{Javoy2014}M. Javoy, and E. Kaminski, \Journal{\em Earth Planet. Sci. Lett.} {407} {1} {2004}

\bibitem{ONeill2008}H.S.C. O'Neill, H. Palme, \Journal{\em Phil. Trans. R. Soc. Lond.} {A366} {4205} {2008}

\bibitem{Caro2010}G. Caro, and B. Bourdon, \Journal{\em Geochim. Cosmochim. Acta} {74} {3333} {2010}

\bibitem{Warren2011}P.H. Warren, \Journal{\em Earth Planet. Sci. Lett.} {311} {93} {2011}

\bibitem{Jackson2013}M.G. Jackson, and A.M. Jellinek, \Journal{\em Earth, Geochem. Geophys. Geosyst.} {14} {2954} {2013}

\bibitem{Campbell2012}I.H. Campbell, H.S.C. O'Neill, \Journal{\em Nature} {483} {553} {2012}

\bibitem{Zhang2012}J. Zhang, N. Dauphas, A.M. Davis, I. Leya, A. Fedkin, \Journal{\em Nat. Geosci.} {5} {251} {2012}

\bibitem{Ringwood1975}A.E. Ringwood, A. E. {\it Composition and Petrology of the Earth's Mantle} (McGraw-Hill, New York) (1975)

\bibitem{Jagoutz1979}E. Jagoutz, H. Palme, H. Baddenhausen, K. Blum, M. Cendales, G. Dreibus, B. Spettel, H. Wanke, and V. Lorenz, {\em Proceedings of the Lunar and Planetary Science Conference} (Pergamon, New York) {2031} (1979)

\bibitem{Wanke1981}H. Wanke, \Journal{\em Philos. Trans. R. Soc.} {A303} {287} {1981}

\bibitem{Hart1986}S.R. Hart, A. Zindler, \Journal{\em Chem. Geol.} {57} {247} {1986} 

\bibitem{Allegre1995}C.J. Allegre, J.-P. Poirier, E. Humler and A.H. Hofmann, \Journal{\em Earth Planet. Sci. Lett.} {134} {515} {1995}

\bibitem{Palme2003}H. Palme, H.S.C. O'Neill, {\it Treatise on Geochemistry, 2nd Edition} (Elsevier, Amsterdam) 1 (2014)

\bibitem{Lyubetskaya2007}T. Lyubetskaya and J. Korenaga, \Journal{\em J. Geophys.Res.} {112} {B03211} {2007}

\bibitem{Wipperfurth2018}S.A. Wipperfurth, M. Guo, O. \v{S}r\'{a}mek, and W.F. McDonough, \Journal{\em Earth Planet. Sci. Lett.} {498} {196} {2018}

\bibitem{Schubert1980}G. Schubert, D. Stevenson, P. Cassen, \Journal{\em J. Geophys. Res.} {85} {2531} {1980}

\bibitem{Turcotte2001}D.L. Turcotte, D. Paul, W.M. White, \Journal{\em J. Geophys. Res.} {106} {4265} {2001}

\bibitem{Turcotte2002}D. Turcotte and G. Schubert, {\it Geodynamics} (Cambridge University Press) (2002)

\bibitem{Davies2010b}G.F. Davies, \Journal{\em Geochem. Geophys. Geosyst.} {11} {Q03} {2010}

\bibitem{Huang2012}Y. Huang, V. Chubakov, F. Mantovani, W. F. McDonough, and R.L. Rudnick, \Journal{\em Journal of Physics: Conference Series} {375} {042041} {2012}

\bibitem{Anderson2007}D.L. Anderson, {\it New Theory of the Earth} {Cambridge University Press} (2007)

\bibitem{Han2016}R. Han, Y.-F. Li, L. Zhan, W.F. McDonough. J. Cao, L. Ludhova, \Journal{\em Chinese Physics C} {40} {033003} {2016}

\bibitem{Beacom2016}J.F. Beacom et al., arXiv:1602.01733 (2016)

\bibitem{Ricci2015}B. Ricci, {\it PoS NEUTEL2015} (2015) 014

\bibitem{Smirnov2003}O. Smirnov, \Journal{\em Instruments and Experimental Techniques} {46} {327} {2003}

\bibitem{Borex2014}G. Bellini, et al., \Journal{\PRD} {89} {112007} {2014}

\bibitem{Wilkinson82}D.H. Wilkinson, \Journal{Nucl. Phys.} {A377} {474} {1982}

\bibitem{PDG}K.A. Olive et al. (Particle Data Group), \Journal{\em Chin. Phys.} {C38} {090001} {2014}

\bibitem{Vogel1984a}P. Vogel, \Journal{\PRD} {29} {1918} {1984}

\bibitem{Fayans1985}S.A. Fayans, 1985, \Journal{\em Sov. J. Nucl. Phys.} {42} {590} {1985}

\bibitem{Vogel1999}P. Vogel and J.F. Beacom, \Journal{\PRD} {60} {053003} {1999}

\bibitem{Kurylov2003}A. Kurylov, M.J. Ramsey-Musolf and P. Vogel, \Journal{\PRC} {67} {035502} {2003}

\bibitem{Zacek1986}G. Zacek et al., \Journal{\PRD} {34} {2621} {1986} 

\bibitem{Vidyakin1994}G. S. Vidyakin et al., \Journal{\em JETP Lett.} {59} {390} {1994}

\bibitem{Declais1994}Y. Declais et al, \Journal{\em Phys. Letters} {338} {383} {1994}

\bibitem{Strumia2003}A. Strumia, F. Vissani, \Journal{\em Physics Letters B} {564} {42} {2003}

\bibitem{Morita1963}M. Morita, \Journal{\em Suppl.Progr.Theor.Phys.} {26} {1} {1963}

\bibitem{Behrens1982}H. Behrens and W. B\"{u}hring, {\it Electron Radial Wave Functions and Nuclear Beta Decay} (Clarendon, Oxford, 1982)

\bibitem{Schopper1966}H.F. Schopper, {\it Weak Interactions and Nuclear Beta Decay} (North-Holland, Amsterdam) 284 (1966)

\bibitem{Hayen2018}L. Hayen, N. Severijns, K. Bodek, D.R. Marian, X. Mougeot, \Journal{\em Reviews of Modern Physics} {90} {015008} {2018}

\bibitem{Bahcall1963}N. Bahcall, \Journal{Phys.Rev.} {129} {2683} {1963}

\bibitem{Harston1992}M.R. Harston and N.C. Pyper, \Journal{\PRA} {45} {6282} {1992}

\bibitem{Mougeot2012}X. Mougeot, M.-M. Bé, C. Bisch, and M. Loidl, \Journal{\PRA} {86} {042506-1} {2012}

\bibitem{Loidl2014}M. Loidl, C. Le-Bret, M. Rodrigues, X. Mougeot, \Journal{\em J.Lowhttps://www.overleaf.com/project/5c939ebdd0dbb3140bd963ce Temp. Phys.} {176} {1040} {2014}

\bibitem{Bisch2013}C. Bisch, X. Mougeot, and M.-M. Bé, {\it 16th International Congress of Metrology} 07004 (2013)

\bibitem{Fiorentini2010}G. Fiorentini, et al., \Journal{\PRC} {81} {034602} {2010}

\bibitem{CTF98a}Alimonti G., et al., Borexino collaboration, \Journal{NIMA} {406} {411} {1998}

\bibitem{CTF98b}Alimonti G., et al., Borexino collaboration, \Journal{\em Astroparticle Physics} {8} {141} {1998}

\bibitem{TOI-1995}Y.A. Alkovali, \Journal{\em Nucl. Data Sheets} {76} {127} {1995}

\bibitem{Fogli2006}G.L. Fogli, E. Lisi, A. Palazzo and A.M. Rotunno, \Journal{\em Earth, Moon, and Planets} {99} {111} {2006}

\bibitem{Takeuchi2019}N. Takeuchi et al., \Journal{\em Physics of the Earth and Planetary Interiors} {288} {37} {2019}

\bibitem{Huang2014}Y. Huang, V. Strati, F. Mantovani, S.B. Shirey, W.F. McDonough, \Journal{\em Geochemistry, Geophysics, Geosystems} {15} {3925} {2014}

\bibitem{Ahrens1954}L.H. Ahrens, \Journal{\em Geochimica et Cosmochimica Acta} {5} {49} {1954}

\bibitem{Ando2004}S. Ando, \Journal{\em The Astrophysical Journal} {607} {20} {2004}

\bibitem{Gaisser2002}T.K. Gaisser and M. Honda, \Journal{\em Annual Review of Nuclear and Particle Science} {52} {53} {2002}

\bibitem{JUNO}Fengpeng An et al., JUNO collaboration, \Journal{\em J.Phys.G: Nucl. Part.Phys.} {43} {030401} {2016}

\bibitem{Kopeikin2004}V. Kopeikin, L. Mikaelyan, V. Sinev, \Journal{\em Phys. At. Nucl.} {67} {1892} {2004}

\bibitem{Fallot2012}M. Fallot, S. Cormon, M. Estienne, A. Algora, V. Bui, et al., \Journal{\PRL} {109} {202504} {2012}

\bibitem{Mueller2011}T.A. Mueller, et al., \Journal{\PRC} {83} {054615} {2011} 

\bibitem{VonFeilitzsch1982}F. Von Feilitzsch, A. Hahn, K. Schreckenbach, \Journal{\PLB} {118} {162} {1982}

\bibitem{Schreckenbach1985}K. Schreckenbach, G. Colvin, W. Gelletly, F. Von Feilitzsch, \Journal{\PLB} {160} {325} {1985}

\bibitem{Hahn1985}A. Hahn, K. Schreckenbach, G. Colvin, B. Krusche, W. Gelletly, et al., \Journal{\PLB} {218} {365} {1989}

\bibitem{Haag2014}N. Haag, A. G\"{u}tlein, M. Hofmann, L. Oberauer, W. Potzel, K. Schreckenbach, F. Wagner, \Journal{\PRL} {112} {122501} {2014}

\bibitem{Huber2011}P. Huber, \Journal{\PRC} {84} {024617} {2011}

\bibitem{Hayes2014}A.C. Hayes, J.L. Friar, G.T. Garvey, G. Jungman, and G. Jonkmans, \Journal{\PRL} {112} {202501} {2014}

\bibitem{Mention2011}G. Mention, M. Fechner, Th. Lasserre, Th.A. Mueller, D. Lhuillier, M. Cribier, and A. Letourneau, \Journal{\PRD} {83} {073006} {2011}

\bibitem{DayaBay2017}F.P.~An et al., (Daya Bay Collaboration),  \Journal{\PRL} {118} {251801} {2017}

\bibitem{DayaBay2017c}F.P.~An et al., (Daya Bay Collaboration),  \Journal{\em Chinese  Physics  C} {41} {013002} {2017}

\bibitem{DoubleChooz2014}Y.~Abe  et al., (Double  Chooz  Collaboration),  \Journal{\em Journal of High Energy Physics} {10} {2014} {086}

\bibitem{RENO2018}M.Y. Pac, et al., \Journal{\em arXiv:1801.04049v1} (2018)

\bibitem{NEOS2018}K.~Siyeon, et al., (NEOS collaboration) \Journal{\em PoS} {ICRC2017} {1024} {2018}

\bibitem{Hayes2016}A.C.~Hayes and P.~Vogel, \Journal{\em Annu. Rev. Nucl. Part. Sci.} {66} {2016} {219}

\bibitem{Huber2004}P. Huber, T. Schwetz, \Journal{\PRD} {70} {053011} {2004}

\bibitem{IAEA}J. Mandula, {\it Nuclear Power Engeneering Section, International Atomic Energy Agency, IAEA-PRIS database} (Report No. IAEA-STI/PUB/1671) (2014)

\bibitem{Baldoncini2015}M. Baldoncini et al., \Journal{\PRD} {91} {065002} {2015}

\bibitem{Bellini2013}G. Bellini, A. Ianni, L. Ludhova, F. Mantovani, and W.F. McDonough, \Journal{\em Prog. Part. Nucl. Phys.} {73} {1} {2013}

\bibitem{Strati2015}Strati et al., \Journal{\em Progress in Earth and Planetary Science} {2:5} {1} {2015}

\bibitem{Kopeikin2006}V.I. Kopeikin, L.A. Mikaelyan, and V.V. Sinev, \Journal{\em Physics of Atomic Nuclei} {69} {185} {2006}

\bibitem{Ma2017}X.B. Ma et al., \Journal{\em Nuclear Physics A} {966} {294} {2017}

\bibitem{Hagner2000}T. Hagner et al., \Journal{\em Astroparticle Physics} {14} {33} {2000}

\bibitem{LSD1989}M. Aglietta et al.,  \Journal{\em Nuovo  Cimento  Soc.  Ital.  Fis.  C} {12}  {467} {1989}

\bibitem{LVD2003}M. Aglietta et al., \Journal{\em Phys.At.Nucl.} {66} {123} {2003}

\bibitem{KamLAND2010}S.~Abe, et al., KamLAND Collaboration, \Journal{\PRC} {81} {025807} {2010}

\bibitem{MUSIC}P. Antonioli, C. Ghetti, E.V. Korolkova, V.A. Kudryavtsev, and G. Sartorelli, \Journal{\em Astropart. Phys.} {7} {357} {1997}

\bibitem{FLUKA}A. Ferrari, P.R. Sala, A. Fas\'so, and J. Ranft, {\em FLUKA: A multi-particle transport code (program version 2005)} CERN, Geneva,2005.

\bibitem{GEANT4}geant4.web.cern.ch

\bibitem{Harissopulos2005}S. Harissopulos, H.W. Becker, J.W. Hammer, A. Lagoyannis, C. Rolfs, and F. Strieder, \Journal{\PRC} {72} {062801} {2005}

\bibitem{Franco2011}D. Franco, G. Consolati, and D. Trezzi, \Journal{\PRC} {83} {015504} {2011} 

\bibitem{Borexino2012b}G. Bellini, et al., \Journal{\PRL} {108} {051302} {2012}

\bibitem{Borexino2015}M. Agostini, et al, \Journal{\PRD} {92} {031101} {2015}

\bibitem{Watanabe2016}Watanabe H., {\em talk at Neutrino Research and Thermal Evolution of the Earth} (2016)
(https://www.tfc.tohoku.ac.jp/wp-content/uploads/2016/10/04\_HirokoWatanabe\_TFC2016.pdf)

\bibitem{Borexino2002}G. Alimonti, et al., Borexino Collaboration, \Journal{\em Astropart.Phys.} {16} {205} {2002}

\bibitem{Borexino2009}G. Alimonti, et al. Borexino Collaboration, \Journal{\em Nucl.Instr. and Methods A} {600} {568} {2009}

\bibitem{Benziger2008}J. Benziger, at al., \Journal{\NIMA} {587} {277} {2008}

\bibitem{Borexino2008}C.~Arpesella et al., Borexino Collaboration, \Journal{\PRL} {101} {091302} {2008}

\bibitem{CTF2008}H. Back, et al., Borexino Collaboration, \Journal{\NIM} {584} {98} {2008}

\bibitem{Borexino2014}G. Bellini et al., Borexino Collaboration, \Journal{\PRD} {89} {112007} {2014}

\bibitem{Borexino2012}H. Back, et al., Borexino Collaboration, \Journal{\JINST} {7} {10018} {2012}


\bibitem{LNGS2017}M.Agostini, et al., Borexino Collaboration, In {\em LNGS annual report}, www.lngs.infn.it/images/REIS/Annual\_Report/AnnualReport\_2017\_low\_def.pdf (2017)

\bibitem{Borexino2018n}M.Agostini, et al., Borexino Collaboration, \Journal{\NATURE} {562} {505} {2018}

\bibitem{Zuzel2004}G. Zuzel, H. Simgen, and G. Heusser,\Journal{\em Appl. Radiat. Isot.} {61} {197} {2004}

\bibitem{Smirnov2016}O. Yu. Smirnov, et al., Borexino Collaboration, \Journal{\em Physics of Particles and Nuclei} {47} {995} {2016}

\bibitem{Benziger2007}J. Benziger, et al., \Journal{\NIMA} {582} {509} {2007}

\bibitem{Borexino2013a}G.Bellini, et al., Borexino collaboration, \Journal{\em Journal of Cosmology and Astroparticle Physics} {08} {049} {2013}

\bibitem{Borexino2013}G.Bellini, et al., Borexino Collaboration, \Journal{\PLB} {722} {295} {2013}

\bibitem{Borexino2018}M. Agostini at al., Borexino Collaboration, \Journal{Astroparticle Physics} {97} {136} {2018}

\bibitem{Borexino2010b}G. Bellini et al., Borexino Collaboration, \Journal{\PRC} {81} {034317} {2010}

\bibitem{Kumaran2019}S. Kumaran, {\it Towards the Updated geo-neutrino Measurement with Borexino} (indico.in2p3.fr/event/18287/contributions/67510/attachments/52191/67338/poster\_appec\_v3.pdf) (2019)

\bibitem{Gando2012}A. Gando, {\it First Results of Neutrinoless Double Beta Decay Search with KamLAND-Zen, Ph.D. Thesis} (2012)

\bibitem{Suzuki2014}A. Suzuki, \Journal{\it Eur. Phys. J. C} {74} {2014} {3094}

\bibitem{Berger2009}B.E. Berger et al, \Journal{\JINST} {4} {04017} {2009}

\bibitem{McKee2008}D.W. McKee, J.K. Busenitz, and I. Ostrovskiy, \Journal{\NIMA} {587} {272} {2008}

\bibitem{KamLAND2008}S.~Abe, et al., KamLAND Collaboration, \Journal{\PRL} {100} {221803} {2008}

\bibitem{KamLAND2003}K.~Eguchi, et al., KamLAND Collaboration, \Journal{\PRL} {90} {021802} {2003}

\bibitem{KamLAND2005a}T.~Araki, et al., KamLAND collaboration, \Journal{\PRL} {94} {081801} {2005}

\bibitem{KamLAND2005b}T.~Araki, et al., KamLAND collaboration, \Journal{\NATURE} {436} {499} {2005}

\bibitem{KamLAND2011a}A.~Gando, et al., KamLAND Collaboration, \Journal{\PRD} {83} {052002} {2011} 

\bibitem{KamLAND2011b}A.~Gando, et al., Kamland Collaboration, \Journal{\em Nature Geoscience} {4} {647} {2011}

\bibitem{KamLAND2013}A.~Gando, et al., KamLAND Collaboration, \Journal{\PRD} {88} {033001} {2013}

\bibitem{KamLAND2015}A.~Gando, et al., Kamland Collaboration, \Journal{\PRC} {92} {055808} {2015}

\bibitem{Keefer2015}G.~Keefer et al., \Journal{\NIMA} {769} {79} {2015}

\bibitem{Enomoto2007}S.~Enomoto, E.~Ohtani, K.~Inoue, and A.~Suzuki, \Journal{\em Earth Planet. Sci. Lett.} {258} {147} {2007}

\bibitem{Inoue2018}K.~Inoue {\it Current status and future prospects of KamLAND-Zen} www.rcnp.osaka-u.ac.jp/dbd18/Data/Prog/W0102\_Inoue.pdf (2018)

\bibitem{Dye2016}S.T.~Dye,  arXiv:1611.03559 [physics.geo-ph] (2016)

\bibitem{Fiorentini2012}G. Fiorentini, G.L. Fogli, E. Lisi, F. Mantovani, and A.M. Rotunno, \Journal{\PRD} {86} {033004} {2012} 

\bibitem{Mantovani2019}F. Mantovani, private communication.

\bibitem{Herndon1996}J.M. Herndon, \Journal{\em Proc. Natl. Acad. Sci. USA} {93} {646} {1996}

\bibitem{Hollenbach2001}D.F. Hollenbach, and J.M. Herndon, \Journal{\em Proc. Natl. Acad. Sci. USA} {98} {11085} {2001}

\bibitem{Degueldre2016}C. Degueldre, C. Fiorina, \Journal{\em Solid Earth Sciences} {1} {49} {2016}

\bibitem{Raghavan2002}R.S. Raghavan, arXiv:hep-ex/0208038 (2002)

\bibitem{Rusov2007}V.D. Rusov, et al., \Journal{\em J. Geophys. Res.} {112} {B09203} {2007}

\bibitem{deMeijer2008}R.J. de Meijer, W. van Westrenen, \Journal{\em S. Afr. J.Sci.} {104} {111} {2008}

\bibitem{Herndon2014}J.M. Herndon, \Journal{\em Curr. Sci.} {104} {528} {2014}

\bibitem{Herndon2005}J.M. Herndon, D.A. Edgerley, arXiv:hep-ph/0501216 (2005)

\bibitem{Dye2008}S.T. Dye, \Journal{\PLB} {679} {15} {2009}

\bibitem{Fields2006}B.D. Fields, K.A. Hochmuth, \Journal{\em Earth, Moon, and Planets} {99} {155} {2006}

\bibitem{Dye2019}S.T.~Dye, private communication.

\bibitem{Lee2003}K.K.M.  Lee,  and R. Jeanloz, \Journal{\em Geophys. Res. Lett.} {30} {2212} {2003}

\bibitem{Rama2003}V. Rama Murthy, W. van Westrenen,  and Y. Fei, \Journal{\NATURE} {423} {163} {2003}

\bibitem{Nimmo2004}F. Nimmo, G.D. Price, J. Brodholt and D. Gubbins, \Journal{\em Geophys. J. Int.} {156} {363} {2004}

\bibitem{Dye2006}S.T. Dye, E. Guillian, J.G. Learned, et al., \Journal{\em Earth, Moon, and Planets} {99} {241} {2006}

\bibitem{Suzuki2006}A.Suzuki, \Journal{\em Earth, Moon, and Planets} {99} {359} {2006}

\bibitem{Apollonio2000}M. Apollonio, et al., \Journal{\PRD} {61} {012001} {2000}

\bibitem{Batygov2006}M. Batygov, \Journal{\em Earth, Moon, and Planets} {99} {183} {2006}

\bibitem{Domogatski2006}G.V. Domogatsky, V.I. Kopeikin, L.A. Mikaelyan, V.V. Sinev, \Journal{\em Physics of Atomic Nuclei} {69} {1894} {2006}

\bibitem{Tanaka2014}H.K.M. Tanaka, H. Watanabe, \Journal{\em Scientific Reports} {4} {4708} {2014} 

\bibitem{Safdi2015}R. Safdi and B. Suerfu, \Journal{\PRL} {114} {071802} {2015}

\bibitem{SuperK2011}K. Abe et al., \Journal{\PRD} {83} {052010} {2011}

\bibitem{Watanabe2009}H. Watanabe, et al., \Journal{\em Astroparticle Physics} {31} {320} {2009}

\bibitem{Dai2008}X. Dai, et al., \Journal{\NIMA} {589} {290} {2008}

\bibitem{Sweany2011}M. Sweany, et al., \Journal{\NIMA} {664} {245} {2012}

\bibitem{Alonso2014}J.R. Alonso, et al., {\it arXiv:1409.5864v3} (2014)

\bibitem{Yeh2011}M. Yeh, et al., \Journal{\NIMA} {660} {51} {2011}

\bibitem{Zuber2003}K. Zuber, \Journal{\PLB} {571} {148} {2003}

\bibitem{Chen2005}M. Chen, {\em Potassium geo-neutrino detection} (www.docstoc.com/docs/82267645/Potassium-Geo-neutrino-Detection) (2005)

\bibitem{Belli2010}P. Belli et al., \Journal{\NIMA} {615} {301} {2010}

\bibitem{Szczerbinska2011}B. Szczerbinska, \Journal{\em Proceedings of the South Dakota Academy of Science} {90} {13} {2011}

\bibitem{Briancon2015}Ch. Briançon, et al., \Journal{\em Physics of Atomic Nuclei} {78} {740} {2015}

\bibitem{Ebert2013}J. Ebert, et al., \Journal{\NIMA} {807} {114} {2016}

\bibitem{Aberle2012}C. Aberle, A. Elagin, H.J. Frisch, M. Wetstein, and L. Winslow, \Journal{\JINST} {9} {P06012} {2014}

\bibitem{Leyton2017}M. Leyton, S. Dye, and J. Monroe, \Journal{\em Nature Communications} {8} {15989} {2017}

\bibitem{MUNU}Z. Daraktchieva, et al., \Journal{\em J. Phys G} {35} {125107} {2008}

\bibitem{Dujmic2008}D. Dujmic, et al., \Journal{\em Astropart. Phys.} {30} {58} {2008}

\bibitem{SNO2018}https://today.lbl.gov/new-beginnings-for-the-sno-neutrino-detector/

\bibitem{Chen2006}M.C. Chen, \Journal{\em Earth, Moon, and Planets} {99} {221} {2006}

\bibitem{Perry2009}H.K.C. Perry, J.-C. Mareschal and C. Jaupart, \Journal{\em Earth and Planetary Science Lett.} {288} {301} {2009}

\bibitem{Strati2017}V. Strati et al., \Journal{\em Geochemistry, Geophysics, Geosystems} {18} {4326} {2017}

\bibitem{Andringa2016}S. Andringa, et al., \Journal{\em Advances in High Energy Physics Volume} {2016} {6194250} {2016}

\bibitem{Baldoncini2016}M. Baldoncini, et al., \Journal{\em Journal of Physics: Conference Series} {718} {062003} {2016}

\bibitem{Lu2017}H. Lu, et al., \Journal{\em Journal of Physics: Conference Series} {888} {012088} {2017}

\bibitem{Yang2017}H. Yang et al, \Journal{\JINST} {12} {T11004} {2017}

\bibitem{Li2013}Y.-F. Li et al., \Journal{\PRD} {88} {013008} {2013}

\bibitem{Forero2017}D.V. Forero, R. Hawkins, P. Huber, {\it arXiv:1710.07378} (2017)

\bibitem{Reguzzoni2019}M. Reguzzoni, L.~Rossi, M.~Baldoncini, I.~Callegari, P.~Poli, D.~Sampietro, V.~Strati, et al., {\it Journal of Geophysical Research: Solid Earth} doi:10.1029/2018jb016681 (2019)

\bibitem{Drinkwater2003}M.R. Drinkwater, R. Floberghagen, R. Haagmans, D. Muzi, and A. Popescu, {\it GOCE: ESA's First Earth Explorer Core Mission} 
(Earth Gravity Field from Space - from Sensors to Earth Sciences. Kluwer Academic Publishers, Dordrecht, Netherlands, ISBN: 1-4020-1408-2) (2003)

\bibitem{Danielson2019}D.L. Danielson, A.C. Hayes, and G.T. Garvey, \Journal{\PRD} {99} {036001} {2019}

\bibitem{Li2015}J. Li, X. Ji, W. Haxton, J.S.Y. Wang, \Journal{\em Physics Procedia} {61} {576} {2015}

\bibitem{Guo2019}Z. Guo, M. Yeh, R. Zhang, D.W. Cao, M. Qi, Z. Wang, S. Chen, \Journal{\em Astroparticle Physics} {109} {33} {2019}

\bibitem{Wang2017}Z. Wang, et al., \Journal{\NIMA} {35} {685} {2012}

\bibitem{Wurm2012}M. Wurm, et al., LENA Collaboration, \Journal{\em Astroparticle Physics} {35} {685} {2012}

\bibitem{Barabanov2017}I.R. Barabanov, et al., \Journal{\em Physics of Atomic Nuclei} {80} {446} {2017}

\bibitem{Learned2008}J.G. Learned, S.T. Dye, S. Pakvasa, {\it arXiv:0810.4975} (2008)

\bibitem{HANOHANO}https://www.phys.hawaii.edu/\textasciitilde sdye/CEROS-report.pdf

\end{thebibliography}
\end{document}